\documentclass[preprint,authoryear,10pt,3p]{elsarticleFTP}

\usepackage{amssymb}
\usepackage{tikz}
\usepackage{graphicx}
\usepackage{caption}
\usepackage{mathtools}
\usepackage{amsmath}
\usepackage{multirow}
\usepackage{colortbl}
\usepackage{color}
\usepackage{tabulary}
\usepackage{bm}
\usepackage{amsmath}
\usepackage{eqparbox}
\usepackage{tabularx}
\usepackage{tabulary}
\usepackage[utf8]{inputenc}
\usepackage{booktabs}
\usepackage{kpfonts}
\usepackage{rotating}
\usepackage{enumitem}
\usepackage{hyperref}
\usepackage{breakurl}
\usepackage{setspace}
\usepackage{epigraph}
\usepackage{longtable}
\usepackage{pdflscape}

\journal{International Journal of Forecasting}

\begin{document}

\begin{frontmatter}
\title{Forecasting: theory and practice}

\author[label1]{Fotios~Petropoulos\corref{cor1}}
\cortext[cor1]{Corresponding author: f.petropoulos@bath.ac.uk; fotios@bath.edu}
\author[label2]{Daniele~Apiletti}
\author[label3]{Vassilios~Assimakopoulos}
\author[label43]{Mohamed~Zied~Babai}
\author[label4]{Devon~K.~Barrow}
\author[label76]{Souhaib~Ben~Taieb}
\author[label47]{Christoph~Bergmeir}
\author[label45]{Ricardo~J.~Bessa}
\author[label77]{Jakub~Bijak}
\author[label33]{John~E.~Boylan}
\author[label5]{Jethro~Browell}
\author[label37]{Claudio~Carnevale}
\author[label38]{Jennifer~L.~Castle}
\author[label72]{Pasquale~Cirillo}
\author[label36]{Michael~P.~Clements}
\author[label41,label71]{Clara~Cordeiro}
\author[label48]{Fernando~Luiz~Cyrino~Oliveira}
\author[label6]{Shari~De~Baets}
\author[label53]{Alexander~Dokumentov}
\author[label77]{Joanne~Ellison}
\author[label7]{Piotr~Fiszeder}
\author[label52]{Philip~Hans~Franses}
\author[label74]{David~T.~Frazier}
\author[label39]{Michael~Gilliland}
\author[label9]{M.~Sinan~G\"on\"ul}
\author[label1]{Paul~Goodwin}
\author[label11]{Luigi~Grossi}
\author[label61]{Yael~Grushka-Cockayne}
\author[label11]{Mariangela~Guidolin}
\author[label12]{Massimo~Guidolin}
\author[label13]{Ulrich~Gunter}
\author[label49]{Xiaojia~Guo}
\author[label11]{Renato~Guseo}
\author[label59]{Nigel~Harvey}
\author[label35]{David~F.~Hendry}
\author[label1]{Ross~Hollyman}
\author[label40]{Tim~Januschowski}
\author[label14]{Jooyoung~Jeon}
\author[label67]{Victor~Richmond~R.~Jose}
\author[label15]{Yanfei~Kang}
\author[label63]{Anne~B.~Koehler}
\author[label16,label33]{Stephan~Kolassa}
\author[label62,label33]{Nikolaos~Kourentzes}
\author[label17]{Sonia~Leva}
\author[label18]{Feng~Li}
\author[label19]{Konstantia~Litsiou}
\author[label20]{Spyros~Makridakis}
\author[label74]{Gael~M.~Martin}
\author[label50,label51]{Andrew~B.~Martinez}
\author[label1]{Sheik~Meeran}
\author[label21]{Theodore~Modis}
\author[label22]{Konstantinos~Nikolopoulos}
\author[label9]{Dilek~\"Onkal}
\author[label23,label66]{Alessia~Paccagnini}
\author[label73]{Anastasios~Panagiotelis}
\author[label57]{Ioannis~Panapakidis}
\author[label42]{Jose~M.~Pav\'ia}
\author[label69,label70]{Manuela~Pedio}
\author[label24]{Diego~J.~Pedregal}
\author[label44]{Pierre~Pinson}
\author[label25]{Patr\'icia~Ramos}
\author[label54]{David~E.~Rapach}
\author[label56]{J.~James~Reade}
\author[label26]{Bahman~Rostami-Tabar}
\author[label27]{Micha\l{}~Rubaszek}
\author[label28]{Georgios~Sermpinis}
\author[label29]{Han~Lin~Shang}
\author[label3]{Evangelos~Spiliotis}
\author[label26]{Aris~A.~Syntetos}
\author[label30]{Priyanga~Dilini~Talagala}
\author[label31]{Thiyanga~S.~Talagala}
\author[label46]{Len~Tashman}
\author[label64]{Dimitrios~Thomakos}
\author[label58]{Thordis~Thorarinsdottir}
\author[label32,label60]{Ezio~Todini}
\author[label24]{Juan~Ram\'on~Trapero~Arenas}
\author[label15]{Xiaoqian~Wang}
\author[label68]{Robert~L.~Winkler}
\author[label33]{Alisa~Yusupova}
\author[label34]{Florian~Ziel}


\address[label1]{School of Management, University of Bath, UK} 
\address[label2]{Politecnico di Torino, Turin, Italy}
\address[label3]{Forecasting and Strategy Unit, School of Electrical and Computer Engineering, National Technical University of Athens, Greece}
\address[label43]{Kedge Business School, France}
\address[label4]{Department of Management, Birmingham Business School, University of Birmingham, UK}
\address[label76]{Big Data and Machine Learning Lab, Université de Mons (UMONS), Belgium}
\address[label47]{Faculty of Information Technology, Monash University, Melbourne, Australia}
\address[label45]{INESC TEC – Institute for Systems and Computer Engineering, Technology and Science, Porto, Portugal}
\address[label77]{Department of Social Statistics and Demography, University of Southampton, UK}
\address[label33]{Centre for Marketing Analytics and Forecasting, Lancaster University Management School, Lancaster University, UK}
\address[label5]{School of Mathematics and Statistics, University of Glasgow, UK}
\address[label37]{Department of Mechanical and Industrial Engineering, University of Brescia, Italy}
\address[label38]{Magdalen College, University of Oxford, UK}
\address[label72]{ZHAW School of Management and Law, Zurich University of Applied Sciences, Switzerland}
\address[label36]{ICMA Centre, Henley Business School, University of Reading, UK}
\address[label41]{Faculdade de Ciências e Tecnologia, Universidade do Algarve, Portugal}
\address[label71]{CEAUL, Faculdade de Ciências, Universidade de Lisboa, Portugal}
\address[label48]{Pontifical Catholic University of Rio de Janeiro (PUC-Rio), Brazil}
\address[label6]{Department of Business Informatics and Operations Management, Faculty of Economics and Business Administration, Universiteit Gent, Belgium}
\address[label53]{Let's Forecast, Australia}
\address[label7]{Faculty of Economic Sciences and Management, Nicolaus Copernicus University in Torun, Poland}
\address[label52]{Econometric Institute, Erasmus School of Economics, Rotterdam, The Netherlands}
\address[label74]{Department of Econometrics and Business Statistics, Monash University, Melbourne, Australia}
\address[label39]{SAS, USA}
\address[label9]{Newcastle Business School, Northumbria University, Newcastle upon Tyne, UK}
\address[label11]{Department of Statistical Sciences, University of Padua, Italy}
\address[label61]{Darden School of Business, University of Virginia, USA}
\address[label12]{Finance Department, Bocconi University and Baffi-CAREFIN Centre, Milan, Italy}
\address[label13]{Department of Tourism and Service Management, MODUL University Vienna, Austria}
\address[label49]{Robert H. Smith School of Business, University of Maryland, USA}
\address[label59]{Department of Experimental Psychology, University College London, UK}
\address[label35]{Nuffield College and Institute for New Economic Thinking at the Oxford Martin School, University of Oxford, UK}
\address[label40]{Amazon Research, Germany}
\address[label14]{Korea Advanced Institute of Science and Technology, South Korea}
\address[label67]{McDonough School of Business, Georgetown University, USA}
\address[label15]{School of Economics and Management, Beihang University, Beijing, China}
\address[label63]{Miami University, Ohio, USA}
\address[label16]{SAP, Switzerland}
\address[label62]{Skövde Artificial Intelligence Lab, School of Informatics, University of Skövde, Sweden}
\address[label17]{Department of Energy, Politecnico di Milano, Italy}
\address[label18]{School of Statistics and Mathematics, Central University of Finance and Economics, Beijing, China}
\address[label19]{Manchester Metropolitan University Business School, UK}
\address[label20]{M Open Forecasting Center \& Institute for the Future, University of Nicosia, Nicosia, Cyprus}
\address[label50]{Office of Macroeconomic Analysis, US Department of the Treasury, Washington DC, USA}
\address[label51]{GWU Research Program on Forecasting, Washington DC, USA}
\address[label21]{Growth Dynamics, Lugano, Switzerland}
\address[label22]{Durham University Business School, Durham University, UK}
\address[label23]{Michael Smurfit Business School, University College Dublin, Ireland}
\address[label66]{Centre for Applied Macroeconomic Analysis, Australia}
\address[label73]{Discipline of Business Analytics, The University of Sydney Business School, Australia}
\address[label57]{Department of Electrical and Computer Engineering, University of Thessaly, Volos, Greece}
\address[label42]{GIPEyOP, UMMICS, Department of Applied Economics, Universitat de Valencia, Spain}
\address[label69]{School of Accounting and Finance, University of Bristol, UK}
\address[label70]{Baffi-CAREFIN Centre, Bocconi University, Italy}
\address[label24]{ETSI Industrial, Universidad de Castilla-La Mancha, Ciudad Real, Spain}
\address[label44]{Department of Technology, Management and Economics, Technical University of Denmark, Denmark}
\address[label25]{Porto Accounting and Business School, Polytechnic of Porto, Portugal}
\address[label54]{Department of Economics, Saint Louis University, USA}
\address[label56]{Department of Economics, School of Politics, Economics and International Relations, University of Reading, UK}
\address[label26]{Cardiff Business School, Cardiff University, UK}
\address[label27]{SGH Warsaw School of Economics, Collegium of Economic Analysis, Poland}
\address[label28]{Adam Smith Business School, University of Glasgow, UK}
\address[label29]{Department of Actuarial Studies and Business Analytics, Macquarie University, Australia}
\address[label30]{Department of Computational Mathematics, University of Moratuwa, Sri Lanka}
\address[label31]{Department of Statistics, Faculty of Applied Sciences, University of Sri Jayewardenepura, Sri Lanka}
\address[label46]{Foresight, International Institute of Forecasters, USA}
\address[label64]{School of Economics and Political Science, National and Kapodistrian University of Athens, Greece}
\address[label58]{Norwegian Computing Center, Oslo, Norway}
\address[label32]{University of Bologna, Italy}
\address[label60]{Italian Hydrological Society, Bologna, Italy}
\address[label68]{Fuqua School of Business, Duke University, Durham, USA}
\address[label34]{House of Energy Markets and Finance, University of Duisburg-Essen, Germany}


\end{frontmatter}


\epigraph{In theory, there is no difference between theory and practice. But, in practice, there is.}{Benjamin Brewster (1882)}

\section*{Abstract}
\label{sec:Abstract}

Forecasting has always been at the forefront of decision making and planning. The uncertainty that surrounds the future is both exciting and challenging, with individuals and organisations seeking to minimise risks and maximise utilities. The large number of forecasting applications calls for a diverse set of forecasting methods to tackle real-life challenges. This article provides a non-systematic review of the theory and the practice of forecasting. We provide an overview of a wide range of theoretical, state-of-the-art models, methods, principles, and approaches to prepare, produce, organise, and evaluate forecasts. We then demonstrate how such theoretical concepts are applied in a variety of real-life contexts.

We do not claim that this review is an exhaustive list of methods and applications. However, we wish that our encyclopedic presentation will offer a point of reference for the rich work that has been undertaken over the last decades, with some key insights for the future of forecasting theory and practice. Given its encyclopedic nature, the intended mode of reading is non-linear. We offer cross-references to allow the readers to navigate through the various topics. We complement the theoretical concepts and applications covered by large lists of free or open-source software implementations and publicly-available databases.

\noindent \textbf{Keywords:} review; encyclopedia; methods; applications; principles; time series; prediction.

\clearpage
\setcounter{tocdepth}{3}
\small
\tableofcontents
\normalsize

\clearpage

\section[Introduction (Robert L. Winkler)]{Introduction\protect\footnote{This subsection was written by Robert L. Winkler.}}
\label{sec:introduction}

Forecasting has come a long way since early humans looked at the sky to see if the weather would be suitable for hunting, and even since hunters could get a forecast such as ``a high of 40 with a chance of rain''. Now a hunter can look at a smartphone to instantly get hour-by-hour forecasts of temperatures and probabilities of rain at multiple locations as well as videos of maps showing forecasted weather patterns over the coming hours. Tailored forecasts of increasing sophistication can be generated to inform important decisions of many different types by managers, public officials, investors, and other decision makers.

In the 15 years since the excellent review paper by \cite{Gooijer2006-hl_GS}, the field of forecasting has seen amazing growth in both theory and practice. Thus, this review is both timely and broad, ranging from the highly theoretical to the very applied.

Rapid advances in computing have enabled the analysis of larger and more complex data sets and stimulated interest in analytics and data science. As a result, the forecaster's toolbox of methods has grown in size and sophistication. Computer science has led the way with methods such as neural networks and other types of machine learning, which are getting a great deal of attention from forecasters and decision makers. Other methods, including statistical methods such as Bayesian forecasting and complex regression models, have also benefited from advances in computing. And improvements have not been limited to those based on computing advances. For example, the literature on judgmental forecasting has expanded considerably, driven largely by the ``wisdom of crowds'' notion.  

The combining, or aggregation, of forecasts, which is not a new idea, has received increased attention in the forecasting community recently and has been shown to perform well. For example, the top-performing entries in the M4 Competition run by Spyros Makridakis combined forecasts from multiple methods. Many models have been developed to forecast the number of deaths that will be caused by COVID-19, and combining the forecasts makes sense because it is hard to know which one will be the most accurate. It is consistent with Bayesian ideas since it can be viewed as updating, with each individual forecast added to the combined forecast (also called an ensemble) contributing some new information. 

Despite the excitement surrounding these new developments, older methods such as ARIMA and exponential smoothing are still valuable too. Exponential smoothing, along with other simple approaches, are quite robust and not as prone to overfitting as more complex methods. In that sense, they are useful not only on their own merits, but as part of an ensemble that also includes more sophisticated methods. Combined forecasts are more valuable if the forecasts come from methods that are diverse so that their forecast errors are not highly correlated.  

The conditions leading to larger, more sophisticated toolboxes for forecasters have also led to larger data sets with denser grids and improved models in areas of application. This has happened with models of the atmosphere, which are important in formulating improved weather forecasts. More detailed information about customers and their preferences allows the development of improved models of customer behaviour for managers. In turn, forecasting methods that can handle all of that information quickly are valuable for decision-making purposes. This process has spurred an explosion in trying to gather information on the internet. 

Risk is an important consideration in decision making, and probability forecasts can quantify such risks. Theoretical work in probability forecasting has been active for some time, and decision makers in many areas of practice have embraced the use of probability forecasts. In the Bayesian approach, inferences and forecasts are probabilistic in nature, and probability forecasts can be generated in many other ways too.

The U.S. National Weather Service began issuing probabilities of precipitation to the public in the 1960s. Yet extensive widespread use and dissemination of probabilities has only developed since the turn of the century. Now probability forecasts are increasingly communicated to the public and used as inputs in decision making. Nate Silver's FiveThirtyEight.com report gives probability forecasts for elections, medicine and science, sporting events, economic measures, and many other areas, often looking at multiple forecasting models individually and also combining them. 

It is natural for people to desire certainty. When probability forecasts of precipitation were first disseminated widely, many were very sceptical about them, with some accusing the forecasters of hedging and saying ``Don’t give me a probability. I want to know if it's going to rain or not''. Of course, point forecasts often are given along with probability forecasts. The current frequent exposure to probabilities helps the general public better understand, appreciate, and feel more comfortable with them. And the current situation in the world with COVID-19, increases in huge fires, big storms, political polarisation, international conflicts, etc., should help them realise that we are living in an age with huge uncertainties, and forecasts that quantify these uncertainties can be important. Where possible, visualisation can help, as indicated by the saying that a picture is worth a thousand words. Examples are the cones of uncertainty on maps in forecasts of the speed, severity, and future path of hurricanes, and the time line of the probability of a team winning a game, updated quickly after each play.

Put simply, this is an exciting time for the field of forecasting with all of the new theoretical developments and forecasting applications in practice. Forecasting is so ubiquitous that it’s not possible to cover all of these developments in a single article. This article manages to cover quite a few, and a good variety. Using short presentations for each one from an expert ``close to the ground'' on that theoretical topic or field of practice works well to provide a picture of the current state of the art in forecasting theory and practice.

\clearpage

\section{Theory}
\label{sec:theory}

\subsection[Introduction to forecasting theory (Anne B. Koehler)]{Introduction to forecasting theory\protect\footnote{This subsection was written by Anne B. Koehler.}}
\label{sec:introduction_to_the_theory_of_forecasting}
The theory of forecasting is based on the premise that current and past knowledge can be used to make predictions about the future. In particular for time series, there is the belief that it is possible to identify patterns in the historical values and successfully implement them in the process of predicting future values. However, the exact prediction of futures values is not expected. Instead, among the many options for a forecast of a single time series at a future time period are an expected value (known as a point forecast), a prediction interval, a percentile and an entire prediction distribution. This set of results collectively could be considered to be ``the forecast''. There are numerous other potential outcomes of a forecasting process. The objective may be to forecast an event, such as equipment failure, and time series may play only a small role in the forecasting process. Forecasting procedures are best when they relate to a problem to be solved in practice. The theory can then be developed by understanding the essential features of the problem. In turn, the theoretical results can lead to improved practice.

In this introduction, it is assumed that forecasting theories are developed as forecasting methods and models. A forecasting method is defined here to be a predetermined sequence of steps that produces forecasts at future time periods. Many forecasting methods, but definitely not all, have corresponding stochastic models that produce the same point forecasts. A stochastic model provides a data generating process that can be used to produce prediction intervals and entire prediction distributions in addition to point forecasts. Every stochastic model makes assumptions about the process and the associated probability distributions. Even when a forecasting method has an underlying stochastic model, the model is not necessarily unique. For example, the simple exponential smoothing method has multiple stochastic models, including state space models that may or may not be homoscedastic (i.e., possess constant variance). The combining of forecasts from different methods has been shown to be a very successful forecasting method. The combination of the corresponding stochastic models, if they exist, is itself a model. Forecasts can be produced by a process that incorporates new and/or existing forecasting methods/models. Of course, these more complex processes would also be forecasting methods/models.

Consideration of the nature of the variables and their involvement in the forecasting process is essential. In univariate forecasting, the forecasts are developed for a single time series by using the information from the historical values of the time series itself. While in multivariate forecasting, other time series variables are involved in producing the forecasts, as in time series regression. Both univariate and multivariate forecasting may allow for interventions (e.g., special promotions, extreme weather). Relationships among variables and other types of input could be linear or involve nonlinear structures (e.g., market penetration of a new technology). When an explicit functional form is not available, methodologies such as simulation or artificial neural networks might be employed. Theories from fields, such as economics, epidemiology, and meteorology, can be an important part of developing these relationships. Multivariate forecasting could also mean forecasting multiple variables simultaneously (e.g., econometric models). 

The data or observed values for time series come in many different forms that may limit or determine the choice of a forecasting method. In fact, there may be no historical observations at all for the item of interest, when judgmental methods must be used (e.g., time taken to complete construction of a new airport). The nature of the data may well require the development of a new forecasting method. The frequency of observations can include all sorts of variations, such as every minute, hourly, weekly, monthly, and yearly (e.g., the electricity industry needs to forecast demand loads at hourly intervals as well as long term demand for ten or more years ahead). The data could be composed of everything from a single important time series to billions of time series. Economic analysis often includes multiple variables, many of which affect one another. Time series for businesses are likely to be important at many different levels (e.g., stock keeping unit, common ingredients, or common size container) and, consequently, form a hierarchy of time series. Some or many of the values might be zero; making the time series intermittent. The list of forms for data is almost endless.

Prior to applying a forecasting method, the data may require pre-processing. There are basic details, such as checking for accuracy and missing values. Other matters might precede the application of the forecasting method or be incorporated into the methods/models themselves. The treatment of seasonality is such a case. Some forecasting method/models require de-seasonalised time series, while others address seasonality within the methods/models. Making it less clear when seasonality is considered relative to a forecasting method/model, some governmental statistical agencies produce forecasts to extend time series into the future in the midst of estimating seasonal factors (i.e., X-12 ARIMA).

Finally, it is extremely important to evaluate the effectiveness of a forecasting method. The ultimate application of the forecasts provides guidance in how to measure their accuracy. The focus is frequently on the difference between the actual value and a point forecast for the value. Many loss functions have been proposed to capture the ``average'' of these differences. Prediction intervals and percentiles can be used to judge the value of a point forecast as part of the forecast. On the other hand, the quality of prediction intervals and prediction distributions can themselves be evaluated by procedures and formulas that have been developed (e.g., ones based on scoring rules). Another assessment tool is judging the forecasts by metrics relevant to their usage (e.g., total costs or service levels).
 
In the remaining subsections of section \S\ref{sec:theory}, forecasting theory encompasses both stochastic modelling and forecasting methods along with related aspects.

\subsection{Pre-processing data}
\label{sec:Preprocessing_time_series_data}

\subsubsection[Box-Cox transformations (Anastasios Panagiotelis)]{Box-Cox transformations\protect\footnote{This subsection was written by Anastasios Panagiotelis.}}
\label{sec:BoxCox_Transformations}

A common practice in forecasting models is to transform the variable of interest $y$ using the transformation initially proposed by \cite{BoxCox1964transformation_AP} as
	\[
	  y^{(\lambda)}=\begin{cases}
	  (y^\lambda-1)/\lambda&\lambda\neq 0\\
	  \log(y)&\lambda = 0
	  \end{cases}\,.
	\]

The range of the transformation will be restricted in a way that depends on the sign of $\lambda$, therefore \cite{BickelDoksum1981transformation_AP} propose the following modification
    \[
	  y^{(\lambda)}=\begin{cases}
	  (|y|^\lambda sign(y_i)-1)/\lambda&\lambda\neq 0\\
	  \log(y)&\lambda = 0
	  \end{cases}\,,
	\]
which has a range from $(-\infty,\infty)$ for any value of $\lambda$. For a recent review of the Box-Cox (and other similar) transformations see \cite{AtkisonRianiCorbellini2021transformation_AP}.
	
The initial motivation for the Box-Cox transformation was to ensure data conformed to assumptions of normality and constant error variance that are required for inference in many statistical models. The transformation nests the log transformation when $\lambda=0$ and the case of no transformation (up to an additive constant) when $\lambda=1$. Additive models for $\log(y)$ correspond to multiplicative models on the original scale of $y$.  Choices of $\lambda$ between $0$ and $1$ therefore provide a natural continuum between multiplicative and additive models. For examples of forecasting models that use either a log or Box-Cox transformation see \S\ref{sec:forecasting_for_multiple_seasonal_cycles} and \S\ref{sec:state_space_models} and for applications see \S\ref{sec:Promotional_forecasting}, \S\ref{sec:Pandemics}, and \S\ref{sec:Call_arrival_forecasting}.
	
The literature on choosing $\lambda$ is extensive and dates back to the original \cite{BoxCox1964transformation_AP} paper - for a review see \cite{Sakia1992transformation_AP}. In a forecasting context, a popular method for finding $\lambda$ is given by \cite{Guerrero1993transformation_AP}. The method splits the data into blocks, computes the coefficient of variation within each block and then computes the coefficent of variation again between these blocks. The $\lambda$ that minimises this quantity is chosen.
	
Since the transformations considered here are monotonic, the forecast quantiles of the transformed data will, when back-transformed, result in the correct forecast quantiles in terms of the original data. As a result finding prediction intervals in terms of the original data only requires inverting the transformation. It should be noted though, that prediction intervals that are symmetric in terms of the transformed data will not be symmetric in terms of the original data. In a similar vein, back-transformation of the forecast median of the transformed data returns the forecast median in terms of the original data. For more on using the median forecast see \S\ref{sec:point_forecast_accuracy_measures} and references therein.
    
The convenient properties that apply to forecast quantiles, do not apply to the forecast mean, something recognised at least since the work of \cite{GrangerNewbold1976transformation_AP}. Back-transformation of the forecast mean of the transformed data does not yield the forecast mean of the original data, due to the non-linearity of the transformation. Consequently forecasts on the original scale of the data will be biased unless a correction is used. For some examples of bias correction methods see \cite{GrangerNewbold1976transformation_AP,Taylor1986retransformation_AP, PankratzDudley1987transformation_AP,Guerrero1993transformation_AP} and references therein. 
	
The issues of choosing $\lambda$ and bias correcting are accounted for in popular forecasting software packages. Notably, the method of \cite{Guerrero1993transformation_AP} both for finding $\lambda$ and bias correcting is implemented in the R packages \textit{forecast} and \textit{fable} (see Appendix B). 
 
\subsubsection[Time series decomposition (Alexander Dokumentov)]{Time series decomposition\protect\footnote{This subsection was written by Alexander Dokumentov.}}
\label{sec:time_series_decomposition}
Time series decomposition is an important building block for various forecasting approaches (see, for example, \S\ref{sec:Theta_method_and_models}, \S\ref{sec:Bagging_for_Time_Series}, and \S\ref{sec:Traffic_flow_forecasting}) and a crucial tools for statistical agencies. Seasonal decomposition is a way to present a time series as a function of other time series, called components. Commonly used decompositions are additive and multiplicative, where such functions are summation and multiplication correspondingly. If logs can be applied to time series, any additive decomposition method can serve as multiplicative after applying log transformation to the data.

The simplest additive decomposition of a time series with single seasonality comprises three components: trend, seasonal component, and the ``remainder''. It is assumed that the seasonal component has a repeating pattern (thus sub-series corresponding to every season are smooth or even constant), the trend component describes the smooth underlying mean and the remainder component is small and contains noise.

The first attempt to decompose time series into trend and seasonality is dated to 1847 when \citet{ad_buys1847changements} performed decomposition between trend and seasonality, modelling the trend by a polynomial and the seasonality by dummy variables. Then, in 1884 \citet{ad_Poynting1884} proposed price averaging as a tool for eliminating trend and seasonal fluctuations. Later, his approach was extended by \citet{ad_hooker1901suspension}, \citet{ad_spencer1904graduation} and \citet{ad_anderson1914nochmals}. \citet{ad_copeland1915statistical} was the first who attempted to extract the seasonal component, and \citet{ad_macaulay1930smooth} proposed a method which is currently considered ``classical''.

The main idea of this method comes from the observation that averaging a time series with window size of the time series seasonal period leaves the trend almost intact, while effectively removes seasonal and random components. At the next step, subtracting the estimated trend from the data and averaging the result for every season gives the seasonal component. The rest becomes the remainder.

Classical decomposition led to a series of more complex decomposition methods such as X-11 \citep{ad_shishkin1967x}, X-11-ARIMA \citep{ad_dagum1988x11arima,ad_ladiray2001seasonal}, X-12-ARIMA \citep{ad_findley1998new}, and X-13-ARIMA-SEATS \citep{ad_findley2005some}; see also \S\ref{sec:autoregressive_integrated_moving_average_models}.

Seasonal trend decomposition using Loess \citep[STL:][]{ad_cleveland1990stl} takes iterative approach and uses smoothing to obtain a better estimate of the trend and seasonal component at every iteration. Thus, starting with an estimate of the trend component, the trend component is subtracted from the data, the result is smoothed along sub-series corresponding to every season to obtain a ``rough'' estimate of the seasonal component. Since it might contain some trend, it is averaged to extract this remaining trend, which is then subtracted to get a detrended seasonal component. This detrended seasonal component is subtracted from the data and the result is smoothed again to obtain a better estimate of the trend. This cycle repeats a certain number of times.

Another big set of methods use a single underlining statistical model to perform decomposition. The model allows computation of confidence and prediction intervals naturally, which is not common for iterative and methods involving multiple models. The list of such methods includes TRAMO/SEATS procedure \citep{ad_monsell2003toward}, the BATS and TBATS models \citep{de2011forecasting_DKB}, various structural time series model approaches \citep{Harvey1990-mt_JRTA,ad_Commandeur2010}, and the recently developed seasonal-trend decomposition based on regression \citep[STR:][]{ad_dokumentov_2017,ad_stR}; see also \S\ref{sec:time_series_regression_models}. The last mentioned is one of the most generic decomposition methods allowing presence of missing values and outliers, multiple seasonal and cyclic components, exogenous variables with constant, varying, seasonal or cyclic influences, arbitrary complex seasonal schedules. By extending time series with a sequence of missing values the method allows forecasting.
 
\subsubsection[Anomaly detection and time series forecasting (Priyanga Dilini Talagala)]{Anomaly detection and time series forecasting\protect\footnote{This subsection was written by Priyanga Dilini Talagala.}}
\label{sec:Anomaly_detection_and_time_series_forecasting}
Temporal data are often subject to uncontrolled, unexpected interventions, from which various types of anomalous observations are produced. Owing to the complex nature of domain specific problems, it is difficult to find a unified definition for an anomaly and mostly application-specific \citep{unwin2019multivariate_PDT}. In time series and forecasting literature, an anomaly is mostly defined with respect to a specific context or its relation to past behaviours. The idea of a context is induced by the structure of the input data and the problem formulation \citep{chandola2007outlier, hand2009mining_PDT, chandola2009anomaly}. Further, anomaly detection in forecasting literature has two main focuses, which are conflicting in nature: one demands special attention be paid to anomalies as they can be the main carriers of significant and often critical information such as fraud activities, disease outbreak, natural disasters, while the other down-grades the value of anomalies as it reflects data quality issues such as missing values, corrupted data, data entry errors, extremes, duplicates and unreliable values \citep{talagala2020anomaly_PDT}. 

In the time series forecasting context, anomaly detection problems can be identified under three major umbrella themes: detection of (\textit{i}) contextual anomalies (point anomalies, additive anomalies) within a given series, (\textit{i}) anomalous sub-sequences within a given series, and (\textit{iii}) anomalous series within a collection of series \citep{gupta2013outlier_PDT,talagala2020anomalyoddstream_PDT}. 
According to previous studies forecast intervals are quite sensitive to contextual anomalies and the greatest impact on forecast are from anomalies occurring at the forecast origin \citep{chen1993forecasting_PDT}. 

The anomaly detection methods in forecasting applications can be categorised into two groups: (\textit{i}) model-based approaches and (\textit{ii}) feature-based approaches. Model-based approaches compare the predicted values with the original data. If the deviations are beyond a certain threshold, the corresponding observations are treated as anomalies \citep{luo2018real_PDT,luo2018benchmarking_PDT,sobhani2020temperature_PDT}. Contextual anomalies and anomalous sub-sequences are vastly covered by model-based approaches. Limitations in the detectability of anomalous events depend on the input effects of external time series. Examples of such effects are included in SARIMAX models for polynomial approaches (see also \S\ref{sec:autoregressive_integrated_moving_average_models}). In nonlinear contexts an example is the generalised Bass model \citep{bass:94_RGuseo} for special life cycle time series with external control processes (see \S\ref{sec:Innovation_diffusion_models}). SARMAX with nonlinear perturbed mean trajectory as input variable may help separating the mean process under control effects  from anomalies in the residual process.
Feature-based approaches, on the other hand, do not rely on predictive models. Instead, they are based on the time series features measured using different statistical operations (see \S\ref{sec:Feature_based_time_series_forecasting}) that differentiate anomalous instances from typical behaviours \citep{Fulcher2014_YK}. Feature-based approaches are commonly used for detecting anomalous time series within a large collection of time series. Under this approach, it first forecasts an anomalous threshold for the systems typical behaviour and new observations are identified as anomalies when they fall outside the bounds of the established anomalous threshold \citep{talagala2019feature_PDT,talagala2020anomalyoddstream_PDT}. Most of the existing algorithms involve a manual anomalous threshold. In contrast, \cite{burridge2006additive_PDT} and \cite{talagala2020anomalyoddstream_PDT} use extreme value theory based data-driven anomalous thresholds. Approaches to the problem of anomaly detection for temporal data can also be divided into two main scenarios: (\textit{i}) batch processing and (\textit{ii}) data streams. The data stream scenario poses many additional challenges, due to nonstationarity, large volume, high velocity, noisy signals, incomplete events and online support \citep{luo2018real_PDT,talagala2020anomalyoddstream_PDT}.

The performance evaluation of the anomaly detection frameworks is typically done using confusion matrices \citep{luo2018real_PDT,sobhani2020temperature_PDT}. However, these measures are not enough to evaluate the performance of the classifiers in the presence of imbalanced data \citep{hossin2015review_PDT}. Following \cite{ranawana2006optimized_PDT} and \cite{talagala2019feature_PDT}, \cite{leigh2019framework_PDT} have used some additional measures such as negative predictive value, positive predictive value and optimised precision to evaluate the performance of their detection algorithms.

\subsubsection[Robust handling of outliers in time series forecasting (Luigi Grossi)]{Robust handling of outliers in time series forecasting\protect\footnote{This subsection was written by Luigi Grossi.}}
\label{sec:Robust_handling_of_outliers}
Estimators of time series processes can be dramatically affected by the presence of few aberrant observations which are called differently in the time series literature: outliers, spikes, jumps, extreme observations (see \S\ref{sec:Anomaly_detection_and_time_series_forecasting}). If their presence is neglected, coefficients could be biasedly estimated. Biased estimates of ARIMA processes will decrease the efficiency of predictions \citep{Bianco2001-ma_LG}. Moreover, as the optimal predictor of ARIMA models (see \S\ref{sec:autoregressive_integrated_moving_average_models}) is a linear combination of observed units, the largest coefficients correspond to observations near the forecast origin and the presence of outliers among these units can severely affect the forecasts. Proper preliminary analysis of possible extreme observations is an unavoidable step, which should be carried out before any time series modelling and forecasting exercise (see \S\ref{sec:Forecasting_with_many_variables}). The issue was first raised in the seminal paper by \cite{Fox1972-da_LG}, who suggests a classification of outliers in time series, separating additive outliers (AO) from innovation outliers (IO). The influence of different types of outliers on the prediction errors in conditional mean models (ARIMA models) is studied by \cite{chen1993forecasting_PDT,Chen1993-vw_LG} and \cite{Ledolter1989-oy_LG,Ledolter1991-er_LG}, while the GARCH context (see also \S\ref{sec:arch_garch_models}) is explored by \cite{Franses1999-gl_LG} and \cite{Catalan2007-bi_LG}. \cite{Abraham1979-dh_LG} propose a Bayesian model which reflects the presence of outliers in time series and allows to mitigate their effects on estimated parameters and, consequently, improve the prediction ability. The main idea is to use a probabilistic framework allowing for the presence of a small group of discrepant units. 

A procedure for the correct specification of models, accounting for the presence of outliers, is introduced by \cite{Tsay1986-yb_LG} relying on iterative identification-detection-removal of cycles in the observed time series contaminated by outliers. The same issue is tackled by \cite{Abraham1989-nj_LG}: in this work non-influential outliers are separated from influential outliers which are observations with high residuals affecting parameter estimation. Tsay's procedure has been later modified \citep{Balke1993-yb_LG} to effectively detect time series level shifts. The impulse- and step-indicator saturation approach is used by \cite{Marczak2016-rj_LG} for detecting additive outliers and level shifts estimating structural models in the framework of nonstationary seasonal series. They find that timely detection of level shifts located towards the end of the series can improve the prediction accuracy. 

All these works are important because outlier and influential observations detection is crucial for improving the forecasting performance of models. The robust estimation of model parameters is another way to improve predictive accuracy without correcting or removing outliers (see \S\ref{sec:Electricity_price_forecasting}, for the application on energy data). \cite{Sakata1998-yp_LG} introduce a new two-stage estimation strategy for the conditional variance based on Hampel estimators and S-estimators. \cite{Park2002-qo_LG} proposes a robust GARCH model, called RGARCH exploiting the idea of least absolute deviation estimation. The robust approach is also followed for conditional mean models by \cite{Gelper2009-ff_LG} who introduce a robust version of the exponential and Holt-Winters smoothing technique for prediction purposes and by \cite{Cheng2015-ly_LG} who propose an outlier resistant algorithm developed starting from a new synthetic loss function. Very recently, \cite{Beyaztas2019-qm_LG} have introduced a robust forecasting procedure based on weighted likelihood estimators to improve point and interval forecasts in functional time series contaminated by the presence of outliers.
 
\subsubsection[Exogenous variables and feature engineering (Jethro Browell)]{Exogenous variables and feature engineering\protect\footnote{This subsection was written by Jethro Browell.}}
\label{sec:Exogenous_variables_and_feature_engineering}
Exogenous variables are those included in a forecasting system because they add value but are not being predicted themselves, and are sometimes called `features' (see \S\ref{sec:Feature_based_time_series_forecasting}). For example, a forecast of county's energy demand may be based on the recent history of demand (an \textit{endogenous} variable), but also weather forecasts, which are exogenous variables. Many time series methods have extensions that facilitate exogenous variables, such as autoregression with exogenous variables (ARX). However, it is often necessary to prepare exogenous data before use, for example so that it matches the temporal resolution of the variable being forecast (hourly, daily, and so on).

Exogenous variables may be numeric or categorical, and may be numerous. Different types of predictor present different issues depending on the predictive model being used. For instance, models based on the variable's absolute value can be sensitive to extreme values or skewness, whereas models based on the variable value's rank, such as tree-based models, are not. Exogenous variables that are correlated with one another also poses a challenge for some models, and techniques such as regularisation and partial leased squares have been developed to mitigate this.

Interactions between exogenous variables my also be important when making predictions. For example, crop yields depend on both rainfall and sunlight: one without the other or both in excess will result in low yields, but the right combination will result in high yields. Interactions may be included in linear models by including product of the two interacting exogenous as a feature in the model. This is an example of feature engineering, the process of creating new features based on domain knowledge or exploratory analysis of available data. In machine learning (see \S\ref{sec:machine_Learning}), many features may be created by combining exogenous variables speculatively and passed to a selection algorithm to identify those with predictive power. Combinations are not limited to products, or only two interacting variables, and where many exogenous variables are available, could include summary statistics (mean, standard deviation, range, quantiles...) of groups of variables.

Where exogenous variables are numerous dimension reduction may be applied to reduce the number of features in a forecasting model (see also \S\ref{sec:Variable_Selection}). Dimension reduction transforms multivariate data into a lower dimensional representation while retaining meaningful information about the original data. Principal component analysis (PCA) is a widely used method for linear dimension reduction, and non-linear alternatives are also available. PCA is useful when the number of candidate predictors is greater than the number of time series observations, as is often the case in macroeconomic forecasting \citep{Stock2002_JB}. It is routinely applied in applications from weather to sales forecasting. In retail forecasting, for example, past sales of thousands of products may be recorded but including them all as exogenous variables in the forecasting model for an individual product may be impractical. Dimension reduction offers an alternative to only using a subset of the available features.

Preparation of data for forecasting tasks is increasingly important as the volume of available data is increasing in many application areas. Further details and practical examples can be found in \cite{MaxKuhn2019_JB} and \cite{Albon2018_JB} among other texts in this area. For deeper technical discussion of a range of non-linear dimension reduction algorithms, see~\cite{Hastie2009_JB}.
 
\subsection{Statistical and econometric models}
\label{sec:Statistical_and_econometric_models}

\subsubsection[Exponential smoothing models (Juan Ramón Trapero Arenas)]{Exponential smoothing models\protect\footnote{This subsection was written by Juan Ramón Trapero Arenas.}}
\label{sec:exponential_smoothing_models}
Exponential smoothing is one of the workhorses of business forecasting. Despite the many advances in the field, it is always a tough benchmark to bear in mind. The development of exponential smoothing dates back to 1944, where Robert G. Brown through a mechanical computing device estimated key variables for fire-control on the location of submarines \citep{Gardner1985-jm_JRTA}. More details about the state of the art of exponential smoothing can be found in \cite{Gardner2006-bv_JRTA}. 

The idea behind exponential smoothing relies on the weighted average of past observations, where that weight decreases exponentially as one moves away from the present observations. The appropriate exponential smoothing method depends on the components that appear in the time series. For instance, in case that no clear trend or seasonal pattern is present, the simplest form of exponential smoothing methods known as Simple (or Single) Exponential Smoothing (SES) is adequate, such as:
$$
f_{t+1} = \alpha y_t + (1-\alpha)f_t
$$

In some references, is also known as Exponentially Weighted Moving Average \citep{Harvey1990-mt_JRTA}. The formula for SES can be obtained from minimising the discounted least squares error function and expressing the resulting equation in a recursive form \citep{Harvey1990-mt_JRTA}. If observations do not have the same weight, the ordinary least squares cannot be applied. On the other hand, the recursive form is very well-suited for saving data storage. 

In order to use SES, we need to estimate the initial forecast ($f_1$) and the exponential smoothing parameter ($\alpha$). Traditionally, the initialisation was done by using either ad hoc values or a heuristic scheme \citep{Hyndman2008-iu_JRTA}, however nowadays it is standard to estimate both the initial forecast and the optimal smoothing parameter by minimising the sum of squares of the one-step ahead forecast errors. The estimation of the smoothing parameter usually is restricted to values between 0 and 1. Once SES is defined, the method only provides point forecasts, i.e., forecasts of the mean. Nonetheless, it is of vital importance for many applications to provide density (probabilistic) forecasts. To that end, \cite{Hyndman2002-jp_JRTA} extended exponential smoothing methods under State Space models using a single source of error (see \S\ref{sec:state_space_models}) to equip them with a statistical framework capable of providing future probability distributions. For example, SES can be expressed in the State Space as a local level model:
\begin{eqnarray}\label{eq:subtasks}
y_t &=& \ell_{t-1} + \epsilon_t, \nonumber \\
\ell_t &=& \ell_{t-1} + \alpha \epsilon_t. \nonumber 
\end{eqnarray}

\noindent where the state is the level ($\ell$) and $\epsilon$ is the Gaussian noise. Note the difference between traditional exponential smoothing methods and exponential smoothing models (under the state space approach). The former only provide point forecasts, meanwhile the latter also offers probabilistic forecasts, which obviously includes prediction intervals. In addition, some exponential smoothing models can be expressed an ARIMA models (see also \S\ref{sec:autoregressive_integrated_moving_average_models}).

So far, we have introduced the main exponential smoothing using SES, however real time series can include other components as trends, seasonal patterns, cycles, and the irregular (error) component. In this sense, the exponential smoothing version capable of handling local trends is commonly known as Holt's method \citep[][originally published in 1957]{Holt2004-zg} and, if it also models a seasonality component, which can be incorporated in an additive or multiplicative fashion, it is called Holt-Winters method \citep{Winters1960-sd}. Exponential smoothing models have been also extended to handle multiple seasonal cycles; see \S\ref{sec:forecasting_for_multiple_seasonal_cycles}.

Fortunately, for various combinations of time series patterns (level only, trended, seasonal, trended and seasonal) a particular exponential smoothing can be chosen. \cite{Pegels1969-jh_JRTA} proposed a first classification of exponential smoothing methods, later extended by \cite{Gardner1985-jm_JRTA} and \cite{Taylor2003-ix_JRTA}. The state space framework mentioned above, developed by \cite{Hyndman2002-jp_JRTA}, allowed to compute the likelihood for each exponential smoothing model and, thus, model selection criteria such as AIC could be used to automatically identify the appropriate exponential smoothing model. Note that the equivalent state space formulation was derived by using a single source of error instead of a multiple source of error \citep{Harvey1990-mt_JRTA}. \cite{Hyndman2008-iu_JRTA} utilised the notation (E,T,S) to classify the exponential smoothing models, where those letters refer to the following components: Error, Trend, and Seasonality. This notation has gained popularity because the widely-used \textit{forecast} package \citep{Hyndman2020forecast_CB}, recently updated to the \textit{fable} package, for R statistical software, and nowadays exponential smoothing is frequently called ETS. 

\subsubsection[Time-series regression models (Vassilios Assimakopoulos)]{Time-series regression models\protect\footnote{This subsection was written by Vassilios Assimakopoulos.}}
\label{sec:time_series_regression_models}
The key idea of linear regression models is that a target (or dependent, forecast, explained, regress) variable, \bm{$y$}, i.e., a time series of interest, can be forecast through other regressor (or independent, predictor, explanatory) variables, \bm{$x$}, i.e., time series or features (see \S\ref{sec:Exogenous_variables_and_feature_engineering}), assuming that a linear relationship exists between them, as follows

\begin{equation*}
    y_t = \beta_{0} + \beta_{1} x_{1_{t}} + \beta_{2} x_{2_{t}} + \dots + \beta_{k} x_{k_{t}} + e_t,
\end{equation*}

\noindent where $e_t$ is the residual error of the model at time $t$, $\beta_{0}$ is a constant, and coefficient $\beta_{i}$ is the effect of regressor \bm{$x_i$} after taking into account the effects of all $k$ regressors involved in the model. For example, daily product sales may be forecast using information related with past sales, prices, advertising, promotions, special days, and holidays (see also \S\ref{sec:Retail_sales_forecasting}).

In order to estimate the model, forecasters typically minimise the sum of the squared errors (ordinary least squares estimation, OLS), $\text{SSE}=\sum_{t=1}^{n} e_t^2$, using the observations available for fitting the model to the data \citep{OrdWessex} and setting the gradient $\frac{\partial \text{SSE}}{\partial \beta_{i}}$ equal to zero. If the model is simple, consisting of a single regressor, then two coefficients are computed, which are the slope (coefficient of the regressor) and the intercept (constant). When more regressor variables are considered, the model is characterised as a multiple regression one and additional coefficients are estimated.

A common way to evaluate how well a linear regression model fits the target series, reporting an average value of $\Bar{y}$, is through the coefficient of determination, $R^2=\frac{\sum_{t=1}^{n} (f_t-\Bar{y})^2}{\sum_{t=1}^{n} (y_t-\Bar{y})^2}$, indicating the proportion of variation in the dependent variable explained by the model. Values close to one indicate sufficient goodness-of-fit, while values close to zero insufficient fitting. However, goodness-of-fit should not be confused with forecastability \citep{HarrellSpringer}. When the complexity of the model is increased, i.e., more regressors are considered, the value of the coefficient will also rise, even if such additions lead to overfitting (see \S\ref{sec:Model_complexity}). Thus, regression models should be evaluated using cross-validation approaches (see \S\ref{sec:Cross_validation_for_time_series_data}), approximating the post-sample accuracy of the model, or measures that account for model complexity, such as information criteria (e.g., AIC, AICc, and BIC) and the adjusted coefficient of determination, $\Bar{R}^2 = 1-(1-R^2)\frac{n-1}{n-k-1}$ \citep{JamesSpringer}. Other diagnostics are the standard deviation of the residuals and the t-values of the regressors. Residual standard error, $\sigma_e = \sqrt{\frac{\sum_{t=1}^{n} (y_t-f_t)^2}{n-k-1}}$, summarises the average error produced by the model given the number of regressors used, thus accounting for overfitting. The t-values measure the impact of excluding regressors from the model in terms of error, given the variation in the data, thus highlighting the importance of the regressors.

To make sure that the produced forecasts are reliable, the correlation between the residuals and the observations of the regressors must be zero, with the former displaying also insignificant autocorrelation. Other assumptions suggest that the residuals should be normally distributed with an average value of zero and that their variability should be equal across time (no heteroscedasticity present). Nevertheless, in practice, it is rarely necessary for residuals to be normally distributed in order for the model to produce accurate results, while the homoscedasticity assumption becomes relevant mostly when computing prediction intervals. If these assumptions are violated, that may mean that part of the variance of the target variable has not been explained by the model and, therefore, that other or more regressors are needed. In case of non-linear dependencies between the target and the regressor variables, data power transformations (see \S\ref{sec:BoxCox_Transformations}) or machine learning approaches can be considered (see \S\ref{sec:machine_Learning}).

Apart from time series regressors, regression models can also exploit categorical (dummy or indicator) variables \citep{hyndman2018Forecasting_CB} which may e.g., inform the model about promotions, special events, and holidays (binary variables), the day of the week or month of the year (seasonal dummy variables provided as one-hot encoded vectors), trends and structural changes, and the number of trading/working days included in the examined period. In cases where the target series is long and displays complex seasonal patterns, additional regressors such as Fourier series and lagged values of both the target and the regressor variables may become useful. Moreover, when the number of the potential regressor variables is significant compared to the observations available for estimating the respective coefficients (see \S\ref{sec:Forecasting_with_Big_Data}), step-wise regression \citep{JamesSpringer} or dimension reduction and shrinkage estimation methods (see \S\ref{sec:Variable_Selection}) can be considered to facilitate training and avoid overfitting. Finally, mixed data sampling (MIDAS) regression models are a way of allowing different degrees of temporal aggregation for the regressors and predictand (see also \S\ref{sec:temporal_aggregation} for further discussions on forecasting with temporal aggregation).
 
\subsubsection[Theta method and models (Dimitrios Thomakos)]{Theta method and models\protect\footnote{This subsection was written by Dimitrios Thomakos.}}
\label{sec:Theta_method_and_models}
In the age of vast computing power and computational intelligence, the contribution of simple forecasting methods is possibly not \textit{en vogue}; the implementation of complicated forecasting systems becomes not only expedient but possibly desirable. Nevertheless forecasting, being a tricky business, does not always favour the complicated or the computationally intensive. Enter the theta method. From its beginnings 20 years back in \cite{Assimakopoulos2000-hj} to recent advances in the monograph of \cite{Nikolopoulos2019-sp_DT}, to other work in-between and recently too, the theta method has emerged as not only a powerfully simple but also enduring method in modern time series forecasting. The reader will benefit by reviewing \S\ref{sec:exponential_smoothing_models}, \S\ref{sec:autoregressive_integrated_moving_average_models}, and \S\ref{sec:Forecasting_with_many_variables} for useful background information. 

The original idea has been now fully explained and understood and, as \cite{Nikolopoulos2019-sp_DT} have shown, even the revered AR(1) model forecast is indeed a theta forecast -- and it has already been shown by \cite{Hyndman2003-ea_DT} that the theta method can represent SES (with a drift) forecasts as well. In its simplest form the method generates a forecast from a linear combination of the last observation and some form of ``trend'' function, be that a constant, a linear trend, a non-parametric trend or a non-linear trend. In summary, and under the conditions outlined extensively in \cite{Nikolopoulos2019-sp_DT}, the theta forecasts can be expressed as functions of the ``theta line'': 
$$
Q_t(\theta) = \theta y_t + (1-\theta)T_{t+1}
$$

\noindent where $T_{t+1}$ is the trend function, variously defined depending on the modelling approach and type of trend one is considering in applications. It can be shown that the, univariate, theta forecasts can given either as
$$
f_{t+1|t} = y_t + \Delta Q_t(\theta)
$$

\noindent when the trend function is defined as $T_{t+1}= \mu t$ and as
$$
f_{t+1|t} = Q_t(\theta) + \theta \Delta \mathbb{E}(T_{t+1})
$$

\noindent when the trend function is left otherwise unspecified. The choice of the weight parameter $\theta$ on the linear combination of the theta line, the choice and number of trend functions and their nature and other aspects on expanding the method have been recently researched extensively. 

The main literature has two strands. The first one details the probabilistic background of the method and derives certain theoretical properties, as in \cite{Hyndman2003-ea_DT}, \cite{Thomakos2012-em_DT,Thomakos2015-ku_DT} and a number of new theoretical results in \cite{Nikolopoulos2019-sp_DT}. The work of Thomakos and Nikolopoulos provided a complete analysis of the theta method under the unit root data generating process, explained its success in the M3 competition \citep{Makridakis2000-ty}, introduced the multivariate theta method and related it to cointegration and provided a number of other analytical results for different trend functions and multivariate forecasting. The second strand of the literature expands and details various implementation (including hybrid approaches) of the method, as in the theta approach in supply chain planning of \cite{Nikolopoulos2012-zy_DT}, the optimised theta models and their relationship with state space models in \cite{Fioruci2015-gf_DT} and \cite{Fiorucci2016-ak_DT}, hybrid approaches as in \cite{Theodosiou2011-mf_DT} and \cite{Spiliotis2019-cd_DT}, to the very latest generalised theta method of \cite{Spiliotis2020-jc_DT}. These are major methodological references in the field, in addition to many others of pure application of the method.

The theta method is also part of the family of adaptive models/methods, and a simple example illustrates the point: the AR(1) forecast or the SES forecast are both theta forecasts but they are also both adaptive learning forecasts, as in the definitions of the recent work by \cite{Kyriazi2019-md_DT}. As such, the theta forecasts contain the basic building blocks of successful forecasts: simplicity, theoretical foundations, adaptability and performance enhancements. Further research on the usage of the theta method within the context of adaptive learning appears to be a natural next step. In the context of this section, see also \S\ref{sec:Robust_equilibrium_correction_forecasting_devices} on equilibrium correcting models and forecasts.

Given the simplicity of its application, the freely available libraries of its computation, its scalability and performance, the theta method should be considered as a critical benchmark henceforth in the literature -- no amount of complexity is worth its weight if it cannot beat a single Greek letter! 

\subsubsection[Autoregressive integrated moving average (ARIMA) models (Philip Hans Franses \& Sheik Meeran)]{Autoregressive integrated moving average (ARIMA) models\protect\footnote{This subsection was written by Philip Hans Franses \& Sheik Meeran.}}
\label{sec:autoregressive_integrated_moving_average_models}
Time series models that are often used for forecasting are of the autoregressive integrated moving average class \citep[ARIMA --][]{Box1976-af_SM}. The notation of an ARIMA($p$, $d$, $q$) model for a time series $y_t$ is
$$
(1 - \phi_1L - \dots - \phi_pL^p)(1-L)^d y_t = c + (1 + \theta_1L + \dots + \theta_qL^q)+\epsilon_t,
$$

\noindent where the lag operator $L$ is defined by $L^k y_t=y_{t-k}$. The $\epsilon_t$ is a zero-mean uncorrelated process with common variance $\sigma_\epsilon^2$. Some exponential smoothing models (see \S\ref{sec:exponential_smoothing_models}) can also be written in ARIMA format, where some ETS models assume that $d=1$ or $d=2$. For example, SES is equivalent to ARIMA(0,1,1) when $\theta_1 = \alpha - 1$. 

The parameters in the ARIMA model can be estimated using Maximum Likelihood, whereas for the ARIMA($p$, $d$, 0) Ordinary Least Squares can be used. The iterative model-building process \citep{Franses2014-kl_PHF} requires the determination of the values of $p$, $d$, and $q$. Data features as the empirical autocorrelation function and the empirical partial autocorrelation function can be used to identify the values of $p$ and $q$, in case of low values of $p$ and $q$. Otherwise, in practice one relies on the well-known information criteria like AIC and BIC (see \S\ref{sec:Model_Selection}). The function \texttt{auto.arima} of the \textit{forecast} package \citep{Hyndman2020forecast_CB} for R statistical software compares models using information criteria, and has been found to be very effective and increasingly being used in ARIMA modelling.

Forecasts from ARIMA models are easy to make. And, at the same time, prediction intervals can be easily computed. Take for example the ARIMA(1,0,1) model:
$$
y_t = c + \phi_1 y_{t-1} + \epsilon_t + \theta_1 \epsilon_{t-1}.
$$
\noindent The one-step-ahead forecast from forecast origin $n$ is $f_{n+1 | n} = c + \phi_1 y_n + \theta_1 \epsilon_n$ as the expected value $E(\epsilon_{n+1}) = 0$. The forecast error is $y_{n+1} - f_{n+1 | n} = \epsilon_{n+1}$ and, hence, the forecast error variance is $\sigma_\epsilon^2$. The two-steps-ahead forecast from $n$ is $f_{n+2 | n} = c + \phi_1 f_{n+1 | n}$ with the forecast error equal to $\epsilon_{n+2}+\phi_1 \epsilon_{n+1}$ and the forecast error variance $(1+\phi_1^2)\sigma_\epsilon^2$. These expressions show that the creation of forecasts and forecast errors straightforwardly follow from the model expressions, and hence can be automated if necessary. 

An important decision when using an ARIMA model is the choice for the value of $d$. When $d=0$, the model is created for the levels of the time series, that is, $y_t$. When $d=1$, there is a model for $(1-L)y_t$, and the data need to be differenced prior to fitting an ARMA model. In some specific but rare cases, $d=2$. The decision on the value of d is usually based on so-called tests for unit roots
\citep{Dickey1979-cj_PHF,Dickey1987-nd_PHF}. Under the null hypothesis that $d=1$, the data are non-stationary, and the test involves non-standard statistical theory \citep{Phillips1987-ak_PHF}. One can also choose to make $d=0$ as the null hypothesis \citep{Kwiatkowski1992-lb_PHF,Hobijn2004-lt_PHF}. The power of unit root tests is not large, and in practice one often finds signals to consider $d=1$ \citep{Nelson1982-hc_PHF}.

For seasonal data, like quarterly and monthly time series, the ARIMA model can be extended to Seasonal ARIMA (SARIMA) models represented by ARIMA($p$, $d$, $q$)($P$, $D$, $Q$)$_s$, where $P$, $D$, and $Q$ are the seasonal parameters and the $s$ is the periodicity. When $D = 1$, the data are transformed as $(1-L^s)y_t$. It can also be that $D = 0$ and $d = 1$, and then one can replace $c$ by $c_1 D_{1,t} + c_2 D_{2,t} + \dots + c_s D_{s,t}$ where the $D_{i,t}$ with $i = 1, 2, \dots, s$ are seasonal dummies. The choice of $D$ is based on tests for so-called seasonal unit roots \citep{Hylleberg1990-wm_PHF,Franses1991-sw_PHF,Ghysels1994-dd_PHF}. 

Another popular extension to ARIMA models is called ARIMAX, implemented by incorporating additional exogenous variables (regressors) that are external to and different from the forecast variable. An alternative to ARIMAX is the use of regression models (see \S\ref{sec:time_series_regression_models}) with ARMA errors.
 
\subsubsection[Forecasting for multiple seasonal cycles (Bahman Rostami-Tabar)]{Forecasting for multiple seasonal cycles\protect\footnote{This subsection was written by Bahman Rostami-Tabar.}}
\label{sec:forecasting_for_multiple_seasonal_cycles}
With the advances in digital data technologies, data is recorded more frequently in many sectors such as energy \citep[][and \S\ref{sec:Energy}]{wang2016electric_BRT}, healthcare \citep[][and \ref{sec:Heathcare}]{whitt2019forecasting_BRT}, transportation \citep{gould2008forecasting_BRT}, and telecommunication \citep{meade2015forecasting_BRT}. This often results in time series that exhibit multiple seasonal cycles (MSC) of different lengths. Forecasting problems involving such series have been increasingly drawing the attention of both researchers and practitioners leading to the development of several approaches.

Multiple Linear Regression (MLR) is a common approach to model series with MSC \citep{kamisan2018load_BRT,rostami2020anticipating_BRT}; for an introduction on time-series regression models, see \S\ref{sec:time_series_regression_models}. While MLR is fast, flexible, and uses exogenous regressors, it does not allow to decompose components and change them over time. Building on the foundation of the regression, Facebook introduced Prophet \citep{taylor2018forecasting_BRT}, an automated approach that utilises the Generalised Additive Model \citep{hastie1990generalized_BRT}. Although the implementation of Prophet may be less flexible, it is easy to use, robust to missing values and structural changes, and can handles outliers well.

Some studies have extended the classical ARIMA (see \S\ref{sec:autoregressive_integrated_moving_average_models}) and Exponential Smoothing (ETS; see \S\ref{sec:exponential_smoothing_models}) methods to accommodate MSC. Multiple/multiplicative Seasonal ARIMA (MSARIMA) model is an extension of ARIMA for the case of MSC \citep{taylor2003short_BRT}. MSARIMA allows for exogenous regressors and terms can evolve over time, however, it is not flexible, and the computational time is high. \cite{svetunkov2020state_BRT} introduced the Several Seasonalities ARIMA (SSARIMA) model which constructs ARIMA in a state-space form with several seasonalities. While SSARIMA is flexible and allows for exogenous regressors, it is computationally expensive, especially for high frequency series.

\cite{taylor2003short_BRT} introduced Double Seasonal Holt-Winters (DSHW) to extend ETS for modelling daily and weekly seasonal cycles. Following that, \cite{taylor2010exponentially_DKB} proposed a triple seasonal model to consider the intraday, intraweek and intrayear seasonalities. \cite{gould2008forecasting_BRT} and \cite{taylor2012forecasting_BRT} instead proposed an approach that combines a parsimonious representation of the seasonal states up to a weekly period in an innovation state space model. With these models, components can change, and decomposition is possible. However, the implementation is not flexible, the use of exogenous regressors is not supported, and the computational time could be high.

An alternative approach for forecasting series with MSC is TBATS \citep[][see also \S\ref{sec:time_series_decomposition}]{de2011forecasting_DKB}. TBATS uses a combination of Fourier terms with an exponential smoothing state space model and a Box-Cox transformation (see \S\ref{sec:BoxCox_Transformations}), in an entirely automated manner. It allows for terms to evolve over time and produce accurate forecasts. Some drawbacks of TBATS, however, are that it is not flexible, can be slow, and does not allow for covariates.

In response to shortcomings in current models, Forecasting with Additive Switching of Seasonality, Trend and Exogenous Regressors (FASSTER) has been proposed by \cite{fasster2020_BRT}. FASSTER is fast, flexible and support the use of exogenous regressors into a state space model. It extends state space models such as TBATS by introducing a switching component to the measurement equation which captures groups of irregular multiple seasonality by switching between states.

In recent years, Machine Learning (ML; see \S\ref{sec:machine_Learning}) approaches have also been recommended for forecasting time series with MSC. MultiLayer Perceptron \citep[MLP:][]{dudek2013forecasting_BRT,zhang2005neural_BRT}, Recurrent Neural Networks \citep[RNN:][]{lai2018modeling_BRT}, Generalised Regression Neural Network \citep[GRNN:][]{dudek2015generalized_BRT}, and Long Short-Term Memory Networks \citep[LSTM][]{zheng2017electric_BRT} have been applied on real data \citep{bandara2020lstm_BRT,xie2020forecasting_BRT} with promising results. These approaches are flexible, allow for any exogenous regressor and suitable when non-linearity exists in series, however interpretability might be an issue for users \citep{MakridakisPLOS}.
 
\subsubsection[State-space models (Diego J. Pedregal)]{State-space models\protect\footnote{This subsection was written by Diego J. Pedregal.}}
\label{sec:state_space_models}
State Space (SS) systems are a very powerful and useful framework for time series and econometric modelling and forecasting. Such systems were initially developed by engineers, but have been widely adopted and developed in Economics as well \citep{Harvey1990-mt_JRTA,koopman2012_Diego}. The main distinguishing feature of SS systems is that the model is formulated in terms of \textit{states} ($\mathbf{\alpha}_t$), which are a set of variables usually unobserved, but which have some meaning. Typical examples are trends, seasonal components or time varying parameters.

A SS system is built as the combination of two sets of equations: (\textit{i}) \textit{state} or \textit{transition} equations which describe the dynamic law governing the states between two adjacent points in time; and (\textit{ii}) \textit{observation} equations which specify the relation between observed data (both inputs and outputs) and the unobserved states. A linear version of such a system is shown in equation (\ref{eq:SSsystem}). 
\begin{equation}\label{eq:SSsystem}
\begin{array}{cc}
	\mathbf{\alpha}_{t+1}=\mathbf{T}_t \mathbf{\alpha}_t+\mathbf{\Gamma}_t+\mathbf{R}_t \mathbf{\eta}_t, &  \mathbf{\eta}_t \sim N(0,\mathbf{Q}_t)\\
	\mathbf{y}_t=\mathbf{Z}_t \mathbf{\alpha}_t+\mathbf{D}_t+\mathbf{C}_t \mathbf{\epsilon}_t, & \mathbf{\epsilon}_t \sim N(0,\mathbf{H}_t)\\
	\mathbf{\alpha}_1 \sim N(\boldsymbol{a}_1,\mathbf{P}_1) & \\
\end{array}
\end{equation}

In this equations $\mathbf{\eta}_t$ and $\mathbf{\epsilon}_t$ are the state and observational vectors of zero mean Gaussian noises with covariance $\mathbf{S}_t$. $\mathbf{T}_t$, $\mathbf{\Gamma}_t$, $\mathbf{R}_t$, $\mathbf{Q}_t$, $\mathbf{Z}_t$, $\mathbf{D}_t$, $\mathbf{C}_t$, $\mathbf{H}_t$ and $\mathbf{S}_t$ are the so-called (time-varying) system matrices, and $\boldsymbol{a}_1$ and $\mathbf{P}_1$ are the initial state and state covariance matrix, respectively. Note that $\mathbf{D}_t$ and $\mathbf{\Gamma}_t$ may be parameterised to include some input variables as linear or non-linear relations to the output variables $\mathbf{y}_t$.

The model in equation (\ref{eq:SSsystem}) is a \textit{multiple error SS model}. A different formulation is the \textit{single error SS model} or the \textit{innovations SS model}.  This latter is similar to (\ref{eq:SSsystem}), but replacing $\mathbf{R}_t \mathbf{\eta}_t$ and $\mathbf{C}_t \mathbf{\epsilon}_t$ by $\mathbf{K}_t \mathbf{e}_t$ and $\mathbf{e}_t$, respectively. Then, naturally, the innovations form may be seen as a restricted version of model (\ref{eq:SSsystem}), but, conversely, under weak assumptions, (\ref{eq:SSsystem}) may also be written as an observationally equivalent \textit{innovations form} \cite[see, for example,][pp. 12-17]{jerez16_Diego}

Once a SS system is fully specified, the core problem is to provide optimal estimates of states and their covariance matrix over time. This can be done in two ways, either by looking back in time using the well-known \textit{Kalman filter} (useful for online applications) or taking into account the whole sample provided by smoothing algorithms (typical of offline applications) \citep{Anderson_Diego}.

Given any set of data and a specific model, the system is not fully specified in most cases because it usually depends on unknown parameters scattered throughout the system matrices that define the SS equations. Estimation of such parameters is normally carried out by Maximum Likelihood defined by prediction error decomposition \citep{Harvey1990-mt_JRTA}.

Non-linear and non-Gaussian models are also possible, but at the cost of a higher computational burden because more sophisticated recursive algorithms have to be applied, like the extended Kalman filters and smoothers of different orders, particle filters \citep{Doucet_Diego}, Unscented Kalman filter and smoother \citep{Uhlmann_Diego}, or simulation of many kinds, like Monte Carlo, bootstrapping or importance sampling \citep{koopman2012_Diego}.

The paramount advantage of SS systems is that they are not a particular model or family of models strictly speaking, but a container in which many very different model families may be implemented, indeed many treated in other sections of this paper. The following is a list of possibilities, not at all exhaustive:
\begin{itemize}[noitemsep]
    \item Univariate models with or without inputs: regression (\S\ref{sec:time_series_regression_models}), ARIMAx (\S\ref{sec:autoregressive_integrated_moving_average_models}), transfer functions, exponential smoothing (\S\ref{sec:exponential_smoothing_models}), structural unobserved components, Hodrick-Prescott filter, spline smoothing.
    \item Fully multivariate: natural extensions of the previous ones plus echelon-form VARIMAx, Structural VAR, VECM, Dynamic Factor models, panel data (\S\ref{sec:Forecasting_with_many_variables}).
    \item Non-linear and non-Gaussian: TAR, ARCH, GARCH (\S\ref{sec:arch_garch_models}), Stochastic Volatility \citep{koopman2012_Diego}, Dynamic Conditional Score \citep{harvey_2013_Diego}, Generalised Autoregressive Score \citep{koopman2013_Diego}, multiplicative unobserved components.
    \item Other: periodic cubic splines, periodic unobserved components models, state dependent models, Gegenbauer long memory processes \citep{Gegenbauer_Diego}.
\end{itemize}

Once any researcher or practitioner becomes acquainted to a certain degree with the SS technology, some important advanced issues in time series forecasting may be comfortably addressed \citep{jerez16_Diego}. It is the case, for example, of systems block concatenation, systems nesting in errors or in variables, treating errors in variables, continuous time models, time irregularly spaced data, mixed frequency models, time varying parameters, time aggregation, hierarchical and group forecasting \citep{pedregal2018_Diego} (time, longitudinal or both), homogeneity of multivariate models (proportional covariance structure among perturbations), etc.

All in all, the SS systems offer a framework capable of handling many modelling and forecasting techniques available nowadays in a single environment. Once the initial barriers are overcome, a wide panorama of modelling opportunities opens up.

\subsubsection[Models for population processes (Jakub Bijak)]{Models for population processes\protect\footnote{This subsection was written by Jakub Bijak.}}
\label{sec:Models_for_population_processes}
Over the past two centuries, formal demography has established its own, discipline-specific body of methods for predicting (or \textit{projecting}\footnote{The demographic literature sometimes makes a distinction between unconditional \textit{forecasts} (or \textit{predictions}) and \textit{projections}, conditional on their underlying assumptions. In this section, we use the former term to refer to statements about the future, and the latter to the result of a numerical exercise of combining assumptions on fertility, mortality and migration in a deterministic model of population renewal.}) populations. Population sciences, since their 17\textsuperscript{th} century beginnings, have been traditionally very empirically focused, with strong links with probability theory \citep{Courgeau2012-wh_JB}. Given the observed regularities in population dynamics, and that populations are somewhat better predictable than many other socio-economic processes, with reasonable horizons possibly up to one generation ahead \citep{Keyfitz1972-ux_JB,Keyfitz1981-mk_JB}, demographic forecasts have become a bestselling product of the discipline \citep{Xie2000-mc_JB}. Since the 20\textsuperscript{th} century, methodological developments in human demography have been augmented by the work carried out in mathematical biology and population ecology \citep{Caswell2019-ba_JB}.

The theoretical strength of demography also lies almost exclusively in the formal mathematical description of population processes \citep{Burch2018-jo_JB}, typically growth functions and structural changes. Historically, such attempts started from formulating the logistic model of population dynamics, inspired by the Malthusian theory \citep{Verhulst1845-ng_JB,Pearl1920-mz_JB}. \citeauthor{Lotka1907-jl_JB}'s (\citeyear{Lotka1907-jl_JB}) work laid the foundations of the stable population theory with asymptotic stability under constant vital rates, subsequently extended to modelling of interacting populations by using differential equations \citep{Lotka1925-xg,Volterra1926-dc_JB}. By the middle of the 20th century, the potential and limitations of demographic forecasting methods were already well recognised in the literature \citep{Hajnal1955-og_JB,Brass1974-zf_JE}.

In the state-of-the-art demographic forecasting, the core engine is provided by matrix algebra. The most common approach relies on the cohort-component models, which combine the assumptions on fertility, mortality and migration, in order to produce future population by age, sex, and other characteristics. In such models, the deterministic mechanism of population renewal is known, and results from the following demographic accounting identity \cite[population balancing equation, see][]{Rees1973-au_JB,Bryant2018-hp_JB}: 
$$
P[x+1, t+1] = P[x, t] - D[(x, x+1), (t, t+1)] + I[(x, x+1), (t, t+1)] - E[(x, x+1), (t, t+1)]
$$

\noindent where $P[x, t]$ denotes population aged $x$ at time $t$, $D[(x, x+1), (t, t+1)]$ refer to deaths between ages $x$ and $x+1$ in the time interval $t$ to $t+1$, with $I$ and $E$ respectively denoting immigration (and other entries) and emigration (and other exits). In addition, for the youngest age group, births $B[(t, t+1)]$ need to be added. The equation above can be written up in the general algebraic form: $\mathbf{P}_{t+1} = \mathbf{G} \mathbf{P}_t$, where $\mathbf{P}_t$ is the population vector structured by age (and other characteristics), and $\mathbf{G}$ is an appropriately chosen growth matrix (Leslie matrix), closely linked with the life table while reflecting the relationship above, expressed in terms of rates rather than events \citep{Leslie1945-eu_JB,Leslie1948-bw_JB,Preston2000-ry_JB,Caswell2019-ba_JB}.

In the cohort-component approach, even though the mechanism of population change is known, the individual components still need forecasting. The three main drivers of population dynamics –- fertility, mortality, and migration -– differ in terms of their predictability \citep{National_Research_Council2000-td_JB}: mortality, which is mainly a biological process moderated by medical technology, is the most predictable; migration, which is purely a social and behavioural process is the least; while the predictability of fertility -– part-biological, part-behavioural -- is in the middle (for component forecasting methods, see \S\ref{sec:Forecasting_mortality_data}, \S\ref{sec:Forecasting_fertility}, and \S\ref{sec:Forecasting_migration}). In practical applications, the components can be either projected deterministically, following judgment-based or expert assumptions \citep[for example,][]{Lutz2017-ib_JB}, or extrapolated by using probabilistic methods, either for the components or for past errors of prediction \citep{Alho1985-xi_JB,AS05,De_Beer2008-zv_JB}. An impetus to the use of stochastic methods has been given by the developments in the UN World Population Prospects \citep{Gerland2014-vo_JB,Azose2016-kt_JB}. Parallel, theoretical advancements included a stochastic version of the stable population theory \citep{Keiding1976-nq_JB}, as well as coupling of demographic uncertainty with economic models \citep{Alho2008-iu_JB}.

Since its original formulation, the cohort-component model has been subject to several extensions \citep[see, for example,][]{Stillwell2011-iu_JB}. The multiregional model \citep{Rogers1975-jy_JB} describes the dynamics of multiple regional populations at the same time, with regions linked through migration. The multistate model  \citep{Schoen1987-kf_JB} generalises the multiregional analysis to any arbitrary set of states (such as educational, marital, health, or employment statuses, and so on; see also state-space models in \S\ref{sec:state_space_models}). The multiregional model can be in turn generalised to include multiple geographic levels of analysis in a coherent way \citep{Kupiszewski2011-ww_JB}. Recent developments include multifocal analysis, with an algebraic description of kinship networks \citep{Caswell2019-si_JB,Caswell2020-lh_JB}. For all these extensions, however, data requirements are very high: such models require detailed information on transitions between regions or states in a range of different breakdowns. For pragmatic reasons, microsimulation-based methods offer an appealing alternative, typically including large-sample Monte Carlo simulations of population trajectories based on available transition rates \citep{Zaidi2009-dl_JB,Belanger2017-ws_JB}.

Aside of a few extensions listed above, the current methodological developments in the forecasting of human populations are mainly concentrated on the approaches for predicting individual demographic components (see \S\ref{sec:Forecasting_mortality_data}, \S\ref{sec:Forecasting_fertility}, and \S\ref{sec:Forecasting_migration}), rather than the description of the population renewal mechanism. Still, the continuing developments in population ecology, for example on the algebraic description of transient and asymptotic population growth \citep{Nicol-Harper2018-zh_JB}, bear substantial promise of further advancements in this area, which can be additionally helped by strengthened collaboration between modellers and forecasters working across the disciplinary boundaries on the formal descriptions of the dynamics of human, as well as other populations. 

\subsubsection[Forecasting count time series (Gael M. Martin)]{Forecasting count time series\protect\footnote{This subsection was written by Gael M. Martin.}}
\label{sec:Forecasting_count_time_series}
Probabilistic forecasts based on predictive mass functions are the most natural way of framing predictions of a variable that enumerates the occurrences of an event over time; i.e. the most natural way of predicting a time series of \textit{counts}. Such forecasts are both \textit{coherent}, in the sense of being consistent with the discrete support of the variable, and capture all distributional -- including tail -- information. In contrast, point forecasts based on summary measures of central location (e.g., a (conditional) mean, median or mode), convey no such distributional information, and potentially also lack coherence as, for example, when the mean forecast of the integer-valued count variable assumes non-integer values. These comments are even more pertinent for \textit{low count} time series, in which the number of \textit{rare} events is recorded, and for which the cardinality of the support is small. In this case, point forecasts of any sort can be misleading, and continuous (e.g., Gaussian) approximations (sometimes adopted for high count time series) are particularly inappropriate. 

These points were first elucidated in \cite{FREELAND2004_GM}, and their subsequent acceptance in the literature is evidenced by the numerous count data types for which discrete predictive distributions are now produced; including counts of: insurance claims \citep{MCCABE2005_GM}, medical injury deaths \citep{BU2008_GM}, website visits \citep{BISAGLIA2016_GM}, disease cases \citep{Rao2016_GM,BISAGLIA2019_GM,Siuli2019_GM}, banking crises \citep{Dungey2020_GM}, company liquidations \citep{Homburg2020_GM}, hospital emergency admissions \citep{Sun2021_GM}, work stoppages \citep{Weib2021_GM}, and the intermittent product demand described in \S\ref{sec:Forecasting_for_intermittent_demands_and_count_data} \citep{SNYDER2012_GM,Kolassa2016_PR,BerryWest2020_GM}.

The nature of the predictive model for the count variable, together with the paradigm adopted (Bayesian or frequentist), determine the form of the probabilistic forecast, including the way in which it does, or does not, accommodate parameter and model uncertainty. As highlighted in \S\ref{sec:Foundations_of_Bayesian_forecasting} and \S\ref{sec:Implementantion_of_Bayesian_forecasting}, the Bayesian approach to forecasting is \textit{automatically} probabilistic, no matter what the data type. It also factors parameter uncertainty into the predictive distribution, plus model uncertainty if Bayesian model averaging is adopted, producing a distribution whose location, shape and degree of dispersion reflect all such uncertainty as a consequence. See \cite{MCCABE2005_GM}, \cite{Neal2015_GM}, \cite{BISAGLIA2016_GM}, \cite{frazier2019approximate_GM}, \cite{BerryWest2020_GM} and \cite{Lu2021_GM}, for examples of Bayesian probabilistic forecasts of counts.

In contrast, frequentist probabilistic forecasts of counts typically adopt a `plug-in' approach, with the predictive distribution conditioned on estimates of the unknown parameters of a given count model. Sampling variation in the estimated predictive (if acknowledged) is quantified in a variety of ways. \cite{FREELAND2004_GM}, for instance, produce confidence intervals for the true (point-wise) predictive probabilities, exploiting the asymptotic distribution of the (MLE-based) estimates of those probabilities. \cite{BU2008_GM} extend this idea to (correlated) estimates of sequential probabilities, whilst \cite{JUNG2006_GM} and \cite{Weib2021_GM} exploit bootstrap techniques to capture point-wise sampling variation in the forecast distribution. \cite{McCabe2011_GM}, on the other hand, use subsampling methods to capture sampling fluctuations in the \textit{full} predictive distribution, retaining the non-negativity and summation to unity properties of the probabilities \citep[see also][for related, albeit non-count data work]{Harris2019_GM}. Model uncertainty is catered for in a variety of ways: via nonparametric \citep{McCabe2011_GM} or bootstrapping \citep{BISAGLIA2019_GM} methods; via (frequentist) model averaging \citep{Sun2021_GM}; or via an informal comparison of predictive results across alternative models \citep{JUNG2006_GM}. Methods designed explicitly for calibrating predictive mass functions to observed count data -- whether those functions be produced using frequentist \textit{or} Bayesian methods -- can be found in \cite{Czado2009_GM} and \cite{Held2014_GM}; see also \S\ref{sec:evaluating_probabilistic_forecasts} and \S\ref{sec:assessing_the_reliability_of_probabilistic forecats}.

Finally, whilst full probabilistic forecasts are increasingly common, point, interval and quantile forecasts are certainly still used. The need for such summaries to be coherent with the discrete nature of the count variable appears to be now well-accepted, with recent work emphasising the importance of this property \citep[for example,][]{BU2008_GM,Homburg2019_GM,Siuli2019_GM,Homburg2020_GM}. 

\subsubsection[Forecasting with many variables (J. James Reade)]{Forecasting with many variables\protect\footnote{This subsection was written by J. James Reade.}}
\label{sec:Forecasting_with_many_variables}
Multivariate models -- regression models with multiple explanatory variables -- are often based on available theories regarding the determination of the variable to be forecast, and are often referred to as \textit{structural models}. In a stationary world without structural change, then it would be anticipated that the best structural model would provide the best forecasts, since it would provide the conditional mean of the data process \citep[see, for example,][]{Clements1998_ABM}. In a non-stationary world of unit roots and structural breaks, however, this need not be the case. In such situations, often simple forecast models can outperform structural models, especially at short forecast horizons \citep[see, for example,][]{hendry01_JJR}. Multivariate forecast models require that explanatory variables also be forecast -- or at least, scenarios be set out for them. These may be simplistic scenarios, for example all explanatory variables take their mean values. Such scenarios can play a useful role in formulating policy making since they illustrate in some sense the outcomes of different policy choices. 

Since the 1980s and \cite{sims1980_JJR}, vector autoregressive (VAR) models have become ubiquitous in macroeconomics, and common in finance \citep[see, for example,][]{hasbrouck95_JJR}. A VAR model is a set of linear regression equations (see also \S\ref{sec:time_series_regression_models}) describing the evolution of a set of endogenous variables. Each equation casts each variable as a function of lagged values of all the variables in the system. Contemporaneous values of system variables are not included in VAR models for identification purposes; some set of identifying restrictions are required, usually based on economic theory, and when imposed the resulting model is known as a structural VAR model. VAR models introduce significantly greater levels of parameterisation of relationships, which increases the level of estimation uncertainty. At the same time VAR models afford the forecaster a straightforward way to generate forecasts of a range of variables, a problem when forecasting with many variables. As with autoregressive methods, VAR models can capture a significant amount of variation in data series that are autocorrelated, and hence VAR methods can be useful as baseline forecasting devices. VAR-based forecasts are often used as a benchmark for complex models in macroeconomics like DSGE models \citep[see, for example,][]{del2006good_JJR}. The curse of dimensionality in VAR models is particularly important and has led to developments in factor-augmented VAR models, with practitioners often reducing down hundreds of variables into factors using principal component analysis \citep[see, for example,][]{bernanke2005measuring_JRR}. Bayesian estimation is often combined with factor-augmented VAR models.

Often, significant numbers of outliers and structural breaks require many indicator variables to be used to model series (see also \S\ref{sec:Anomaly_detection_and_time_series_forecasting} and \S\ref{sec:Robust_handling_of_outliers}). Indicator saturation is a method of detecting outliers and structural breaks by saturating a model with different types of indicator, or deterministic variables \citep{johansen2009analysis_JJR,castle2015detecting_JJR}. The flexibility of the approach is such that it has been applied in a wide variety of contexts, from volcanic eruptions \citep{pretis2016detecting_JJR} to prediction markets and social media trends \citep{vaughanwilliams2016prediction_JJR}. 

A particularly important and ever-expanding area of empirical analysis involves the use of panel data sets with long time dimensions: panel time series \citep{Eberhardt2012-hc}. The many variables are then extended across many cross sectional units, and a central concern is the dependence between these units, be they countries, firms, or individuals. At the country level one approach to modelling this dependence has been the Global VAR approach of, for example, \cite{Dees2007-nv}. In more general panels, the mean groups estimator has been proposed to account for cross-section dependence \citep{Pesaran1999-wo}.

Outliers, structural breaks, and split trends undoubtedly also exist in panel time series. The potential to test for common outliers and structural changes across cross sectional units would be useful, as would the ability to allow individual units to vary individually, e.g., time-varying fixed effects. 
\cite{nymoen2015equilibrium_JJR} is the first application of indicator saturation methods in a panel context, looking at equilibrium unemployment dynamics in a panel of OECD countries, but applications into the panel context are somewhat constrained by computer software packages designed for indicator saturation (\S\ref{sec:Forecasting_UK_unemployment} discusses further the case of forecasting unemployment). The \textit{gets} R package of \cite{gets_JJR,pretis2018general_JJR} can be used with panel data.
 
\subsubsection[Functional time series models (Han Lin Shang)]{Functional time series models\protect\footnote{This subsection was written by Han Lin Shang.}}
\label{sec:Functional_time_series_models}
Functional time series consist of random functions observed at regular time intervals. Functional time series can be classified into two categories depending on if the continuum is also a time variable. On the one hand, functional time series can arise from measurements obtained by separating an almost continuous time record into consecutive intervals \citep[e.g., days or years, see][]{HK12}. We refer to such data structure as sliced functional time series, examples of which include daily precipitation data \citep{Gromenko2017-ln_HLS}. On the other hand, when the continuum is not a time variable, functional time series can also arise when observations over a period are considered as finite-dimensional realisations of an underlying continuous function \citep[e.g., yearly age-specific mortality rates, see][]{LRS20}.

Thanks to recent advances in computing storage, functional time series in the form of curves, images or shapes is common. As a result, functional time series analysis has received increasing attention. For instance, \cite{Bosq00} and \cite{BB07} proposed the functional autoregressive of order 1 (FAR(1)) and derived one-step-ahead forecasts that are based on a regularised Yule-Walker equations. FAR(1) was later extended to FAR($p$), under which the order $p$ can be determined via \citeauthor{KR13}'s \citeyearpar{KR13} hypothesis testing. \cite{Horvath2020-ey_HLS} compared the forecasting performance between FAR(1), FAR($p$), and functional seasonal autoregressive models of \cite{Chen2019-zs_HLS}.

To overcome the curse of dimensionality (see also \S\ref{sec:Exogenous_variables_and_feature_engineering}, \S\ref{sec:Model_complexity} and \S\ref{sec:Variable_Selection}), a dimension reduction technique, such as functional principal component analysis (FPCA), is often used. \cite{ANH15} showed asymptotic equivalence between a FAR and a VAR model (for a discussion of VAR models, see \S\ref{sec:Forecasting_with_many_variables}). Via an FPCA, \cite{ANH15} proposed a forecasting method based on the VAR forecasts of principal component scores. This approach can be viewed as an extension of \cite{HS09}, in which principal component scores are forecast via a univariate time series forecasting method. With the purpose of forecasting, \cite{Kargin2008-jz_HLS} proposed to estimate the FAR(1) model by using the method of predictive factors. \cite{KK17} proposed a functional moving average process and introduced an innovations algorithm to obtain the best linear predictor. \cite{KKW16} extended the VAR model to the vector autoregressive moving average model and proposed the functional autoregressive moving average model. The functional autoregressive moving average model can be seen as an extension of autoregressive integrated moving average model in the univariate time series literature (see \S\ref{sec:autoregressive_integrated_moving_average_models}).

Extending short-memory to long-memory functional time series analysis, \cite{LRS20b,LRS20} considered local Whittle and rescale-range estimators in a functional autoregressive fractionally integrated moving average model. The models mentioned above require stationarity, which is often rejected. \cite{HKR14} proposed a functional KPSS test for stationarity. \cite{CKP16} studied nonstationarity of the time series of state densities, while \cite{BSS17} considered a cointegrated linear process in Hilbert space. \cite{NSS19} proposed a variance ratio-type test to determine the dimension of the nonstationary subspace in a functional time series. \cite{LRS20c} studied the estimation of the long-memory parameter in a functional fractionally integrated time series, covering the functional unit root.

From a nonparametric perspective, \cite{BCS00} proposed a functional kernel regression method to model temporal dependence via a similarity measure characterised by semi-metric, bandwidth and kernel function. \cite{AV08} introduced a semi-functional partial linear model that combines linear and nonlinear covariates. Apart from conditional mean estimation, \cite{HHR13} considered a functional autoregressive conditional heteroscedasticity model for estimating conditional variance. 
\cite{Rice2020-zl_HLS} proposed a conditional heteroscedasticity test for functional data.
\cite{KRS17} proposed a portmanteau test for testing autocorrelation under a functional generalised autoregressive conditional heteroscedasticity model.

\subsubsection[ARCH/GARCH models (Jooyoung Jeon)]{ARCH/GARCH models\protect\footnote{This subsection was written by Jooyoung Jeon.}}
\label{sec:arch_garch_models}
Volatility has been recognised as a primary measure of risks and uncertainties \citep{Markowitz1952a_JJ,Sharpe1964a_JJ,Taylor2009_JJ,Gneiting2011a_SK}; for further discussion on uncertainty estimation, see \S\ref{sec:Estimation_and_representation_of_uncertainty}. Estimating future volatility for measuring the uncertainty of forecasts is imperative for probabilistic forecasting. Yet, the right period in which to estimate future volatility has been controversial as volatility based on too long a period will make irrelevant the forecast horizon of our interests, whereas too short a period results in too much noise \citep{Engle2004_JJ}. An alternative to this issue is the dynamic volatility estimated through the autoregressive conditional heteroscedasticity (ARCH) proposed by \cite{Engle1982_JJ}, and the generalised autoregressive conditional heteroscedasticity (GARCH) model proposed by \cite{Bollerslev1987_JJ}. The ARCH model uses the weighted average of the past squared forecast error whereas the GARCH model generalises the ARCH model by further adopting past squared conditional volatilities. The GARCH model is the combination of (\textit{i}) a constant volatility, which estimates the long-run average, (\textit{ii}) the volatility forecast(s) in the last steps, and (\textit{iii}) the new information collected in the last steps. The weightings on these components are typically estimated with maximum likelihood. The models assume a residual distribution allowing for producing density forecasts. One of the benefits of the GARCH model is that it can model heteroscedasticity, the volatility clustering characteristics of time series \citep{Mandelbrot1963_JJ}, a phenomenon common to many time series where uncertainties are predominant. Volatility clustering comes about as new information tends to arrive time clustered and a certain time interval is required for the time series to be stabilised as the new information is initially recognised as a shock.

The GARCH model has been extended in the diverse aspects of non-linearity, asymmetry and long memory. Among many such extensions, the Exponential GARCH (EGARCH) model by \cite{Nelson1991_JJ} uses log transformation to prevent negative variance; the Threshold GARCH (TGARCH) model by \cite{Zakoian1994_JJ} allows for different responses on positive and negative shocks. A small piece of information can have more impact when the time series is under stress than under a stable time series \citep{Engle2004_JJ}. Another pattern often observed in the volatility time series is slowly decaying autocorrelation, also known as a long memory pattern, which \cite{Baillie1996_JJ} capture using a slow hyperbolic rate of decay for the ARCH terms in the fractionally integrated GARCH (FIGARCH) model. Separately, in a further approach to directly estimating long term volatility, the GARCH-MIDAS (Mixed Data Sampling) model proposed by \cite{Engle2013_JJ} decomposes the conditional volatility into the short-term volatility, as captured by the traditional GARCH, and the long-term volatility represented by the realised volatilities. The Heterogeneous Autoregressive (HAR) model by \cite{Corsi2014} considers the log-realised volatility as a linear function of the log-realised volatility of yesterday, last week and last month to reflect traders' preferences on different horizons in the past. This model is extended by \cite{WILMS2021484} to incorporate information about future stock market volatility by further including option-implied volatility. A different approach to volatility modelling, discussed in \S\ref{sec:Low_and_high_prices_in_volatility_models}, is the use of low and high prices in the range-based volatility models.

The univariate GARCH models surveyed so far have been exended to multivariate versions, in order to model changes in the conditional covariance in multiple time series, resulting in such examples as the VEC \citep{Bollerslev1987_JJ} and BEKK \citep{Engle1995_JJ}, an acronym derived from Baba, Engle, Kraft, and Kroner. The VEC model, a direct generalisation of the univariate GARCH, requires more parameters in the covariane matrices and provides better fitness at the expense of higher estimation costs than the BEKK. The VEC model has to ensure the positivity of the covariance matrix with further constraints, whereas the BEKK model and its specific forms, e.g., factor models, avoid this positivity issue directly at the model specification stage. In an effort to further reduce the number of parameters to be estimated, the linear and non-linear combinations of the univariate GARCH models, such as the constant conditional correlation model of \cite{Bollerslev1990_JJ} and the dynamic conditional correlation models of \cite{Tse2002_JJ} and of \cite{Engle2002_JJ}, were investigated.

\subsubsection[Markov switching models (Massimo Guidolin)]{Markov switching models\protect\footnote{This subsection was written by Massimo Guidolin.}}
\label{sec:Markov_switching_models}
Since the late 1980s, especially in macroeconomics and finance, the applications of dynamic econometric modelling and forecasting techniques have increasingly relied on a special class of models that accommodate regime shifts, Markov switching (MS) models. The idea of MS is to relate the parameters of otherwise standard dynamic econometric frameworks (such as systems of regressions, vector autoregressions, and vector error corrections) to one or more unobserved state variables (see \S\ref{sec:state_space_models} for a definition), say, $S_t$, that can take $K$ values and capture the notion of systems going through phases or ``regimes'', which follow a simple, discrete stochastic process and are independent of the shocks of the model. 

For instance, an otherwise standard AR(1) model can be extended to $y_t= \phi_{0,S_t}+\phi_{1,S_t} y_(t-1)+\sigma_{S_t} \epsilon_t$, where all the parameters in the conditional mean as well as the variance of the shocks may be assumed to take different, estimable values as a function of $S_t$. Similarly, in a $K$-regime MS VAR($p$), the vector of intercepts and the $p$ autoregressive matrices may be assumed to depend on $S_t$. Moreover, the covariance matrix of the system shocks may be assumed to depend on some state variable, either the same as the mean parameters ($S_t$) or an additional, specific one ($V_t$), which may depend on lags of $S_t$. When a MS VAR model is extended to include exogenous regressors, we face a MS VARX, of which MS regressions are just a special case.

Even though multivariate MS models may suffer from issues of over-parameterisations that must be kept in check, their power of fitting complex non-linearities is unquestioned because, as discussed by \cite{Marron1992-ga_MG}, mixtures of normal distributions provide a flexible family that can be used to approximate many distributions. Moreover, MS models are known \citep{Timmermann2000-hf_MG} to capture key features of many time series. For instance, differences in conditional means across regimes enter the higher moments such as variance, skewness, and kurtosis; differences in means in addition to differences in variances across regimes may generate persistence in levels and squared values of the series.

The mainstream literature (see, e.g., \citeauthor{Hamilton1990-fs_MG} \citeyear{Hamilton1990-fs_MG}, or the textbook treatments by \citeauthor{Kim1999-ef_MG} \citeyear{Kim1999-ef_MG}, and \citeauthor{Guidolin2018-cj_MG} \citeyear{Guidolin2018-cj_MG}) initially focused on time-homogeneous Markov chains (where the probabilities of the state transitions are constant). However, the finance and business cycles literatures \citep{Gray1996-vg_MG} has moved towards time-heterogeneous MS models, in which the transition matrix of the regimes may change over time, reacting to lagged values of the endogenous variables, to lagged exogenous variables, or to the lagged values of the state (in a self-exciting fashion).

MS models may be estimated by maximum likelihood, although other estimation methods cannot be ruled out, like GMM \citep{Lux2008-nx_MG}. Typically, estimation and inference are based on the Expectation-Maximisation algorithm proposed by \cite{Dempster1977-eh_MG}, a filter that allows the iterative calculation of the one-step ahead forecast of the state vector given the information set and a simplified construction of the log-likelihood of the data. However, there is significant evidence of considerable advantages offered by Bayesian approaches based on Monte Carlo Markov chain techniques to estimating multivariate MS models \citep[see, for example,][]{Hahn2010-bd_MG}.

Notably, MS models have been recently generalised in a number of directions, such as including regimes in conditional variance functions, for example of a GARCH or DCC type \citep[see][and \S\ref{sec:arch_garch_models}]{Pelletier2006-qb_MG}.
 
\subsubsection[Threshold models (Manuela Pedio)]{Threshold models\protect\footnote{This subsection was written by Manuela Pedio.}}
\label{sec:threshold_models}
It is a well-known fact that financial and economic time series often display non-linear patterns, such as structural instability,  which may appear in the form of recurrent regimes in model parameters. In the latter case, such instability is stochastic, it displays structure, and as such, it can be predicted. Accordingly, modelling economic and financial instability has become an essential goal for econometricians since the 1970s.

One of the first and most popular models is the threshold autoregressive (TAR) model developed by \cite{Tong1978-md_MP}. A TAR model is an autoregressive model for the time series $y_t$ in which the parameters are driven by a state variable $S_t$ (see \S\ref{sec:state_space_models} for a definition), which is itself a random variable taking $K$ distinct integer values (i.e., $S_t=k$, $k=1,\dots,K$). In turn, the value assumed by $S_t$ depends on the value of the threshold variable $q_t$ when compared to $K-1$ threshold levels, $q_k^*$. For instance, if only two regimes exists, it is $S_t=1$ if $q_t \leq q_1^*$ and $S_t=2$ otherwise. The threshold variable $q_t$ can be exogenous or can be a lagged value of $y_t$. In the latter case, we speak of self-exciting threshold autoregressive (SETAR) models. Other choices of $q_t$ include linear \citep{Chen2003-sq_MP,Chen2006-rz_MP,Gerlach2006-ff_MP} or non-linear \citep{Chen1995-yr_MP,Wu2007-qz_MP} combinations of the lagged dependent variable or of exogenous variables. 

The TAR model has also been extended to account for different specifications of the conditional mean function leading to the development of the threshold moving average \citep[TMA -- see, for example,][]{Tong1990-jv_MP,Gooijer1998-eh_MP,Ling2007-ou_MP} and the threshold autoregressive moving average \citep[TARMA -- see, for example,][]{Ling1999-ra_MP,Amendola2006-eh_MP} models. Those models are similar to the ones described in \S\ref{sec:autoregressive_integrated_moving_average_models}, but their parameters depend on the regime $K$.

A criticism of TAR models is that they imply a conditional moment function that fails to be continuous. To address this issue, \cite{Chan1986-xi_MP} proposed the smooth transition autoregressive (STAR) model. The main difference between TAR and STAR models is that, while a TAR imposes an abrupt shift from one regime to the others at any time that the threshold variable crosses above (or below) a certain level, a STAR model allows for gradual changes among regimes.

In its simplest form, a STAR is a two-regime model where the dependent variable $y_t$ is determined as the weighted average of two autoregressive (AR) models, i.e., 
$$
y_t = \sum+{j=1}^p{\phi_{j,1} y_{t-j} P(S_t=1; g(x_t))} + \sum_{j=1}^p{\phi_{j,2} y_{t-j} P(S_t=2; g(x_t))} + \epsilon_t,
$$

\noindent where $x_t$ is the transition variable and $g$ is some transformation of the transition variable $x_t$. Regime probabilities are assigned through the transition function $F(k; g(x_t))$, with $F$ being a cumulative density function of choice. The transition variable $x_t$ can be the lagged endogenous variable, $y_{t-d}$ for $d \geq 1$ \citep{Terasvirta1994-uw_MP}, a (possibly non-linear) function of it, or an exogenous variable. The transition variable can also be a linear time trend ($x_t=t$), which generates a model with smoothly changing parameters \citep{Lin1994-lv_MP}. Popular choices for the transition function $F$ are the logistic function (which gives rise to the LSTAR model) and the exponential function (ESTAR). Notably, the simple STAR model we have described can be generalised to have multiple regimes \citep{Van_Dijk1999-oo_MP}.

Threshold models are also applied to modelling and forecasting volatility; for instance, the GJR-GARCH model of \cite{Glosten1993-ef_MP} can be interpreted as a special case of a threshold model. A few multivariate extensions of threshold models also exist, such as vector autoregressive threshold models, threshold error correction models \citep{Balke1997-py_MP}, and smooth transition vector error correction models \citep{Granger1996-ae_MP}. 

\subsubsection[Low and high prices in volatility models (Piotr Fiszeder)]{Low and high prices in volatility models\protect\footnote{This subsection was written by Piotr Fiszeder.}}
\label{sec:Low_and_high_prices_in_volatility_models}
Volatility models of financial instruments are largely based solely on closing prices (see \S\ref{sec:arch_garch_models}); meanwhile, daily low and high (LH) prices significantly increase the amount of information about the variation of returns during a day. LH prices were used for the construction of highly efficient estimators of variance, so called the range-based (RB) estimators \citep[e.g.,][]{Parkinson1980_PF,Garman1980_PF,Rogers1991_PF,Yang2000_PF,Magdon2003,Fiszeder2013_PF}. Recently, \cite{Riedel2021} analysed how much additional information about LH reduces the time averaged variance in comparison to knowing only open and close. RB variance estimators, however, have a fundamental drawback, as they neglect the temporal dependence of returns (like conditional heteroscedasticity) and do not allow for the calculation of multi-period dynamic volatility forecasts.

In the last dozen or so years, numerous univariate dynamic volatility models have been constructed based on LH prices. Some of them were presented in the review paper of \cite{Chou2015_PF}. These models can be divided into four groups. The first one comprises simple models, used traditionally to describe returns, but they are based on the price range or on the mentioned earlier RB variance estimators. They include such models as random walk, moving average, exponentially weighted moving average (EWMA), autoregressive (AR), autoregressive moving average (ARMA; see \S\ref{sec:autoregressive_integrated_moving_average_models}), and heterogeneous autoregressive (HAR). The second group contains models which describe the conditional variance (or standard deviation) of returns. It comprises models like GARCH-PARK-R \citep{Mapa2003_PF}, GARCH-TR \citep{Fiszeder2005_PF}, REGARCH \citep{Brandt2006_PF}, RGARCH \citep{Molnar2016_PF}. The third group includes models which describe the conditional mean of the price range. It means that in order to forecast variance of returns the results have to be scaled. This group contains models like RB SV \citep{Alizadeh2002_PF}, CARR \citep{Chou2005_PF}, TARR \citep{Chen2008_PF}, CARGPR \citep{Chan2012_PF}, STARR \citep{Lin2012_PF}, and MSRB \citep{Miao2013_PF}. The last group is methodologically different because the estimation of model parameters is based on the sets of three prices, i.e., low, high and closing. This approach comprises the GARCH models \citep{Lildholdt2002_PF,Venter2005_PF,Fiszeder2016_PF} and the SV model \citep{Horst2012_PF}. 

The development of multivariate models with LH prices has taken place in the last few years. They can be divided into three groups. The first one includes models, used traditionally to describe returns or prices, but they are based on the price range or RB variance estimators. They comprise such models like multivariate EWMA, VAR, HVAR, and vector error correction (VEC). It is a simple approach, however most models omit modelling the covariance of returns. The second group is formed by the multivariate RB volatility models like RB-DCC \citep{Chou2009a_PF}, DSTCC-CARR \citep{Chou2009_PF}, RR-HGADCC \citep{Asai2013_PF}, RB-MS-DCC \citep{Su2014_PF}, DCC-RGARCH \citep{Fiszeder2019a_PF}, RB-copula \citep{Chiang2011_PF,Wu2011_PF}. The third group includes the multivariate co-range volatility models like multivariate CARR \citep{Fernandes2005_PF}, BEKK-HL \citep{Fiszeder2018_PF} and co-range DCC \citep{Fiszeder2019_PF}. These models apply LH prices directly not only for the construction of variances of returns but also for covariances. \S\ref{sec:Financial_time_series_forecasting_with_range_based_volatility_models} discusses the use of the range-based volatility models in financial time series forecasting.

\subsubsection[Forecasting with DSGE models (Alessia Paccagnini)]{Forecasting with DSGE models\protect\footnote{This subsection was written by Alessia Paccagnini.}}
\label{sec:Forecasting_with_DSGE_Models}
Dynamic Stochastic General Equilibrium (DSGE) models are the workhorse of modern macroeconomics employed by monetary and fiscal authorities to explain and forecast comovements of aggregate time series over the business cycle and to perform quantitative policy analysis. These models are studied in both academia and policy-making institutions 
\citep[for details, see:][]{MPC_Deln13,Paccagnini17_Alessia,CET18_Alessia}. 
For example, the European Central Bank uses the New Area-Wide Model introduced by \cite{WCC10_Alessia} and the Federal Reserve Board has created the Estimated, Dynamic, Optimisation-based model (FRB/EDO) as discussed in \cite{EDO10_Alessia}. For an application on forecasting GDP and inflation, see \S\ref{sec:Forecasting_GDP_and_Inflation}.
Developed as a response to \cite{Lucas76_Alessia} critique of structural macroeconometrics models, DSGEs introduced microfundations to describe business cycle fluctuations.
Initially calibrated, estimated DSGEs have been employed in shocks identification and forecasting horseraces for the last 15 years. Estimation became possible thanks to computational progress and adoption of Bayesian techniques \citep[for technical details, see:][]{AS07_Alessia,HS16_Alessia,FG20_Alessia}. Bayesian estimation allows for attributing prior distributions, instead of calibrating, and computing the posterior distribution for selected model parameters as well as drawing from predictive density.
The \cite{SW07_Alessia} DSGE is the most popular framework referred to in both research and policy literature. Proposed for the US economy, this medium-scale model is a closed economy composed of households, labor unions, a productive sector, and a monetary policy authority that sets the short-term interest rate according to a Taylor rule. These ingredients are mathematically represented by a system of linear rational expectation equations. Using a solution algorithm  \citep[for example,][]{Blanchard1980-bw,Sims02_Alessia}, researchers can write the model using the state-space representation composed by the transition equation and the measurement equation. The latter matches the observed data (in the Smets and Wouters: output growth rate, consumption, investment, wages, worked hours, inflation, and short-term interest rate)
with the model latent variables. The solved model is employed for quantitative policy analysis and to predict and explain the behavior of macroeconomic and financial indicators.

DSGE models forecasting performance is investigated along two dimensions: point forecast and density forecast (see \S\ref{sec:point_forecast_accuracy_measures} and \S\ref{sec:evaluating_probabilistic_forecasts} for discussions on their evaluation).

The point forecast is implemented by conducting both static and dynamic analysis, as described in \cite{CPV19_Alessia}.
If the static analysis provides a unique forecast value, the dynamic analysis describes the evolution of the prediction along the time dimension to investigate possible time-varying effects.
Usually, point predictions are compared using the \cite{DM95_Alessia} and the \cite{CW06_Alessia} tests that compare predictions from two competing models
The accuracy of the static analysis is based mainly on Mean Absolute Error (MAE) and Root Mean Square Error (RMSE). 
MAE and RMSE are used to provide a relative forecasting evaluation compared to other competitors.
Following \cite{Clements1998_ABM}, \cite{KRS12_Alessia} apply the standard forecast unbiased test to assess if DSGEs are good forecasters in the absolute sense.
The accuracy of the dynamic analysis is based on the Fluctuation Test \citep[for some  DSGE applications, see:][]{GR16_Alessia,CPV19_Alessia,BFMW19_Alessia}. This test is based on the calculation of RMSEs that are assessed to investigate if the forecasting performance can be influenced by instabilities in the model parameters.

The density forecast is based on the uncertainty derived by the Bayesian estimation and it is commonly evaluated using the probability integral transform and the log predictive density scores \citep[as main references,][]{Wolters15_Alessia,KR15_Alessia}. The statistical significance of these predictions is evaluated using the \cite{AG07_Alessia} test that compares log predictive density scores from two competing models.

\subsubsection[Robust equilibrium-correction forecasting devices (Andrew B. Martinez)]{Robust equilibrium-correction forecasting devices\protect\footnote{This subsection was written by Andrew B. Martinez.}}
\label{sec:Robust_equilibrium_correction_forecasting_devices}
The use of equilibrium-correction models is ubiquitous in forecasting. \citet{hendry2010_ABM} notes that this class commonly includes models with explicit equilibrium-correction mechanisms such as vector equilibrium-correction models (VEqCM) as well as models with implicit equilibrium-correction (or long-run mean reversion) mechanisms such as vector auto-regressions (VARs; see \S\ref{sec:Forecasting_with_many_variables}), dynamic factor models (DFMs), dynamic stochastic general-equilibrium (DSGE) models (see \S\ref{sec:Forecasting_with_DSGE_Models}), most models of the variance (see \S\ref{sec:arch_garch_models}), and almost all regression equations (see \S\ref{sec:time_series_regression_models} and \S\ref{sec:autoregressive_integrated_moving_average_models}). This class of forecast model is prevalent across most disciplines. For example, \citet{Pretis2020_ABM} illustrates that there is an equivalence between physical energy balance models, which are used to explain and predict the evolution of climate systems, and VEqCMs.

Despite their wide-spread use in economic modeling and forecasting, equilibrium-correction models often produce forecasts that exhibit large and systematic forecast errors. \citet{Clements1998_ABM,Clements1999_ABM} showed that forecasts from equilibrium-correction models are not robust to abrupt changes in the equilibrium. These types of regime changes are very common in macroeconomic time series \citep[see][ as well as \S\ref{sec:Markov_switching_models}]{Hamilton2016_ABM} and can cause the forecasts from many models to go off track. Therefore, if for example, there is a change in the equilibrium towards the end of the estimation sample, forecasts from this class of models will continue to converge back to the previous equilibrium. 

In general, the forecasts from equilibrium-correction models can be robustified by estimating all model parameters over smaller or more flexible sub-samples. Several studies have proposed general procedures that allow for time-variation in the parameters; see, for example, \citet{pesaran2013_ABM}, \citet{Giraitis2013_ABM} and \citet{inoue2017_ABM}. This allows for an optimal or more adaptive selection of model estimation windows in order generate forecasts after structural breaks have occurred.

An alternative approach for robustifying forecasts from equilibrium-correction models is to focus on the formulation and estimation of the equilibrium. \citet{hendry2006_ABM} shows that differencing the equilibrium-correction mechanism can improve the forecasts by removing aspects which are susceptible to shifts. However, differencing the equilibrium also induces additional forecast-error variance. \citet{Castle2010_ABM} show that it is beneficial to update the equilibrium or to incorporate the underlying structural break process. Alternatively, \citet{Castle2015_ABM} show that there can be large forecast gains from smoothing over estimates of the transformed equilibrium. Building on this, \citet{Martinez2019_ABM} show that there are many possible transformations of the equilibrium that can improve the forecasts. Several of these transformations imply that the equilibrium-correction model collapses to different naive forecast devices whose forecasts are often difficult to beat. By replacing the equilibrium with smooth estimates of these transformations, it is possible to outperform the naive forecasts at both short and long forecast horizons while retaining the underlying economic theory embedded within the equilibrium-correction model. Thus, it is possible to dramatically improve forecasts from equilibrium-correction models using targeted transformations of the estimated equilibrium so that it is less susceptible to the shifts which are so damaging to the model forecasts.
 
\subsubsection[Forecasting with data subject to revision (Michael P. Clements)]{Forecasting with data subject to revision\protect\footnote{This subsection was written by Michael P. Clements.}}
\label{sec:Forecasting_with_Data_Subject_to_Revision}
When a forecast is made today of the future value of a variable, the forecast is necessarily `real time' -- only information available at the time the forecast is made can be used. The forecasting ability of a model can be evaluated by mimicking this setup -- generating forecasts over some past period (so outcomes known) only using data known at each forecast origin. As noted by \cite{MPC_MPCDFHcho}, out-of-sample forecast performance is the gold standard. Sometimes the analysis is pseudo real time. At a given forecast origin $t$, forecasts are constructed only using data up to the period $t$, but the data are taken from the latest-available vintage at the time the study is undertaken. Using revised data to estimate the forecasting model -- instead of the data available at the time the forecast was made -- may exaggerate forecast performance, and present a misleading picture of how well the model might perform in real time. The improved availability of real-time databases has facilitated proper real-time studies.\footnote{For example, the Federal Reserve Bank of Philadelphia maintain a real-time data set covering a number of US macro-variables, at:
https://www.philadelphiafed.org/research-and-data/real-time-center/real-time-data/, and see \cite{MPC_Crou01}.} At time $t$ the data are taken from the vintage available at time $t$. Data revisions are often important, and occur because statistical
agencies strive to provide timely estimates which are based on incomplete source data \citep[see, for example,][]{MPC_Fixl05,MPC_Fixl08,MPC_Zwij15}.

There are a number of possible real-time approaches. The conventional approach is to estimate the forecasting model using the latest vintage of data available at time $t$. Suppose the vintage-$t$ contains data for time periods up to $t-1$, denoted $\ldots ,y_{t-3}^{t},y_{t-2}^{t},y_{t-1}^{t}$.
The observation for time $t-1$ is a first estimate, for $t-2$ a second estimate, and so on, such that data for earlier periods will have been revised many times. Hence the model will be estimated on data of different maturities, much of which will have been revised many times. But the forecast will typically be generated by feeding into the model `lightly-revised' data for the most recent time periods. The accuracy of the resulting forecasts can be improved upon (in principle) by taking into
account data revisions \cite[see, for example,][]{MPC_Koen03,MPC_Kish05,MPC_CGAR}. In the following two paragraphs, we consider alternative real-time approaches which solve the problem of estimating the model on mainly revised data, and feeding in mainly unrevised forecast origin data.

\cite{MPC_Koen03} suggest using real-time-vintage (RTV) data to estimate the model. The idea is to use early estimates of the data to estimate the model, so that the model is estimated on `lightly-revised' data that matches the maturity of the forecast-origin data that the forecast is conditioned on.

Other approaches seek to model the data revisions process along with the fully-revised true values of the data, as in \cite{MPC_Kish05}, \cite{MPC_Cunn07}, and \cite{MPC_Jaco06}. Reduced form models that avoid the necessity of estimating unobserved components have adapted the vector autoregression
(VAR; see also \S\ref{sec:Forecasting_with_many_variables}) of \cite{sims1980_JJR} to jointly model different observed vintages of data. Following \cite{MPC_Patt95}, \cite{MPC_Garr08} work in terms of the level of the log
of the variable, $Y_{t}^{t+1}$, and model the vector given by $\mathbf{Z}^{t+1}=\left(Y_{t}^{t+1}-Y_{t-1}^{t},Y_{t-1}^{t+1}-Y_{t-1}^{t},Y_{t-2}^{t+1}-Y_{t-2}^{t}\right) ^{\prime }$. \cite{MPC_CGVARog,MPC_CGVARf} and \cite{MPC_CCGbay} minimise the effects of benchmark revisions and re-basing by modelling
`same-vintage-growth rates', namely $\mathbf{Z}^{t+1}=\left(
y_{t}^{t+1},y_{t-1}^{t+1},\ldots ,y_{t-q+1}^{t+1}\right) ^{\prime }$, where $y_{t}^{t+1}=Y_{t}^{t+1}-Y_{t-1}^{t+1}$, and $q$ denotes the greatest data maturity. 

\cite{MPC_GalvDSGE} shows how forecasts of fully-revised data can be generated for dynamic stochastic general equilibrium (DSGE; \S\ref{sec:Forecasting_with_DSGE_Models}) models \citep[for example,][]{MPC_Deln13}, by applying the approach of \cite{MPC_Kish05}. \cite{MPC_ClemPIRT} argues that improvements in forecast accuracy might be expected to be greater for interval or density forecasts than point
forecasts, and this is further explored by \cite{MPC_CGDRRT}.

Surveys on data revisions and real-time analysis, including forecasting, are provided by \cite{MPC_CrouHB,MPC_Crou11,MPC_CrouOUPHB} and \cite{MPC_CGencyc}; see also \S\ref{sec:Macroeconomic_survey_expectations}.

\subsubsection[Innovation diffusion models (Mariangela Guidolin)]{Innovation diffusion models\protect\footnote{This subsection was written by Mariangela Guidolin.}}
\label{sec:Innovation_diffusion_models}
Forecasting the diffusion of innovations is a broad field of research, and influential reviews on the topic have highlighted its importance in many disciplines for strategic or anticipative reasons \citep{mahajan:90_MG,meade:06_MG,peres:10_MG}. Large-scale and fast diffusion processes of different nature, ranging from the spread of new epidemics to the adoption of new technologies and products, from the fast diffusion of news to the wide acceptance of new trends and social norms, are demanding a strong effort in terms of forecasting and control, in order to manage their impact into socio-economic, technological and ecological systems.

The formal representation of diffusion processes is often based on epidemic models, under the hypothesis that an innovation spreads in a social system through communication among people just like an epidemics does through contagion. The simplest example is represented by the (cumulative) logistic equation that describes a pure epidemic process in a homogeneously mixing population \citep{Verhulst1838}. The most famous and employed evolution of the logistic equation is the Bass model, \cite{bass:69_RGuseo}, developed in the field of quantitative marketing and soon become a major reference, due to its simple and powerful structure. 

The Bass model (BM) describes the life-cycle of an innovation, depicting its characterising phases of launch, growth/maturity, and decline, as result of the purchase decisions of a given cohort of potential adopters. 
Mathematically, the model is represented by a first order differential equation, describing a diffusion process by means of three parameters: the maximum market potential, $m$, assumed to be constant along the whole diffusion process, and parameters $p$ and $q$, referring respectively to two distinct categories of consumers, the \textit{innovators}, identified with parameter $p$, adopting for first, and the \textit{imitators}, adopting at a later stage by imitating others' behaviour and thus responsible for \textit{word-of-mouth} dynamics. In strategic terms, crucial forecasts are referred to the point of maximum growth of the life cycle, the \textit{peak}, and the point of market saturation. For a general description of new product forecasting please refer to \S\ref{sec:New_product_forecasting}.

Innovation diffusion models may be also used for \textit{post-hoc} explanations, helping understand the evolution of a specific market and its response to different external factors. Indeed, one of the most appealing characteristics of this class of models is the possibility to give a simple and nice interpretation to all the parameters involved. 
In this perspective, a valuable generalisation of the BM was proposed in \cite{bass:94_RGuseo} with the Generalised Bass Model (GBM). 
The GBM enlarges the BM by multiplying its hazard rate by a very general intervention function $x(t)$, assumed to be non-negative, which may account for exogenous shocks able to change the temporal dynamics of the diffusion process, like marketing strategies, incentive mechanisms, change in prices and policy measures.

Another generalisation of the BM and the GBM, relaxing the assumption of a constant market potential was proposed in \cite{guseoguidolin:09_RGuseo} with the GGM. This model postulates a time-dependent market potential, $m(t)$, which is function of the spread of knowledge about the innovation, and thus assumes that a diffusion process is characterised by two separate phases, information and adoption.
The GGM allows a significant improvement in forecasting over the simpler BM, especially through a more efficient description of the first part of the time series, often characterised by a slowdown pattern, as noticed by \cite{guseo:11_RGuseo}.

Other generalisations of innovation diffusion models, considering competition between products, are treated in \S\ref{sec:Synchronic_and_diachronic_competition}. Applications of innovation diffusion models are presented in \S\ref{sec:New_product_forecasting} and \S\ref{sec:Forecasting_renewable_energy_technologies}.
 
\subsubsection[The natural law of growth in competition (Theodore Modis)]{The natural law of growth in competition\protect\footnote{This subsection was written by Theodore Modis.}}
\label{sec:The_natural_law of growth_in_competition_Logistic_growth}
As early as in 1925 Alfred J. Lotka demonstrated that manmade products diffuse in society along S-shaped patterns similar to those of the populations of biological organisms \citep{Lotka1925-xg}. Since then S curve logistic descriptions have made their appearance in a wide range of applications from biology, epidemiology and ecology to industry, competitive substitutions, art, personal achievement and others \citep{Fisher1971-tb,Marchetti1983-rt,Meade1984-sl,Modis1992-ax}. The reader is also referred to \S\ref{sec:Innovation_diffusion_models} and \S\ref{sec:Forecasting_renewable_energy_technologies}. In fact, logistic growth can be detected whenever there is growth in competition, and competition can be generalised to a high level of abstraction, e.g. diseases competing for victims and all possible accidents competing for the chance to be materialised.

S-curves enter as modular components in many intricate natural patterns. One may find S curves inside S curves because logistics portray a fractal aspect; a large S curve can be decomposed in a cascade of smaller ones \citep{Modis1994-wp}. One may also find chaos by rendering the logistic equation discrete \citep{Modis1992-sg}. Finally, logistics sit in the heart of the Lotka-Volterra equations, which describe the predator–prey relations and other forms of competition. In its full generality, the logistic equation, in a discrete form, with cross terms to account for all interrelations between competing species, would give a complete picture in which growth in competition, chaos, self-organisation, complex adaptive systems, autopoiesis, and other such academic formulations, all ensue as special cases \citep{Modis1997-rf}.

\begin{figure}[ht!]
\begin{center}
\includegraphics[trim=150 490 150 115, clip, width=3.2in ]{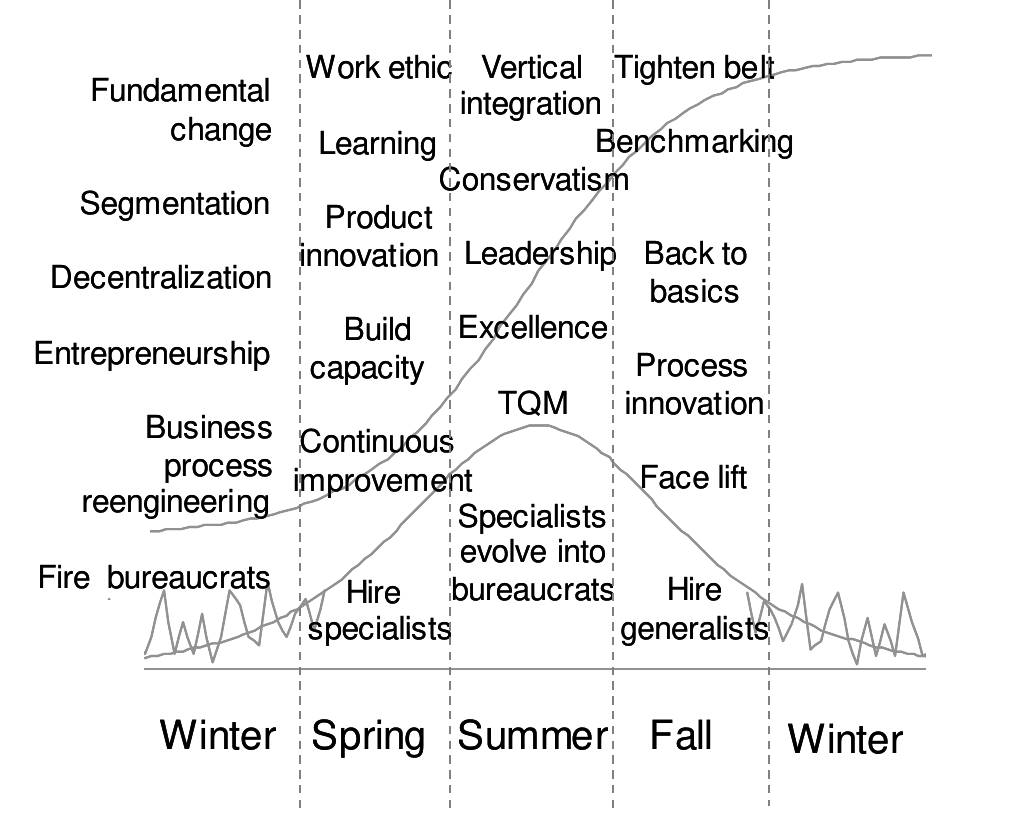}
\caption{Typical attributes of a growth cycle's ``seasons''. Adopted from \cite{Modis1998-ci} with the permission from the author.}
\label{fig:modis_theory_1}
\end{center}
\end{figure}

Each S curve has its own life cycle, undergoing good and bad ``seasons'' (see figure \ref{fig:modis_theory_1}). A large set of behaviours have been tabulated, each one best suited for a particular season \citep{Modis1998-ci}. Becoming conservative -- seeking no change -- is appropriate in the summer when things work well. But excellence drops in second place during the difficult times of winter -- characterised by chaotic fluctuations -- when fundamental change must take place. Learning and investing are appropriate for spring, but teaching, tightening the belt, and sowing the seeds for the next season's crop belong in the fall.

\begin{figure}[ht!]
\begin{center}
\includegraphics[trim=70 350 70 75, clip, width=2.6in ]{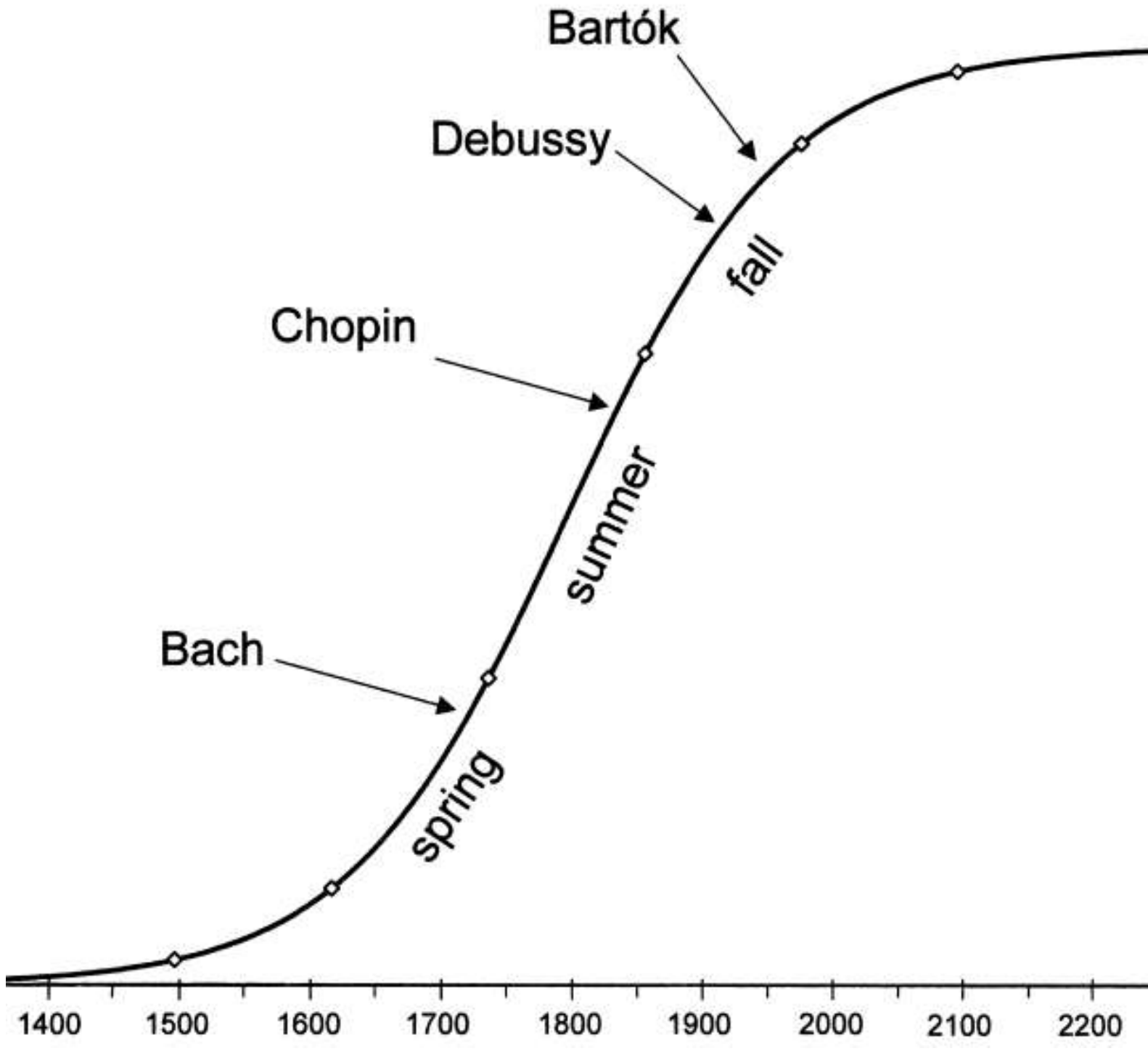}
\caption{The evolution of classical music. The vertical axis could be something like ``importance'', ``public appreciation'', or ``public preoccupation with music'' (always cumulative). Adopted from \cite{Modis2013-to} with the permission from the author.} 
\label{fig:modis_theory_2}
\end{center}
\end{figure}

\begin{figure}[ht!]
\begin{center}
\includegraphics[width=3.2in ]{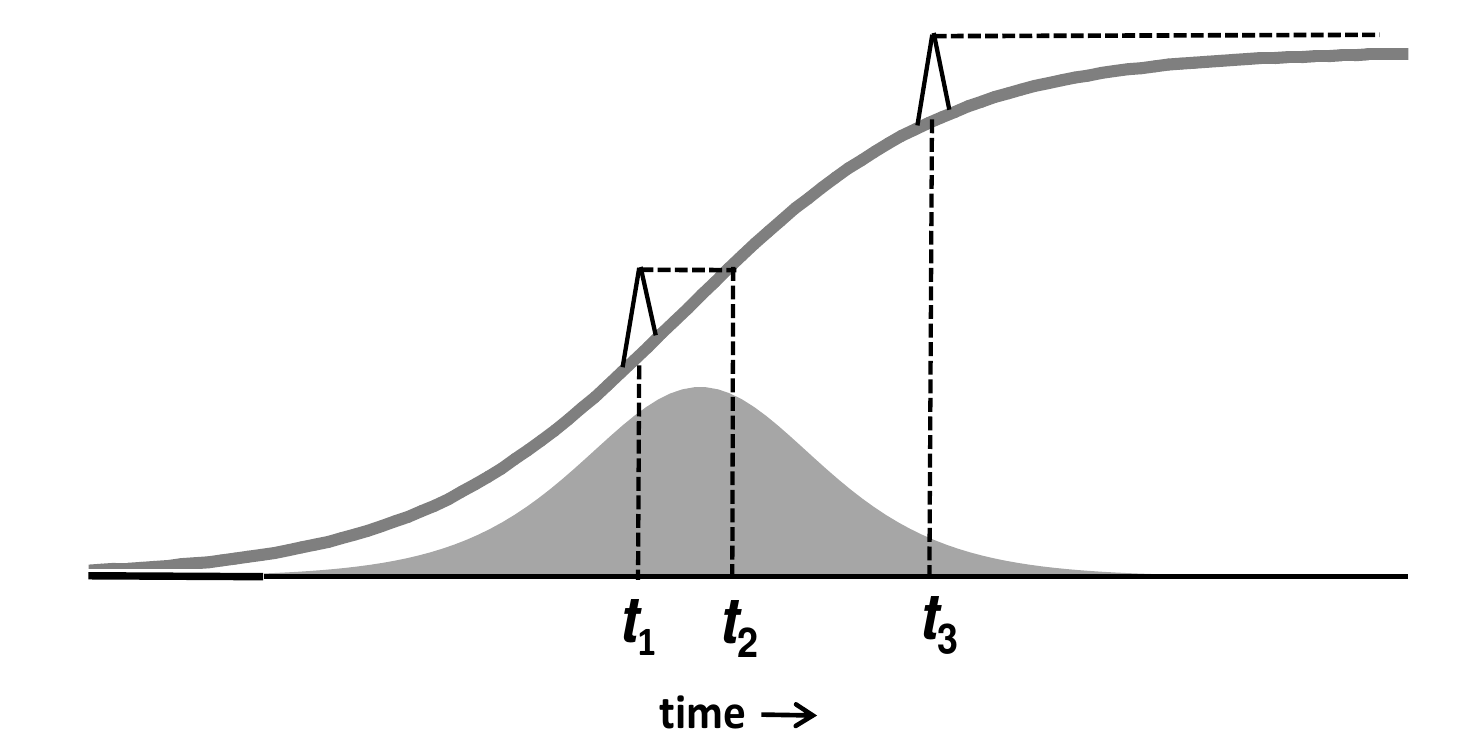}
\caption{An upward excursion at $t_1$ reaches the same level as the logistic curve at $t_2$ and can be considered as a ``natural'' deviation. The same-size excursion at time $t_3$ has no corresponding point on the curve. The grey life cycle delimits the position and size of all ``natural'' deviations. Adapted from \cite{Modis2007-ur} with permission from the author.} 
\label{fig:modis_theory_3}
\end{center}
\end{figure}

Focusing on \textit{what} to do is appropriate in spring, whereas in fall the emphasis shifts to the \textit{how}. For example, the evolution of classical music followed a large-timeframe S curve beginning in the fifteenth century and reaching a ceiling in the twentieth century; see figure \ref{fig:modis_theory_2} \citep{Modis2013-to}. In Bach's time composers were concerned with \textit{what} to say. The value of their music is in its architecture and as a consequence it can be interpreted by any instrument, even by simple whistling. But two hundred years later composers such as Debussy wrote music that depends crucially on the interpretation, the \textit{how}. Classical music was still ``young'' in Bach's time but was getting ``old'' by Debussy's time. No wonder Chopin is more popular than Bartók. Chopin composed during the ``summer'' of music's S curve when public preoccupation with music grew fastest. Around that time composers were rewarded more handsomely than today. The innovations they made in music -- excursions above the curve -- were assimilated by the public within a short period of time because the curve rose steeply and would rapidly catch up with each excursion/innovation. But today the curve has flattened and composers are given limited space. If they make an innovation and find themselves above the curve, there won't be any time in the future when the public will appreciate their work; see figure \ref{fig:modis_theory_3} \citep{Modis2007-ur}. On the other hand, if they don't innovate, they will not be saying anything new. In either case today's composers will not be credited with an achievement. 

S curves constructed only qualitatively can be accurate, informative, and insightful. Practical challenges of applying S curves are discussed in \S\ref{sec:Dealing_with_logistic_forecasts_in_practice}.

\subsubsection[Synchronic and diachronic competition (Renato Guseo)]{Synchronic and diachronic competition\protect\footnote{This subsection was written by Renato Guseo.}}
\label{sec:Synchronic_and_diachronic_competition}
Synchronic and diachronic competition models account for critical life cycle perturbations due to interactions, not captured by univariate innovation diffusion models (see \S\ref{sec:Innovation_diffusion_models}) or other time series models, such as ARIMA and VAR. This is important especially in medium and long-term prediction.

Competition in a natural or socio-economic system generally refers to the presence of antagonists that contend for the same resource. This typically occurs in nature, where multiple species struggle with each other to survive, or in socio-economic contexts where products, technologies and ideas concur to be finally accepted within a market and compete for the same market potential. These competition dynamics are reflected in separate time series -- one for each concurrent -- characterised by a locally oscillatory behaviour with nonlinear trends, unexpected deviations and saturating effects.

The analytic representation of competition has followed different approaches. A first approach has been based on complex systems analysis \citep{boccara:04_RGuseo}, which refers to a class of agents (see also \S\ref{sec:Agent_based_models}) that interact, through local transition rules, and produce competition as an emergent behaviour. This approach may be frequently reduced to a system of differential equations, with suitable mean field approximations. A second approach, systems analysis, has been based on systems of ordinary differential equations (ODE).

In this domain, competition may be a synchronic process, if competitors are present in the environment at the same time; for example, two products may enter the market in the same period. Instead, it is diachronic if competitors come at a later stage; for example, one product enters a market in a given period and just subsequently other similar products enter the same market and start to compete. Pioneering contributions of this competition modelling are due to \citet{lotka:20_RGuseo} and \citet{volterra:26_RGuseo}, who independently obtained a class of synchronic predator-prey models; see also \S\ref{sec:The_natural_law of growth_in_competition_Logistic_growth}. A generalised version of the Lotka-Volterra (LV) model has been provided by \citet{abramson:98_RGuseo}.

\citet{morris:03_RGuseo} proposed an extended LV model for a duopolistic situation, by making explicit the role of carrying capacities or market potentials, and the inhibiting strength of the competitors in accessing the residual resource. LV models typically do not have closed form solutions. In such cases, a staked form of equations allows a first-stage inference based on nonlinear least squares (NLS), with no strong assumptions on the stochastic distributions of error component. Short-term refining may be grounded on a Seasonal Autoregressive Moving Average with exogenous input (SARMAX) representation. Outstanding forecasts may be obtained including the estimated first-stage solution as `exogenous' input (see \S\ref{sec:Anomaly_detection_and_time_series_forecasting} and \S\ref{sec:Exogenous_variables_and_feature_engineering}).

A different synchronic model, termed Givon-Bass (GB) model, extending the univariate innovation diffusion models described in \S\ref{sec:Innovation_diffusion_models} \citep{bass:69_RGuseo, givonmamu:95_RGuseo}, has been presented in \citet{bonaldo:91_RGuseo}, introducing parametric components of global interaction. In this model, the residual market (or carrying capacity) is completely accessible to all competitors, and the rate equations introduce distributed seeding effects. The GB model has a closed form solution that was independently published by \citet{krishnanbasskumar:00_RGuseo}. The more general model by \citet{savin:05_RGuseo} and related advances by \citet{guseomortarino:10_RGuseo} were extended to the diachronic case in \citet{guseomortarino:12_RGuseo}, defining a competition and regime change diachronic (CRCD) model with a closed form solution. A relevant improvement of CRCD has been proposed in \citet{guseomortarino:14_RGuseo}, by introducing within-brand and cross-brand word-of-mouth effects, not present in standard LV models. The unrestricted unbalanced diachronic competition (unrestricted UCRCD) model, is defined with these new factors. The model assumes, among other specifications, a constant market potential. In \citet{guseomortarino:15_RGuseo} this assumption is relaxed, by introducing a dynamic market potential \citep{guseoguidolin:09_RGuseo, guseo:11_RGuseo}. Some applications are summarised in \S\ref{sec:Competing_products}.
 
\subsubsection[Estimation and representation of uncertainty (Ricardo Bessa)]{Estimation and representation of uncertainty\protect\footnote{This subsection was written by Ricardo Bessa.}}
\label{sec:Estimation_and_representation_of_uncertainty}
Forecasting uncertainty consists in estimating the possible range of forecast errors (or true values) in the future and the most widely adopted representation is a forecast interval~\citep{Patel1989}. The forecast interval indicates a range of values and the respective probability, which is likely to contain the true value (which is yet to be observed) of the response variable. Since for a specific lead-time the forecast interval only encompasses information from a marginal distribution, it can also be named marginal forecast interval (MFI). A MFI can be obtained from: parametric distribution, such as a Gaussian distribution with conditional mean and variance estimated with a Generalised ARCH model~\citep{Baillie1992}; non-parametric distribution, e.g., obtained with conditional kernel density estimation~\citep{Hyndman1996}; directly estimated with statistical learning methods, such as quantile regression~\citep{Taylor1999} or bootstrapping~\citep{Masarotto1990}, or with machine learning algorithms like quantile random forests~\citep{Meinshausen2006}. For a combination of density forecasts from different models, see \S\ref{sec:Density_forecast_combinations}.

For multi-step ahead forecasting problems (see also \S\ref{sec:multi_step_ahead_forecasting}), in particular when information about forecast uncertainty is integrated in multi-period stochastic optimisation \citep{Dantzig1993}, information about the temporal dependency structure of forecast errors (or uncertainty) is a fundamental requirement. In this case, the concept of simultaneous forecast intervals (SFI) can be found in the statistical literature \citep{Chew1968}. SFI differ from MFI since take into account the temporal interdependency of forecast errors and are constructed to have the observed temporal trajectory of the response variable fully contained inside the forecast intervals during all lead-times of the time horizon. The number of works that cover SFI is lower when compared to the MFI, but some examples are: methods based on Bonferroni- and product-type inequalities applied time series forecasting models likes ARIMA and Holt-Winters \citep{Ravishanker1991}; combination of bootstrap replications and an heuristic optimisation procedure to find an envelope of the temporal trajectories closest to the deterministic forecast \citep[][see also \S\ref{sec:Forecasting_with_bootstrap}]{Staszewska‐Bystrova2011}; sampling forecast errors at different horizons and estimate the SFI with the empirical Mahalanobis distance \citep{Jorda2013}. 

Advances in Operations Research for decision-making problems under uncertainty imposed new requirements in forecast uncertainty estimation and representation. On one hand, stochastic optimisation requires a scenario representation for forecast uncertainty \citep{Powell2019}. This motivated research in methods that generate uncertainty forecasts represented by random vectors (term used in statistics) or path forecasts (term used in econometrics), such as parametric copula combined with MFI \citep{Pinson2009}, parametric dynamic stochastic model \citep{Li2011} or epi-spline basis functions \citep{Rios2015}. On the other hand, robust optimisation does not make specific assumptions on probability distributions and the uncertain parameters are assumed to belong to a deterministic uncertainty set. Hence, some authors proposed new methods to shape forecast uncertainty as polyhedral or ellipsoidal regions to enable a direct integration of forecasts in this type of optimisation problem \citep{Bertsimas2008, Golestaneh2019}.

Finally, communication of forecast uncertainty (e.g., MFI, SFI, random vectors) to decision-makers requires further attention since it remains as a major bottleneck for a wide adoption by industry, particularly in uses cases with multivariate time series \citep{Akram2015} and adverse weather events \citep{Ramos2010}. Please also see \S\ref{sec:Communicating_forecast_uncertainty}.
 
\subsubsection[Forecasting under fat tails (Pasquale Cirillo)]{Forecasting under fat tails\protect\footnote{This subsection was written by Pasquale Cirillo.}}
\label{sec:Forecasting_under_fat_tails}
A non-negative continuous random variable $X$ is fat-tailed, if its survival function $S(x)=P(X\geq x)$ is regularly varying, that is to say if $S(x)=L(x) x^{-\alpha}$, where $L(x)$ is a slowly varying function, for which $\lim_{x \to \infty}\frac{L(tx)}{L(x)}=1$ for $t>0$ \citep{Embrechts2013-ne_PC}. The parameter $\alpha$ is known as the tail parameter, and it governs the thickness of the tail -- the smaller $\alpha$ the fatter the tail -- and the existence of moments, so that $E[X^p]<\infty$ if and only if $\alpha>p$. Often $\alpha$ is re-parametrised as $\xi=1/\alpha$.

Fat tails are omnipresent in nature, from earthquakes to floods, and they are particularly common in human-related phenomena like financial markets, insurance, pandemics, and wars \citep[see, for example,][and references therein]{Mandelbrot1983-eh_PC,Taleb2020-ow_PC}. 

Forecasting fat-tailed random variables is therefore pivotal in many fields of life and science. However, while a basic coverage of the topic is available in most time series and risk management manuals \citep[e.g.,][]{Shumway2017-am_PC,McNeil2015-lj_PC}, the profound implications of fat tails are rarely taken into consideration, and this can generate substantial errors in forecasts. 

As observed in \cite{Taleb2020-ki_PC}, any statistical forecasting activity about the mean -- or another quantity -- of a phenomenon needs the law of large numbers (LLN), which guarantees the convergence of the sample mean at a given known rate, when the number of observations $n$ grows. 

Fat-tailed phenomena with tail parameter $\alpha \leq 1$ are trivially not predictable. Since their theoretical mean is not defined, the LLN does not work, for there is nothing the sample mean can converge to. This also applies to apparently infinite-mean phenomena, like pandemics and wars, i.e., extremely tricky objects of study, as discussed in \cite{Cirillo2016-lr_PC}. In similar situations, one can rely on extreme value theory to understand tail risk properties, but should refrain from standard forecasting.

For random variables with $1<\alpha \leq 2$, the LLN can be extremely slow, and an often unavailable number of observations is needed to produce reliable forecasts. Even for a well-behaved and non-erratic phenomenon, we all agree that a claim about the fitness or non-fitness of a forecasting approach, just on the basis of one single observation ($n=1$), would be considered unscientific. The fact is that with fat-tailed variables that ``$n=1$'' problem can be made with $n=10^6$ observations \citep{Embrechts2013-ne_PC,Taleb2020-ow_PC}. In the case of events like operational losses, even a larger $n\to \infty$ can still be just anecdotal \citep{Cirillo2016-lr_PC}.

According to \cite{Taleb2020-ow_PC}, owing to preasymptotics, a conservative heuristic is to manage variables with $\alpha\leq 2.5$ as practically unpredictable (for example, see \S\ref{sec:Pandemics} and \S\ref{sec:Forecasting_risk_for_violence_and_wars}). Their sample average is indeed too unstable and needs too many observations for forecasts to be reliable in a reasonable period of time. For $\alpha >2.5$, conversely, forecasting can take place, and the higher $\alpha$ the better. In any case, more research is strongly needed and desirable (see also \S\ref{sec:conclusions}).

Observe that even discussing the optimality of any alarm system \citep[see, for example,][]{Turkman1990-uz_PC,Svensson1996-cw_PC} based on average forecasts would prove meaningless under extremely fat tails ($\alpha \leq 2$), when the LLN works very slowly or does not work. In fact, even when the expected value is well-defined (i.e., $1<\alpha<2$), the non-existence of the variance would affect all the relevant quantities for the verification of optimality \citep{De_Mare1980-mn_PC}, like for instance the chance of undetected events. For all these quantities, the simple sample estimates commonly used would indeed be misleading. 

\subsection{Bayesian forecasting}
\label{sec:bayesian_forecasting}

\subsubsection[Foundations of Bayesian forecasting (David T. Frazier \& Gael M. Martin)]{Foundations of Bayesian forecasting\protect\footnote{This subsection was written by David T. Frazier \& Gael M. Martin.}}
\label{sec:Foundations_of_Bayesian_forecasting}
The Bayesian approach to forecasting produces, by default, a probabilistic forecast (see also \S\ref{sec:Density_forecast_combinations} and \S\ref{sec:evaluating_probabilistic_forecasts}) describing the uncertainty about future values of the phenomenon of interest, conditional on all \textit{known quantities}; and with uncertainty regarding all \textit{unknown quantities} having been integrated out. In order to produce such forecasts, the Bayesian approach requires (\textit{i}) a predictive model for the future value of the relevant phenomenon, conditional on the observed data and all model unknowns; (\textit{ii}) a model for the observed data; and (\textit{iii}) a distribution describing the (subjective or objective) prior beliefs regarding the model unknowns. Using these quantities, the standard calculus of probability distributions, and Bayes' theorem, then yield the Bayesian predictive (equivalently, forecast) density function (where density is used without loss of generality).

Stated more formally, given observed data up to time $n$, $\mathbf{y}=(y_{1},\dots,y_{n})'$, and denoting the model unknowns by $\theta$, Bayesian forecasting describes the behaviour of the future {random variable} $Y_{n+1}$ via the predictive density: 
\begin{equation}
p(y_{n+1}|\mathbf{y})=\int p(y_{n+1}|\theta,\mathbf{y})p(\theta|\mathbf{y})d\theta\mathbf{,} \label{eq:exact}
\end{equation}

\noindent where $y_{n+1}$ denotes a value in the support of $Y_{n+1}$ and $p(y_{n+1}|\theta,\mathbf{y})$ is the predictive model for $Y_{n+1}$ conditional on $\mathbf{y}$ and the model unknowns $\theta$. Critically, and in
contrast with frequentist approaches to forecasting, parameter uncertainty has been factored into $p(y_{n+1}|\mathbf{y})$ via the process of integration with respect to the posterior probability density function (pdf) for $\theta$, $p(\theta|\mathbf{y})$. The posterior pdf is given, by Bayes' theorem, as $p(\theta|\mathbf{y})\propto p(\mathbf{y}|\theta)\times p(\theta),$ where $p(\mathbf{y}|\theta)$ defines the assumed model for $\mathbf{y}$ (equivalently, the likelihood function), and the prior pdf $p(\theta)$ captures prior beliefs about $\theta.$ Moreover, uncertainty about the assumed predictive model itself can be easily accommodated using Bayesian model averaging, which involves taking a weighted average of \textit{model-specific} predictives, with posterior model probabilities (also obtained via by Bayes' theorem) serving as the weights. See \cite{koop2003bayesian_GM}, \cite{ohagan2004_GM} and \cite{greenberg2008_GM} for textbook illustrations of all of these steps.

No matter what the data type, the form of predictive model, or the dimension of the unknowns, the basic manner in which all Bayesian forecast problems are framed is the same. What differs however, from problem to problem, is the way in which the forecasting problem is \textit{solved}. To understand why, it is sufficient to recognise that in order to obtain the predictive density $p(y_{T+1}|\mathbf{y})$ we must be able to (somehow) perform the integration that defines this quantity. In almost any practical setting, this integration is {infeasible analytically} and we must rely on \textit{computational methods} to access the predictive density. Therefore, the evolution of the practice of Bayesian forecasting has gone hand in hand with developments in Bayesian computation \citep{martin2020computing_GM}. Through the lens of computation, in \S\ref{sec:Implementantion_of_Bayesian_forecasting} we briefly describe the methods of implementing Bayesian forecasting.

\subsubsection[Implementation of Bayesian forecasting (David T. Frazier \& Gael M. Martin)]{Implementation of Bayesian forecasting\protect\footnote{This subsection was written by David T. Frazier \& Gael M. Martin.}}
\label{sec:Implementantion_of_Bayesian_forecasting}
If the posterior is accessible via methods of \textit{exact simulation} -- e.g., Monte Carlo simulation, importance sampling, Markov chain Monte Carlo (MCMC) sampling, pseudo-marginal MCMC \citep{andrieu:roberts:2009_GM,andrieu:doucet:holenstein:2010_GM} -- an estimate of the predictive density $p(y_{n+1}|\mathbf{y})$ in (\ref{eq:exact}) can be produced using draws of the unknown $\theta$ from the posterior pdf, $p(\theta|\mathbf{y})$. In most cases, this simulation-based estimate of $p(y_{n+1}|\mathbf{y})$ can be rendered arbitrarily accurate by choosing a very large number of posterior draws; hence the use of the term `exact predictive' to reference this estimate. See \cite{gewekewhiteman2006_GM} for an early review of Bayesian forecasting implemented using exact simulation, \S\ref{sec:Forecasting_with_DSGE_Models}, \S \ref{sec:bayesian_forecasting_with_copulas}, \S\ref{sec:Variable_Selection}, \S\ref{sec:Forecasting_GDP_and_Inflation}, and \S\ref{sec:Copula_forecasting_with_dependent_financial_times_series} for further discussions and a range of relevant applications, and Chapters 3, 7, and 9 in \cite{geweke2011handbook_GM} for applications in (general) state space, macroeconometric and finance settings respectively. In addition, a 2008 special issue of \textit{International Journal of Forecasting} on Bayesian Forecasting in Economics provides coverage of forecasting applications (and methods of computation) that exploit exact simulation methods, as do selected chapters in \cite{ohagan2010handbook_GM} and \cite{brooks:etal:2011_GM}.

In cases where the posterior is not readily accessible, due to either the intractability of the likelihood function or the high dimension of the model unknowns, or both, methods of \textit{approximation} are required. \cite{frazier2019approximate_GM}, for instance, produce an `approximate predictive' by replacing the exact posterior in (\ref{eq:exact}), $p(\theta|\mathbf{y})$, with an `approximate posterior' constructed using
\textit{approximate Bayesian computation} \citep[ABC --][]{sisson2018handbook_GM}. In large samples, this approximate predictive is shown to be equivalent to the exact predictive. The approximate and exact predictives are also shown to be numerically indistinguishable in finite samples, in the cases investigated; see also \cite{canale2016_GM}, and \cite{konkamking2019_GM}. Related work produces an approximate predictive by exploiting \textit{variational Bayes} \citep{blei2017variational_GM} approximations of the posterior \citep{tran2017variational_GM,quiroz2018gaussian_GM,koop2018variational_GM,chan2020fast_GM,loaiza2020fast_GM}. The flavour of this work is broadly similar to that of \cite{frazier2019approximate_GM}; that is, computing the predictive $p(y_{n+1}|\mathbf{y})$ via an approximation to the posterior does not \textit{significantly} reduce predictive accuracy.

The Bayesian paradigm thus provides a very natural and coherent approach to prediction that can be implemented successfully via one of any number of computational methods. Inherent to the approach, however, is the assumption that the model we are using to make predictions is an accurate description of the data generating process (DGP) that has generated the observed data; or if we are averaging across models using Bayesian model averaging, we must assume that this average contains the DGP for the observed data. In response to this limitation, allied with the desire to drive prediction by user-specified measures of predictive loss, new approaches to Bayesian prediction have recently been proposed, and we briefly discuss two such classes of methods.

First are methods for combining predictives in which the weights are not equated to the posterior model probabilities (as in standard Bayesian model averaging) but, rather, are updated via problem-specific predictive criteria, or via predictive calibration \citep{Dawid1982_GM,Dawid1985_GM,Gneiting&2007_TT}; see \cite{BILLIO2013213_AY}, \cite{casarin2015_GM}, \cite{Pett2016_GM}, \cite{Bassetti2018_GM}, \cite{BASTURK2019_GM}, \cite{MCALINN2019155_AY}, and \cite{McAlinn2020_GM}, for a selection of approaches, and \S\ref{sec:Density_forecast_combinations} for related discussion. Importantly, these methods do not assume the true model is spanned by the constituent model set. Second are methods in which the standard Bayesian posterior, which is itself based on a potentially misspecified model, is replaced by a generalised version that is designed for the specific predictive task at hand (e.g., accurate prediction of extreme values); with the overall goal then being to produce predictions that are accurate according to the particular measure of interest. See \cite{syring2020gibbs_GM}, \cite{loaiza2019focused_GM}, and \cite{frazierloss_GM} for specific examples of this methodology, as well as proofs of its theoretical validity.

\subsubsection[Bayesian forecasting with copulas (Feng Li)]{Bayesian forecasting with copulas\protect\footnote{This subsection was written by Feng Li.}}
\label{sec:bayesian_forecasting_with_copulas}
Copulas provide intrinsic multivariate distribution structure that allows for modelling multivariate dependence with marginal distributions as the input, making it possible to forecast dependent time series and time series dependence. This review focuses on the Bayesian approach for copula forecasting. For rigorous topics on copula introduction \citep{nelsen2006introduction_Feng,trivedi2007copula_Feng,joe1997multivariate_Feng}, copula modelling techniques \citep{durante2015principles_Feng}, vine copulas \citep{joe2014dependence_Feng}, review on frequentist approaches for copula-based forecasting \citep{patton2013copula_Feng}, see the aforementioned references and the references therein.

The advantages of the Bayesian copula approach compared to the frequentist treatments are (\textit{i}) the Bayesian approach allows for jointly modelling the marginal models and the copula parameters, which improves the forecasting efficiency \citep{joe2005asymptotic_Feng,li2018improving_Feng}, (\textit{ii}) probabilistic forecasting is naturally
implied with the Bayesian predictive density, and (\textit{iii}) experts information can be seamlessly integrated into forecasting via the priors' setting.

Forecasting with copulas involves selecting between a different class of copulas. Common approaches include Bayesian hypothesis testing (see \S\ref{sec:Foundations_of_Bayesian_forecasting}) where copula parameters are treated as nuisance variables \citep{huard2006bayesian_Feng}, or parsimonious modelling of covariance structures using Bayesian selection and model averaging
\citep{pitt2006efficient_Feng,smith2010modeling_Feng,min2011bayesian_Feng}.

One particular interest in copulas is to forecast the dynamic dependencies between multivariate time series. Time-varying copula construction is possible via (\textit{i}) an autoregressive or GARCH form (see \S\ref{sec:arch_garch_models}) of dependence parameters
\citep{patton2006modelling_Feng,lucas2014conditional_Feng}, (\textit{ii}) factor copula construction \citep{oh2018time_Feng,tan2019bayesian_Feng} that simplifies the computation, (\textit{iii}) the stochastic copula autoregressive model \citep{almeida2012efficient_Feng} that dependence is modelled by a real-valued latent variable, and (\textit{iv}) covariate-dependent copulas approach by parameterising the dependence as a function of possible time-dependent covariates
\citep{li2018improving_Feng} that also improves the forecasting interpretability. ARMA-like and GARCH-like dependences in the tail can be considered as special cases of \cite{li2018improving_Feng}.

In multivariate time series forecasting (see also \S\ref{sec:Forecasting_with_many_variables}), unequal length of data is a common issue. One possible approach is to partition the copula parameters into elements relating only to the marginal distributions and elements only relating to the copula \citep{patton2006estimation_Feng}. For mixed frequency data, it is possible to decompose the copula dependence structure into linear and nonlinear components. Then the high and low frequency data is used to model the linear and nonlinear dependencies, respectively \citep{oh2016high_Feng}. Bayesian data augmentation is also used to forecast multivariate time series with mixed discrete and continuous margins \citep{smith2012estimation_Feng}. For other treatments for discrete and continuous time series \citep[see, for example,][]{panagiotelis2012pair_Feng,panagiotelis2017model_Feng}.

Bayesian approach for lower dimensional copula forecasting ($d < 10$) is straightforward with traditional Gaussian copulas, Student's-$t$ copulas, Archimedean copulas, or pair copula combinations. In higher dimensional settings, special considerations are required to save the computational burden, such as low rank approximation of covariates matrix \citep{salinas2019high_Feng} or factor copula models with stochastic loadings \citep{creal2015high_Feng}.

In the Bayesian setup, forecasting model performance is typically evaluated based on a $K$-fold out-of-sample log predictive score \citep[LPS:][]{geweke2010comparing_Feng}, and out-of-sample Value-at-Risk (VaR) or Expected Shortfall (ES) are particularly used in financial applications. The LPS is an overall forecasting evaluation tool based on predictive densities, serving out-of-sample probabilistic forecasting. LPS is ideal for decision makers \citep{geweke2001bayesian_Feng,geweke2010comparing_Feng}. The VaR gives the percentile of the conditional distribution, and the corresponding ES is the expected value of response variable conditional on it lying below its VaR.
 
\subsection{Variable and model selection}
\label{sec:Variable_and_model_selection}

\subsubsection[Leading indicators and Granger causality (Ulrich Gunter)]{Leading indicators and Granger causality\protect\footnote{This subsection was written by Ulrich Gunter.}}
\label{sec:Leading_indicators_and_Granger_causality}
Leading (economic) indicators are variables that try to capture changes in the development of economic activity before those changes materialise. Typically, market participants, policy makers, the general public, etc. are interested in which direction the economy of a country is developing ahead of the publication date of the respective quarterly GDP figures (see also \S\ref{sec:Forecasting_GDP_and_Inflation}), for which these leading indicators play a crucial role. This is of particular relevance for short-term forecasting and nowcasting of such a (low-frequency) variable characterised by a publication lag.

For some leading indicators, single variables are combined into a composite index such as The Conference Board Leading Economic Index published by The Conference Board for the United States. It consists of averages of ten variables, including average weekly hours in manufacturing, manufacturers' new orders, private housing building permits, and average consumer expectations \citep{The_Conference_Board2020-yi_Ulrich}. Other leading indicators consist of one variable only and are purely survey-based, such as the monthly ifo Business Climate Index published by the ifo Institute for Germany, which covers the appraisal of current and future expected business of executives of approximately 9,000 German firms across industries \citep{Ifo_Institute2020-df_Ulrich}.

If a specific leading indicator is able to capture certain economic changes before they happen, such a leading indicator $x_t$ is said to Granger-cause (or predictively cause) a forecast variable $y_t$ \citep{Granger1969-uk_UG}, implying that a ``cause'' cannot happen after an effect. In (theoretical) econometric terms and following the notation of \cite{Lutkepohl2005-xj_UG}, $y_t$ is Granger-caused by $x_t$ if for at least one forecast horizon $h=1, 2, \dots$ the following inequality holds: 
$$
\sum_{y}(h \mid \Omega_i) < \sum_{y}(h \mid \Omega_i \setminus \{x_t \mid t \leq i \}).
$$

In other words, the mean square error of the optimal $h$-step-ahead predictor of the forecast variable, which includes the information contained in the leading indicator in the set of all relevant information to the forecaster available at the forecast origin $\sum_{y}(h \mid \Omega_i)$, must be smaller than the mean square error of that $h$-step-ahead predictor of the forecast variable without the information contained in said leading indicator $\sum_{y}(h \mid \Omega_i \setminus \{x_t \mid t \leq i \})$ \citep{Lutkepohl2005-xj_UG}.

Nonetheless, the concept of Granger causality also has its limitations. The notion that a ``cause'' cannot happen after an effect is only a necessary, yet not a sufficient condition for causality, hence the reason why ``cause'' has been written with quotation marks in this section. If one ignored this fact and equated causality with Granger causality, they would commit an informal logical fallacy called Post hoc ergo propter hoc \citep[i.e., after this, therefore because of this;][]{Walton2008-uv_Ulrich}. A bold example of this fallacy would be, ``Because the birds migrate, winter is coming''. This is a fallacy, as winter would come at about the same time every year, no matter if there were migrating birds or not. Moreover, hardly any economic activity is monocausal.

There are also different types of formal statistical Granger causality tests available for different data structures that are implemented in typical statistics/econometrics software packages. For the simple case of two variables (Granger, 1969), say the forecast variable and the leading indicator, the null hypothesis of a bivariate Granger causality test is that the leading indicator does not Granger-cause the forecast variable. Under this null hypothesis, the $F$-test statistic on the joint impact of the coefficients of the past realisations of the leading indicator employed as explanatory variables in a multiple linear regression with the forecast variable as dependent variable and its past realisations as additional explanatory variables will not be statistically significantly different from zero. The maximum lag for past realisations would be optimally determined, for instance, by some information criterion (e.g., AIC, BIC; see also \S\ref{sec:Model_Selection}). Practical applications of Granger causality and leading indicators in tourism demand forecasting can be found, for instance, in \S\ref{sec:Tourism_demand_forecasting}.

\subsubsection[Model complexity (Michał Rubaszek)]{Model complexity\protect\footnote{This subsection was written by Michał Rubaszek.}}
\label{sec:Model_complexity}
A simple model must be easily understood by decision-makers. On the contrary, relationships in a complex model are opaque for its users (see also \S\ref{sec:Trust_in_forecasts}). In this context, complexity is not measured solely by the number of parameters, but also by the functional form of the relationships among variables.

Complex models are commonly believed to deliver better forecasts, as they well describe sophisticated economic structures and offer good fit the data. Consequently, they are favoured by researchers, decision-makers and academic journals. However, empirical studies provide little evidence about their forecasting superiority over simple models. This can be explained using the bias-variance trade-off framework, in which the mean squared error can be decomposed into
\[
MSE = noise + variance + bias^2.
\]
Noise is driven by the random term in the model. It cannot be avoided, even if we know the true DGP. Variance is caused by the need to estimate the model parameters, hence its value increases with model complexity and declines with the sample size. Bias is predominantly related to model mis-specification, which is most likely to occur for simple methods. The implications of the above framework are twofold: (\textit{i}) the relationship between model complexity and MSE is U-shaped and (\textit{ii}) the optimal model complexity increases with the sample size.

The illustration of the bias-variance trade-off for a simple autoregressive model (see \S\ref{sec:autoregressive_integrated_moving_average_models}) is provided by \cite{CaZorziEtAl:2016OER_MR}. They present analytical proof that for any persistent DGP of the form $y_t=c+\phi y_{t-1}+\epsilon_t$ characterised by half-life of over two years, the accuracy of forecasts from the random walk or the AR(1) model with parameter $\phi$ fixed at an arbitrary value consistent with half-life of five years tend to be higher than that from the estimated AR(1) model. This result explains why in numerous studies the random walk is a tough benchmark as well as why a simple, calibrated AR(1) model can be successful in forecasting inflation \citep{FaustWright:2013HEF_MR}, exchange rates \citep{CaZorziEtAl:2017JIE_MR} or oil prices \citep{Rubaszek:2020IJF_MR} compared to a number of complex competitors.

Wide support to the view that model simplicity improves forecasting performance is presented by \citet{GreenArmstrong:2015JBR_SI_MR}, in an introductory article to the special issue of Journal of Business Research ``Simple versus complex forecasting'', as well as the results of M1 and M2 competitions \citep{Makridakis1982-co, Makridakis1993-bw}. \citet{StockWatson:1998_MR} also show that for most US monthly series complex non-linear autoregressive models deliver less accurate forecasts than their linear counterparts. On the contrary, the results of M4 competition tend to favour more complex models \citep[][and \S\ref{sec:Forecasting_competitions}]{Makridakis2020-mm}.

Why are then complex models preferred to simple ones, if the latter deliver more accurate forecasts? \citet{BrightonGigerenzer:2015JBR_SI_MR} claim that there is a tendency to overemphasize the bias component and downplay the role of variance. This behaviour is leading to an implicit preference towards more complex forecasting methods, which is called by the authors as ``bias bias''. To avoid it, one can follow the golden rules of forecasting, which says: be conservative in the choice of over-ambitious models and be wary of the difficulty to forecast with complex methods \citep{ArmstrongEtAl:2015JBR_GoldenRule_MR}. The alternative is to use methods, which explicitly account for the bias-variance trade-off, e.g. machine learning (see \S\ref{sec:machine_Learning}).

\subsubsection[Variable selection (Ross Hollyman)]{Variable selection\protect\footnote{This subsection was written by Ross Hollyman.}}
\label{sec:Variable_Selection}
References to `big data' have become somewhat ubiquitous both in the media and in academic literature in recent years (see \S\ref{sec:Forecasting_with_Big_Data} but also \S\ref{sec:Exogenous_variables_and_feature_engineering}). Whilst in some disciplines (for example Finance) it has become possible to observe time series data at ever higher frequency, it is in the cross section that the amount of data available to analysts has seen exponential growth. 

Ordinary Least Squares (OLS) is the standard tool for data analysis and prediction, but is well known to perform poorly when there many potential explanatory variables; \cite{Bishop2006_HR} sets out clearly why this is so. In situations where there is no obvious underlying model to suggest which of a potentially large set of candidate variables to focus on, the researcher needs to add new tools to the tool kit.

There are two principal sets of approaches to this problem. The first seeks to reduce the model dimensions by summarising the predictors in to a much smaller number of aggregated indices or factors. The second is to employ a regularisation strategy to reduce the effective dimension of the problem by shrinking the size of the regression coefficients. Some strategies reduce a subset of these coefficients to zero, removing some variables from the model entirely and these can hence be described as variable selection procedures.

Such procedures may be applicable either when a problem is believed to be truly sparse, with only a small number of variables having an effect, or alternatively when a sparse model can provide effective forecasts, even though the underling system is in fact more complex (see also \S\ref{sec:Model_complexity}). 

In the frequentist framework, the Least Absolute Shrinkage and Selection Operator (LASSO) procedure of \cite{JLC_Tibshir96} has proven effective in many applications. The LASSO requires choice of an additional regularisation parameter (usually selected by some statistical criteria or via a cross validation process). Various refinements to the original LASSO procedure have been developed, see in particular \cite{Rapach2018_HR} for a recent forecasting application. An alternative frequentist approach to variable selection and shrinkage is the Complete Subset Regression (CSR) model of \cite{Elliott2015_HR}, where separate OLS regressions are run on all possible combinations of potential regressors, with forecasts generated by averaging across the entire set of models. \cite{Kotchoni2019_HR} combine CSR with LASSO in a comprehensive empirical economic forecasting exercise. 

Where the underlying process is not properly sparse \citep[see][for a discussion]{Giannone2017_HR}, it is perhaps more natural to work in a Bayesian framework where samples can be drawn from the variable selection part of model reflecting the estimated probability of inclusion of each variable. The appropriate degree of regularisation can also be selected in a similar way. Forecasts are then constructed as weighted averages over several sparse models. This approach has proven to be very effective in practice, for example in fitting Bayesian Vector Auto Regressions to forecast economic time series. Early examples include \cite{George1993_HR} and \cite{Mitchell1988_HR}, which use binary indicators to select variables in to the model. More recent approaches use continuous random variables to achieve a similar effect, making computation more tractable. Examples include the Horeshoe Prior \citep{Carvalho2010_HR,Piironen2017_HR} and the LN-CASS prior of \cite{Thomson2019_HR}. \cite{Cross2020_HR} is a recent example of an economic forecasting exercise using several such models.
 
\subsubsection[Model selection (David F. Hendry)]{Model selection\protect\footnote{This subsection was written by David F. Hendry.}}
\label{sec:Model_Selection}
Taxonomies of all possible sources of forecast errors from estimated models that do not coincide with their  data generating process (DGP) have revealed that two mistakes determine forecast failures (i.e., systematic departures between forecasts and later outcomes), namely mis-measured forecast origins and unanticipated location shifts \citep{Clements1999_ABM,DFH_HendMizOpen11}. The former can be addressed by nowcasting designed to handle breaks \citep{CastleJoF09,ReichOxHB11,NowCast18}. It is crucial to capture the latter as failure to do so will distort estimation and forecasts, so must be a focus when selecting models for forecasting facing  an unknown number of in-sample contaminating outliers and multiple breaks at unknown times.

Consequently, selection must be jointly over both observations and variables, requiring computer learning methods \citep[][see also \S\ref{sec:Pandemics}]{JLC_HendryDoornikModelDiscovery,DFH_CDHRobustSel20}. \textit{Autometrics}, a multiple-path block-search algorithm (see \citealp{JLC_Door07Auto}), uses impulse (IIS: \citealp{JLC_HendJoha07} and \citealp{johansen2009analysis_JJR}) and step (SIS: \citealp{JLC_CDHPSIS15}) indicator saturation for discovering outliers and breaks, as well as selecting over other variables. Although knowing the in-sample DGP need not suffice for successful forecasting after out-of-sample shifts, like financial crises and pandemics, `robust variants' of selected models can then avoid systematic mis-forecasts \citep[see, for example,][and \S\ref{sec:Forecasting_productivity}]{Martinez2019_ABM,JLC_DoornikCastleHendry20}.

Saturation estimation has approximately $K=k+2n$ candidates for $k$ regressors, with $2^K$ possible models, requiring selection with more candidate variables, $K$, than observations, $n$ (also see \S\ref{sec:data_driven_approaches} and \S\ref{sec:Forecasting_with_many_variables}). Selection criteria like AIC \citep{Akaike1973-hg_ASJBZB}, BIC \citep{Schwarz1978_DR}, and HQ \citep{DFH_Hann79q} are insufficient in this setting. For saturation estimation, we select at a tight significance level, $\alpha=1/K$, retaining subject-theory variables and other regressors. When forecasting is the aim, analyses and simulations suggest loose significance for then selecting other regressors, close to the 10\% to 16\% implied significance level of AIC, regardless of location shifts at or near the forecast origin \citep{DFH_CastDoorHend18}. At loose levels, \textit{Autometrics} can find multiple undominated terminal models across paths, and averaging over these, a univariate method and a robust forecasting device can be beneficial, matching commonly found empirical outcomes. The approach applies to systems, whence selection significance of both indicators
and variables is judged at the system level. Capturing in-sample location shifts remains essential \citep{Doornik2020-es_KN}.

There are costs and benefits of selection for forecasting (also see \S\ref{sec:Variable_Selection} and \S\ref{sec:judgmental_model_selection}). Selection at a loose significance level implies excluding fewer relevant variables that contribute to forecast accuracy, but retaining more irrelevant variables that are adventitiously significant, although fewer than by simply averaging over all sub-models \citep{DFH_HoetBMA99}. Retaining irrelevant variables that are subject to location shifts worsens forecast performance, but their coefficient estimates are driven towards zero when updating estimates as the forecast origin moves forward. Lacking omniscience about future values of regressors that then need to be forecast, not knowing the DGP need not be costly relative to selecting from a general model that nests it \citep{DFH_CastDoorHend18}. Overall, the costs of model selection for forecasting are small compared to the more fundamental benefit of finding location shifts that would otherwise induce systematic forecast failure.
 
\subsubsection[Cross-validation for time-series data (Christoph Bergmeir)]{Cross-validation for time-series data\protect\footnote{This subsection was written by Christoph Bergmeir.}}
\label{sec:Cross_validation_for_time_series_data}
When building a predictive model, its purpose is usually not to predict well the already known samples, but to obtain a model that will generalise well to new, unseen data. To assess the out-of-sample performance of a predictive model we use a test set that consists of data not used to estimate the model (for a discussion of different error measures used in this context see \S\ref{sec:point_forecast_accuracy_measures}).
However, as we are now using only parts of the data for model building, and other parts of the data for model evaluation, we are not making the best possible use of our data set, which is a problem especially if the amount of data available is limited. Cross-validation (CV), first introduced by \cite{stone1974cross_CB}, is a widely used standard technique to overcome this problem \citep{Hastie2009_JB} by using all available data points for both model building and testing, therewith enabling a more precise estimation of the generalisation error and allowing for better model selection. The main idea of ($k$-fold) cross-validation is to partition the data set randomly into $k$ subsets, and then use each of the $k$ subsets to evaluate a model that has been estimated on the remaining subsets. An excellent overview of different cross-validation techniques is given by \citet{arlot2010survey_CB}.

Despite its popularity in many fields, the use of CV in time series forecasting is not straightforward. Time series are often non-stationary and have serial dependencies (see also \S\ref{sec:autoregressive_integrated_moving_average_models}). Also, many forecasting techniques iterate through the time series and therewith have difficulties dealing with missing values (withheld for testing). Finally, using future values to predict data from the past is not in accordance with the normal use case and therewith seems intuitively problematic. Thus, practitioners often resort to out-of-sample evaluation, using a subset from the very end of a series exclusively for evaluation, and therewith falling back to a situation where the data are not used optimally. 

To overcome these problems, the so-called time-series cross-validation \citep{hyndman2018Forecasting_CB} extends the out-of-sample approach from a fixed origin to a rolling origin evaluation \citep{tashman2000out_CB}. Data is subsequently moved from the out-of-sample block from the end of the series to the training set. Then, the model can be used (with or without parameter re-estimation) with the newly available data. The model re-estimation can be done on sliding windows with fixed window sizes or on expanding windows that always start at the beginning of the series \citep{Bell2018Forecasting_CB}.

However, these approaches extending the out-of-sample procedure make again not optimal use of the data and may not be applicable when only small amounts of data are available. Adapting the original CV procedure, to overcome problems with serial correlations, blocked CV approaches have been proposed in the literature, where the folds are chosen in blocks \citep{Racine2000Consistent_CB,bergmeir2012use_CB} and/or data around the points used for testing are omitted \citep{Racine2000Consistent_CB,Burman1994Cross_CB}. Finally, it has been shown that with purely autoregressive models, CV can be used without modifications, i.e., with randomly choosing the folds \citep{bergmeir2018note_CB}. Here, CV estimates the generalisation error accurately, as long as the model errors from the in-sample fit are uncorrelated. This especially holds when models are overfitting. Underfitting can be easily detected by checking the residuals for serial correlation, e.g., with a Ljung-Box test \citep{ljung1978measure_CB}. This procedure is implemented in the \textit{forecast} package \citep{Hyndman2020forecast_CB} in R \citep{RCT2020R_CB}, in the function \texttt{CVar}.

\subsection{Combining forecasts}
\label{sec:combining_forecasts}

\subsubsection[Forecast combination: a brief review of statistical approaches (Devon K. Barrow)]{Forecast combination: a brief review of statistical approaches\protect\footnote{This subsection was written by Devon K. Barrow.}}
\label{sec:Forecast_combination_a_brief_review_of_statistical_approaches}
Given $N$ forecasts of the same event, forecast combination involves estimation of so called combination weights assigned to each forecast, such that the accuracy of the combined forecast generally outperforms the accuracy of the forecasts included. Early statistical approaches adopted a range of strategies to estimate combination weights including (\textit{i}) minimising in-sample forecast error variance among forecast candidates \citep{BatesGranger1969_DR,newbold1974experience_DKB,min1993bayesian_DKB}, (\textit{ii}) formulation and estimation via ordinary least squares regression \citep{granger1984improved_DKB,macdonald1994combining_DKB}, (\textit{iii}) use of approaches based on Bayesian probability theory \citep[][and \S\ref{sec:bayesian_forecasting}]{bunn1975bayesian_DKB,bordley1982combination_DKB,clemen1986combining_DKB,diebold1990use_DKB,Raftery1993-wj_ET}, (\textit{iv}) and the use of regime switching and time varying weights recognising that weights can change over time \citep[][and \S\ref{sec:Markov_switching_models}]{diebold1987structural_DKB,elliott2005estimation_DKB,lutkepohl2011forecasting_DKB,tian2014forecast_DKB}. \cite{barrow2016distributions_DKB} contains a very good documentation and empirical evaluation of a range of these early approaches, while \cite{de2000review_DKB} and \cite{armstrong2001principles_DKB} contain guidelines on their use.

Recent statistical approaches use a variety of techniques to generate forecasts and/or derive weights. \cite{kolassa2011combining_DKB} apply so called Akaike weights based on Akaike Information Criterion \citep{sakamoto1986akaike_DKB}, while bootstrapping has been used to generate and combine forecast from exponential smoothing \citep[][but also \S\ref{sec:Forecasting_with_bootstrap} and \S\ref{sec:Bagging_for_Time_Series}]{cordeiro2009forecasting_CC,barrow2020automatic_DKB,Bergmeir2016-su}, artificial neural networks \citep[][and \S\ref{sec:neural_networks}]{barrow2016comparison_DKB,barrow2016cross_DKB}, and other forecast methods \citep{athanasopoulos2018bagging_DKB,hillebrand2010benefits_DKB,inoue2008useful_DKB}. \cite{barrow2016cross_DKB} developed cross-validation and aggregating (Crogging) using cross-validation to generate and average multiple forecasts, while more recently, combinations of forecasts generated from multiple temporal levels has become popular \citep[][and \S\ref{sec:temporal_aggregation}]{Kourentzes2014-jq,Athanasopoulos2017-ta,kourentzes2019cross_DKB}. These newer approaches recognise the importance of forecast generation in terms of uncertainty reduction \citep{Petropoulos2018-cl}, the creation of diverse forecasts \citep{brown2005diversity_DKB,lemke2010meta_DKB}, and the pooling of forecasts \citep{kourentzes2019another_DKB,lichtendahl2020some_DKB}.

Now nearly 50 years on from the seminal work of \cite{BatesGranger1969_DR}, the evidence that statistical combinations of forecasts improves forecasting accuracy is near unanimous, including evidence from competitions \citep[][and \S\ref{sec:Forecasting_competitions}]{Makridakis1982-co,Makridakis2000-ty,Makridakis2020-mm}, and empirical studies \citep{elliott2005estimation_DKB,jose2008simple_DKB,Andrawis2011-bo,kourentzes2019another_DKB}. Still, researchers have tried to understand why and when combinations improve forecast accuracy \citep{palm1992combine_DKB,Petropoulos2018-cl,Timmermann2006_DR,ATIYA2020197}, and the popularity of the simple average \citep{chan2018some_DKB,smith2009simple_DKB,claeskens2016forecast_DKB}. Others have investigated properties of the distribution of the forecast error beyond accuracy considering issues such as normality, variance, and in out-of-sample performance of relevance to decision making \citep{makridakis1989sampling_DKB,chan1999value_DKB,barrow2016distributions_DKB}.

Looking forward, evidence suggests that the future lies in the combination of statistical and machine learning generated forecasts \citep{Makridakis2020-mm}, and in the inclusion of human judgment \citep[][but also \S\ref{sec:judgmental_forecasting}]{gupta1994managerial_DKB,wang2016select_DKB,Petropoulos2018-mt}. Additionally, there is need to investigate such issues as decomposing combination accuracy gains, constructing prediction intervals \citep{koenker2005quantile_TJ,grushka2020combining_DKB}, and generating combined probability forecasts \citep[][and \S\ref{sec:Density_forecast_combinations}]{Raftery1997-em_ET,ranjan2010combining_DKB,HALL20071_AY,clements2011combining_DKB}. Finally, there is need for the results of combined forecasts to be more interpretable and suitable for decision making \citep{bordignon2013combining_DKB,graefe2014combining_DKB,barrow2016distributions_DKB,Todini2018-dj_ET}.

\subsubsection[Density forecast combinations (Alisa Yusupova)]{Density forecast combinations\protect\footnote{This subsection was written by Alisa Yusupova.}}
\label{sec:Density_forecast_combinations}
Density forecasts provide an estimate of the future probability distribution of a random variable of interest. Unlike point forecasts (and point forecasts supplemented by prediction intervals) density forecasts provide a complete measure of forecast uncertainty. This is particularly important as it allows decision makers to have full information about the risks of relying on the forecasts of a specific model. Policy makers like the Bank of England, the European Central Bank and Federal Reserve Banks in the US routinely publish density forecasts of different macroeconomic variables such as inflation, unemployment rate, and GDP. In finance, density forecasts find application, in particular, in the areas of financial risk management and forecasting of stock returns \citep[see, for example,][\textit{inter alia}]{TayWallis_2000_AY,Berkowitz_2001_AY,GUIDOLIN2006285_AY,SHACKLETON20102678_AY}. The reader is referred to \S\ref{sec:Economics_and_finance} for a discussion of relevant applications.

Initial work on forecast combination focused on the combination of point forecasts (see \S\ref{sec:Forecast_combination_a_brief_review_of_statistical_approaches}). In recent years attention has shifted towards evaluation, comparison and combination of density forecasts with empirical applications that are mostly encountered in the areas of macroeconomics and finance. The improved performance of combined density forecasts stems from the fact that pooling of forecasts allows to mitigate potential misspecifications of the individual densities when the true population density is non-normal. Combining normal densities yields a flexible mixture of normals density which can accommodate heavier tails (and hence skewness and kurtosis), as well as approximate non-linear specifications \citep{HALL20071_AY,Jore_2010_AY}. 

The predictive density combination schemes vary across studies and range from simple averaging of individual density forecasts to complex approaches that allow for time-variation in the weights of prediction models, also called experts \citep[see][for a comprehensive survey of density forecast combination methods]{AastveitMitchellRavazzolo_2019_AY}. A popular approach is to combine density forecasts using a convex combination of experts' predictions, so called `linear opinion pools' \citep[see, for example,][]{HALL20071_AY,Kascha_2010_AY,GEWEKE2011130_AY}. In order to determine the optimal combination weights this method relies on minimising Kullback-Leibler divergence of the true density from the combined density forecast. These linear approaches have been extended by \citet{GneitingRanjan2013_TT} to allow non-linear transformations of the aggregation scheme and by \citet{KAPETANIOS2015150_AY}, whose `generalised pools' allow the combination weights to depend on the (forecast) value of the variable of interest.

\citet{BILLIO2013213_AY} developed a combination scheme that allows the weights associated with each predictive density to be time-varying, and propose a general state space representation of predictive densities and
combination schemes. The constraint that the combination weights must be non-negative and sum to unity implies that linear and Gaussian state-space models cannot be used for inference and instead Sequential Monte Carlo methods (particle filters) are required. More recently, \citet{MCALINN2019155_AY} developed a formal Bayesian framework for forecast combination (Bayesian predictive synthesis) which generalises existing methods. Specifying a dynamic linear model for the synthesis function, they develop a time-varying (non-convex/nonlinear) synthesis of predictive densities, which forms a dynamic latent (agent) factor model. 

For a discussion on methods for evaluating probabilistic forecasts, see \S\ref{sec:evaluating_probabilistic_forecasts} and \S\ref{sec:assessing_the_reliability_of_probabilistic forecats}.

\subsubsection[Ensembles and predictive probability post processors (Ezio Todini)]{Ensembles and predictive probability post processors\protect\footnote{This subsection was written by Ezio Todini.}}
\label{sec:Ensembles_and_predictive_probability_post_ processors}
Improved rational decisions are the final objective of modelling and forecasting. Relatively easy decisions among a number of alternatives with predefined and known outcomes become hard when they are conditioned by future unknown events. This is why one resorts to modelling and forecasting, but this is insufficient. To be successful, one must account for the future conditioning event uncertainty to be incorporated into the decision-making process using appropriate Bayesian approaches (see also \S\ref{sec:bayesian_forecasting}), as described by the decision theory literature \citep{Berger1985-sm_ET,Bernardo1994-is_ET,DeGroot2004-xh_ET}. The reason is that taking a decision purely based on model forecasts is equivalent to assuming the future event (very unlikely, as we know) to equal the forecasted value. Therefore, the estimation of the predictive probability density is the essential prerequisite to estimating the expected value of benefits (or of losses) to be compared and traded-off in the decision-making process \citep{Draper2013-tx_ET}. This highly increases the expected advantages together with the likelihood of success and the robustness of the decision \citep{Todini2017-jm_ET,Todini2018-dj_ET}.

In the past, the assessment of the prediction uncertainty was limited to the evaluation of the confidence limits meant to describe the quality of the forecast. This was done using continuous predictive densities, as in the case of the linear regression, or more frequently in the form of predictive ensembles. These probabilistic predictions, describing the uncertainty of the model forecasts given (knowing) the observations can be used within the historical period to assess the quality of the models \citep{Todini2016-lv_ET}. When predicting into the future, observations are no more available and what we look for (known as predictive probability) is the probability of occurrence of the unknown ``future observations'' given (knowing) the model forecasts. This can be obtained via Bayesian inversion, which is the basis of several uncertainty post-processors used in economy \citep[][and \S\ref{sec:Economics_and_finance}]{Diebold1998-ti_ET}, hydrology \citep[][and \S\ref{sec:Forecasting_and_decision_making_for_floods}]{Krzysztofowicz1999-rm_ET,Todini1999-xc_ET,Todini2008-zr_ET,Schwanenberg2015-eu_ET}, meteorology \citep[][see also \S\ref{sec:Weather_Forecasting}]{Granger2000-oe_ET,Katz2011-ee_ET,Economou2016-ax_ET,Reggiani2019-ik_ET}, etc. Accordingly, one can derive a predictive probability from a single model forecast to estimate the expected value of a decision by integrating over the entire domain of existence all the possible future outcomes and theirs effects, weighted with their appropriate probability of occurrence.

When several forecasts are available to a decision maker, the problem of deciding on which of them one should rely upon becomes significant. It is generally hard to choose among several forecasts because one model could be best under certain circumstances but rather poor under others. Accordingly, to improve the available knowledge on a future unknown event, predictive densities are extended to multiple forecasts to provide decision makers with a single predictive probability, conditional upon several model's forecasts \citep[][see also \S\ref{sec:Forecast_combination_a_brief_review_of_statistical_approaches} and \S\ref{sec:Density_forecast_combinations}]{Raftery1997-em_ET,Coccia2011-oy_ET}.

A number of available uncertainty post processors can cope with multi-model approaches, such as Bayesian model averaging \citep{Raftery1993-wj_ET,Raftery1997-em_ET}, model output statistics \citep{Glahn1972-cc_ET,Wilkd2005-ez_ET}, ensemble model output statistics \citep{Gneiting2005-zr_ET}, and model conditional processor \citep{Todini2008-zr_ET,Coccia2011-oy_ET}. 

Finally, important questions such as: (\textit{i}) ``what is the probability that an event will happen within the next $x$ hours?'' and (\textit{ii}) ``at which time interval it will most likely occur?'' can be answered using a multi-temporal approach \citep[][see also \S\ref{sec:temporal_aggregation}]{Krzysztofowicz2014-hv_ET,Coccia2011-fx_ET} and results of its applications were presented in \cite{Coccia2011-fx_ET}, \cite{Todini2017-jm_ET}, and \cite{Barbetta2017-sd_ET}. 

\subsubsection[The wisdom of crowds (Yael Grushka-Cockayne)] {The wisdom of crowds\protect\footnote{This subsection was written by Yael Grushka-Cockayne.}}
\label{sec:Wisdom_of_crowds}
Multiple experts' forecasts are collected in a wide variety of situations: medical diagnostics, weather prediction, forecasting the path of a hurricane, predicting the outcome of an election, macroeconomic forecasting, and more. One of the central findings from the forecasting literature is that there is tremendous power in combining such experts' forecasts into a single forecast. The simple average, or what Surowiecki refers to as `the wisdom of crowds' \citep{Surowiecki2005-ty_YGC}, has been shown to be a surprisingly robust combined forecast in the case of point forecasting \citep[][and \S\ref{sec:Forecast_combination_a_brief_review_of_statistical_approaches}]{clemen1986combining_DKB,Clemen1989-fo_YGC,Scott_Armstrong2001-bl_YGC}. The average forecast is more accurate than choosing a forecast from the crowd at random and is sometimes even more accurate than nearly all individuals \citep{Mannes2012-gw_YGC}. The average point forecast also often outperforms more complicated point aggregation schemes, such as weighted combinations \citep{Smith2009-lm_YGC,Soule2020-kg_YGC}.

\cite{Mannes2012-gw_YGC} highlight two crucial factors that influence the quality of the average point forecast: individual expertise and the crowd's diversity. Of the two: ``The benefits of diversity are so strong that one can combine the judgments from individuals who differ a great deal in their individual accuracy and still gain from averaging'' \citep[page 234]{Mannes2012-gw_YGC}.

\cite{Larrick2006-rc_YGC} define the idea of `bracketing': In the case of averaging, two experts can either bracket the realisation (the truth) or not. When their estimates bracket, the forecast generated by taking their average performs better than choosing one of the two experts at random; when the estimates do not bracket, averaging performs equally as well as the average expert. Thus, averaging can do no worse than the average expert, and with some bracketing, it can do much better. Modern machine learning algorithms such as the random forest exploit this property by averaging forecasts from hundreds of diverse experts \citep[here, each ``expert'' is a regression tree;][]{Grushka-Cockayne2017-cy_YGC}.

Only when the crowd of forecasts being combined has a high degree of dispersion in expertise, some individuals in the crowd might stand out, and in such cases, there could be some benefits to chasing a single expert forecaster instead of relying on the entire crowd. \cite{Mannes2014-yn_YGC} suggest that combining a small crowd can be especially powerful in practice, offering some diversity among a subset of forecasters with an minimum level of expertise.

When working with probabilistic forecasting (see also \S\ref{sec:Density_forecast_combinations}, \S\ref{sec:Ensembles_and_predictive_probability_post_ processors}, and \S\ref{sec:evaluating_probabilistic_forecasts}), averaging probabilities is the most widely used probability combination method \citep{Cooke1991-en_YGC,Hora2004-dv_YGC,Clemen2008-te_YGC}. \cite{Stone1961-it_YGC} labelled such an average the linear opinion pool. \cite{OHagan2006-hl_YGC} claimed that the linear opinion pool is: ``hard to beat in practice''.

Although diversity benefits the average point forecast, it can negatively impact the average probability forecast. As the crowd's diversity increases, the average probability forecast becomes more spread out, or more underconfident \citep{Dawid1995-zz_YGC,Hora2004-dv_YGC,ranjan2010combining_DKB}. Averaging quantiles, instead of probabilities, can offer sharper and better calibrated forecasts \citep{Lichtendahl2013-ci_YGC}. Trimmed opinion pools can be applied to probability forecasts, also resulting in better calibrated forecasts \citep[][see also \S\ref{sec:assessing_the_reliability_of_probabilistic forecats}]{Jose2014-gm_YGC}.

The ubiquity of data and the increased sophistication of forecasting methods results in more use of probabilistic forecasts. While probabilities are more complex to elicit, evaluate, and aggregate compared to point estimates, they do contain richer information about the uncertainty of interest. The wisdom of combining probabilities, however, utilises diversity and expertise differently than combining point forecasts. When relying on a crowd, eliciting point forecasts versus eliciting probabilities can significantly influence the type of aggregation one might choose to use.

\subsection{Data-driven methods}
\label{sec:data_driven_approaches}

\subsubsection[Forecasting with big data (Jennifer L. Castle)]{Forecasting with big data\protect\footnote{This subsection was written by Jennifer L. Castle.}}
\label{sec:Forecasting_with_Big_Data}
The last two decades have seen a proliferation of literature on forecasting using big data \citep{JLC_Varian14,JLC_SwanXiong18,JLC_HassaniSilva15} but the evidence is still uncertain as to whether the promised improvements in forecast accuracy can systematically be realised for macroeconomic phenomena. In this section we question whether big data will significantly increase the forecast accuracy of macroeconomic forecasts. \cite{JLC_NBERc14009} argues that machine learning methods are seen as an efficient approach to dealing with big data sets, and we present these methods before questioning their success at handling non-stationary macroeconomic data that are subject to shifts. \S\ref{sec:Forecasting_on_distributed_systems} discusses big data in the context of distributed systems, and \S\ref{sec:machine_Learning_with_very_noisy_data} evaluates a range of machine learning methods frequently applied to big data.

The tools used to analyse big data focus on regularization techniques to achieve dimension reduction, see \cite{JLC_KimSwanson14} for a summary of the literature. This can be achieved through selection \citep[such as \emph{Autometrics},][but also see \S\ref{sec:Variable_Selection} and \S\ref{sec:Model_Selection}]{JLC_Door07Auto}, shrinkage (including Ridge Regression, LASSO, and Elastic Nets, see \S\ref{sec:machine_Learning_with_very_noisy_data} but also \S\ref{sec:Forecasting_stock_returns} for an applied example), variable combination (such as Principal Components Analysis and Partial Least Squares), and machine learning methods (including Artificial Neural Networks, see \S\ref{sec:neural_networks}). Many of these methods are `black boxes' where the algorithms are not easily interpretable, and so they are mostly used for forecasting rather than for policy analysis. 

Big data has been effectively used in nowcasting, where improved estimates of the forecast origin lead to better forecasts, absent any later shifts. Nowcasting can benefit from large data sets as the events have happened and the information is available, see \cite{NowCast18} for a nowcasting application, and \S\ref{sec:Leading_indicators_and_Granger_causality} on leading indicators. However, the benefits of big data are not as evident in a forecasting
context where the future values of all added variables also need to be forecast and are as uncertain as the variable(s) of interest.

Macroeconomic time series data are highly non-stationary, with stochastic trends and structural breaks. The methods of cross-validation and hold-back, frequently used to handle bid data, often assume that the data generating process does not change over time. Forecasting models that assume the data are drawn from a stationary distribution (even after differencing) do not forecast well \emph{ex ante}. So while there seems to be lots of mileage in improving forecasts using big data, as they allow for more flexible models that nest wider information sets, more general dynamics and many forms of non-linearity, the statistical problems facing `small' data forecasting models do not disappear \citep{JLC_Harford14,JLC_DoornikHendry15}. \cite{JLC_CastleDoornikHendry20} do not find improvements in forecasting from big data sets over small models. It is essential to keep in mind the classical statistical problems of mistaking correlation for causation, ignoring sampling biases, finding excess numbers of false positives and not handling structural breaks and non-constancies both in- and out-of-sample, in order to guard against these issues in a data abundant environment.

\subsubsection[Forecasting on distributed systems (Xiaoqian Wang)]{Forecasting on distributed systems\protect\footnote{This subsection was written by Xiaoqian Wang.}}
\label{sec:Forecasting_on_distributed_systems}
Big data is normally accompanied by the nature that observations are indexed by timestamps, giving rise to big data time series characterised by high frequency and long-time span. Processing big data time series is obstructed by a wide variety of complications, such as significant storage requirements, algorithms' complexity and high computational cost \citep{l2017machine_XW,wang2018novel_XW,galicia2018novel_XW,wang2020distributed_XW}. These limitations accelerate the great demand for scalable algorithms. Nowadays, increasing attention has been paid to developing data mining techniques on distributed systems for handling big data time series, including but not limited to processing \citep{mirko2013hadoop_XW}, decomposition \citep{bendre2019time_XW}, clustering \citep{ding2015yading_XW}, classification \citep{triguero2015mrpr_XW}, and forecasting \citep{galicia2018novel_XW}. For forecasting problems based on big data sets and/or large sets of predictors, please refer to \S\ref{sec:Forecasting_with_Big_Data} and \S\ref{sec:Forecasting_stock_returns}.

Distributed systems, initially designed for independent jobs, do not support to deal with dependencies among observations, which is a critical obstacle in time series processing \citep{li2014rolling_XW,wang2020distributed_XW}. Various databases (e.g., InfluxDB\footnote{Available at \url{https://www.influxdata.com/time-series-database/}},
OpenTSDB\footnote{Available at \url{http://opentsdb.net/}}, RRDtool\footnote{Available at \url{https://oss.oetiker.ch/rrdtool/}}, and Timely\footnote{Available at \url{https://code.nsa.gov/timely/}}) can function as storage platforms
for time series data. However, none of these databases supports advanced analysis, such as modelling, machine learning algorithms and forecasting. Additional considerations are therefore required in further processing time series on distributed systems. \citet{mirko2013hadoop_XW} developed the Hadoop.TS library for processing large-scale time series by creating a time series bucket. \citet{li2014rolling_XW} designed an index pool serving as a data structure for assigning index keys to time series entries, allowing time series data to be sequentially stored on HDFS \citep[Hadoop Distributed File System:][]{shvachko2010hadoop_XW} for MapReduce \citep{dean2008mapreduce_XW} jobs. \citet{chen2019periodicity_XW} proposed a data compression and abstraction method for large-scale time series to facilitate the periodicity-based time series prediction in a parallel manner.

The evolution of the algorithms for efficiently forecasting big data time series on distributed systems is largely motivated by a wide range of applications including meteorology, energy, finance, transportation and farming \citep{galicia2018novel_XW,chen2019periodicity_XW,hong2019energy_XW,sommer2020online_XW}. Researchers have made several attempts to make machine learning techniques available for big data time series forecasting on distributed systems \citep{li2014rolling_XW,talavera2016nearest_XW,galicia2019multi_XW,xu2020distributed_XW}.
\citet{talavera2016nearest_XW} presented a nearest neighbours-based algorithm implemented for Apache Spark \citep{zaharia2016apache_XW} and achieved satisfactory forecasting performance. \citet{galicia2018novel_XW} proposed a scalable methodology which enables Spark's MLlib \citep{meng2016mllib_XW} library to conduct multi-step forecasting by splitting the multi-step forecasting problem into $h$ sub-problems ($h$ is the forecast horizon).

Another strand of the literature on forecasting big data time series is to improve time-consuming estimation methods using a MapReduce framework. \citet{sheng2013extended_XW} learned the parameters of echo state networks for time series forecasting by designing a parallelised extended Kalman filter involving two MapReduce procedures. Recently, \citet{sommer2020online_XW} accurately estimated coefficients of a high-dimensional ARX model by designing two online distributed learning algorithms. \citet{wang2020distributed_XW} resolved challenges associated with forecasting ultra-long time series from a new perspective that global estimators are approximated by combining the local estimators obtained from subseries by minimising a global loss function. Besides, inspired by the \emph{no-free-lunch} theorem \citep{wolpert1997no_XW}, model selection (see \S\ref{sec:Model_Selection}) and model combination (see \S\ref{sec:combining_forecasts}) are involved in finalisation of algorithms for forecasting on distributed systems \citep[e.g.,][]
{li2014rolling_XW,galicia2019multi_XW,bendre2019time_XW,xu2020distributed_XW}.

\subsubsection[Agent-based models (Thiyanga S. Talagala)]{Agent-based models\protect\footnote{This subsection was written by Thiyanga S. Talagala.}}
\label{sec:Agent_based_models}
Time series forecasting involves use of historical data to predict values for a specific period time in future. This approach assumes that recent and historical patterns in the data will continue in the future. This assumption is overly ingenuous. However, this is not reliable in some situations. For example, (\textit{i})~forecasting COVID-19 cases (see also \S\ref{sec:Pandemics}) where, due to interventions and control measures taken by the governments and due to the change in personal behaviour, the disease transmission pattern changes rapidly, and (\textit{ii})~forecasting sales of a new product (see also \S\ref{sec:New_product_forecasting}): external factors such as advertisement, promotions (see \S\ref{sec:Promotional_forecasting}), social learning, and imitation of other individuals change the system behaviour.

In such circumstances to make reliable forecasts it is important to take into account all information that might influence the variable that is being forecast. This information includes a variety of environmental-level and individual-level factors. An agent-based modelling is a powerful tool to explore such complex systems. Agent-based modelling approach is useful when, (\textit{i})~data availability is limited, (\textit{ii})~uncertainty of various interventions in place and a rapidly changing social environment, and (\textit{iii})~limited understanding of the dynamics of the variable of interest.

Agent-based modelling disaggregates systems into individual level and explores the aggregate impact of individual behavioural changes on the system as a whole. In other words, the key feature of agent-based modelling is the bottom-up approach to understand how a system's complexity arises, starting with individual level (see also \S\ref{sec:Cross_sectional_hierarchical_forecasting}). As opposed to this, the conventional time series forecasting approaches are considered top-down approaches.

Agent-based models have two main components: (\textit{i})~Agents, and (\textit{ii})~Rules, sometimes referred as procedures and interactions. Agents are individuals with autonomous behaviour. Agents are heterogeneous. Each agent individually assesses on the basis of a set of rules. An agent-based modelling approach simulates how heterogeneous agents interact and behave to assess the role of different activities on the target variable. According to \cite{farmer2009economy}, ``An agent-based model is a computerised simulation of a number of decision-makers (agents) and institutions, which interact through prescribed rules''. Their paper highlights the importance of adopting agent-based models as a better way to help guide financial policies.

\begin{figure}[ht!]
\begin{center}
\includegraphics[width=0.65\linewidth]{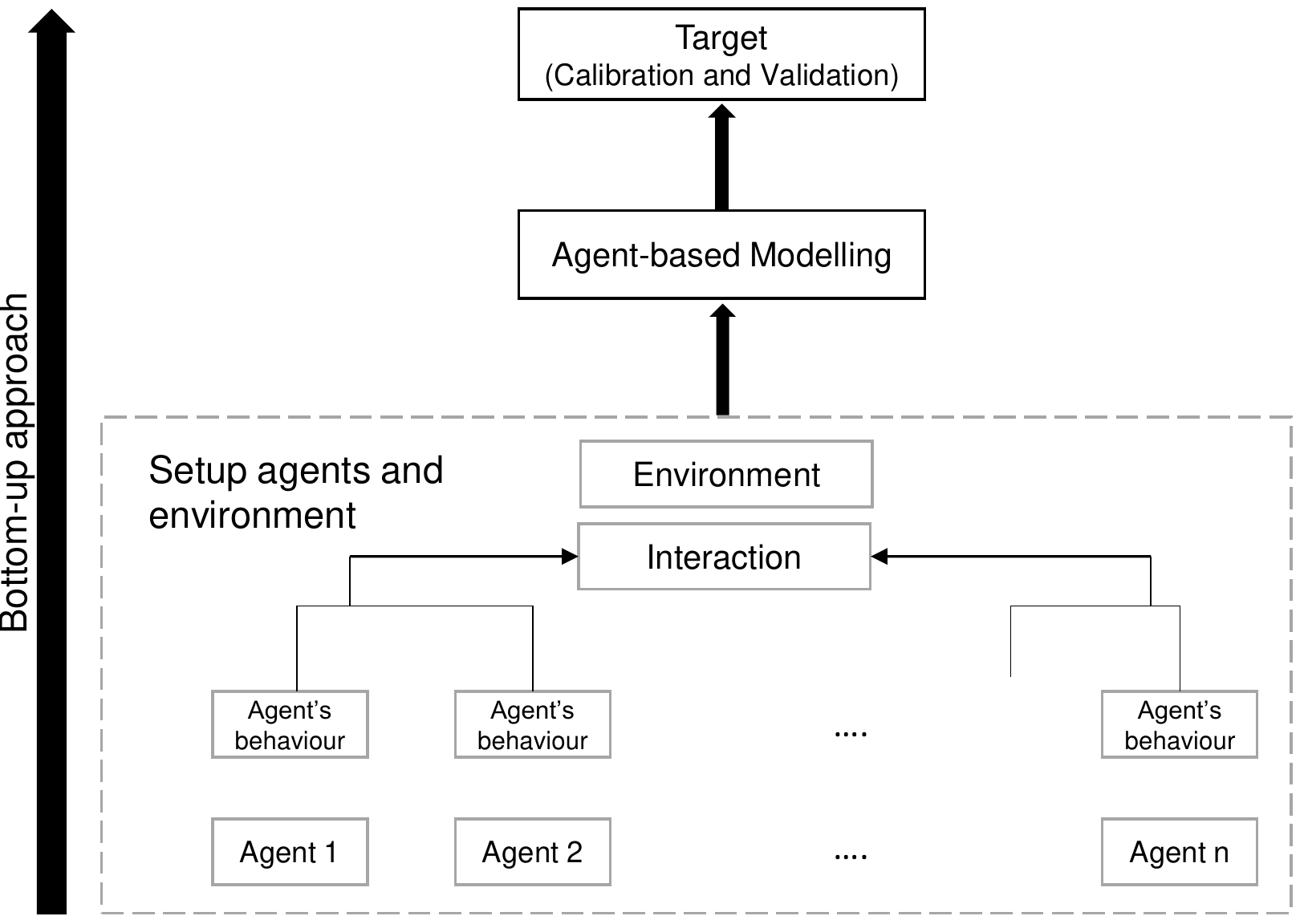}
\caption{Framework for Agent-based modelling.}
\label{fig:ABM_Thiyanga}
\end{center}
\end{figure}

A general framework for Agent-based modelling involves three main stages (See Figure~\ref{fig:ABM_Thiyanga}): (\textit{i})~setup environments and agents, (\textit{ii})~agent-based modelling, and (\textit{iii})~calibration and validation. The first two steps are self-explanatory. The final step involves calibration of the model with empirical data and then evaluates whether the agent-based model mirrors the real-world system/target. The validation step involves testing the significance of the difference between agent-based model results and real data collected about the target. One of the main challenges in designing an agent-based model is finding a balance between model simplicity and model realism (see also \S\ref{sec:Model_complexity}). The KISS principle (keep it simple, stupid), introduced by \cite{axelrod1997advancing_Thiyanga} is often cited as an effective strategy in agent-based modelling. A high level of expertise in the area of the subject is necessary when developing an agent-based model.

Despite these limitations and challenges, agent-based modelling has been used extensively to model
infectious disease transmission and forecasting \citep{tracy2018agent_Thiyanga,venkatramanan2018using_Thiyanga}.
Agent-based modelling approaches have been widely used in early phases of the COVID-19 outbreak, to assess the impact of different interventions on disease spread and forecasts \citep{wallentin2020covid_Thiyanga}. In a review paper, \cite{weron2014electricity_FZ} states some applications of agent-based models for electricity demand forecasting. \cite{xiao2016forecasting_Thiyanga} use agent-based models to forecast new product diffusion. Furthermore, thinking the other way around, \cite{hassan2013asking_Thiyanga} explain how forecasting principles can be applied in agent-based modelling.
 
\subsubsection[Feature-based time series forecasting (Yanfei Kang)]{Feature-based time series forecasting\protect\footnote{This subsection was written by Yanfei Kang.}}
\label{sec:Feature_based_time_series_forecasting}
A time series \emph{feature} can be any statistical representation of time series characteristics. A vast majority of time series mining tasks are based on similarity quantification using their feature representations, including but not limited to time series clustering \citep[][and \S\ref{sec:clustering_based_forecasting}]{Wang2006_YK,Kang2014-ik_YK,Kang2015-yj_YK}, classification \citep{Nanopoulos2001-ky_YK,Fulcher2013_YK}, anomaly detection \citep[][and \S\ref{sec:Anomaly_detection_and_time_series_forecasting}]{Yanfei_Kang2012-sr_YK,talagala2020anomaly_PDT}, and forecasting \citep[][see also \S\ref{sec:Exogenous_variables_and_feature_engineering}]{Kang2017_YK,Montero-Manso2020-et}. The choice of features depends on the nature of the data and the application context. The state-of-the-art time series feature representation methods quantify a wide range of time series characteristics, including simple summary statistics, stationarity \citep{Montero-Manso2020-et,Wang2020_YK}, model fits \citep{Fulcher2014_YK,Christ2018_YK}, time series imaging \citep{Li2020_YK}, and others. In the forecasting community, two lines of forecasting approaches have been developed using time series features, namely feature-based model selection and feature-based model combination. The motivation behind them is no single model always performs the best for all time series. Instead of choosing one model for all the data, features can be used to obtain the most appropriate model or the optimal combination of candidate models, per series. 

As early as in 1972, \citet{Reid1972_YK} argues that time series characteristics provide valuable information in forecast model selection, which is further echoed by \citet{makridakis1979accuracy_YK}. One way to forecast an extensive collection of time series is to select the most appropriate method per series according to its features. Pioneer studies focus on rule-based methods \citep[for example,][]{Arinze1994_YK,Wang2009_YK} to recommend the ``best'' forecasting model per series based on its features. Another line of approaches apply regression to study how useful features are in predicting which forecasting method performs best \citep[for example,][]{Meade2000_YK,Petropoulos2014_YK}. With the advancement of machine learning (see also \S\ref{sec:machine_Learning}), more recent literature uses ``meta-learning'' to describe the process of automatically acquiring knowledge for forecast model selection. The first such study is by \citet{prudencio2004meta_YK}, who apply decision trees for forecast model selection. \citet{lemke2010meta_DKB} compare different meta-learning approaches to investigate which model works best in which situation. \citet{Kang2017_YK} propose using feature spaces to visualise the strengths and weaknesses of different forecasting methods. Other algorithms such as neural networks (see also \S\ref{sec:neural_networks}) and random forecasts are also applied to forecast model selection \citep{kuck2016meta_YK,Talagala2018_YK}.

One of the pioneering studies in feature-based forecast combination is the rule-based approach by \citet{Collopy1992_YK}, who develop 99 rules for forecast combination based on 18 features. Recently, \citet{Kang2020a_YK} use 26 features to predict the performances of nine forecasting methods with nonlinear regression models, and obtain the combination weights as a tailored softmax function of the predicted forecasting errors. The feature-based forecast model averaging (FFORMA) framework proposed by \citet{Montero-Manso2020-et} employ 42 features to estimate the optimal combination weights of nine forecasting methods based on extreme gradient boosting \citep[XGBoost, ][]{Chen2016_YK}. \citet{Li2020_YK} first transform time series into images, and use features extracted from images to estimate the optimal combination weights. For feature-based interval forecasting, \citet{Wang2020_YK} investigate how time series features affect the relative performances of prediction intervals from different methods, and propose a general feature-based interval forecasting framework to provide reliable forecasts and their uncertainty estimation.

\subsubsection[Forecasting with bootstrap (Clara Cordeiro)]{Forecasting with bootstrap\protect\footnote{This subsection was written by Clara Cordeiro.}}
\label{sec:Forecasting_with_bootstrap}
The bootstrap methodology has been widely applied in many areas of research, including time series analysis. The bootstrap procedure \citep{Efron1979-mr} is a very popular methodology for independent data because of its simplicity and nice properties. It is a computer-intensive method that presents solutions in situations where the traditional methods fail or are very difficult to apply. However, Efron's bootstrap (iid bootstrap) has revealed itself inefficient
in the context of dependent data, such as in the case of time series, where the dependence structure arrangement has to be kept during the resampling scheme. 

Most of the resampling for dependent data consider segments of the data to define blocks, such that the dependence structure within each block can be kept. Different versions of blocking differ in the way as blocks are constructed: the nonoverlapping block bootstrap \citep{carlstein1990resampling_CC}, the moving block bootstrap \citep{kunsch1989jackknife_CC}, the circular block bootstrap \citep{politis1992circular_CC}, and the stationary block bootstrap \citep{politis1994stationary_CC}. But, if the time series process is driven from iid innovations, another way of resampling can be used. 

The iid Bootstrap can then be easily extended to a dependent setup. That was the spirit of sieve bootstrap proposed by \cite{buhlmann1997sieve_CC}. This method is based on the idea of fitting parametric models first and resampling from the residuals. Such models include, for example, the linear regression \citep{freedman1981bootstrapping_CC} and autoregressive time series \citep{efron1986bootstrap_CC}. This approach is different from the previous bootstrap methods for dependent data; the sample bootstrap is (conditionally) stationary and does not present a structure of dependence. Another different feature is that the sieve bootstrap sample is not a subsample from the original data, as in the previous methods. Observe that even if the sieve bootstrap is based on a parametric model, it is nonparametric in its spirit. The AR model (see \S\ref{sec:autoregressive_integrated_moving_average_models}) here is just used to filter the residuals series. 

A few years ago, the sieve bootstrap was used for estimating forecast intervals \citep{zagdanski2001prediction_CC, andre2002forecasting_CC}. Motivated by these works, \cite{cordeiro2006bootstrap_CC,cordeiro2009forecasting_CC,cordeiro2010boot_CC} developed a procedure to estimate point forecasts. The idea of these authors was to fit an exponential smoothing model (see \S\ref{sec:exponential_smoothing_models}) to the time series, extract the residuals and then apply the sieve bootstrap to the residuals. Further developments of this procedure include the estimation of forecast intervals \citep{cordeiro2014forecast_CC} and also the detection, estimation and imputation of missing data \citep{cordeiro2013predicting_CC}. In a recent work \citep{Bergmeir2016-su} a similar approach was also consider, the residuals were extracted and resampled using moving block bootstrap (see \S\ref{sec:Bagging_for_Time_Series} for further discussion).

\cite{bickel1981some_CC} and later in \cite{angus1992asymptotic_CC} showed that in extreme value theory, the bootstrap version for the maximum (or minimum) does not converge to the extremal limit laws. \cite{zelterman1993semiparametric_CC} pointed out that ``to resample the data for approximating the distribution of the $k$ largest observations would not work because the `pseudo-samples' would never have values greater than $X_{n:n}$''\footnote{max$(X_1,\cdots,X_n)$.}. A method considering to resample a smaller size than the original sample was proposed in \cite{hall1990using_CC}. Recently, \cite{neves2020modelling_CC} used this idea and developed a preliminary work in modelling and forecasting extremes in time series.

\subsubsection[Bagging for time series forecasting (Fernando Luiz Cyrino Oliveira)]{Bagging for time series forecasting\protect\footnote{This subsection was written by Fernando Luiz Cyrino Oliveira.}}
\label{sec:Bagging_for_Time_Series}
The term \textit{bagging} was proposed by \cite{Breiman1996-mg} to describe the generation of several versions of a predictor, via Bootstrap procedures (introduced in \S\ref{sec:Forecasting_with_bootstrap}), with a final aggregation stage. Thus, ``\textbf{b}ootstrap \textbf{agg}regat\textbf{ing}'' was established as bagging. The main idea is to improve predictors' accuracy once the data sets, draw randomly with replacement, will approximating the original distribution. The author argues that bagging works for unstable procedures, but it was not tested for time series. Years after, \cite{Kilian2004-he} suggested the first attempts for temporal dependent data. For data-driven methods, to forecasting and simulation time series and deal with predictors ensembles, bagging has shown as a powerful tool.

A general framework for ensemble forecasting methods involves four main stages: (\textit{i})~data treatment, (\textit{ii})~resampling, (\textit{iii})~forecasting, and (\textit{iv})~aggregation. However, for time series, bootstrap should be done carefully, as the serial dependence and non-stationarity must be considered. 

As mentioned in \S\ref{sec:Forecasting_with_bootstrap}, this led \cite{Bergmeir2016-su} to propose a bagging version for exponential smoothing, the Bagged ETS. As pre-treatment, after a Box-Cox transformation, the series is decomposed into trend, seasonal, and remainder components via STL decomposition \citep{ad_cleveland1990stl}. The resampling stage uses moving block bootstrap \citep[MBB:][]{Lahiri2003-ne}, applied to the remainder. There are several discussions in the literature about this procedure, mainly regarding the size of the blocks. MBB resampling the collection of overlapping (consecutive) blocks of observations. The idea is to keep the structure still present in the remainder. The forecasts are obtained via ETS methods (see \S\ref{sec:exponential_smoothing_models}) and, for the final aggregation, the authors adopted the median. Their method is evaluated on the M3 data set and outperformed the original benchmarks. The work of \cite{Bergmeir2016-su} inspired many others: \cite{Dantas2017-su} applied the idea for air transport demand data and \cite{De_Oliveira2018-bg} for energy consumption, proposing the so-called remainder sieve bootstrap (RSB).

\cite{Dantas2018-lv} proposed an extension to the Bagged ETS where bagging and exponential smoothing are combined with clustering methods (clustering-based forecasting methods are discussed in \S\ref{sec:clustering_based_forecasting}). The approach aims to consider and reduce the covariance effects among the ensemble time series, creating clusters of similar forecasts – since it could impact the variance of the group. A variety of forecasts are selected from each cluster, producing groups with reduced variance. 

In light of the aforementioned, there are several possibilities for each stage of the mentioned framework. In this context, to investigate the reasons why bagging works well for time series forecasting, \cite{Petropoulos2018-cl} explored three sources of uncertainty: model form, data, and parameter. While arguably bagging can handle all of them, \cite{Petropoulos2018-cl} showed that simply tackling model uncertainty is enough for achieving a superior performance, leading to the proposal of a Bootstrap Model Combination (BMC) approach, where different model forms are identified in the ensemble and fitted to the original data.

Finally, \cite{De_Oliveira2020} proposed ``treating and pruning'' strategies to improve the performance of prediction intervals for both model selection and forecast combinations. Testing over a large set of real time series from the M forecasting competitions (see also \S\ref{sec:Forecasting_competitions}), their results highlighted the importance of analysing the prediction intervals of the ensemble series before the final aggregation.

\subsubsection[Multi-step ahead forecasting (Souhaib Ben Taieb)]{Multi-step ahead forecasting\protect\footnote{This subsection was written by Souhaib Ben Taieb.}}
\label{sec:multi_step_ahead_forecasting}

Given a univariate time series comprising $n$ observations, $y_1, y_2,\dots, y_n$, multi-step ahead point forecasting involves producing point estimates of the $H$ future values $y_{n+1}, y_{n+2},\dots, y_{n+H}$, where $H > 1$, is the forecast horizon \citep{Ben_Taieb2014-kb_SBT}.

The (naive) recursive strategy estimates a one-step-ahead autoregressive model to predict $y_{t+1}$ from $y_t, y_{t-1}, \dots$, by minimising the one-step-ahead forecast errors. Each forecast is then obtained dynamically by iterating the model $H$ times, and by plugging in the missing lagged values with their respective forecasts. The direct strategy builds separate $h$-step-ahead models to predict $y_{t+h}$ from $y_t, y_{t-1}, ...$ for $h = 1, 2, \dots, H$, by minimising $h$-step-ahead forecast errors, and forecasts are computed directly by the estimated models.

In theory, with linear models, model misspecification plays an important role in the relative performance between the recursive and direct strategy \citep{Chevillon2007-dr_SBT}. If the model is correctly speciﬁed, the recursive strategy benefits from more efficient parameter estimates, while the direct strategy is more robust to model misspecification. With nonlinear models, recursive forecasts are known to be asymptotically biased \citep{Lin1994-td_SBT, Fan2005-hs_SBT, Terasvirta2010-ea_SBT}, and the direct strategy is often preferred over the recursive strategy since it avoids the accumulation of forecast errors. In practice, the results are mixed \citep{Atiya1999-vg_SBT, Kline2004-ms_SBT, Marcellino2006-bf_SBT,Sorjamaa2007-ry_SBT, Pesaran2011-px_SBT}, and depend on many interacting factors including the model complexity (see also \S\ref{sec:Model_complexity}), the (unknown) underlying data generating process, the number of observations, and the forecast horizon (see also \S\ref{sec:Feature_based_time_series_forecasting}).

Hybrids and variants of both recursive and direct strategies have been proposed in the literature. For example, one of the hybrid strategies \citep{Zhang1994-ie_SBT, Sorjamaa2006-cc_SBT, Zhang2013-zx_SBT} first produce recursive forecasts, then adjust these forecasts by modelling the multi-step forecast errors using a direct strategy \citep{Ben_Taieb2014-vl_SBT}. Variants of the recursive strategy match the model estimation and forecasting loss functions by minimising the implied $h$-step-ahead recursive forecast errors \citep{McNames1998-ih_SBT,Bontempi1999-li_SBT,Bhansali2002-fy_SBT}.  Variants of the direct strategy exploit the fact that the errors of different models are serially correlated \citep{Lee2003-xa_SBT, Chen2004-cg_SBT, Franses2009-yc_SBT, Pesaran2011-px_SBT}. The idea is to reduce the forecast variance of independently selected models by exploiting the relatedness between the forecasting tasks, as in multi-task learning \citep{Caruana1997-tg_SBT}. For example, a multi-horizon strategy will measure forecast accuracy (see \S\ref{sec:point_forecast_accuracy_measures}) by averaging the forecast errors over multiple forecast horizons \citep{Kline2004-ms_SBT, Bontempi2011-lb_SBT}. Different multi-horizon strategies can be speciﬁed, with different formulation of the objective function \citep{Ben_Taieb2010-jh_SBT}. One particular case is the multi-output strategy which estimates a single model for all horizons by minimising the average forecast error over the entire forecast horizon \citep{Bontempi2011-lb_SBT}.

Forecasting strategies are often model-dependent, especially with machine learning models (see \S\ref{sec:machine_Learning}). Furthermore, model architecture and parameters are often trained by taking into account the chosen forecasting strategy. For example, we can naturally implement and train recursive forecasting models using recurrent neural networks (see also \S\ref{sec:neural_networks}). Also, different specifications of the decoder in sequence-to-sequence models will induce different forecasting strategies, including variants of direct and multi-horizon strategies. For more details, we refer the reader to \cite{Hewamalage2019Recurrent_CB} and Section 4.2 in \cite{Benidis2020-tn_SBT}.

Which forecasting strategy is best is an empirical question since it involves a tradeoff between forecast bias and variance \citep{Ben_Taieb2015-fm_SBT, Ben_Taieb2012-aq_SBT}. Therefore, the forecasting strategy should be part of the design choices and the model selection procedure of any multi-step-ahead forecasting model.

\subsubsection[Neural networks (Georgios Sermpinis)]{Neural networks\protect\footnote{This subsection was written by Georgios Sermpinis.}}
\label{sec:neural_networks}
Neural Networks (NNs) or Artificial Neural Networks (ANNs) are mathematical formulations inspired by the work and functioning of biological neurons. They are characterized by their ability to model non-stationary, nonlinear and high complex datasets. This property along with the increased computational power have put NNs in the frontline of research in most fields of science \citep{Gooijer2006-hl_GS,Zhang1998-bv_GS}.

A typical NN topology is consisted by three types of layers (input, hidden and output) and each layer is consisted by nodes. The first layer in every NN, is the input layer and the number of its nodes corresponds to the number of explanatory variables (inputs). The last layer is the output layer and the number of nodes corresponds to the number of response variables (forecasts). Between the input and the output layer, there is one or more hidden layers where the nodes define the amount of complexity the model is capable of fitting. Most NN topologies in the input and the first hidden layer contain an extra node, called the bias node. The bias node has a fixed value of one and serves a function similar to the intercept in traditional regression models. Each node in one layer has connections (weights) with all or a subset (for example, for the convolutional neural network topology) of the nodes of the next layer.

NNs process the information as follows: the input nodes contain the explanatory variables. These variables are weighted by the connections between the input and the first hidden nodes, and the information reaches to the hidden nodes as a weighted sum of the inputs. In the hidden nodes, there is usually a non-linear function (such as the sigmoid or the RelU) which transform the information received. This process is repeated until the information reaches the output layer as forecasts. NNs are trained by adjusting the weights that connect the nodes in a way that the network maps the input value of the training data to the corresponding output value. This mapping is based on a loss function, the choice of which depends on the nature of the forecasting problem. The most common NN procedure, is the back-propagation of errors (for additional details on training, see \S\ref{sec:machine_Learning_with_very_noisy_data}). 

The simpler and most common NN topology, is the Multilayer Forward Perceptron (MLP). In MLP, the hidden nodes contain the sigmoid function and the information moves in forward direction (from the inputs to the output nodes). An another NN topology where the information moves also only in a forward direction is the Radial Basis Function NN (RBF). Now the hidden neurons compute the Euclidean distance of the test case from the neuron's centre point and then applies the Gaussian function to this distance using the spread values. Recurrent Neural Networks (RNNs) are NN topologies that allow previous outputs to be used as inputs while having hidden states. The information moves both forwards and backwards. RNNs have short-term memory and inputs are taken potentially from all previous values. MLPs, RBFs and RNNs are universal function approximators \citep{Hornik1991-ju_GS,Schafer2006-wc_GS,Park1991-uv_GS}. However, the amount of NN complexity in terms of hidden layers and nodes to reach this property, might make the NN topology computationally unfeasible to train (see also the discussion in \S\ref{sec:machine_Learning_with_very_noisy_data}). For the interaction of NNs with the probability theory, we refer the reader to last part of \S\ref{sec:Deep_Probabilistic_Forecasting_models}.
 
\subsubsection[Deep probabilistic forecasting models (Tim Januschowski)]{Deep probabilistic forecasting models\protect\footnote{This subsection was written by Tim Januschowski.}}
\label{sec:Deep_Probabilistic_Forecasting_models}
Neural networks (\S\ref{sec:neural_networks}) can be equipped to provide not only a single-valued forecast, but rather the entire range of values possible in a number of ways (see also \S\ref{sec:Density_forecast_combinations} and \S\ref{sec:Ensembles_and_predictive_probability_post_ processors}). We will discuss three selected approaches in the following, but remark that this is a subjective selection and is by far not comprehensive.\footnote{For example, contemporary topics in machine learning such as generative adversarial networks can be naturally lifted to forecasting and similarly, more traditional probabilistic machine learning approaches such as Gaussian Processes \citep{maddix2018deep_TJ}. We ignore the important area of Bayesian deep learning \citep[see][for a survey]{wang2016survey_TJ} entirely here for lack of space.}

\begin{enumerate}[noitemsep]
 \item Analogously to linear regression and Generalised Linear Models, obtaining probabilistic forecasts can be achieved by the neural network outputting not the forecasted value itself but rather parameters of probability distribution or density \citep{Bishop2006_HR}. In forecasting, a prominent example is the DeepAR model \citep{flunkert2017deepar_TJ}, which uses a recurrent 
neural network architecture and assumes the probability distribution to be from a standard probability density function (e.g., negative binomial or Student's $t$). Variations are possible, with either non-standard output distributions in forecasting such as the multinomial distribution \citep{rabanser2020effectiveness_TJ} or via representing the probability density as cumulative distribution function \citep{salinas2019high_Feng} or the quantile function \citep{gasthaus2019probabilistic_TJ}. 
 \item An alternative approach is to apply concepts for quantile regression \citep{koenker2005quantile_TJ} to neural networks, e.g., by making the neural network produce values for selected quantiles directly \citep{wen2017multi_TJ}. 
 \item It is possible to combine neural networks with existing probabilistic models. For example, neural networks can parametrise state space models \citep{koopman2012_Diego} as an example for another class of approaches \citep{rangapuram2018deep_TJ}, dynamic factor models \citep{geweke1977dynamic_TJ} with neural networks \citep{wang2019deepfactors_TJ} or deep temporal point processes \citep{turkmen19_TJ}. 
\end{enumerate}

The appeals of using neural networks for point forecasts carry over to probabilistic forecasts, so we will only comment on the elegance of modern neural network programming frameworks. To the forecasting model builder, the availability of auto gradient computation, the integration with highly-tuned optimisation algorithms and scalability considerations built into the frameworks, means that the time from model idea to experimental evaluation has never been shorter. 
In the examples above, we brushed over the need to have loss functions with which we estimate the parameters of the neural networks. Standard negative log-likelihood based approaches are easily expressible in code as are approaches based on non-standard losses such as the continuous ranked probability score \citep[][and \S\ref{sec:evaluating_probabilistic_forecasts}]{Gneiting&2007_TT}. With open source proliferating in the deep learning community, most of the above examples for obtaining probabilistic forecasts can readily be test-driven \citep[see, for example,][]{alexandrov2019gluonts_TJ}.

For the future, we see a number of open challenges. Most of the approaches mentioned above are univariate, in the following sense. If we are interested in forecasting values for all time series in a panel, we may be interested in modelling the 
relationship among these time series. The aforementioned approaches mostly assume independence of the time series. In recent years, a number of multivariate probabilistic forecasting models have been proposed \citep{salinas2019high_Feng,rangapuram2020nkf_TJ}, but much work remains to obtain a better understanding. Another counter-intuitive challenge for neural networks is to scale them down. Neural networks are highly parametrised, so in order to estimate parameters correctly, panels with lots of time series are needed. However, a large part of the forecasting problem landscape \citep{janusch18_TJ} consists of forecasting problems with only a few time series. Obtaining good uncertainty estimates with neural networks in these settings is an open problem.

\subsubsection[Machine learning (Evangelos Spiliotis)]{Machine learning\protect\footnote{This subsection was written by Evangelos Spiliotis.}}
\label{sec:machine_Learning}
Categorising forecasting methods into statistical and machine learning (ML) is not trivial, as various criteria can be considered for performing this task \citep{januschowski2020criteria_CB}. Nevertheless, more often than not, forecasting methods are categorised as ML when they do not prescribe the data-generating process, e.g., through a set of equations, thus allowing for data relationships to be automatically learned \citep{BARKER2020150}. In this respect, methods that build on unstructured, non-linear regression algorithms (see also \S\ref{sec:time_series_regression_models}), such as Neural Networks (NN), Decision Trees, Support Vector Machines (SVM), and Gaussian Processes, are considered as ML \citep{MakridakisPLOS}.

Since ML methods are data-driven, they are more generic and easier to be adapted to forecast series of different characteristics \citep{SPILIOTIS202037}. However, ML methods also display some limitations. First, in order for ML methods to take full advantage of their capacity, sufficient data are required. Thus, when series are relatively short and display complex patterns, such as seasonality and trend, ML methods are expected to provide sub-optimal forecasts if the data are not properly pre-processed \citep{Zhang1998-bv_GS,MakridakisPLOS}. On the other hand, when dealing with long, high-frequency series, typically found in energy \citep[][but also \S\ref{sec:Energy}]{CHAE2016184}, stock market \citep[][and \S\ref{sec:Economics_and_finance}]{MOGHADDAM201689}, and demand \citep[][but also \S\ref{sec:Operations_and_Supply_Chain_Management}]{CARMO2004137} related applications, ML methods can be applied with success. Second, computational intensity may become relevant \citep{Makridakis2020-mm}, especially when forecasting numerous series at the weekly and daily frequency \citep{SEAMAN2018822} or long-term accuracy improvements over traditional methods are insignificant \citep{NIKOLOPOULOS2018322}. Third, given that the effective implementation of ML methods strongly depends on optimally determining the values of several hyper-parameters, related both with the forecasting method itself and the training process, considerable complexity is introduced, significant resources are required to set up the methods, and high experience and a strong background in other fields than forecasting, such as programming and optimisation, are needed.

In order to deal with these limitations, ML methods can be applied in a cross-learning (CL) fashion instead of a series-by-series one \citep{Makridakis2020-mm}, i.e., allow the methods to learn from multiple series how to accurately forecast the individual ones (see also \S\ref{sec:Forecasting_competitions}). The key principle behind CL is that, although series may differ, common patterns may occur among them, especially when data are structured in a hierarchical way and additional information, such as categorical attributes and exogenous/explanatory variables (see \S\ref{sec:Exogenous_variables_and_feature_engineering}), is provided as input \citep{FRY2020156}. The CL approach has several advantages. First, computational time can be significantly reduced as a single model can be used to forecast multiple series simultaneously \citep{SEMENOGLOU20211072}. Second, methods trained in a particular dataset can be effectively used to provide forecasts for series of different datasets that display similar characteristics (transfer-learning), thus allowing the development of generalised forecasting methods \citep{Oreshkin2020MetalearningFW}. Third, data limitations are mitigated and valuable information can be exploited at global level, thus allowing for patterns shared among the series, such as seasonal cycles \citep{DEKKER2004151} and special events \citep{HUBER2020}, to be effectively captured. 

Based on the above, CL is currently considered the most effective way of applying ML for batch time series forecasting. Some state-of-the-art implementations of CL include long short-term memory NNs \citep{Smyl2020-dv}, deep NNs based on backward and forward residual links \citep{Oreshkin2020NBEATSNB}, feature-based XGBoost \citep{Montero-Manso2020-et}, and gradient boosted decision trees \citep{Bojer2020-yk}.
 
\subsubsection[Machine learning with (very) noisy data (David E. Rapach)]{Machine learning with (very) noisy data\protect\footnote{This subsection was written by David E. Rapach.}}
\label{sec:machine_Learning_with_very_noisy_data}
With the advent of big data, machine learning now plays a leading role in forecasting.\footnote{We do not make a sharp distinction between statistical learning and machine learning. For brevity, we use the latter throughout this subsection.} There are two primary reasons for this. First, conventional ordinary least squares (OLS) estimation is highly susceptible to \textit{overfitting} in the presence of a large number of regressors (or features); see also \S\ref{sec:Model_complexity} and \S\ref{sec:Variable_Selection}. OLS maximises the fit of the model over the estimation (or training) sample, which can lead to poor out-of-sample performance; in essence, OLS over-responds to \textit{noise} in the data, and the problem becomes magnified as the number of features grows. A class of machine-learning techniques, which includes the popular least absolute shrinkage and selection operator \citep[LASSO,][]{JLC_Tibshir96} and elastic net \citep[ENet,][]{JLC_ZouHastie2005}, employs \textit{penalised regression} to improve out-of-sample performance with large numbers of features. The LASSO and ENet guard against overfitting by \textit{shrinking} the parameter estimates toward zero.

Very noisy data -- data with a very low signal-to-noise ratio -- exacerbate the overfitting problem. In such an environment, it is vital to induce adequate shrinkage to guard against overfitting and more reliably uncover the predictive signal amidst all the noise. For LASSO and ENet estimation, a promising strategy is to employ a stringent information criterion, such as the Bayesian information criterion \citep[BIC,][]{Schwarz1978_DR}, to select (or tune) the regularisation parameter governing the degree of shrinkage (often denoted by $\lambda$). \cite{WangLiLeng2009_DR} and \cite{FanTang2013_DR} modify the BIC penalty to account for a diverging number of features, while \cite{HuiWartonFoster2015_DR} refine the BIC penalty to include the value of $\lambda$. These BIC variants induce a greater degree of shrinkage by strengthening the BIC's penalty term, making them useful for implementing the LASSO and ENet in noisy data environments; see \cite{FilippouRapachTaylorZhou2020_DR} for a recent empirical application.

A second reason for the popularity of machine learning in the era of big data is the existence of powerful tools for accommodating complex predictive relationships. In many contexts, a linear specification appears overly restrictive, as it may neglect important nonlinearities in the data that can potentially be exploited to improve forecasting performance. Neural networks (NNs; see \S\ref{sec:neural_networks}) are perhaps the most popular machine-learning device for modelling nonlinear predictive relationships with a large number of features. Under a reasonable set of assumptions, a sufficiently complex NN can approximate any smooth function \citep[for example,][]{Cybenko1989_DR, Funahashi1989_DR, HornikStinchcombeWhite1989_DR, Barron1994_DR}.

By design, NNs are extremely flexible, and this flexibility means that a large number of parameters (or weights) need to be estimated, which again raises concerns about overfitting, especially with very noisy data. The weights of a NN are typically estimated via a stochastic gradient descent (SGD) algorithm, such as \textit{Adam} \citep{KingmaBa2015_DR}. The SGD algorithm itself has some regularising properties, which can be strengthened by adjusting the algorithm's hyperparameters. We can further guard against overfitting by shrinking the weights via LASSO or ENet penalty terms, as well as imposing a dropout rate \citep{HintonSrivastavaKrizhevskySutskeverSalakhutdinov2012_DR,SrivastavaHintonKrizhevskySutskeverSalakhutdinov2014_DR}.

Perhaps the quintessential example of a noisy data environment is forecasting asset returns, especially at short horizons (e.g., monthly). Because many asset markets are reasonably efficient, most of the fluctuations in returns are inherently unpredictable -- they reflect the arrival of new information, which, by definition, is unpredictable. This does not mean that we should not bother trying to forecast returns, as even a seemingly small degree of return predictability can be economically significant \citep[e.g.,][]{CampbellThompson2008_DR}. Instead, it means that we need to be particularly mindful of overfitting when forecasting returns in the era of big data. \S\ref{sec:Forecasting_stock_returns} discusses applications of machine-learning techniques for stock return forecasting.

\subsubsection[Clustering-based forecasting (Ioannis Panapakidis)]{Clustering-based forecasting\protect\footnote{This subsection was written by Ioannis Panapakidis.}}
\label{sec:clustering_based_forecasting}
The robustness of the forecasting process depends mainly on the characteristics of the target variable. In cases of high nonlinear and volatile time series, a forecasting model may not be able to fully capture and simulate the special characteristics, a fact that may lead to poor forecasting accuracy \citep{Pradeepkumar2017-ek_IP}. Contemporary research has proposed some approaches to increase the forecasting performance \citep{Sardinha-Lourenco2018-no_IP}. Clustering-based forecasting refers to the application of unsupervised machine learning in forecasting tasks. The scope is to increase the performance by employing the information of data structure and of the existing similarities among the data entries \citep{Goia2010-jp_IP}; see also \S\ref{sec:Feature_based_time_series_forecasting} and \S\ref{sec:machine_Learning}. Clustering is a proven method in pattern recognition and data science for deriving the level of similarity of data points within a set. The outputs of a clustering algorithm are the centroids of the clusters and the cluster labels, i.e., integer numbers that denote the number of cluster that a specific data entry belongs to \citep{Xu2005-yr_IP}.

There are two approaches in clustering-based forecasting: (\textit{i}) Combination of clustering and supervised machine learning, and (\textit{ii}) solely application of clustering. In the first case, a clustering algorithm is used to split the training set into smaller sub-training sets. These sets contain patterns with high similarity. Then for each cluster a dedicated forecaster is applied \citep{Chaouch2014-ek_IP,Fan2008-rh_IP}. Thus, the number of forecasting algorithms is equal to the number of clusters. This approach enables to train forecasters with more similar patterns and eventually achieve better training process. The forecasting systems that involve clustering are reported to result in lower errors \citep{Fan2006-dv_IP,Mori2001-vf_IP}. The combination of clustering and forecasting has been presented in the literature earlier than the sole application of clustering. One of the first articles in the literature on combining clustering and forecasting sets up the respective theoretical framework \citep{Kehagias1997}.

In the second case, a clustering algorithm is used to both cluster the load data set and perform the forecasting \citep{Lopez2012-sv_IP}. In the sole clustering applications, either the centroids of the clusters can be utilised or the labels. Pattern sequence-based forecasting is an approach that employs the cluster labels. In this approach, a clustering of all days prior to the test day is held and this results in sequences of labels of a certain length. Next, the similarity of the predicted day sequence with the historical data sequences is examined. The load curve of the predicted day is the average of the curves of the days following the same sequences \citep{Martinez_Alvarez2011-jh_IP,Kang2020-xq}. 

There is variety of clustering algorithms that have been proposed in forecasting such as the $k$-means, fuzzy C-means (FCM), self-organising map (SOM), deterministic annealing (DA), ant colony clustering ACC, and others. Apart from the clustering effectiveness, a selection criterion for an algorithm is the complexity. $k$-means and FCM are less complex compared to the SOM that needs a large number of variables to be calibrated prior to its application. Meta-heuristics algorithms, such as ACC, strongly depend on the initialisation conditions and the swarm size. Therefore, a comparison of clustering algorithms should take place to define the most suitable one for the problem under study \citep{Mori2001-vf_IP,Li2008-ka_IP,Elangasinghe2014-nu_IP,Wang2015-wy_IP}.

The assessment of clustering-based forecasting is held via common evaluation metrics for forecasting tasks (see also \S\ref{sec:evaluation_and_validation}). The optimal number of clusters, which is a crucial parameter of a clustering application, is selected via trial-and-error, i.e., the optimal number corresponds to the lowest forecasting error \citep{Nagi2011-eh_IP}. 
 
\subsubsection[Hybrid methods (Sonia Leva)]{Hybrid methods\protect\footnote{This subsection was written by Sonia Leva.}}
\label{sec:Hybrid_methods}
Hybrid approaches combine two or more of the above-mentioned advanced methods. In general, when methods based on AI-based techniques, physical, and statistical approaches are combined together, the result is often improved forecasting accuracy as a benefit from the inherent integration of the single methods. The idea is to mix diverse methods with unique features to address the limitations of individual techniques, thus enhancing the forecast performance \citep[][see also \S\ref{sec:combining_forecasts}]{Nespoli2019-vg_SL,Mandal2012-ll_SL}. The performance of the hybrid methods depends on the performance of the single methods, and these single methods should be specifically selected for the problem that has to be addressed.

Hybrid methods can be categorised based on the constituent methods, but also considering that these base methods may not necessarily act only on the forecasting stage but also on data treatment and parameters identification stages. In data pre-processing combining approaches (see also \S\ref{sec:Preprocessing_time_series_data}), different methods can be used for decomposing the time series into subseries \citep{Son2019-yj_SL} or the signal into different frequencies \citep{Zang2018-aj_SL}, and for classifying the historical data \citep{Huang2015-vy_SL}. An advantage of such hybrid methods is robustness against sudden changes in the values of the main parameters. However, they require additional knowledge and understanding of the base methods, and have the disadvantage of slow response time to new data.

The purpose of the parameter selection stage is to optimise the parameters of the model, in terms of extracting nonlinear features and invariant structures \citep{Behera2018-co_SL,Ogliari2018-pz_SL} but also in terms of estimation of the parameter adopted for the prediction; for example, meteorological factors such as temperature, humidity, precipitation, snowfall, cloud, sunshine, wind speed, and wind direction \citep{Qu_Xiaoyun2016-mx_SL}. Hybrid methods feature straightforward determination of the parameters with relatively basic structures. However, the implementation is sometimes challenging, and depends on the knowledge and expertise of the designer.

Finally, the data post-processing hybrid approaches forecast the residual errors resulted from the forecasting model. Since these hybrid methods consider residual errors from the model, they aim in further improving the predictions of the base methods by applying corrections in the forecasts. However, a disadvantage of these hybrid methods is the increased calculation time, as the residual errors must also be estimated. Also, such hybrid methods are not general and will depend on the field of application. 
In many cases, hybrids approaches outperform other (single) approaches such as $k$NN, NN, and ARIMA-based models \citep{Mellit2020-eb_SL}. A great example is the hybrid method by \citep{Smyl2020-dv}, which achieved the best performance in the M4 forecasting competition (see also \S\ref{sec:Forecasting_competitions}). In particular, in energy applications (see \S\ref{sec:Energy}), a combination of physical and AI-based techniques can lead to improved forecasting performance. Furthermore, machine learning methods (see \S\ref{sec:machine_Learning}) based on historical data of meteorological variables combined with an optimal learning algorithm and weather classification can further improve the forecasting accuracy of single methods. However, in general, the weak point of such hybrid approaches is that they underperform when meteorological conditions are unstable \citep{Chicco2015-ig_SL}.

\subsection{Methods for intermittent demand}
\label{sec:Forecasting_for_intermittent_demands_and_count_data}

\subsubsection[Parametric methods for intermittent demand forecasting (Aris A. Syntetos)]{Parametric methods for intermittent demand forecasting\protect\footnote{This subsection was written by Aris A. Syntetos.}}
\label{sec:Parametric_methods_for_intermittent_demand_forecasting}

Demand forecasting is the basis for most planning and control activities in any organisation. Demand will typically be accumulated in some pre-defined `time buckets' (periods), such as a day, a week or a month. On many occasions, demand may be observed in every time period, resulting in what is sometimes referred to as `non-intermittent demand'. Alternatively, demand may appear sporadically, with no demand at all in some periods, leading to an intermittent appearance of demand occurrences. Intermittent demand items monopolise the stock bases in the after sales industry and are prevalent in many other industries, including the automotive, IT, and electronics sectors. Their inventory implications are dramatic and forecasting their requirements is a very challenging task.

Methods to forecast intermittent demand may broadly be classified as parametric and non-parametric. The former suppose that future demand can be well represented by a statistical distribution (say Poisson or Negative Binomial) which has parameters that are unknown but may be forecasted using past data. These methods are discussed in this sub-section. In the latter, the data are not assumed to follow any standard probability distribution. Instead, direct methods are used to assess the distributions required for inventory management (see also \S\ref{sec:Forecasting_for_inventories}). Such methods are discussed in \S\ref{sec:non_parametric_intermittent_demand_methods}.

Simple Exponential Smoothing (SES; see \S\ref{sec:exponential_smoothing_models}) is often used in practice to forecast intermittent demand series. However, SES fails to recognise that intermittent demand is built from two constituent elements: (\textit{i}) the inter-demand intervals, which relate to the probability of demand occurring, and (\textit{ii}) the demand sizes, when demand occurs. The former indicates the degree of intermittence, whereas the latter relates to the behaviour of the positive demands. \cite{Croston1972-zw_ASJBZB} showed that this inherent limitation leads to SES being (positively) biased after a demand occurring period; this is sometimes referred to as an `issue point' bias. Subsequently, he proposed a method that forecasts separately the sizes of demand, when demand occurs, and the inter-demand intervals. Both forecasts are produced using SES, and the ratio of the former over the latter gives a forecast of the mean demand per period. Croston's method was shown by \cite{Syntetos2001-ih_ASJBZB} to suffer from another type of bias (inversion bias) and the same researchers \citep{Syntetos2005-ic_ASJBZB} proposed a modification to his method that leads to approximately unbiased estimates. This method is known in the literature as the Syntetos-Boylan Approximation (SBA). It has been found repeatedly to account for considerable empirical inventory forecasting improvements \citep{Eaves2004-ce_ASJBZB,Gutierrez2008-iy_ASJBZB,Van_Wingerden2014-hy_ASJBZB,Nikolopoulos2016-wd_ASJBZB} and, at the time of writing, it constitutes the benchmark against which other (new) proposed methodologies in the area of intermittent demand forecasting are assessed. 

Croston's method is based upon the assumption of a Bernoulli demand arrival process. Alternatively, demand may be assumed to arrive according to a Poisson process. It is also possible to adapt Croston's method so that sizes and intervals are updated based on a simple moving average (SMA) procedure instead of SES. \cite{Boylan2003-xz_ASJBZB}, \cite{Shale2006-lt_ASJBZB}, and \cite{Syntetos2015-ic_ASJBZB} presented correction factors to overcome the bias associated with Croston's approach under a Poisson demand arrival process and/or estimation of demand sizes and intervals using an SMA. 

For a detailed review of developments in intermittent demand forecasting interested readers are referred to \cite{Boylan2021-fo_ASJBZB}.

\subsubsection[Non-parametric intermittent demand methods (Mohamed Zied Babai)]{Non-parametric intermittent demand methods\protect\footnote{This subsection was written by Mohamed Zied Babai.}}
\label{sec:non_parametric_intermittent_demand_methods}
Two main non-parametric forecasting approaches have dominated the intermittent demand literature: the bootstrapping approach and the Overlapping/Non-Overlapping aggregation Blocks approach \citep{Boylan2021-fo_ASJBZB}.

The bootstrapping approach relies upon a resampling (with or without replacement) of the historical demand data to build the empirical distribution of the demand over a specified interval. As discussed in \S\ref{sec:Forecasting_with_bootstrap}, this approach was initially introduced by \cite{Efron1979-mr}. Since then, it has been developed by \cite{Willemain2004-cc_ASJBZB} and \cite{Zhou2011-sr_ASJBZB} to deal with intermittent demand items \citep{Babai2020-mz_ASJBZB}. \cite{Willemain2004-cc_ASJBZB} have proposed a method that resamples demand data by using a Markov chain to switch between no demand and demand periods. The empirical outperformance of this method has been shown when compared to Simple Exponential Smoothing (SES) and Croston's method (see also \S\ref{sec:Parametric_methods_for_intermittent_demand_forecasting}). However, the findings of \cite{Willemain2004-cc_ASJBZB}'s work have been challenged by \cite{Gardner2005-xw_ASJBZB} and some limitations have been addressed by \cite{Syntetos2015-my_ASJBZB}. \cite{Zhou2011-sr_ASJBZB} have developed an alternative bootstrapping method. Their method samples separately demand intervals and demand sizes and it has been shown to be associated with a good performance for long lead-times. \cite{Teunter2009-lx_ASJBZB} and \cite{Hasni2019-cv_ASJBZB} have developed adjustments of the bootstrapping methods, where the lead-time demand forecast is adjusted by assuming that the first period in the lead-time bucket corresponds to a non-zero demand. They have demonstrated the outperformance of the adjusted bootstrapping methods in a periodic order-up-to-level inventory control system. A review of the bootstrapping methods in the context of intermittent demand is provided by \cite{Hasni2019-dg_ASJBZB}.

\cite{Porras2008-rk_ASJBZB} were the first to consider aggregation with overlapping and non-overlapping blocks (OB and NOB) approach in forecasting the demand of spare parts. In the NOB approach, a demand series is divided into consecutive non-overlapping blocks of time, whereas in OB, at each period the oldest observation is dropped and the newest is included \citep{Rostami-Tabar2013-nf_ASJBZB}. \cite{Boylan2016-er_ASJBZB} have compared the statistical and inventory performance of the OB and NOB methods. They found that, unless the demand history is short, there is a clear advantage of using OB instead of NOB. More recently, based on extreme value theory (EVT), \cite{Zhu2017-ax_ASJBZB} have proposed an improvement of the OB method that models better the tail of lead-time demand. They have shown that the empirical-EVT method leads to higher achieved target cycle service levels when compared to the original method proposed by \cite{Porras2008-rk_ASJBZB}. Temporal aggregation is further discussed in \S\ref{sec:temporal_aggregation}.

\subsubsection[Classification methods (John E. Boylan)]{Classification methods\protect\footnote{This subsection was written by John E. Boylan.}}
\label{sec:classification_methods}
In many application areas, forecasts are required across a wide collection of products, services or locations. In this situation, it is convenient to introduce classification rules that allow subsets of time series to be forecasted using the same approaches and methods. Categorisation rules, such as the ABC inventory classification, serve the forecasting function only coincidentally. They do not necessarily align to the selection of the best forecasting method.

Within certain modelling frameworks, classification of time series is well established. For example, within an ARIMA framework \citep[][and \S\ref{sec:autoregressive_integrated_moving_average_models}]{Box2008-mh}, or within a state-space framework for exponential smoothing \citep[][and \S\ref{sec:exponential_smoothing_models}]{Hyndman2002-jp_JRTA}, series may be classified, for example based on the AIC \citep{Akaike1973-hg_ASJBZB}. It is more challenging to classify series according to their recommended forecasting method if some of the candidate methods, such as Croston's method (see \S\ref{sec:Parametric_methods_for_intermittent_demand_forecasting}), lack a fully satisfactory model base. In the field of intermittent demand forecasting, \cite{Syntetos2005-qb_ASJBZB} proposed the SBC classification scheme, enabling time series to be classified according to their length of average demand interval and coefficient of variation of demand sizes (when demand occurs). These rules were based on assumptions of independent and identically distributed (iid) demand, and a comparison of expected mean square error between methods. The scheme has been extended by \cite{Kostenko2006-py_ASJBZB} and by \cite{Petropoulos2015-vq}. In an empirical case-study, \cite{Boylan2008-tq} examined series not necessarily conforming to iid assumptions and found the rules to be robust to inexact specification of cut-off values. \cite{Moon2013-kc_ASJBZB} used logistic regression to classify time series of demand for spare parts in the South Korean Navy. The classification was designed to identify superior performance (accuracy and inventory costs) of direct and hierarchical forecasting methods, based on the serial correlation of demands, the coefficient of variation of demand volume of spare parts (see also \S\ref{sec:Spare_parts_forecasting}), and the functionality of the naval equipment. 

\cite{Bartezzaghi1999-xb_ASJBZB} identified five factors that contribute towards intermittence and `lumpiness' (intermittence with highly variable demand sizes): number of potential customers, frequency of customers' requests, heterogeneity of customers, variety of individual customer's requests, and correlations between customers' requests. These may contribute towards useful classifications, for example by the number of customers for an item. When this number is low and some of the customers are large, then direct communication with the customers can inform judgmental forecasts. Similarly, if a customer's requests are highly variable, then `advance demand information' from customers can help to improve judgmental estimates. These strategies can be very useful in a business-to-business environment, where such strategies are feasible.

An alternative approach to classification is combination of forecasts (see \S\ref{sec:combining_forecasts} for a review, and \S\ref{sec:Forecast_combination_a_brief_review_of_statistical_approaches} in particular).  \cite{Petropoulos2015-vq} investigated combining standard forecasting methods for intermittent demand (e.g., SES, Croston, Syntetos-Boylan Approximation; see also \S\ref{sec:Parametric_methods_for_intermittent_demand_forecasting}). They did not find this to improve accuracy directly, but obtained good results from the use of combinations of forecasts at different temporal  frequencies, using methods selected from the extended SBC classification scheme.

\subsubsection[Peak over the theshold (Konstantinos Nikolopoulos)]{Peak over the theshold\protect\footnote{This subsection was written by Konstantinos Nikolopoulos.}}
\label{sec:peak_over_the_threshold_POT}
In the forecasting literature, \cite{Nikolopoulos2020_YK} argues that great attention has been given to modelling fast-moving time series with or without cues of information available \citep{Nikolopoulos2007-fx_KN}. Less attention has been given to intermittent/count series (see \S\ref{sec:Parametric_methods_for_intermittent_demand_forecasting}, \S\ref{sec:non_parametric_intermittent_demand_methods}, and \S\ref{sec:Forecasting_count_time_series}), which are more difficult to forecast given the presence of two sources of uncertainty: demand volume, and timing.

Historically there have been few forecasting methods developed specifically for such data \citep{Syntetos2015-my_ASJBZB}. We believe that through a time series decomposition approach \'a la \cite{Leadbetter1991-xv_KN} we can isolate `peaks over threshold' (POT) data points, and create new intermittent series from any time series of interest. The derived series present almost identical characteristics with the series that \cite{Croston1972-zw_ASJBZB} analysed. In essence one could use such decomposition forecasting techniques to tackle much more difficult phenomena and problems coming from finance, politics, healthcare, humanitarian logistics, business, economics, and social sciences.

Any time series can be decomposed into two sub-series: one containing the baseline (\textit{white swans}) and one containing the extreme values over an arbitrary-set or rule-based-set threshold  (\textit{grey} and \textit{black swans}) as proposed by \cite{Taleb2008-jd_PG}; see also \S\ref{sec:Forecasting_under_fat_tails}. Unfortunately, major decision-related risks and most of the underlying uncertainty lay with these extremes. So, it is very important to be able to effectively model and forecast them. 

It is unlikely that any forecasting approach will accurately give the exact timing of the forthcoming extreme event, but it will instead provide a satisfactory cumulative forecast over a long period of time. The question still stands what can one do with this forecast? For earthquake data, although even if we know that a major earthquake is going to hit a region, it is almost impossible to decide to evacuate cities, but still we can influence and legislate the way structures are built and increase the awareness, training, preparedness and readiness of the public; and also ensure enough capital on hold to cope with the aftermath of the major event. For epidemics/pandemics (see \S\ref{sec:Pandemics}) there are clear implications, as we have evidenced with COVID-19, on how proactively we can source and secure human resources, medical supplies, etc.

What is the current doctrine when forecasting in such a context: advanced probabilistic models. These methods typically require a lot of data and reconstruct the distributions of the underlying phenomena. These come with common successes and a plethora of constraints: \textit{big data sets} needed for training the models, high mathematical \textit{complexity}, and invisibility to practitioners how these methods do actually work and thus \textit{less acceptance in practice}. Yet again, forecasting accuracy is the name of the game and thus these forecasting methods are serious contenders for the task in hand.

Extreme Value Theory (EVT) analyses extreme deviations from statistical measures of central location to estimate the probability of events that are more extreme than anything observed in the time series. This is usually done in the following two ways \citep{Nikolopoulos2020_YK}: (\textit{i}) deriving maxima and/or minima series as a first step and then having the Generalised Extreme Value Distribution fitted (often the number of extreme events is limited), and (\textit{ii}) isolating the values that exceed a threshold (point over threshold) that can also lead to only a few instances extracted -- so a very intermittent series in nature. The analysis involves fitting a Poisson distribution for the number of events in a basic time period and a second distribution -- usually a Generalised Pareto Distribution -- for the size of the resulting POT values.

\subsection{Reasoning and mining}
\label{sec:Reasining_and_mining}

\subsubsection[Fuzzy logic (Claudio Carnevale)]{Fuzzy logic\protect\footnote{This subsection was written by Claudio Carnevale.}}
\label{sec:Fuzzy_logic}
The ``classical'' Boolean logic is not able to handle for uncertainties and/or vagueness that are necessary when dealing with many real world problems. This is in part due to the fact that the Boolean logic is based on only two values (i.e., a statement can only be true or false). Fuzzy logic tries to overcome this issue by admitting that a statement/variable could be partially true or partially false. Mathematically, the fuzzy logic framework is based on the work of \cite{ZADEH1965338_CC} who introduced the theory of \textit{fuzzy sets}.
The main point of this theory is the definition of two kinds of sets:
\begin{enumerate}[noitemsep]
  \item \textit{Crisp sets} are the ``classical'' sets in the Boolean logic. An element can belong (or not) to a certain set.
  \item \textit{Fuzzy sets}, where an element can belong to the sets with a certain \textit{membership grade}, with a value that varies in the interval $[0,1]$. 
\end{enumerate}

The definition of the fuzzy sets allows the framework to take into account the uncertainty and vagueness of information. An extension of this approach is related to the fact that a certain variable can assume a crisp value (classical theory) or can belong to different fuzzy sets with different membership grade. For example, in a system implemented to forecast the daily pollutant concentration in atmosphere, one of the inputs could relate to the weather conditions, such as the wind speed. In the classical approach, the system must have as an input the value of the wind speed at a certain day. In the fuzzy approach, the input of the system could be the membership grade of the input variable to three different fuzzy sets: (\textit{i})~``Not windy'', (\textit{ii})~``average windy'', and (\textit{iii})~``strong windy''. On the other hand, the user of the forecasting system may be only interested in a classification of the output variable instead of the crisp value. In this case, the fuzzy approach is applied to the pollutant concentration which could belong with a certain degree to the fuzzy sets (\textit{i})~``not polluted day'', (\textit{ii})~``medium polluted day'', (\textit{iii})~``high polluted day'', and (\textit{iv})~``critical polluted day''.

In fuzzy theory, each fuzzy set is characterised by a (generally nonlinear) function, called the \textit{membership function}, linking crisp values to the membership of the different sets. The association of a crisp value to its membership for a set is called \textit{fuzzyfication}, while the inverse operation (from a membership value to a crisp value) is called \textit{defuzzification}. As with the logic theory, the \textit{inference system} assumes a key role in the fuzzy theory. A Fuzzy Inference System (FIS) allows the interpretation of the membership grades of the input variable(s) and, given some sets of fuzzy rules, assigns the corresponding values to the output variable(s). In the literature, two main fuzzy inference systems are presented: 
\begin{enumerate}[noitemsep]
  \item Mamdani system \citep{MAMDANI1975_CC}, where both the input and output of the inference system are membership functions.
  \item Sugeno system \citep{SUGENO1985_CC}, where the output of the inference system is a crisp value, usually obtained by applying a linear function to the defuzzified value of the input.
\end{enumerate}
 
\subsubsection[Association rule mining (Daniele Apiletti)]{Association rule mining\protect\footnote{This subsection was written by Daniele Apiletti.}}
\label{sec:Association_rule_mining}
Association rule mining is an exploratory data-driven approach which is able to automatically and exhaustively extract all existing correlations in a data set of categorical features. It is a powerful but computationally intensive technique, successfully applied in different forecasting contexts \citep{acquaviva2015enhancing_DA, ap2020correlating_DA, di2018metatech_DA}. Its results are in a human-readable form.

The data set must be in the form of transactions, i.e., a collection of events, each described by categorical features. If the phenomena under analysis are modelled by continuous-valued variables, discretisation can be applied to obtain a suitable data set.

Association rule mining core task is the frequent itemset extraction, which consists in finding frequently-occurring relationships among items in a data set \citep{han2011data_DA}. 
Given a data set of records characterised by several attributes, an item refers to a pair of (attribute $=$ value), while a set of items is called itemset. The support count of an itemset is the number of records $r$ containing that itemset. The support of an itemset is the percentage of records containing it with respect to the total number of records in the data set. An itemset is frequent when its support is greater than or equal to a minimum support threshold.

An association rule is an implication in the form $A \rightarrow B$, where $A$ and $B$ are disjoint itemsets (i.e., $A \cap B = \emptyset $) \citep{KumarDMBook_DA}. $A$ is called rule body or antecedent and $B$ rule head or consequent. 

To evaluate the quality of an association rule, the support, confidence, and lift metrics are commonly exploited \citep{han2011data_DA}. \textit{Rule support} is the fraction of records containing both $A$ and $B$, indicating the probability that a record contains every item in these itemsets. The support of the rule is computed as the support of the union of $A$ and $B$. 

\textit{Rule confidence} represents the strength of the implication, and is the conditional probability that a transaction containing $A$ also contains $B$, $P(B|A)$, i.e., the proportion of records that contain $A$ with respect to those that also contain $B$.

Finally, the \textit{lift} of a rule measures the correlation between antecedent and consequent. It is defined as the ratio between the rule $A \rightarrow B$ confidence and the support of $B$. A lift ratio equal to 1.0 implies that itemsets $A$ and $B$ are not correlated. A lift higher than 1.0 indicates a positive correlation, meaning that the occurrence of $A$ likely leads to the occurrence of $B$ with the given confidence. The greater the lift, the stronger the association. Finally, a lift lower than 1.0 indicates a negative correlation between $A$ and $B$.

The problem of association rule mining consists in the extraction of all the association rules having rule support and confidence greater than the respective support and confidence thresholds, $MinConf$ and $MinSup$, defined as parameters of the mining process \citep{KumarDMBook_DA}. These thresholds allow to control the statistical relevance of the extracted rules.

The process of rule mining consists of two steps. The first step is the computation of frequent itemsets, i.e., itemsets with support greater or equal to $MinSup$. The second step is the extraction of association rules from frequent itemsets. Let be $F$ a frequent itemset, hence having a support higher than $MinSup$, pairs $A$ and $B=F-A$ are derived so that the confidence of $A \rightarrow B$ is higher than $MinConf$. The first step of the process is the most computationally expensive. Thus, several algorithms have been proposed to solve the problem of frequent itemset extraction \citep{zaki2000scalable_DA}, some specifically addressing high-dimensionality issues \citep{ap2017parallel_DA, ap2015pampa_DA}. Despite being computationally demanding, association rule mining is an exhaustive approach, i.e., all and only statistically relevant correlations are extracted. \S\ref{sec:Forecasting_food_beverage} offers an example of applying association rule mining to forecast the quality of beverages.
 
\subsubsection[Forecasting with text information (Xiaojia Guo)]{Forecasting with text information\protect\footnote{This subsection was written by Xiaojia Guo.}}
\label{sec:Text_based_forecasting}
Text data, such as social media posts, scholar articles and company reports, often contains valuable information that can be used as predictors in forecasting models \citep{aggarwal2012mining_XG}. Before extracting useful features from the data, a text document needs to be cleaned and normalised for further processing. The first step of preparing the data is to filter out \textit{stop words} -- the words that do not add much meaning to a sentence, e.g., ``a'', ``is'', and ``me''. For grammatical reasons, it is necessary for documents to use different forms of words. Stemming and lemmatisation can be applied to reduce inflectional forms or relate different forms of a word to a common base form. Stemming often chops off the end of a word, while lemmatisation uses vocabularies and morphological analysis to return the base form (or the \textit{lemma}) of a word \citep{lovins1968development_XG,manning2008introduction_XG}. For example, the word ``industries" will be turned into ``industri'' or ``industry'' if stemming or lemmatisation is applied. 

To model and analyse text data, we need to transform it into numerical representations so the forecasting models can process them as predictors. One way of transforming the data is through sentiment analysis. Sentiment analysis is often applied to detect polarity within customer materials such as reviews and social media posts \citep{das2007yahoo_XG,archak2011deriving_XG}. An easy way to obtain the sentiment score of a word is to look it up in a happiness dictionary \citep[for example, the hedonometer dictionary,][]{hedonometer2020_XG}. Another common way of representing the sentiment is to use a vector of numeric values that denote the word's positivity, negativity and neutrality based on existing lexical databases such as the \textit{WordNet} \citep{godbole2007large_XG,baccianella2010sentiwordnet_XG}. Once the sentiment of each word is calculated, we can apply an aggregation algorithm (e.g., simple average) to measure the sentiment of an entire sentence or paragraph. 

In scholar articles and company reports, context features might be more important than sentiments. The bag-of-words model and word embeddings are often applied to generate numeric representations of such text. A bag-of-words model simply returns a matrix that describes the occurrence of words within a document \citep{goldberg2017neural_XG}. When we use this matrix as input to a forecasting model, each word count can be considered as a feature. The Word2Vec method is a widely used embedding method that is built based on the context of a word. Specifically, it trains a two-layer neural network that takes each word as an input to predict its surrounding words (see \S\ref{sec:neural_networks} for a discussion of neural networks for forecasting). The weights from the input layer to the hidden layer are then utilised as the numerical representation for the input word \citep{le2014distributed_XG}. Once the text is turned into numeric representations, they can be used as predictors in any forecasting models. The most challenging part in this process is to find the right technique to extract features from the text. 

In terms of software implementation, the Natural Language Toolkit (NLTK) and SpaCy library in Python can be applied to remove stop words and stem or lemmatise text \citep{loper2002nltk_XG,honnibal2015spacy_XG}. The bag-of-words technique is also available in NLTK. A particular implementation of the Word2Vec model is available on \cite{googlecode2013_XG}. Moreover, a public data set of movie reviews that is commonly studied in literature is available from the \cite{standordgroup2013_XG}. 
 
\subsection{Forecasting by aggregation}
\label{sec:forecasting_by_aggregation}

\subsubsection[Cross-sectional hierarchical forecasting (Patr\'icia Ramos)]{Cross-sectional hierarchical forecasting\protect\footnote{This subsection was written by Patr\'icia Ramos.}}
\label{sec:Cross_sectional_hierarchical_forecasting}
In many applications time series can be aggregated at several levels of aggregation based on geographic or logical reasons to form hierarchical structures. These are called hierarchical time series. In the retail industry (see also \S\ref{sec:Retail_sales_forecasting}), for example, individual sales of products at the bottom-level of the hierarchy can be grouped in categories and families of related products at increasing aggregation levels, with the total sales of the shop or distribution centre at the top level \citep{Pennings17_PR,Oliveira2019_PR,pedregal2018_Diego}. Similarly, cross-sectional hierarchies can be used for spatial aggregation to help model housing prices or traffic in transportation networks, or otherwise formed geographical demarcations \citep[for example,][]{Athanasopoulos09_PR,kourentzes2019cross_DKB}. The forecasts of hierarchical time series produced independently of the hierarchical structure generally will not add up according to the aggregation constrains of the hierarchy, i.e., they are not coherent. Therefore, hierarchical forecasting methods that generate coherent forecasts should be considered to allow appropriate decision-making at the different levels. Actually, by taking advantage of the relationships between the series across all levels these methods have shown to improve forecast accuracy \citep{Athanasopoulos09_PR,Shang2017_PR,Yagli2019_PR}. One of the main reasons behind this improved performance is that forecast reconciliation is effectively a special case of forecast combinations \citep{Hollyman2021-wj}; see also \S\ref{sec:combining_forecasts}.

The most common approaches to hierarchical forecasting are bottom-up and top-down. In the bottom-up approach forecasts for each time series at the bottom-level are first produced and then these are added up to obtain forecasts for all other series at the hierarchy \citep{Dunn76_PR}. Since forecasts are obtained at the bottom-level no information is lost due to aggregation. In the top-down approach forecasts for the top-level series are first generated and then these are disaggregated generally using historical proportions to obtain forecasts for the bottom-level series, which are then aggregated \citep{Gross90_PR}. \cite{Hyndman11_PR} claim that this approach introduces bias to the forecasts, however \cite{Hollyman2021-wj} showed that it is possible to calculate unbiased top-down forecasts.

Recent research on hierarchical forecasting tackles the problem using a two-stage approach. Forecasts for the series at all levels of the hierarchy are first obtained independently without considering any aggregation constrains (we refer to these as base forecasts). Then, base forecasts are adjusted so that they become coherent (we refer to these as reconciled forecasts). This adjustment is achieved by a matrix that maps the base forecasts into new bottom-level forecasts which are then added up \citep{Hyndman11_PR}.

\cite{Wickramasuriya2019_PR} found the optimal solution for this matrix, which minimises the trace of the covariance matrix of the reconciled forecast errors (hence MinT reconciliation). This optimal solution is based on the covariance matrix of the base forecast errors which incorporates the correlation structure of the hierarchy. \cite{Wickramasuriya2019_PR} presented several alternative estimators for this covariance matrix: (\textit{i})~proportional to the identity which is optimal only when base forecast errors are uncorrelated and equivariant (referred to as OLS), (\textit{ii})~proportional to the sample covariance estimator of the in-sample one-step-ahead base forecast errors with off-diagonal elements null accounts for the differences in scale between the levels of the hierarchy (referred to as WLS), (\textit{iii})~proportional to the previous estimator unrestricted also accounts for the relationships between the series (referred to as MinT-Sample), and (\textit{iv})~proportional to a shrinkage estimator based on the two previous estimators, parameterising the shrinkage in terms of variances and correlations, accounts for the correlation across levels (referred as MinT-Shrink). Other researchers focus on simple (equal-weighted) combinations of the forecasts produced at different hierarchical levels \citep{Abouarghoub2018-wz,Hollyman2021-wj}. \cite{PRITULARGA2021108221} showed that more complex reconciliation schemes result in more variability in the forecasts, due to the estimation of the elements in the covariance matrix, or the implicit combination weights. They provide approximations for the covariance matrix that balance this estimation uncertainty with the benefits of more finely tuned weights. 

More recently these techniques were extended to probabilistic forecasting \citep{Taieb2020_PR}. When base forecasts are probabilistic forecasts characterised by elliptical distributions, \cite{Panagiotelis2020_PR} showed that reconciled probabilistic forecasts also elliptical can be obtained analytically. When it is not reasonable to assume elliptical distributions, a non-parametric approach based on bootstrapping in-sample errors can be used.
 
\subsubsection[Temporal aggregation (Fotios Petropoulos)]{Temporal aggregation\protect\footnote{This subsection was written by Fotios Petropoulos.}}
\label{sec:temporal_aggregation}
Temporal aggregation is the transformation of a time series from one frequency to another of lower frequency. As an example, a time series of length $n$ that is originally sampled at a monthly frequency can be transformed to a quarterly series of length $n/3$ by using equally-sized time buckets of three periods each. It is usually applied in an non-overlapping manner, but overlapping aggregation can also be considered. The latter is preferred in the case when the original series is short, but has the disadvantage of applying lower weights on the few first and last observations of the series and introducing autocorrelations \citep[][and \S\ref{sec:non_parametric_intermittent_demand_methods}]{Boylan2016-er_ASJBZB}.

Temporal aggregation is appealing as it allows to investigate the original series through different lenses. By changing the original frequency of the data, the apparent series characteristics also change. In the case of slow-moving series, temporal aggregation leads to decrease of intermittence \citep[][see also \S\ref{sec:Forecasting_for_intermittent_demands_and_count_data}]{Nikolopoulos2011-bt}. In the case of fast-moving series, higher levels of aggregation (i.e., lower frequencies) allow for better modelling of trend patterns, while lower aggregation levels (i.e., higher frequencies) are more suitable for capturing seasonal patterns \citep{Spithourakis2014-vf,Kourentzes2014-jq}.

Research has found evidence of improved forecasting performance with temporal aggregation for both slow \citep{Nikolopoulos2011-bt} and fast \citep{Spithourakis2011-df} moving time series. This led to characterise temporal aggregation as a ``self-improving mechanism''. The good performance of temporal aggregation was reconfirmed by \cite{Babai2012-ye}, who focused on its utility performance rather than the forecast error. However, one challenge with single levels of aggregation is the choice of a suitable aggregation level for each series \citep{Kourentzes2017-vi}.

Instead of focusing on a single aggregation level, \cite{Andrawis2011-bo}, \cite{Kourentzes2014-jq}, \cite{Petropoulos2014-ad}, and \cite{Petropoulos2015-vq} suggested the use of multiple levels of aggregation, usually abbreviated as MTA (multiple temporal aggregation). This not only tackles the need to select a single aggregation level, but also partly addresses the issue of model uncertainty, instead of relying on model selection and parametrisation at a single aggregation level. Using this property, \cite{Kourentzes2017-vi} showed that MTA will typically lead to more accurate forecasts, even if in theory suboptimal. Different frequencies allow for better identification of different series patterns, so it is intuitive to consider multiple temporal levels and benefit from the subsequent forecast combination across frequencies. \cite{Kourentzes2016-bf} showed how multiple temporal aggregation can be extended to incorporate exogenous variables (see also \S\ref{sec:Exogenous_variables_and_feature_engineering}). However, forecasting at a single level of aggregation can still result in better performance when the seasonal pattern is strong \citep{Spiliotis2019-hg,Spiliotis2020-hj}.

\cite{Athanasopoulos2017-ta} expressed multiple temporal aggregation within the hierarchical forecasting framework (see \S\ref{sec:Cross_sectional_hierarchical_forecasting}) using the term ``temporal hierarchies''. Temporal hierarchies allow for the application of established hierarchical reconciliation approaches directly to the temporal dimension. \cite{Jeon2019-xo} show how temporal hierarchies can be used to obtain reconciled probabilistic forecasts, while \cite{Spiliotis2019-hg} explored empirical bias-adjustment strategies and a strategy to avoid excessive seasonal shrinkage. \cite{Nystrup2020-vz} proposed estimators for temporal hierarchies suitable to account for autocorrelation in the data. Finally, \cite{Kourentzes2020-mc} applied temporal hierarchies on intermittent data, and showed that higher aggregation levels may offer structural information which can improve the quality of the forecasts. 

\subsubsection[Cross-temporal hierarchies (Nikolaos Kourentzes)]{Cross-temporal hierarchies\protect\footnote{This subsection was written by Nikolaos Kourentzes.}}
\label{sec:crosstemporal_hierarchies}
In the last two subsections (\S\ref{sec:Cross_sectional_hierarchical_forecasting} and \S\ref{sec:temporal_aggregation}), we saw two complimentary hierarchical structures, cross-sectional and temporal.
Although the machinery behind both approaches is similar, often relying on the hierarchical framework by \cite{Hyndman11_PR} and \cite{Athanasopoulos09_PR}, and work that followed from these \citep[particularly][]{Wickramasuriya2019_PR}, they address different forecasting problems. Cross-sectional hierarchies change the unit of analysis but are fixed in the period of analysis. For example, a manufacturer may operate using a hierarchy across products. The different nodes in the hierarchy will correspond to different products, product groups, super-groups, and so on, but will all refer to the same period, for example, a specific week. Temporal hierarchies do the opposite, where the unit of analysis is fixed, but the period is not. For example, we may look at the sales of a specific Stock Keeping Unit (SKU) at a daily, weekly, monthly, quarterly and annual levels. However, one can argue that having annual forecasts at the SKU level may not be useful. Similarly, having aggregate sales across an organisation at a weekly frequency is also of little value. 

In connecting these to organisational decisions, we can observe that there is only a minority of problems that either cross-sectional or temporal hierarchies are natural, as typically decisions can differ across both the unit and the period (planning horizon) of analysis. In the latter case, both hierarchical approaches are more akin to statistical devices that can improve forecast accuracy through the use of forecast combinations, rather than satisfy the motivating argument behind hierarchical forecasting that is to provide coherent predictions for decisions at different levels of the hierarchy. 

Cross-temporal hierarchies attempt to overcome this limitation, providing coherent forecasts across all units and periods of analysis, and therefore a common outlook for the future across decision-makers at different functions and levels within an organisation. The literature remains sparse on how to construct cross-temporal forecasts, as the size of the hierarchy can easily become problematic. \cite{kourentzes2019cross_DKB} propose a heuristic approach to overcome the ensuing estimation issues. The approach works by compartmentalising the estimation. First, they obtain estimates of the cross-sectional reconciliation weights for each temporal level of the hierarchy. Then, these are combined across temporal levels, to a unique set that satisfies all coherency constraints. Using these combined weights, they obtain the reconciled bottom level forecasts, which can be aggregated as needed. Although they recognise that their approach can result in suboptimal results in terms of reconciliation errors, it guarantees coherent forecasts. Cross-temporal forecasts are more accurate than either temporal or cross-sectional hierarchical forecasts and provide a holistic view of the future across all planning levels and demarcations. \cite{Spiliotis2020-hj} also identify the problem, however, they do not focus on the coherency of forecasts and propose a sequential reconciliation across the two dimensions. This is shown to again be beneficial, but it does not achieve coherency. Arguably one can adapt the iterative correction algorithm by \cite{Kourentzes2020-mc} to enforce coherency in this approach as well. 
 
\subsubsection[Ecological inference forecasting (Jose M. Pav\'ia)]{Ecological inference forecasting\protect\footnote{This subsection was written by Jose M. Pav\'ia.}}
\label{sec:ecological_inference_forecasting}
Ecological inference forecasting (EIF) aims to predict the inner-cells values of a set of contingency tables when only the margins are known. It defines a fundamental problem in disciplines such as political science, sociology and epidemiology \citep{Salway2004-yp_JMP}. \cite{Cleave1995-eg_JMP}, \cite{Greiner2007-tz_JMP} and \cite{Pavia2009-ic_JMP} describe other areas of application. The fundamental difficulty of EIF lies in the fact that this is a problem with more unknowns than observations, giving rise to concerns over identifiability and indeterminacy: many sets of substantively different internal cell counts are consistent with a given marginal table. To overcome this issue, a similarity hypothesis (and, sometimes, the use of covariates) is routinely assumed. The basic hypothesis considers that either conditional row (underlying) probabilities or fractions are similar (related) among contingency tables \citep{Greiner2010-eg_JMP}. The covariations among row and column margins of the different tables are then used to learn about the internal cells.

The above hypothesis is not a cure-all to the main drawback of this approach. EIF is exposed to the so-called ecological fallacy \citep{Robinson1950-sz_JMP}: the presence of inconsistencies in correlations and association measures across different levels of aggregation. This is closely related to the well-known Simpson's Paradox \citep{Simpson1951-xf_JMP}. In this setting, the ecological fallacy is manifested through aggregation bias \citep{Wakefield2004-br_JMP} due to contextual effects and/or spatial autocorrelation \citep{Achen1995-ht_JMP}. This has led many authors to disqualify ecological inference forecasts \citep[see, for example,][]{Freedman1998-og_JMP,Tam_Cho1998-tf_JMP,Anselin2002-ea_JMP,Herron2004-uj_JMP} and many others to study under which circumstances ecological inference predictions would be reliable \citep{Firebaugh1978-va_JMP,Gelman2001-ev_JMP,Forcina2019-kz_JMP,Guseo2010-ym}.
Despite the criticisms, many algorithms for solving the EIF problem can be found in the literature, mainly from the ecological regression and mathematical programming frameworks (some of them available in functions of the R statistical software). 

The ecological regression literature has been prolific since the seminal papers of \cite{Goodman1953-lp_JMP,Goodman1959-gl_JMP} and \cite{Duncan1953-ip_JMP} and is undergoing a renaissance after \cite{King1997-ju_JMP}: new methods generalised from $2 \times 2$ tables to $R \times C$ tables have been proposed \citep{King1999-nt_JMP,Rosen2001-ec_JMP}, the geographical dimension of the data is being explicitly considered \citep{Calvo2003-lb_JMP,Puig2015-sm_JMP}, and new procedures combining aggregated and individual level data, including exit polls (see also \S\ref{sec:Elections_forecasting}), are introduced \citep{Glynn2010-se_JMP,Greiner2010-eg_JMP,Klima2019-gm_JMP}. See \cite{King2004-rj_JMP} for a wide survey and \cite{Klima2016-pg_JMP} and \cite{Plescia2018-lh_JMP} for an extensive evaluation of procedures. In mathematical programming exact and inequality constraints for the inner-cell values are incorporated in a natural way. Hence, this approach has shown itself to be a proper framework for generating ecological inference forecasts. The proposals from this approach can be traced back to \cite{Hawkes1969-gt_JMP} and \cite{Irwin1969-hn_JMP}. After them, some key references include \cite{McCarthy1977-jx_JMP}, \cite{Tziafetas1986-qt_JMP}, \cite{Corominas2015-lu_JMP}, \cite{Romero2020-no_JMP}, and \cite{PaviaRomero2021}. Solutions based on other strategies, for instance, entropy maximization, have been also suggested \citep[see, for example,][]{Johnston2000-mg_JMP,Bernardini_Papalia2020-ra_JMP}.

\subsection{Forecasting with judgment}
\label{sec:forecasting_with_judgment}

\subsubsection[Judgmental forecasting (Nigel Harvey)]{Judgmental forecasting\protect\footnote{This subsection was written by Nigel Harvey.}}
\label{sec:judgmental_forecasting}
People may use judgment alone to make forecasts or they may use it in combination with statistical methods. Here the focus is on pure judgmental forecasting (for judgmental adjustments, see \S\ref{sec:judgmental_adjustments_of_computer_based_forecasts}). Different types of judgment heuristic (mental `rules of thumb') can be used to make forecasts. The heuristic used depends on the nature of the information available to the forecaster \citep{Harvey2007-xo_NG}.

Consider cases where the only relevant information is held in the forecaster's memory. For example, someone might be asked whether Manchester United or Burnley will win next week's match. Here one memory-based heuristic that might be applicable is the recognition heuristic: if one recognises one object but not the other, then one should infer that the recognised object has higher value \citep{Goldstein2002-jm_NG}. In the above example, most people who recognise just one of the teams would be likely to make a correct forecast that Manchester United will win \citep{Ayton2011-k_NGt}. The availability heuristic is another memory-based heuristic that may be applicable: objects that are most easily brought to mind are those which are more likely. Thus, if we are asked which team is likely to come top of the premier league, we would say Manchester United if that is the one that most easily comes to mind. The availability heuristic is often effective because more likely events are encountered more often and more recently and are hence better remembered. However, it can be disrupted by, for example, greater media coverage of more unlikely (and hence more interesting) events. 

Consider next cases in which forecasters possess information about values of one or more variables correlated with the variable to be forecast. For example, teachers may wish to forecast the grades of their students in a final examination on the basis of past records of various other measures. \cite{Kahneman1973-fx_NG} suggested that people use the representativeness heuristic to deal with this type of situation. Forecasters first select a variable that they think is able to represent the one that must be predicted. For example, a teacher may consider that frequency in attending voluntary revision classes represents a student's ability in the final examination. Thus, if a student attended 15 of the 20 revision classes, they are likely to obtain 75\% in the final examination.

Finally, consider situations in which people forecast future values of a variable on the basis of a record of previous values of that variable. There is some evidence that, when forecasting from time series, people use anchor-and-adjustment heuristics \citep{Hogarth1981-ft_NG,Lawrence1992-vo_NG}. For example, (\textit{i}) when forecasting from an upward trended series, they anchor on the last data point and then make an upward adjustment to take the trend into account and (\textit{ii}) when forecasting from an untrended series containing autocorrelation, they anchor on the last data point and make an adjustment towards the mean to take the autocorrelation into account. 

\cite{Kahneman_2011_NG} and others have divided cognitive processes into those which are intuitive (System 1) and those which are deliberative (System 2). We have discussed only intuitive processes underlying judgmental forecasting \citep{Gigerenzer2007-cd_NG}. However, they can be supplemented by deliberative (System 2) processes \citep{Theocharis2019-tn_NG} in some circumstances.

\subsubsection[Judgmental adjustments of computer-based forecasts (Paul Goodwin)]{Judgmental adjustments of computer-based forecasts\protect\footnote{This subsection was written by Paul Goodwin.}}
\label{sec:judgmental_adjustments_of_computer_based_forecasts}
Judgmental adjustments to algorithmic computer-based forecasts can enhance accuracy by incorporating important extra information into forecasts \citep{McNees1990-zr_PG,Fahimnia2020-zj_PG,Perera2019-xi_PG}. 
However, cognitive factors (see, for example, \S\ref{sec:judgmental_forecasting}), and motivational biases (see \S\ref{sec:Forecasting_in_the_supply_chain}), can lead to the inefficient use of information \citep{Fildes2019-uj_PG}, unwarranted adjustments and reductions in accuracy \citep{Fildes2009-tc,Franses2009-ar_PG}. 

People may `egocentrically discount' a computer's forecasts when its rationale is less clear than their own reasoning \citep{Bonaccio2006-vy_PG}. They can also be less tolerant of errors made by algorithms than those made by humans \citep[][and \S\ref{sec:Trust_in_forecasts}]{Onkal2009-nm,Dietvorst2015-fy}. The random errors associated with algorithmic forecasts, and the salience of rare large errors, can therefore lead to an unjustified loss of trust in computer forecasts. Adjustments may also give forecasters a sense of ownership of forecasts or be used to justify their role \citep{Onkal2005-nr_PG}. 

Computer-based forecasts are designed to filter randomness from time-series. In contrast, humans tend to perceive non-existent systematic patterns in random movements \citep[][and \S\ref{sec:Judgmental_forecasting_in_practice}]{OConnor1993-mq_PG,Reimers2011-qn_NG} and apply adjustments to reflect them. This can be exacerbated by the narrative fallacy \citep{Taleb2008-jd_PG}, where people invent stories to explain these random movements, and hindsight bias \citep{Fischhoff2007-cg_PG}, where they believe, in retrospect, that these movements were predictable. Recent random movements, and events, are particularly likely to attract undue attention, so long-run patterns identified by the computer are given insufficient weight \citep{Bolger1993-zl_PG}. Damaging interventions are also probable when they result from political interference \citep{Oliva2009-nz_PG} or optimism bias \citep{Fildes2009-tc}, or when they reflect information already factored into the computer’s forecast, leading to double counting \citep{Van_den_Broeke2019-os_PG}.

How can interventions be limited to occasions when they are likely to improve accuracy? Requiring people to document reasons justifying adjustments can reduce gratuitous interventions \citep{Goodwin2000-xp}. Explaining the rationale underlying statistical forecasts also improved adjustment behaviour when series had a simple underlying pattern in a study by \cite{Goodwin1999-my_PG}. However, providing guidance on when to adjust was ineffective in an experiment conducted by \cite{Goodwin2011-mx}, as was a restriction preventing people from making small adjustments.

When determining the size of adjustments required, decomposing the judgment into a set of easier tasks improved accuracy in a study by \cite{Webby2005-rb_PG}. Providing a database of past outcomes that occurred in circumstances analogous to those expected in the forecast period also improved adjustments in a study by \cite{Lee2007-wm_PG}. Outcome feedback, where the forecaster is informed of the most recent outcome is unlikely to be useful since it contains noise and exacerbates the tendency to over-focus on recent events \citep{Goodwin1999-my_PG,Petropoulos2016-uk_PG}. However, feedback on biases in adjustments over several recent periods may improve judgments \citep{Petropoulos2017-dj_PG}. Feedback will be less useful where interventions are rare so there is insufficient data to assess performance. 

The evidence for this section is largely based on laboratory-based studies of adjustment behaviour. \S\ref{sec:Judgmental_forecasting_in_practice} gives details of research into forecast adjustment in practice and discusses the role of forecasting support systems in improving the effectiveness of judgmental adjustments.

\subsubsection[Judgmental model selection (Shari De Baets)]{Judgmental model selection\protect\footnote{This subsection was written by Shari De Baets.}}
\label{sec:judgmental_model_selection}
Forecasters -- practitioners and researchers alike -- use Forecast Support Systems (FSS) in order to perform their forecasting tasks. Usually, such an FSS allows the forecaster to load their historical data and they can then apply many different types of forecasting techniques to the selected data. The idea is that the forecaster selects the method which leads to highest forecast accuracy. Yet, there is no universally `best' method, as it depends on the data that is being forecasted (see also \S\ref{sec:Model_Selection}). Thus, selection is important in achieving high accuracy. But how does this selection occur? 

Research on judgmental selection of statistical forecasts is limited in quantity. \cite{Lawrence2002-wv} found that participants were not very adept at selecting good forecasting algorithms from a range offered to them by an FSS and had higher error than those who were presented with the optimal algorithm by an FSS. \cite{Petropoulos2018-mt} compared judgmental selection of forecasting algorithms with automatic selection based on predetermined information criteria. They found that judgmental selection was better than automatic selection at avoiding the `worst' models, but that automatic selection was superior at choosing the `best' ones. In the end, overall accuracy of judgmental selection was better than that of algorithmic selection. If their experiment had included more variation of the data (trends, fractals, different autoregressive factors) and variation of proposed models, this could possibly have led to better algorithmic than judgmental performance \citep{Harvey2019foresight}. Time series that are more complex will place a higher cognitive load on judgmental selection. This was confirmed in a study by \cite{Han2019-ln}, who used an electroencephalogram (EEG) for the comparison of judgmental forecast selection versus (judgmental) pattern identification. They found that pattern identification outperformed forecast selection, as the latter required a higher cognitive load, which in turn led to a lower forecasting accuracy.

It is likely that, in practice, judgmental selection is much more common than automatic selection. This preference for human judgment over advice from an algorithm has been shown in an experiment by \cite{Onkal2009-nm}. But how apt are forecasters in distinguishing ‘good’ models from `bad' models? This was investigated by \cite{De_Baets2020-du} in an experiment. People were asked to select the best performing model out of a choice of two different qualities (accuracies) of models (different combinations of good versus medium versus bad). People’s choice outperformed forecasts made by averaging the model outputs, lending credence to the views of \cite{Fific2014-zk}. The performance of the participants improved with a larger difference in quality between models and a lower level of noise in the data series. In a second experiment, \cite{De_Baets2020-du} found that participants adjusted more towards the advice of what they perceived to be a good quality model than a medium or bad quality one.

Importantly, in selecting an algorithm and seeing it err, people are quick to abandon it. This phenomenon is known as `algorithm aversion' \citep[][see also \S\ref{sec:Trusting_model_and_expert_forecasts}]{Dietvorst2015-fy} and is due to a `perfection schema' we have in our heads where algorithms are concerned \citep{Madhavan2007-hj}. We do not expect them to `fail' and thus react strongly when they do. While a model may not perform as it should for a particular dataset and may thus elicit algorithm aversion for that particular method, one should not abandon it for all datasets and future forecasts. 

\subsubsection[Panels of experts (Konstantia Litsiou)]{Panels of experts\protect\footnote{This subsection was written by Konstantia Litsiou.}}
\label{sec:panels_of_experts}
Panels of experts are often used in practice to produce judgmental forecasts (see, for example, \S\ref{sec:New_product_forecasting} and \S\ref{sec:Elections_forecasting}). This is especially true in cases with limited available quantitative data and with the level of uncertainty being very high. In this section, three methods for eliciting judgmental forecasts from panels of experts are presented: the Delphi method, interaction groups (IG), and structured analogies (SA).

The Delphi method is centred around organising and structuring group communication \citep{Rao2010-kd_Litsiou}, which aims to achieve a convergence of opinion on a specific real-world issue. It is a multiple-round survey in which experts participate anonymously to provide their forecasts and feedback \citep{Rowe2001-jg_Litsiou}. At the end of each round, the facilitator collects and prepares statistical summaries of the panel of experts’ forecasts. These summaries are presented as feedback to the group, and may be used towards revising their forecasts. This loop continues until a consensus is reached, or the experts in the panel are not willing to revise their forecasts further. In some implementations of the Delphi method, justification of extreme positions (forecasts) is also part of the (anonymous) feedback process. The Delphi method results in a more accurate outcome in the decision-making process \citep{Dalkey1969-dp_Litsiou,Steurer2011-yz_Litsiou}. \cite{Rowe2001-jg_Litsiou} mentioned that, by adopting the Delphi method, groups of individuals can produce more accurate forecasts than simply using unstructured methods. A drawback of the Delphi method is the additional cost associated with the need to run multiple rounds, extending the forecasting process as well as increasing the potential drop-out rates. On the other hand, the anonymity in the Delphi method eliminates issues such as groupthink and the `dominant personalities' effects \citep{Van_de_Ven1971-vd_Litsiou}. 

The IG method suggests that the members of the panel of experts actively interact and debate their points to the extent they have to reach an agreement on a common forecast \citep{Litsiou2019-gy_Litsiou}. \cite{Sniezek1989-gc_Litsiou} found that members of interacting groups provide more accurate judgments compared to individuals. However, there is mixed evidence about the forecasting potential of IG \citep{Scott_Armstrong2006-cp_Litsiou,Boje1982-qk_Litsiou,Graefe2011-ha_Litsiou}. Besides, the need for arranging and facilitating meetings for the IG makes it a less attractive option.

Another popular approach to judgmental forecasting using panels of experts is SA, which refers to the recollection of past experiences and the use analogies \citep{Green2007-ts_Litsiou}. In the SA method, the facilitator assembles a panel of experts. The experts are asked to recall and provide descriptions, forecasts, and similarities/differences for cases analogous to the target situation, as well as a similarity ranking for each of these analogous cases. The facilitator gathers the lists of the analogies provided by the experts, and prepares summaries, usually using weighted averages of the recalled cases based on their similarity to the target situation (see also \S\ref{sec:Wisdom_of_crowds}). Semi-structured analogies (sSA) have also been proposed in the literature, where the experts are asked to provide a final forecasts based on the analogous cases they recalled, which essentially reduces the load for the facilitator \citep{Nikolopoulos2015-mz_Litsiou}.
\cite{Nikolopoulos2015-mz_Litsiou} supported that the use of SA and IG could result to forecasts that are 50\% more accurate compared to unstructured methods (such as unaided judgment). One common caveat of using panels of experts is the difficulty to identify who a real expert is. Engaging experts with high level of experience, and encouraging the interaction of experts are also supported by \cite{Armstrong2018-cf_Litsiou}. 

\subsubsection[Scenarios and judgmental forecasting (M. Sinan G\"on\"ul)]{Scenarios and judgmental forecasting\protect\footnote{This subsection was written by M. Sinan G\"on\"ul.}}
\label{sec:scenarios_forecasting}
Scenarios provide exhilarating narratives about conceivable futures that are likely to occur. Through such depictions they broaden the perspectives of decision makers and act as mental stimulants to think about alternatives. Scenarios enhance information sharing and provide capable tools for communication within organisations. By virtue of these merits, they have been widely used in corporate planning and strategy setting since 1960's \citep{Godet1982-sw_SG,Schoemaker1991-su_SG,Wright1999-zd_SG,Wright2009-ro_SG,Goodwin2010-ma_SG}. Even though utilisation of scenarios as decision advice to judgmental forecasting has been proposed earlier \citep{Schnaars1987-py_SG,Bunn1993-wq_SG}, the research within this domain remained limited until recently when the interest in the subject has rekindled \citep{Onkal2013-ek_DILEK,Goodwin2019-ak_SG,Wicke2019-tq_SG}.

The recent research has used behavioural experimentation to examine various features of scenarios and their interactions with judgmental adjustments (see \S\ref{sec:judgmental_adjustments_of_computer_based_forecasts}) of model-based forecasts. \cite{Onkal2013-ek_DILEK} explored the `content' effects of scenarios where through the narration either a bleak/negative future (a pessimistic scenario) or a bright/positive future (an optimistic scenario) was portrayed. On a demand forecasting context for mobile phones, the participants first received time-series data, model-based forecasts and then asked to generate point and interval forecasts as well as provide a confidence measure. With respect to the existence of scenarios, there were four conditions where the participants may receive: (\textit{i}) no scenarios, (\textit{ii}) optimistic scenarios, (\textit{iii}) pessimistic scenarios, and (\textit{iv}) both scenarios. Findings indicate that decision makers respond differently to optimistic and pessimistic scenarios. Receiving optimistic scenarios resulted in making larger adjustments to the model-based forecasts. At the same time, led to an increased confidence of the participants in their predictions. On the other hand, participants who received negative scenarios tend to lower their predictions the most among the four groups. An intriguing finding was the balancing effect of scenarios on the interval forecast symmetry. The lower interval bounds were adjusted upwards the most towards the centre-point of the interval (i.e., model-based predictions) when optimistic scenarios were received. Similarly, the upper bounds were adjusted downwards the most towards the centre-point of the interval in the presence of pessimistic scenarios. 

The prospects of receiving a single scenario versus multiple scenarios were further explored in \cite{Goodwin2019-ak_SG}. The researchers investigated whether assimilation or contrast effects will occur when decision makers see optimistic (pessimistic) forecasts followed by pessimistic (optimistic) ones compared against receiving a single scenario in solitude. In case of assimilation, a scenario presenting an opposing world view with the initial one would cause adjustments in the opposite direction creating an offset effect. On the other hand, in case of contrast, the forecasts generated after the initial scenarios would be adjusted to more extremes when an opposing scenario is seen. In two experiments conducted in different contexts the researchers found resilient evidence for contrast effects taking place. Interestingly, seeing an opposing scenario also increased the confidence of the forecasters in their initial predictions.

In terms of the effects of scenario presence on the forecasting performance, however, the experimental evidence indicates the benefits are only circumstantial. \cite{Goodwin2019-qh_SG} found that providing scenarios worsened forecast accuracy and shifted the resultant production order decisions further away from optimality. Despite this performance controversy, the decision makers express their fondness in receiving scenarios and belief in their merits \citep{Onkal2013-ek_DILEK,Goodwin2019-ak_SG}. Therefore, we need more tailored research on scenarios and judgmental forecasting to reveal the conditions when scenarios can provide significant improvements to the forecasting accuracy.

\subsubsection[Trusting model and expert forecasts (Dilek \"Onkal)]{Trusting model and expert forecasts\protect\footnote{This subsection was written by Dilek \"Onkal.}}
\label{sec:Trusting_model_and_expert_forecasts}
Defined as ``firm belief in the reliability, truth, and ability of someone/something'' (Oxford English Dictionary), trust entails accepting vulnerability and risk \citep{Rousseau1998-ix_DILEK}. Given that forecasts are altered or even discarded when distrusted by users, examining trust is a central theme for both forecasting theory and practice. 

Studies examining individual's trust in model versus expert forecasts show that individuals often distrust algorithms \citep{Meehl2013-is_DILEK,Burton2020-nr_DILEK} and place higher trust on human advice \citep[but also \S\ref{sec:judgmental_adjustments_of_computer_based_forecasts}, \S\ref{sec:judgmental_model_selection}, and \S\ref{sec:Trust_in_forecasts}]{Diab2011-ab_DILEK,Eastwood2012-cg_DILEK}. We live in an era where we are bombarded with news about how algorithms get it wrong, ranging from COVID-19 forecasts affecting lockdown decisions to algorithmic grade predictions affecting university admissions. Individuals appear to prefer forecasts from humans over those from statistical algorithms even when those forecasts are identical \citep{Onkal2009-nm}. Furthermore, they lose trust in algorithms quicker when they see forecast errors \citep{Dietvorst2015-fy,Prahl2017-ry_DILEK}. Such `algorithm aversion' and error intolerance is reduced when users have opportunity to adjust the forecasting outcome, irrespective of the extent of modification allowed \citep{Dietvorst2018-no_DILEK}. Feedback appears to improve trust, with individuals placing higher trust in algorithms if they can understand them \citep{Seong2008-ac_DILEK}. Overuse of technical language may reduce understanding of the forecast/advice, in turn affecting perceptions of expertise and trustworthiness \citep{Joiner2002-ij_DILEK}. Explanations can be helpful \citep{Goodwin2013-na_DILEK}, with their packaging affecting judgments of trustworthiness \citep{Elsbach2000-cm_DILEK}. Algorithmic appreciation appears to easily fade with forecasting expertise \citep{Logg2019-zd_DILEK}, emphasising the importance of debiasing against overconfidence and anchoring on one’s own predictions.

Trusting experts also presents challenges \citep{Hendriks2015-vh_DILEK,Hertzum2014-fm_DILEK,Maister2012-qb_DILEK}. Expert forecasts are typically seen as predisposed to group-based preconceptions \citep{Brennan2020-od_DILEK,Vermue2018-ab_DILEK}, along with contextual and motivational biases \citep{Burgman2016-of_DILEK}. Misinformed expectations, distorted exposures to `forecast failures', and over-reliance on one's own judgments may all contribute to distrusting experts as well as algorithms.

Credibility of forecast source is an important determinant in gaining trust \citep{Onkal2019-uq_DILEK}. Studies show that the perceived credibility of system forecasts affects expert forecasters' behaviours and trust \citep{Alvarado-Valencia2014-eb_DILEK}, while providing information on limitations of such algorithmic forecasts may reduce biases \citep{Alvarado-Valencia2017-ix_DILEK}. Previous experience with the source appears to be key to assessing credibility \citep{Hertzum2002-es_DILEK} and trust \citep{Cross2004-uf_DILEK}. Such `experienced' credibility appears to be more influential on users' acceptance of given forecasts as opposed to `presumed' credibility \citep{Onkal2017-wd_DILEK}. Source credibility can be revised when forecast (in)accuracy is encountered repetitively \citep{Jiang1996-ow_DILEK}, with forecaster and user confidence playing key roles \citep{Sah2013-lm_DILEK}. 

Trust is critical for forecasting efforts to be translated into sound decisions \citep{Choi2020-rr_DILEK,Ozer2011-jx_DILEK}. Further work on fostering trust in individual/collaborative forecasting will benefit from how trusted experts and models are selected and combined to enhance decision-making.

\subsection{Evaluation, validation, and calibration}
\label{sec:evaluation_and_validation}

\subsubsection[Benchmarking (Anastasios Panagiotelis)]{Benchmarking\protect\footnote{This subsection was written by Anastasios Panagiotelis.}}
\label{sec:benchmarking}
When a new forecasting model or methodology is proposed, it is common for its performance to be benchmarked according to some measure of forecast accuracy against other forecasting methods using a sub-sample of some particular time series. In this process, there is the risk that either the measures of accuracy, competing forecasting methods or test data, are chosen in a way that exaggerates the benefits of a new method. This possibility is only exacerbated by the phenomenon of publication bias \citep{Dickersin1990Benchmark}. 
	
A rigorous approach to benchmarking new forecasting methods should follow the following principles:	
\begin{enumerate}[noitemsep]
\item New methods should always be compared to a larger number of suitable benchmark methods. These should at a minimum include na\"ive methods such as a random walk and also popular general purpose forecasting algorithms such as ARIMA models, Exponential Smoothing, Holt Winters and the Theta method (see \S\ref{sec:Statistical_and_econometric_models} and references therein).
\item Forecasts should be evaluated using a diverse set of error metrics for point, interval and probabilistic forecasts (see \S\ref{sec:point_forecast_accuracy_measures}). Where the forecasting problem at hand should be tailored to a specific problem, then appropriate measures of forecast accuracy must be used. As an example, the literature on Value at Risk forecasting has developed a number of backtesting measures for evaluating the quality of quantile forecasts \citep[see][and references therein]{YangNadarajah2018Benchmark}.
\item Testing should be carried out to discern whether differences between forecasting methods are statistically significant. For discussion see \S\ref{sec:Statistical_tests_of_forecast_performance}. However, there should also be a greater awareness of the debate around the use of hypothesis testing both in forecasting \citep{Armstrong2007Benchmark} and more generally in statistics \citep{WassersteinLazar2016Benchmark}.
\item Sample sizes for rolling windows should be chosen with reference to the latest literature on rolling window choice \citep[see][and references therein]{inoue2017_ABM}.
\item All code used to implement and benchmark new forecasting methods should, where possible, be written in open source programming languages (such as C, Python and R). This is to ensure replicability of results \citep[for more on the replicablity crisis in research see][and references therein]{Peng2015Benchmark}
\item Methods should be applied to appropriate benchmark datasets.
\end{enumerate}

Regarding the last of these points there are some examples in specific fields, of datasets that already play a de facto role as benchmarks. In macroeconomic forecasting, the U.S. dataset of \cite[][see \S\ref{sec:Forecasting_with_Big_Data}]{JLC_StockWatson12} is often used to evaluate forecasting methods that exploit a large number of predictors, with \cite{ForniEtAl2003Benchmark} and \cite{PanagiotelisEtAl2019Benchmark} having constructed similar datasets for the EU and Australia respectively. In the field of energy, the GEFCom data \citep{hong2016probabilistic_FZ} discussed in \S\ref{sec:Load_forecasting} and the IEEE 118 Bus Test Case data \citep{PenaEtAl2018Benchmark} are often used as benchmarks. Finally, the success of the M Forecasting competitions \citep{Makridakis2020-mm} provide a benchmark dataset for general forecasting methodologies (see \S\ref{sec:Forecasting_competitions} and references therein).

A recent trend that has great future potential is the publication of websites that demonstrate the efficacy of different forecasting methods on real data. The Covid-19 Forecast Hub\footnote{https://viz.covid19forecasthub.org/} and the Business Forecast Lab\footnote{https://business-forecast-lab.com/} provide notable examples in the fields of epidemiology and macroeconomics and business respectively.

\subsubsection[Point, interval, and pHDR forecast error measures (Stephan Kolassa)]{Point, interval, and pHDR forecast error measures\protect\footnote{This subsection was written by Stephan Kolassa.}}
\label{sec:point_forecast_accuracy_measures}
\textit{Point forecasts} are single number forecasts for an unknown future quantity also given by a single number. \textit{Interval forecasts} take the form of two point forecasts, an upper and a lower limit. Finally, a less common type of forecast would be a \textit{predictive Highest Density Region} (pHDR), i.e., an HDR \citep{Hyndman1996_SK} for the conditional density of the future observable. pHDRs would be interesting for multimodal (possibly implicit) predictive densities, e.g., in scenario planning. Once we have observed the corresponding realisation, we can evaluate our point, interval and pHDR forecasts.

There are many common point forecast error measures (PFEMs), e.g., the mean squared error (MSE), mean absolute error (MAE), mean absolute scaled error (MASE), mean absolute percentage error (MAPE) or many others \citep[see section 3.4 in][]{hyndman2018Forecasting_CB}. Which one is most appropriate for our situation, or should we even use multiple different PFEMs? 

There are many common point forecast error measures (PFEMs), e.g., the MSE, MAE, MASE, (s)MAPE, the quantile score or pinball loss or many others 
\citep[e.g., sections 5.8 and 5.9 in][]{HyndmanAthanasopoulos2021-09-27}.
Assuming $n$ historical periods with observations $y_1, \dots, y_n$ and a forecasting horizon $H$ with observations $y_{n+1}, \dots, y_{n+H}$ and point forecasts $f_{n+1}, \dots, f_{n+H}$, we have:
\begin{gather*}
    \text{MSE} = \sum_{t=n+1}^{n+H}(y_t-f_t)^2, \quad
	\text{MAE} = \sum_{t=n+1}^{n+H}|y_t-f_t|, \quad
	\text{MASE} = \frac{\sum_{t=n+1}^{n+H}|y_t-f_t|}{\sum_{t=2}^n|y_t-y_{t-1}|} \\
    \text{MAPE} = \sum_{t=n+1}^{n+H}\frac{|y_t-f_t|}{y_t}, \quad
	\text{sMAPE} = \sum_{t=n+1}^{n+H}\frac{|y_t-f_t|}{\frac{1}{2}(y_t+f_t)} \\
    Q_\alpha = \sum_{t=n+1}^{n+H}(1-\alpha)(f_t-y_t)1_{y_t<f_t} + \alpha(y_t-f_t)1_{y_t \geq f_t}.
\end{gather*}

Let us take a step back. Assume we have a full density forecast and wish to ``condense'' it to a point forecast that will minimise some PFEM in expectation. The key observation is that \textit{different PFEMs will be minimised by different point forecasts derived from the same density forecast} \citep{Kolassa2020_SK}. 
\begin{itemize}[noitemsep]
\item The MSE is minimised by the expectation.
\item The MAE and MASE are minimised by the median \citep{HanleyJosephPlattEtAl2001_SK}.
\item The MAPE is minimised by the $(-1)$-median \citep[][p.~752 with $\beta=-1$]{Gneiting2011a_SK}.
\item The sMAPE is minimised by an unnamed functional that would need to be minimised numerically \citep{Goncalves2015}.
\item The hinge/tick/pinball $Q_p$ loss is minimised by the appropriate $p$-quantile \citep{Gneiting2011_SK}.
\item In general, there is no loss function that is minimised by the mode \citep{Heinrich2014_SK}.
\end{itemize}

We note that intermittent demand (see \S\ref{sec:Forecasting_for_intermittent_demands_and_count_data}) poses specific challenges. On the one hand, the MAPE is undefined if there are zeros in the actuals. On the other hand, the point forecasts minimising different PFEMs will be very different. For instance, the conditional median (minimising the MAE) may well be a flat zero, while the conditional mean (minimising the MSE) will usually be nonzero. Our forecasting algorithm may not output an explicit density forecast. It is nevertheless imperative to think about which functional of the implicit density we want to elicit \citep{Gneiting2011a_SK}, and tailor our error measure -- and forecasting algorithm! -- to it. It usually makes no sense to evaluate a point forecast with \textit{multiple} PFEMs \citep{Kolassa2020_SK}.

Interval forecasts can be specified in multiple ways. We can start with a probability coverage and require two appropriate quantiles -- e.g., we could require a 2.5\% and a 97.5\% quantile forecast, yielding a symmetric or equal-tailed 95\% interval forecast. Interval forecasts ($\ell_t,u_t$) of this form can be evaluated by the interval score \citep{Winkler1972_SK,BrehmerGneiting2020_SK}, a proper scoring rule \citep[section 6.2 in][]{GneitingRaftery2007_SK}:
\begin{equation}
	\text{IS}_\alpha = \sum_{t=n+1}^{n+H} (u_t-\ell_t) + \frac{2}{\alpha}(\ell_t-y_t)1_{y_t<\ell_t} +
		\frac{2}{\alpha}(y_t-u_t)1_{y_t>u_t}. 
\end{equation}

We can also use the hinge loss to evaluate the quantile forecasts separately. 

Alternatively, we can require a shortest interval subject to a specified coverage. This interval is not elicitable relative to practically relevant classes of distributions \citep{BrehmerGneiting2020_SK,FisslerFrongilloHlavinovaEtAl2020}.

Yet another possibility is to maximise the interval forecast's probability coverage, subject to a maximum length $\ell$. This modal interval forecast $(f_t,f_t+\ell)$ is elicitable by an appropriate $\ell$-zero-one-loss \citep{BrehmerGneiting2020_SK}
\begin{equation}
	L_\ell = \sum_{t=n+1}^{n+H}1_{f_t<y_t<f_t+\ell} =
	\#\big\{t\in n+1, \dots, n+H\,\big|\,f_t<y_t<f_t+\ell\big\}.
\end{equation}

The pHDR is not elicitable even for unimodal densities \citep{BrehmerGneiting2020_SK}. In the multimodal case, the analysis is likely difficult. Nevertheless, a variation of the Winkler score has been proposed to evaluate pHDRs on an ad hoc basis \citep{Hyndman2020a_SK}. One could also compare the achieved to the nominal coverage, e.g., using a binomial test -- which disregards the volume of the pHDR \citep{Kolassa2020a_SK}.

In conclusion, there is a bewildering array of PFEMs, which require more thought in choosing among than is obvious at first glance. The difficulties involved in evaluating interval and pHDR forecasts motivate a stronger emphasis on full density forecasts \citep[cf.][and \S\ref{sec:evaluating_probabilistic_forecasts}]{AskanaziDieboldSchorfheideEtAl2018_SK}.

\subsubsection[Scoring expert forecasts (Yael Grushka-Cockayne)]{Scoring expert forecasts\protect\footnote{This subsection was written by Yael Grushka-Cockayne.}}
\label{sec:Scoring_expert_forecasts_for_decision_making}
Evaluating forecasting capabilities can be a difficult task. One prominent way to evaluate an expert's forecast is to score the forecast once the realisation of the uncertainty is known. Scoring forecasts using the outcome's realisations over multiple forecasts offers insights into an individual's expertise. Experts can also use scoring information to identify ways to improve future forecasts. In addition, scoring rules and evaluation measures can be designed to match decision-making problems, incentivising forecasts that are most useful in a specific situation \citep{Winkler2019-yi_YGC}.

Scoring rules were first suggested for evaluate meteorological forecasts in work by \cite{Brier1950-ps_YGC}. Scoring rules have since been used in a wide variety of settings, such as business and other applications. When forecasting a discrete uncertainty with only two possible outcomes (e.g., a loan with be defaulted on or not, a customer will click on an ad or not), the Brier score assigns a score of $-(1-p)^2$, where $p$ is the probability forecast reported that the event will occurs. The greater the probability reported for an event that occurs, the higher the score the forecast receives. Over multiple forecasts, better forecasters will tend to have higher average Brier scores. For discrete events with more than two outcomes, a logarithmic scoring rule can be used.

The scoring rules are attractive to managers in practice since they are considered proper. Proper scoring rules (see also \S\ref{sec:evaluating_probabilistic_forecasts}) incentivise honest forecasts from the experts, even prior to knowing the realisation of an uncertainty, since ex ante the expected score is maximised only when reported probabilities equals true beliefs \citep{Winkler1996-tk_YGC,OHagan2006-hl_YGC,Bickel2007-uv_YGC,GneitingRaftery2007_SK,Merkle2013-up_YGC}. Examples of a scoring rule that is not proper yet still commonly used are the linear score, which simply equals the reported probability or density for the actual outcome, or the skill score, which is the percentage improvement of the Brier score for the forecast relative to the Brier score of some base line naive forecast \citep{Winkler2019-yi_YGC}.

For forecasting continuous quantities, forecasts could be elicited by asking for an expert's quantile (or fractile) forecast rather than a probability forecast. For instance, the 0.05, 0.25, 0.50, 0.75 and 0.95 quantiles are often elicited in practice, and in some cases every 0.01 quantile, between 0-1 are elicited
\citep[e.g., the 2014 Global Energy Forecasting Competition,][]{hong2016probabilistic_FZ}. Proper scoring rules for quantiles are developed in \cite{Jose2009-ju_YGC}.

When forecasts are used for decision-making, it is beneficial if the scoring rule used relates in some manner to the decision problem itself.  In certain settings, the connection of the scoring rule to the decision context is straight forward. For example, \cite{Jose2008-jk_YGC} develop scoring rules that can be mapped to decision problems based on the decision maker’s utility function. \cite{Johnstone2011-ta_YGC} develop tailored scoring rules aligning the interest of the forecaster and the decision maker.  \cite{Grushka-Cockayne2017-za_YGC} link quantile scoring rules to business profit-sharing situations. 
 
\subsubsection[Evaluating probabilistic forecasts (Florian Ziel)]{Evaluating probabilistic forecasts\protect\footnote{This subsection was written by Florian Ziel.}}
\label{sec:evaluating_probabilistic_forecasts}
Probabilistic forecasting is a term that is not strictly defined,
but usually refers to everything beyond point forecasting \citep{Gneiting2011a_SK}. However,
in this section we consider only the evaluation of full predictive distributions or equivalent characterisations. For the evaluation of prediction of quantiles, intervals and related objects, see \S\ref{sec:point_forecast_accuracy_measures}.

One crucial point for evaluating probabilistic forecasts is the reporting, which is highly influenced from meteorologic communities. From the theoretical point of view, we should always report the predicted cumulative distribution function $\widehat{F}$ of our prediction target $F$. Alternatively for continuous data, reporting the probability density function is a popular choice. For univariate prediction problems a common alternative is to report quantile forecast on a dense grid of probabilities, as it approximates the full distribution \citep{hong2016probabilistic_FZ}. For multivariate forecasts, it seems to become standard to report a large ensemble (a set of simulated trajectories/paths) of the full predictive distribution. 
The reason is that the reporting of a multivariate distribution (or an equivalent characterisation) of sophisticated prediction models is often not feasible or practicable, especially for non-parametric or copula-based forecasting methods.

In general, suitable tools for forecasting evaluation are proper scoring rules as they address calibration and sharpness simultaneously \citep{GneitingRaftery2007_SK,gneiting2014_VRRJ}. Preferably, we consider strictly proper scoring rules which can identify the true predicted distribution among a set of forecast candidates that contains the true model.

In the univariate case the theory is pretty much settled and there is quite some consensus about the evaluation of probabilistic forecasts \citep{gneiting2014_VRRJ}. The continuous ranked probability score (CRPS) and logarithmic scores (log-score) are popular strictly proper scoring rules, while the quadratic and pseudospherical score remain strictly proper alternatives. The CRPS can be well approximated by averaging across quantile forecasts on an equidistant grid of probabilities \citep{nowotarski2018recent_FZ}.

For multivariate forecast evaluation the situation is more complicated and many questions remain open \citep{GneitingRaftery2007_SK, meng2020scoring_FZ}. The multivariate version of the log-score is a strictly proper scoring rule, but it requires the availability of a multivariate density forecast. This makes it impracticable for many applications. \cite{GneitingRaftery2007_SK} discuss the energy score, a multivariate generalisation of the CRPS, that is strictly proper. Still, it took the energy score more than a decade to increase its popularity in forecasting. A potential reason is the limited simulation study of \cite{pinson2013discrimination_FZ} that concludes that the energy score can not discriminate well differences in the dependency structure. In consequence other scoring rules were proposed in literature, e.g., the variogram score \citep{scheuerer2015variogram_FZ} which is not strictly proper. \cite{ziel2019multivariate_FZ} consider a strictly proper scoring method for continuous variables using copula techniques. In contrast to \cite{pinson2013discrimination_FZ}, recent studies \citep{ziel2019multivariate_FZ,lerch2020simulation_FZ} show that the energy score discriminates well when used together with significance tests like the Diebold-Mariano (DM) test. 
In general, we recommended scoring be applied with
reliability evaluation (see \S\ref{sec:assessing_the_reliability_of_probabilistic forecats}) and significance tests (see \S\ref{sec:Statistical_tests_of_forecast_performance}). 
Additionally, if we want to learn about the performance of our forecasts it is highly recommended to consider multiple scoring rules and evaluate on lower-dimensional subspaces. For multivariate problems, this holds particularly for the evaluation of univariate and bivariate marginal distributions.
 
\subsubsection[Assessing the reliability of probabilistic forecasts (Thordis Thorarinsdottir)]{Assessing the reliability of probabilistic forecasts\protect\footnote{This subsection was written by Thordis Thorarinsdottir.}}
\label{sec:assessing_the_reliability_of_probabilistic forecats}
Probabilistic forecasts in the form of predictive distributions are central in risk-based decision making where reliability, or calibration, is a necessary condition for the optimal use and value of the forecast. A probabilistic forecast is calibrated if the observation cannot be distinguished from a random draw from the predictive distribution or, in the case of ensemble forecasts, if the observation and the ensemble members look like random draws from the same distribution. Additionally, to ensure their utility in decision making, forecasts should be sharp, or specific, see \S\ref{sec:point_forecast_accuracy_measures} and \S\ref{sec:evaluating_probabilistic_forecasts} as well as \cite{Gneiting&2007_TT}.

In the univariate setting, several alternative notions of calibration exist for both a single forecast \citep{Gneiting&2007_TT, Tsyplakov2013_TT} and a group of forecasts \citep{StraehlZiegel2017_TT}. The notion most commonly used in applications is probabilistic calibration \citep{Dawid1984_TT}; the forecast system is probabilistically calibrated if the probability integral transform (PIT) of a random observation, that is, the value of the predictive cumulative distribution function in the observation, is uniformly distributed. If the predictive distribution has a discrete component, a randomised version of the PIT should be used \citep{GneitingRanjan2013_TT}. 

Probabilistic calibration is assessed visually by plotting the histogram of the PIT values over a test set. A calibrated forecast system will return a uniform histogram, a $\cap$-shape indicates overdispersion and a $\cup$-shape indicates underdispersion, while a systematic bias results in a biased histogram \citep[e.g.][]{ThorarinsdottirSchuhen2018_TT}. The discrete equivalent of the PIT histogram, which applies to ensemble forecasts, is the verification rank histogram \citep{Anderson1996_TT, HamillColucci1997_TT}. It shows the distribution of the ranks of the observations within the corresponding ensembles and has the same interpretation as the PIT histogram. 

For small test sets, the bin number of a PIT/rank histogram must be chosen with care. With very few bins, the plot may obscure miscalibration while with many bins, even perfectly calibrated forecasts can yield non-uniformly appearing histograms \citep{ThorarinsdottirSchuhen2018_TT, Heinrich2020_TT}. The bin number should be chosen based on the size of the test set, with the bin number increasing linearly with the size of the test set \citep{Heinrich2020_TT}. More specifically, the uniformity of PIT/rank values can be assessed with statistical tests \citep{DelleMonache&2006_TT, Taillardat&2016_TT, Wilks2019_TT}, where the test statistics can be interpreted as a distance between the observed and a flat histogram \citep{Wilks2019_TT, Heinrich2020_TT}. Testing predictive performance is further discussed in \S\ref{sec:Statistical_tests_of_forecast_performance}.

Calibration assessment of multivariate forecasts is complicated by the lack of a unique ordering in higher dimensions and the many ways in which the forecasts can be miscalibrated \citep{Wilks2019_TT}. \citet{Gneiting&2008_TT} propose a general two-step approach where an ensemble forecast and the corresponding observation are first mapped to a single value by a pre-rank function. Subsequently, the pre-rank function values are ranked in a standard manner. The challenge here is to find a pre-rank function that yields informative and discriminative ranking \citep{Wilks2004_TT, Gneiting&2008_TT, Thorarinsdottir&2016_TT}, see \citet{Thorarinsdottir&2016_TT} and \citet{Wilks2019_TT} for comparative studies. Alternatively, \citet{ZiegelGneiting2014_TT} propose a direct multivariate extension of the univariate setting based on copulas.
 
\subsubsection[Statistical tests of forecast performance (Victor Richmond R. Jose)]{Statistical tests of forecast performance\protect\footnote{This subsection was written by Victor Richmond R. Jose.}}
\label{sec:Statistical_tests_of_forecast_performance}
A natural consequence of growth in forecasting methodologies was the development of statistical tests for predictive ability in the last thirty years. These tests provided forecasters some formal reassurance that the predictive superiority of a leading forecast is statistically significant and is not merely due to random chance.

One of the early papers that undoubtedly sparked growth in this field was Diebold and Mariano (\citeyear{DM95_Alessia}, DM hereafter). In their seminal paper, DM provided a simple yet general approach for testing equal predictive ability, i.e., if two forecasting sources ($f_{1,t}$ and $ f_{2,t}$, $t = 1,\ldots,h$) are equally accurate on average. Mathematically, if we denote the error $e_{i,t} =  y_t - f_{i,t}$ for $i = 1$, $2$ and $t = 1,\ldots,h$, the hypotheses for this DM test is $H_0$: $E[L(-e_{1,t}) - L(-e_{2,t})] = 0$ for all $t$ versus $H_1$: $E[L(-e_{1,t}) - L(-e_{2,t})] \neq 0$ under a loss function $L$.  Their population-level predictive ability test has very few assumptions (e.g., covariance stationary loss differential) and is applicable to a wide range of loss functions, multi-period settings, and wide class of forecast errors (e.g., non-Gaussian, serially and/or contemporaneously correlated). This test though not originally intended for models has been widely used by others to test forecasting models' accuracy \citep{diebold2015_VRRJ}.

Modifications were later introduced by \citet{harvey1998_VRRJ} to improve small sample properties of the test. Generalisations and extensions have emerged to address issues that DM tests encountered in practice such as nested models \citep{clark2001_VRRJ,clark2009_VRRJ}, parameter estimation error \citep{west1996_VRRJ}, cointegrated variables \citep{corradi2001_VRRJ}, high persistence \citep{rossi2005_VRRJ}, and panel data \citep{timmermann2019_VRRJ}. Finite-sample predictive ability tests also emerged from the observation that models may have equal predictive ability in finite samples, which generated a class called conditional predictive accuracy tests \citep{giacomini2006_VRRJ,clark2013_VRRJ}.

An alternative approach to comparing forecast accuracy is through the notion of forecast encompassing, which examines if a forecast encompasses all useful information from another with respect to predictions \citep{chong1986_VRRJ,harvey1998_VRRJ,clark2001_VRRJ}. Though it has a few more assumptions, forecast encompassing tests in certain contexts might be preferable to the mean square prediction error tests \`{a} la Diebold-Mariano \citep{busetti2013_VRRJ}.

Another stream of available statistical tests looks at multiple forecasts simultaneously instead of pairs. Addressing a need for a reality check on ``data snooping'', \citet{white2000_VRRJ} later modified by \citet{hansen2005_VRRJ} developed a multiple model test that uses a null hypothesis of ``superior predictive ability'' instead of the equal predictive ability used in DM tests. These have also been generalised to deal with issues such as cointegrated variables \citep{corradi2001_VRRJ} and multi-horizon forecasts \citep{quaedvlieg2019_VRRJ}. Recently, \citet{li2020_VRRJ} proposed a conditional superior predictive ability test similar to \citet{giacomini2006_VRRJ}'s innovation to the DM test. A different approach for studying performance of multiple forecasting models is through the use of multiple comparison tests such as multiple comparison with a control and multiple comparison with the best \citep{hsu1981_VRRJ,edwards1983_VRRJ,horrace2000_VRRJ}. These tests often are based on jointly estimated confidence intervals that measure the difference between two parameters of interest such as the forecast accuracies of a model and a benchmark. \citet{koning2005_VRRJ} illustrates how they can be ex post used to analyse forecasting performance in the M3 forecasting competition \citep{Makridakis2000-ty} using model ranking instead of forecast accuracy scores as its primitives. The multiple comparison of the best was used in the analysis of the subsequent M4 and M5 Competitions \citep[and \S\ref{sec:Forecasting_competitions}]{Makridakis2020-mm,Makridakis2020-wq}.

\subsubsection[Forecasting competitions (Fotios Petropoulos)]{Forecasting competitions\protect\footnote{This subsection was written by Fotios Petropoulos.}}
\label{sec:Forecasting_competitions}
Forecasting competitions provide a ``playground'' for academics, data scientists, students, practitioners, and software developers to compare the forecasting performance of their methods and approaches against others. Organisers of forecasting competitions test the performance of the participants' submissions against some hidden data, usually the last window of observations for each series. The benefits from forecasting competitions are multifold. Forecasting competitions (\textit{i})~motivate the development of innovative forecasting solutions, (\textit{ii})~provide a deeper understanding of the conditions that some methods work and others fail, (\textit{iii})~promote knowledge dissemination, (\textit{iv})~provide a much-needed, explicit link between theory and practice, and (\textit{v})~leave as a legacy usable and well-defined data sets. Participation in forecasting competitions is sometimes incentivised by monetary prizes. However, the stakes are usually much higher, including reputational benefits.

The most famous forecasting competitions are the ones organised by Spyros Makridakis. Initially, the research question focused on the relative performance of simple versus complex forecast. M and M3 competitions \citep{Makridakis1982-co,Makridakis2000-ty} empirically showed that simple methods (such as exponential smoothing; see \S\ref{sec:exponential_smoothing_models}) are equally good compared to other more complex methods and models (such as ARIMA and neural networks; see \S\ref{sec:autoregressive_integrated_moving_average_models} and \S\ref{sec:neural_networks} respectively) in point-forecast accuracy -- if not better. Moreover, the early Makridakis competitions showed the importance of forecast combinations in increasing predictive accuracy. For example, the winner of the M3 competition was the Theta method (see \S\ref{sec:Theta_method_and_models}), a simple statistical method that involved the combination of linear regression and simple exponential smoothing forecasts \citep{Assimakopoulos2000-hj}.

The M4 competition \citep{Makridakis2020-mm} challenged researchers and practitioners alike with a task of producing point forecasts and prediction intervals for 100 thousand time series of varied frequencies. This time, the main hypothesis focused on the ability of machine learning and neural network approaches in the task of time series forecasting. Machine learning approaches (see \S\ref{sec:machine_Learning}) that focused on each series independently performed poorly against statistical benchmarks, such as Theta, Damped exponential smoothing or simple averages of exponential smoothing models. However, the best two performing submissions in the M4 competition \citep{Smyl2020-dv,Montero-Manso2020-et} used neural network and machine learning algorithms towards utilising cross-learning. So, the main learning outcome from the M4 competition is that, if utilised properly, machine learning can increase the forecasting performance. Similarly to previous competitions, M4 demonstrated again the usefulness of combining across forecasts, with five out of the top six submissions offering a different implementation of forecast combinations.

Several other forecasting competitions focused on specific contexts and applications. For example, M2 competition \citep{Makridakis1993-bw} suggested that the benefits from additional information (domain expertise) are limited; see also \S\ref{sec:panels_of_experts}. The tourism forecasting competition \citep{Athanasopoulos2011-lw} also showed that exogenous variables do not add value, while naive forecasts perform very well on a yearly frequency (for a discussion on tourism forecasting applications, see \S\ref{sec:Tourism_demand_forecasting}). The NN3 competition \citep{Crone2011-si} confirmed the superior performance of statistical methods, but noted that neural network approaches are closing the distance. Tao Hong's series of energy competitions \citep{Hong2014-sz,hong2016probabilistic_FZ,Hong2019} demonstrated best practices for load, price, solar, and wind forecasting, with extensions to probabilistic and hierarchical forecasts (for energy forecasting applications, see \S\ref{sec:Energy}). Finally, many companies have hosted forecasting challenges through the Kaggle platform. \cite{Bojer2020-yk} reviewed the Kaggle competitions over the last five years, and concluded that access to hierarchical information, cross-learning, feature engineering, and combinations (ensembles) can lead to increased forecasting performance, outperforming traditional statistical methods. These insights were a forerunner to the results of the M5 competition, which focused on hierarchically organised retail data  \citep{Makridakis2020-wq,Makridakis2020-bj}.

\cite{Makridakis2021-xa} provide a list of design attributes for forecasting competitions and propose principles for future competitions.

\subsection[The future of forecasting theory (Pierre Pinson)]{The future of forecasting theory\protect\footnote{This subsection was written by Pierre Pinson.}}
\label{sec:the_theory_of_forecasting_the_future}

The theory of forecasting appears mature today, based on dedicated developments at the interface among a number of disciplines, e.g., mathematics and statistics, computer sciences, psychology, etc. A wealth of these theoretical developments have originated from specific needs and challenges in different application areas, e.g., in economics, meteorology and climate sciences, as well as management science among others. In this section, many aspects of the theory of forecasting were covered, with aspects related to data, modelling and reasoning, forecast verification. Now, the fact that forecasting is mature does not mean that all has been done -- we aim here at giving a few pointers at current and future challenges.

First of all, it is of utmost importance to remember that forecasting is a process that involves both quantitative aspects (based on data and models) and humans, at various levels, i.e., from the generation of forecasts to their use in decision-making. A first consequence is that we always need to find, depending on the problem at hand, an optimal trade-off between data-driven approaches and the use of expert judgment. In parallel, forecasting is to be thought of in a probabilistic framework in a systematic manner \citep{gneiting2014_VRRJ}. This allows us to naturally convey uncertainty about the future, while providing the right basis to make optimal decisions in view of the characteristics of the decision problem, as well as the loss (or utility) function and risk aversion of the decision maker. 
Another consequence is that using forecasts as input to decision-making often affects the outcome to be predicted itself -- a problem known as self-negating forecasts (possibly also self-fulfilling) or the prophet dilemma. With advances in the science of dynamic systems and game theory, we should invest in modelling those systems as a whole (i.e., forecasting and decision-making) in order to predict the full range of possible outcomes, based on the decisions that could be made.

In parallel, it is clear that today, the amount of data being collected and possibly available for forecasting is growing at an astounding pace. This requires re-thinking our approaches to forecasting towards high-dimensional models, online learning, etc. Importantly, the data being collected is distributed in terms of ownership. And, due to privacy concerns and competitive interests, some may not be ready to share their data. Novel frameworks to learning and forecasting ought to be developed with that context in mind, for instance focusing on distributed and privacy-preserving learning -- an example among many others is that of Google pushing forward federated learning \citep{Abadi2016_PP}, an approach to deep learning where the learning process is distributed and with a privacy layer. Eventually the access and use of data, as well as the contribution to distributed learning (and collaborative analytics, more generally), may be monetised, bringing a mechanism design component to the future theory of forecasting. A simple and pragmatic example is that of forecast reconciliation: if asking various agents to modify their forecasts to make them coherent within a hierarchy, such modifications could be monetised to compensate for accuracy loss.

A large part of today's modelling and forecasting approaches uses a wealth of data to identify and fit models, to be eventually used to forecast based on new data and under new conditions. Different approaches have been proposed to maximise the generalisation ability of those models, to somewhat maximise chances to do well out-of-sample. At the root of this problem is the effort to go beyond correlation only, and to identify causality (see, e.g., \citet{Pearl2009_PP} for a recent extensive coverage). While causality has been a key topic of interest to forecasters for a long time already, new approaches and concepts are being pushed forward for identification of and inference in causal models \citep{Peters2017_PP}, which may have a significant impact on the theory of forecasting.

Eventually, the key question of \emph{what a good forecast is} will continue to steer new developments in the theory of forecasting in the foreseeable future. The nature of goodness of forecasts (seen from the meteorological application angle) was theorised a few decades ago already \citep{Murphy1993_PP}, based on consistency, quality and value. We still see the need to work further on that question -- possibly considering these 3 pillars, but possibly also finding other ways to define desirable properties of forecasts. This will, in all cases, translates to further developing frameworks for forecast verification, focusing on the interplay between forecast quality and value, but also better linking to psychology and behavioural economics. In terms of forecast verification, some of the most pressing areas most likely relate to (multivariate) probabilistic forecasting and to the forecasting of extreme events. When it comes to forecast quality and value, we need to go beyond the simple plugging of forecasts into decision problems to assess whether this yields better decisions, or not. Instead, we ought to propose suitable theoretical frameworks that allow assessing whether certain forecasts are fundamentally better (than others) for given classes of decision problems. Finally, the link to psychology and behavioural economics should ensure a better appraisal of how forecasts are to be communicated, how they are perceived and acted upon.

Most of the advances in the science of forecasting have come from the complementarity between theoretical developments and applications. We can then only be optimistic for the future since more and more application areas are relying heavily on forecasting. Their specific needs and challenges will continue fuelling upcoming developments in the theory of forecasting.

\clearpage

\section{Practice}
\label{sec:practice}

\subsection[Introduction to forecasting practice (Michael Gilliland)]{Introduction to forecasting practice\protect\footnote{This subsection was written by Michael Gilliland.}}
\label{sec:introduction_to_the_practice_of_forecasting}
The purpose of forecasting is to improve decision making in the face of uncertainty. To achieve this, forecasts should provide an unbiased guess at what is most likely to happen (the point forecast), along with a measure of uncertainty, such as a prediction interval (PI). Such information will facilitate appropriate decisions and actions.

Forecasting should be an objective, dispassionate exercise, one that is built upon facts, sound reasoning, and sound methods. But since forecasts are created in social settings, they are influenced by organisational politics and personal agendas. As a consequence, forecasts will often reflect aspirations rather than unbiased projections. 

\begin{figure}[ht!]
\centering
\includegraphics[trim=72 725 71 70, clip, width=5in]{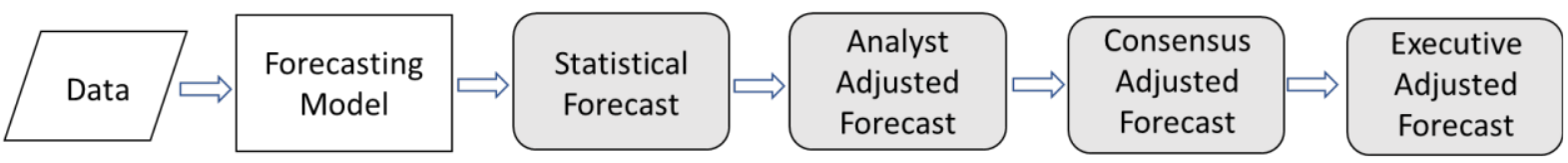}
\caption{Multi-stage forecasting process.}
\label{fig:fig_MG_process}
\end{figure}

In organisations, forecasts are created through processes that can involve multiple steps and participants. The process can be as simple as executive fiat (also known as evangelical forecasting), unencumbered by what the data show. More commonly, the process begins with a statistical forecast (generated by forecasting software), which is then subject to review and adjustment, as illustrated in figure \ref{fig:fig_MG_process}.

In concept, such an elaborate multi-stage process allows ``management intelligence'' to improve forecast quality, incorporating information not accounted for in the statistical model. In reality, however, benefits are not assured. \cite{Lawrence2006-ey_SG} reviewed more than 200 studies, concluding that human judgment can be of significant benefit but is also subject to significant biases. Among the many papers on this subject, there is general agreement on the need to track and review overrides, and the need to better understand the psychological issues around judgmental adjustments.

The underlying problem is that each human touch point subjects the forecast to the interests of the reviewers – and these interests may not align with creating an accurate, unbiased forecast. To identify where such problems are occurring, Forecast Value Added (FVA) analysis is an increasingly popular approach among practitioners.

FVA is defined as the change in a forecasting performance metric that can be attributed to a particular step or participant in the forecasting process \citep{Gilliland2002-qx_MG}. Any activity that fails to deliver positive FVA (i.e., fails to improve forecast quality) is considered process waste.

Starting with a naive forecast, FVA analysis seeks to determine whether each subsequent step in the process improves upon the prior steps. The ``stairstep report'' of table \ref{tab:FVA} is a familiar way of summarising results, as in this example from Newell Rubbermaid \citep{Schubert2011-ie_MG}.

\begin{table}[t]
\centering
\textbf{\caption{Stairstep report showing FVA results.}\label{tab:FVA}}
\medskip
\begin{tabular}{cccc}
 \hline 
 \textbf{Process Step} & \textbf{Forecast accuracy} & \textbf{FVA vs. Naive} & \textbf{FVA vs. Statistical}\\
 & (100\%$-$MAPE) \\
 \hline
Naive forecast & 60\% \\
Statistical forecast & 65\% & 5\% \\
Adjusted forecast & 62\% & 2\% & -3\% \\ 
\hline
\end{tabular}
\end{table}

Here, averaged across all products, naive (random walk) achieved forecast accuracy of 60\%. The company’s statistical forecast delivered five percentage points of improvement, but management review and adjustment delivered negative value. Such findings – not uncommon – urge further investigation into causes and possible process corrections (such as training reviewers or limiting adjustments). Alternatively, the management review step could be eliminated, providing the dual benefits of freeing up management time spent on forecasting and, on average, more accurate forecasts.

\cite{Morlidge2014-yd_MG} expanded upon FVA analysis to present a strategy for prioritising judgmental adjustments, finding the greatest opportunity for error reduction in products with high volume and high relative absolute error. \cite{Chase2021-tn_MG} described a machine learning (ML) method to guide forecast review, identifying which forecasts are most likely to benefit from adjustment along with a suggested adjustment range. \cite{Baker2021-gi_MG} used ML classification models to identify characteristics of non-value adding overrides, proposing the behavioural economics notion of a ``nudge'' to prompt desired forecaster behaviour. Further, \cite{Goodwin2017-io_MG} derived upper bounds for FVA relative to naive forecasts. And \cite{De_Kok2017-pw_MG} created a Stochastic Value Added (SVA) metric to assess the difference between actual and forecasted distributions, knowledge of which is valuable for inventory management.

Including an indication of uncertainty around the point forecast remains an uncommon practice. Prediction intervals in software generally underestimate uncertainty, often dramatically, leading to unrealistic confidence in the forecast. And even when provided, PIs largely go unused by practitioners. \cite{Goodwin2014-jv_MG} summarised the psychological issues, noting that the generally poor calibration of the PIs may not explain the reluctance to utilise them. Rather, ``an interval forecast may accurately reflect the uncertainty, but it is likely to be spurned by decision makers if it is too wide and judged to be uninformative'' \citep[page 5]{Goodwin2014-jv_MG}. 

It has long been recognised \citep{Chatfield1986-zf_MG,Lawrence2000-os_MG} that the practice of forecasting falls well short of the potential exhibited in academic research, and revealed by the M forecasting competitions. In the M4, a simple benchmark combination method (the average of Single, Holt, and Damped exponential smoothing) reduced the overall weighted average (OWA) error by 17.9\% compared to naive. The top six performing methods in M4 further reduced OWA by over 5\% compared to the combination benchmark \citep{Makridakis2020-mm}. But in forecasting practice, just bettering the accuracy of naive has proven to be a surprising challenge. \citeauthor{Morlidge2014-kd_MG}'s (\citeyear{Morlidge2014-kd_MG}) study of eight consumer and industrial businesses found 52\% of their forecasts failed to do so. And, as shown, Newel Rubbermaid beat naive by just two percentage points after management adjustments.

Ultimately, forecast accuracy is limited by the nature of the behaviour being forecast. But even a highly accurate forecast is of little consequence if overridden by management and not used to enhance decision making and improve organisational performance.

Practitioners need to recognise limits to forecastability and be willing to consider alternative (non-forecasting) approaches when the desired level of accuracy is not achievable \citep{Gilliland2010-iw_MG}. Alternatives include supply chain re-engineering – to better react to unforeseen variations in demand, and demand smoothing – leveraging pricing and promotional practices to shape more favourable demand patterns. 

Despite measurable advances in our statistical forecasting capabilities \citep{Makridakis2020-eu_MG}, it is questionable whether forecasting practice has similarly progressed. The solution, perhaps, is what \citeauthor{Morlidge2014-qs_MG} (\citeyear{Morlidge2014-qs_MG}, page 39) suggests that ``users should focus less on trying to optimise their forecasting process than on detecting where their process is severely suboptimal and taking measures to redress the problem''. This is where FVA can help. 

For now, the challenge for researchers remains: To prompt practitioners to adopt sound methods based on the objective assessment of available information, and avoid the ``worst practices'' that squander resources and fail to improve the forecast.

\subsection{Operations and supply chain management}
\label{sec:Operations_and_Supply_Chain_Management}

\subsubsection[Demand management (Yanfei Kang)]{Demand management\protect\footnote{This subsection was written by Yanfei Kang.}}
\label{sec:Demand_management}
Demand management is one of the dominant components of supply chain management \citep{Fildes2006-km}. Accurate demand estimate of the present and future is a first vital step for almost all aspects of supply chain optimisation, such as inventory management, vehicle scheduling, workforce planning, and distribution and marketing strategies \citep{kolassa2016demand_YK}. Simply speaking, better demand forecasts can yield significantly better supply chain management, including improved inventory management and increased service levels. Classic demand forecasts mainly rely on qualitative techniques, based on expert judgment and past experience \citep[e.g.,][]{weaver1971delphi_YK}, and quantitative techniques, based on statistical and machine learning modelling \citep[e.g.,][]{taylor2003short_BRT,Bacha1992_YK}. A combination of qualitative and quantitative methods is also popular and proven to be beneficial in practice by, e.g., judgmental adjustments \citep[][and \S\ref{sec:judgmental_adjustments_of_computer_based_forecasts}]{turner1990role_YK,Onkal2005-nr_PG,syntetos2016effects_YK}, judgmental forecast model selection \citep[][and \S\ref{sec:judgmental_model_selection}]{Petropoulos2018-mt,Han2019-ln}, and other advanced forecasting support systems \citep[][see also \S\ref{sec:Suport_systems}]{Baecke2017-wj,Arvan2019-xy_SG}.

The key challenges that demand forecasting faces vary from domain to domain. They include:
\begin{enumerate}[noitemsep]
 \item The existence of intermittent demands, e.g., irregular demand patterns of fashion products. According to \citet{Nikolopoulos2020_YK}, limited literature has focused on intermittent demand. The seminal work by \citet{Croston1972-zw_ASJBZB} was followed by other representative methods such as the SBA method by \citet{Syntetos2001-ih_ASJBZB}, the aggregate–disaggregate intermittent demand approach (ADIDA) by \citet{Nikolopoulos2011-bt}, the multiple temporal aggregation by \citet{Petropoulos2015-vq}, and the $k$ nearest neighbour ($k$NN) based approach by \citet{Nikolopoulos2016-wd_ASJBZB}. See \S\ref{sec:Forecasting_for_intermittent_demands_and_count_data} for more details on intermittent demand forecasting and \S\ref{sec:temporal_aggregation} for a discussion on temporal aggregation.
 \item The emergence of new products. Recent studies on new product demand forecasting are based on finding analogies \citep{Wright2015_YK,Hu2019_YK}, leveraging comparable products \citep{baardman2018leveraging_YK}, and using external information like web search trends \citep{Kulkarni2012_YK}. See \S\ref{sec:New_product_forecasting} for more details on new product demand forecasting.
 \item The existence of short-life-cycle products, e.g., smartphone demand \citep[e.g.,][]{szozda2010analogous_YK,chung2012sales_YK,shi2020block_YK}.
 \item The hierarchical structure of the data such as the electricity demand mapped to a geographical hierarchy \citep[e.g.,][but also \S\ref{sec:Cross_sectional_hierarchical_forecasting}]{Athanasopoulos09_PR,Hyndman11_PR,Hong2019}. 
\end{enumerate}

With the advent of the big data era, a couple of coexisting new challenges have drawn the attention of researchers and practitioners in the forecasting community: the need to forecast a large volume of related time series \citep[e.g., thousands or millions of products from one large retailer:][]{salinas2019high_Feng}, and the increasing number of external variables that have significant influence on future demand (e.g., massive amounts of keyword search indices that could impact future tourism demand \citep{law2019tourism_YK}). Recently, to deal with these new challenges, numerous empirical studies have identified the potentials of deep learning based global models, in both point and probabilistic demand forecasting \citep[e.g.,][]{wen2017multi_TJ,rangapuram2018deep_TJ,salinas2019high_Feng,Bandara2020_YK}. With the merits of cross-learning, global models have been shown to be able to learn long memory patterns and related effects \citep{montero2020principles_YK}, latent correlation across multiple series \citep{Smyl2020-dv}, handle complex real-world forecasting situations such as data sparsity and cold-starts \citep{Chen2020_YK}, include exogenous covariates such as promotional information and keyword search indices \citep{law2019tourism_YK}, and allow for different choices of distributional assumptions \citep{salinas2019high_Feng}. 

\subsubsection[Forecasting in the supply chain (Paul Goodwin)]{Forecasting in the supply chain\protect\footnote{This subsection was written by Paul Goodwin.}}
\label{sec:Forecasting_in_the_supply_chain}
A supply chain is `a network of stakeholders (e.g., retailers, manufacturers, suppliers) who collaborate to satisfy customer demand' \citep{Perera2019-xi_PG}. Forecasts inform many supply chain decisions, including those relating to inventory control, production planning, cash flow management, logistics and human resources (also see \S\ref{sec:Demand_management}). Typically, forecasts are based on an amalgam of statistical methods and management judgment \citep{Fildes2007-od_PG}. \cite{Hofmann2018-iz_PG} have investigated the potential for using big data analytics in supply chain forecasting but indicate more research is needed to establish its usefulness.

In many organisations forecasts are a crucial element of Sales and Operations Planning (S\&OP), a tool that brings together different business plans, such as those relating to sales, marketing, manufacturing and finance, into one integrated set of plans \citep{Thome2012-gs_PG}. The purposes of S\&OP are to balance supply and demand and to link an organisation's operational and strategic plans. This requires collaboration between individuals and functional areas at different levels because it involves data sharing and achieving a consensus on forecasts and common objectives \citep{Mello2010-la_PG}. Successful implementations of S\&OP are therefore associated with forecasts that are both aligned with an organisation's needs and able to draw on information from across the organisation. This can be contrasted with the `silo culture' identified in a survey of companies by \cite{Moon2003-tm_PG} where separate forecasts were prepared by different departments in `islands of analysis'. Methods for reconciling forecasts at different levels in both cross-sectional hierarchies (e.g., national, regional and local forecasts) and temporal hierarchies (e.g., annual, monthly and daily forecasts) are also emerging as an approach to break through information silos in organisations (see \S\ref{sec:Cross_sectional_hierarchical_forecasting}, \S\ref{sec:temporal_aggregation}, and \S\ref{sec:crosstemporal_hierarchies}). Cross-temporal reconciliation provides a data-driven approach that allows information to be drawn from different sources and levels of the hierarchy and enables this to be blended into coherent forecasts \citep{kourentzes2019cross_DKB}.

In some supply chains, companies have agreed to share data and jointly manage planning processes in an initiative known as Collaborative Planning, Forecasting, and Replenishment (CPFR) \citep[][also see \S\ref{sec:Forecasting_for_inventories}]{Seifert2003-db_PG}. CPFR involves pooling information on inventory levels and on forthcoming events, like sales promotions. Demand forecasts can be shared, in real time via the Internet, and discrepancies between them reconciled. In theory, information sharing should reduce forecast errors. This should mitigate the `bullwhip effect' where forecast errors at the retail-end of supply chains cause upstream suppliers to experience increasingly volatile demand, forcing them to hold high safety stock levels \citep{Lee2007-wm_PG}. 
Much research demonstrating the benefits of collaboration has involved simulated supply chains \citep{Fildes2017-os_PG}. Studies of real companies have also found improved performance through collaboration \citep[e.g.,][]{Boone2008-eq_PG,Hill2018-rm_PG,Eksoz2019-mh_PG}, but case study evidence is still scarce \citep{Syntetos2016-wd_PG}. The implementation of collaborative schemes has been slow with many not progressing beyond the pilot stage \citep{Panahifar2015-zy_PG,Galbreth2015-ge_PG}. Barriers to successful implementation include a lack of trust between organisations, reward systems that foster a silo mentality, fragmented forecasting systems within companies, incompatible systems, a lack of relevant training and the absence of top management support \citep{Fliedner2003-ha_PG,Thome2014-lb_PG}.

Initiatives to improve supply chain forecasting can be undermined by political manipulation of forecasts and gaming. Examples include `enforcing': requiring inflated forecasts to align them with sales or financial goals, `sandbagging': underestimating sales so staff are rewarded for exceeding forecasts, and `spinning': manipulating forecasts to garner favourable reactions from colleagues \citep{Mello2009-sv_PG}. \cite{Pennings2019-uw_PG} discuss schemes for correcting such intentional biases.

For a discussion of the forecasting of returned items in supply chains, see \S\ref{sec:Reverse_logistics}, while \S\ref{sec:The_practice_of_forecasting_the_future} offers a discussion of possible future developments in supply chain forecasting.

\subsubsection[Forecasting for inventories (John E. Boylan)]{Forecasting for inventories\protect\footnote{This subsection was written by John E. Boylan.}}
\label{sec:Forecasting_for_inventories}
Three aspects of the interaction between forecasting and inventory management have been studied in some depth and are the subject of this review: the bullwhip effect, forecast aggregation, and performance measurement.

The `bullwhip effect' occurs whenever there is amplification of demand variability through the supply chain \citep{Lee2004-iv_PG}, leading to excess inventories. This can be addressed by supply chain members sharing downstream demand information, at stock keeping unit level, to take advantage of less noisy data. Analytical results on the translation of ARIMA (see \S\ref{sec:autoregressive_integrated_moving_average_models}) demand processes have been established for order-up-to inventory systems \citep{Gilbert2005-ey_ASJBZB}. There would be no value in information sharing if the wholesaler can use such relationships to deduce the retailer’s demand process from their orders \citep[see, for example,][]{Graves1999-jq_ASJBZB}. Such deductions assume that the retailer's demand process and demand parameters are common knowledge to supply chain members. \cite{Ali2011-ti_ASJBZB} showed that, if such common knowledge is lacking, there is value in sharing the demand data itself and \cite{Ali2012-pt_ASJBZB} established relationships between accuracy gains and inventory savings. Analytical research has tended to assume that demand parameters are known. \cite{Pastore2020-er_ASJBZB} investigated the impact of demand parameter uncertainty, showing how it exacerbates the bullwhip effect.

Forecasting approaches have been developed that are particularly suitable in an inventory context, even if not originally proposed to support inventory decisions. For example, \cite{Nikolopoulos2011-bt} proposed that forecasts could be improved by aggregating higher frequency data into lower frequency data (see also \S\ref{sec:temporal_aggregation}; other approaches are reviewed in \S\ref{sec:Demand_management}). Following this approach, Forecasts are generated at the lower frequency level and then disaggregated, if required, to the higher frequency level. For inventory replenishment decisions, the level of aggregation may conveniently be chosen to be the lead time, thereby taking advantage of the greater stability of data at the lower frequency level, with no need for disaggregation.

The variance of forecast errors over lead time is required to determine safety stock requirements for continuous review systems. The conventional approach is to take the variance of one-step-ahead errors and multiply it by the lead time. However, this estimator is unsound, even if demand is independent and identically distributed, as explained by \cite{Prak2017-vl}. A more direct approach is to smooth the mean square errors over the lead time \citep{Syntetos2006-gi_ASJBZB}.

\cite{Strijbosch2005-ou} showed that unbiased forecasts will not necessarily lead to achievement, on average, of target cycle service levels or fill rates. \cite{Wallstrom2010-ss} proposed a `Periods in Stock' measure, which may be interpreted, based on a `fictitious stock', as the number of periods a unit of the forecasted item has been in stock or out of stock. Such measures may be complemented by a detailed examination of error-implication metrics \citep{Boylan2006-rz_ASJBZB}. For inventory management, these metrics will typically include inventory holdings and service level implications (e.g., cycle service level, fill rate). Comparisons may be based on total costs or via `exchange curves', showing the trade-offs between service and inventory holding costs. Comparisons such as these are now regarded as standard in the literature on forecasting for inventories and align well with practice in industry.

\subsubsection[Forecasting in retail (Stephan Kolassa \& Patr\'icia Ramos)]{Forecasting in retail\protect\footnote{This subsection was written by Stephan Kolassa \& Patr\'icia Ramos.}}
\label{sec:Retail_sales_forecasting}
Retail companies depend crucially on accurate demand forecasting to manage their supply chain and make decisions concerning planning, marketing, purchasing, distribution and labour force. Inaccurate forecasts lead to unnecessary costs and poor customer satisfaction. Inventories should be neither too high (to avoid waste and extra costs of storage and labour force), nor too low (to prevent stock-outs and lost sales, \citealp{Ma2017_PR}).

Forecasting retail demand happens in a three-dimensional space \citep{Syntetos2016-wd_PG}: the position in the supply chain hierarchy (store, distribution centre, or chain), the level in the product hierarchy (SKU, brand, category, or total) and the time granularity (day, week, month, quarter, or year). In general, the higher is the position in the supply chain, the lower is the time granularity required, e.g., retailers need daily forecasts for store replenishment and weekly forecasts for DC distribution/logistics activities at the SKU level \citep{FildesMaKolassa2019_SK}. Hierarchical forecasting (see \S\ref{sec:Cross_sectional_hierarchical_forecasting}) is a promising tool to generate coherent demand forecasts on multiple levels over different dimensions \citep{Oliveira2019_PR}.

Several factors affect retail sales, which often increase substantially during holidays, festivals, and other special events. Price reductions and promotions on own and competitors' products, as well as weather conditions or pandemics, can also change sales considerably \citep{Huang2019_PR}.

Zero sales due to stock-outs or low demand occur very often at the SKU~$\times$~store level, both at weekly and daily granularity. The most appropriate forecasting approaches for intermittent demand are Croston's method \citep{Croston1972-zw_ASJBZB}, the Syntetos-Boylan approximation \citep[SBA;][]{Syntetos2005-ic_ASJBZB}, and the TSB method \citep{Teunter2011-dy_ASJBZB}, all introduced in \S\ref{sec:Parametric_methods_for_intermittent_demand_forecasting}. These methods have been used to forecast sales of spare parts in automotive and aerospace industries but have not yet been evaluated in the retail context. 

Univariate forecasting models are the most basic methods retailers may use to forecast demand. They range from simple methods such as simple moving averages or exponential smoothing to ARIMA and ETS models (discussed in \S\ref{sec:Statistical_and_econometric_models}). These are particularly appropriate to forecast demand at higher aggregation levels \citep{Ramos15_PR,Ramos16_PR}. The main advantage of linear causal methods such as multiple linear regression is to allow the inclusion of external effects discussed above. There is no clear evidence yet that nonlinear models and novel machine learning methods can improve forecast accuracy \citep{FildesMaKolassa2019_SK}.

To be effective, point estimates should be combined with quantile predictions or prediction intervals for determining safety stock amounts needed for replenishment. However, to the best of our knowledge this is an under-investigated aspect of retail forecasting \citep{Taylor2007_PR,Kolassa2016_PR}.

The online channel accounts for an ever-increasing proportion of retail sales and poses unique challenges to forecasting, beyond the characteristics of brick and mortar (B\&M) retail stores. First, there are multiple \emph{drivers or predictors} of demand that could be leveraged in online retail, but not in B\&M:
\begin{itemize}[noitemsep]
\item Online retailers can fine-tune customer interactions, e.g., through the landing page, product recommendations, or personalised promotions, leveraging the customer's purchasing, browsing or returns history, current shopping cart contents, or the retailer's stock position, in order to tailor the message to one specific customer in a way that is impossible in B\&M. 
\item Conversely, product reviews are a type of interaction between the customer and the retailer and other customers which drives future demand.
\end{itemize}

Next, there are differences in forecast \emph{use}:
\begin{itemize}[noitemsep]
\item Forecast use strongly depends on the retailer's omnichannel strategy \citep{Armstrong2017_SK,SopadjievaDholakiaBenjamin2017_SK,MelaciniPerottiRasiniEtAl2018_SK}: e.g., for ``order online, pick up in store'' or ``ship from store'' fulfillment, we need separate but related forecasts for both total online demand and for the demand fulfilled at each separate store.
\item Online retailers, especially in fashion, have a much bigger problem with product returns. They may need to forecast how many products are returned overall \citep[e.g.,][]{ShangMcKieFergusonEtAl2020_SK}, or whether a \emph{specific} customer will return a \emph{specific} product.
\end{itemize}

Finally, there are differences in the forecasting \emph{process}:
\begin{itemize}[noitemsep]
\item B\&M retailers decouple pricing/promotion decisions and optimisation from the customer interaction, and therefore from forecasting. Online, this is not possible, because the customer has total transparency to competitors' offerings. Thus, online pricing needs to react much more quickly to competitive pressures -- faster than the forecasting cycle.
\item Thus, the specific value of predictors is often not known at the time of forecasting: we don't know yet which customer will log on, so we don't know yet how many people will see a particular product displayed on their personalised landing page. (Nor do we know today what remaining stock will be displayed.) Thus, changes in drivers need to be ``baked into'' the forecasting algorithm.
\item Feedback loops between forecasting and other processes are thus even more important online: yesterday's forecasts drive today's stock position, driving today's personalised recommendations, driving demand, driving today's forecasts for tomorrow. Overall, online retail forecasting needs to be more agile and responsive to the latest interactional decisions taken in the web store, and more tightly integrated into the retailer's interactional tactics and omnichannel strategy. 
\end{itemize}

Systematic research on demand forecasting in an online or omnnichannel context is only starting to appear \citep[e.g.][who use basket data from online sales to improve omnichannel retail forecasts]{OmarKlibiBabaiEtAlmanuscript}.
 
\subsubsection[Promotional forecasting (Nikolaos Kourentzes)]{Promotional forecasting\protect\footnote{This subsection was written by Nikolaos Kourentzes.}}
\label{sec:Promotional_forecasting}
Promotional forecasting is central for retailing (see \S\ref{sec:Retail_sales_forecasting}), but also relevant for many manufacturers, particularly of Fast Moving Consumer Goods (FMCG). In principle, the objective is to forecast sales, as in most business forecasting cases. However, what sets promotional forecasting apart is that we also make use of information about promotional plans, pricing, and sales of complementary and substitute products \citep{bandyopadhyay2009dynamic_NK, zhang2008joint_NK}. Other relevant variables may include store location and format, variables that capture the presentation and location of a product in a store, proxies that characterise the competition, and so on \citep{van2002promotions_NK, andrews2008estimating_NK}. 

Three modelling considerations guide us in the choice of models. First, promotional (and associated effects) are proportional. For instance, we do not want to model the increase in sales as an absolute number of units, but instead, as a percentage uplift. We do this to not only make the model applicable to both smaller and larger applications, for example, small and large stores in a retailing chain, but also to gain a clearer insight into the behaviour of our customers. Second, it is common that there are synergy effects. For example, a promotion for a product may be offset by promotions for substitute products. Both these considerations are easily resolved if we use multiplicative regression models. However, instead of working with the multiplicative models, we rely on the logarithmic transformation of the data (see \S\ref{sec:BoxCox_Transformations}) and proceed to construct the promotional model using the less cumbersome additive formulation (see \S\ref{sec:time_series_regression_models}). Third, the objective of promotional models does not end with providing accurate predictions. We are also interested in the effect of the various predictors: their elasticity. This can in turn provide the users with valuable information about the customers, but also be an input for constructing optimal promotional and pricing strategies \citep{zhang2008joint_NK}. 

Promotional models have been widely used on brand-level data \citep[for example,][]{divakar2005practice_NK}. However, they are increasingly used on Stock Keeping Unit (SKU) level data \citep{trapero2015identification_NK, ma2016demand_NK}, given advances in modelling techniques. Especially at that level, limited sales history and potentially non-existing examples of past promotions can be a challenge. \cite{trapero2015identification_NK} consider this problem and propose using a promotional model that has two parts that are jointly estimated. The first part focuses on the time series dynamics and is modelled locally for each SKU. The second part tackles the promotional part, which pools examples of promotions across SKUs to enable providing reasonable estimates of uplifts even for new SKUs. To ensure the expected heterogeneity in the promotional effects, the model is provided with product group information. Another recent innovation is looking at modelling promotional effects both at the aggregate brand or total sales level, and disaggregate SKU level, relying on temporal aggregation \citep[][and \S\ref{sec:temporal_aggregation}]{Kourentzes2016-bf}. \cite{ma2016demand_NK} concern themselves with the intra-and inter-category promotional information. The challenge now is the number of variables to be considered for the promotional model, which they address by using sequential LASSO (see also \S\ref{sec:Variable_Selection}). Although the aforementioned models have shown very promising results, one has to recognise that in practice promotions are often forecasted using judgmental adjustments, with inconsistent performance \citep{ trapero2013analysis_NK}; see also \S\ref{sec:judgmental_adjustments_of_computer_based_forecasts} and \S\ref{sec:Judgmental_forecasting_in_practice}.
 
\subsubsection[New product forecasting (Sheik Meeran)]{New product forecasting\protect\footnote{This subsection was written by Sheik Meeran.}}
\label{sec:New_product_forecasting}
Forecasting the demand for a new product accurately has even more consequence with regards to well-being of the companies than that for a product already in the market. However, this task is one of the most difficult tasks managers must deal with simply because of non-availability of past data \citep{Wind1981-sh_SM}. Much work has been going on for the last five decades in this field. Despite his Herculean attempt to collate the methods reported, \cite{Assmus1984-yg_SM} could not list all even at that time. The methods used before and since could be categorised into three broad approaches \citep{Goodwin2013-ih_SM} namely management judgment, consumer judgment and diffusion/formal mathematical models. In general, the hybrid methods combining different approaches have been found to be more useful \citep{hyndman2018Forecasting_CB,peres:10_MG}. Most of the attempts in New product Forecasting (NPF) have been about forecasting `adoption' (i.e., enumerating the customers who bought at least one time) rather than `sales', which accounts for repeat purchases also. In general, these attempts dealt with point forecast although there have been some attempts in interval and density forecasting \citep{Meade2001-hp_SM}.

Out of the three approaches in NPF, management judgment is the most used approach \citep{Kahn2002-fg_SM,Gartner1993-vz_SM,Lynn1999-kq_SM} which is reported to have been carried out by either individual managers or group of them. \cite{Ozer2011-xj_SM} and \cite{Surowiecki2005-ty_YGC} articulated their contrasting benefits and deficits. The Delphi method (see \S\ref{sec:panels_of_experts}) has combined the benefits of these two modes of operation \citep{Rowe1999-vs_SM} which has been effective in NPF. Prediction markets in the recent past offered an alternative way to aggregate forecasts from a group of Managers \citep{Wolfers2004-mq_JMP,Meeran_undated-wk_SM} and some successful application of prediction markets for NPF have been reported by \cite{Plott2002-es_SM} and \cite{Karniouchina2011-ed_SM}.

In the second category, customer surveys, among other methods, are used to directly ask the customers the likelihood of them purchasing the product. Such surveys are found to be not very reliable \citep{Morwitz1997-va_SM}. An alternative method to avoid implicit bias associated with such surveys in extracting inherent customer preference is conjoint analysis, which makes implicit trade off customers make between features explicit by analysing the customers' preference for different variants of the product. One analysis technique that attempts to mirror real life experience more is Choice Based Conjoint analysis (CBC) in which customers choose the most preferred product among available choices. Such CBC models used together with the analysis tools such as Logit \citep{McFadden1977-tx_SM} have been successful in different NPF applications \citep{Meeran2017-iw_SM}.

In the third approach, mathematical/formal models known as growth or diffusion curves (see \S\ref{sec:Innovation_diffusion_models} and \S\ref{sec:The_natural_law of growth_in_competition_Logistic_growth}) have been used successfully to do NPF \citep{Hu2019_YK}. The non-availability of past data is mitigated by growth curves by capturing the generic pattern of the demand growth of a class of products, which could be defined by a limited number of parameters such as saturation level, inflexion point, etc. For a new product a growth curve can be constituted from well-estimated parameters using analogous products, market intelligence or regression methods. Most extensively used family of growth curves for NPF has started with Bass model \citep{bass:69_RGuseo} that has been extended extensively \citep{Easingwood1983-ii_SM,Simon1987-ig_SM,Bass2001-uz_SM,Islam2000-mt_SM,peres:10_MG}. A recent applications of NPF focused on consumer electronic goods using analogous products \citep{Goodwin2013-ih_SM}.

\subsubsection[Spare parts forecasting (Mohamed Zied Babai)]{Spare parts forecasting\protect\footnote{This subsection was written by Mohamed Zied Babai.}}
\label{sec:Spare_parts_forecasting}
Spare parts are ubiquitous in modern societies. Their demand arises whenever a component fails or requires replacement. Demand for spare parts is typically intermittent, which means that it can be forecasted using the plethora of parametric and non-parametric methods presented in \S\ref{sec:Forecasting_for_intermittent_demands_and_count_data}. In addition to the intermittence of demand, spare parts have two additional characteristics that make them different from Work-In-Progress and final products, namely: (\textit{i}) they are generated by maintenance policies and part breakdowns, and (\textit{ii}) they are subject to obsolescence \citep{Bacchetti2012-zk_ASJBZB,Kennedy2002-lj_ASJBZB}.

The majority of forecasting methods do not link the demand to the generating factors, which are often related to maintenance activities. The demand for spare parts originates from the replacement of parts in the installed base of machines (i.e., the location and number of products in use), either preventively or upon breakdown of the part \citep{Kim2017-gf_ASJBZB}. \cite{Fortuin1984-rj_ASJBZB} claims that using installed base information to forecast the spare part demand can lead to stock reductions of up to 25\%. An overview of the literature that deals with spare parts forecasting with installed base information is given by \cite{Van_der_Auweraer2019-lq_JRTA}. Spare parts demand can be driven by the result of maintenance inspections and, in this case, a maintenance-based forecasting model should then be considered to deal with this issue. Such forecasting models include the Delay Time (DT) model analysed in \cite{Wang2011-mo_ASJBZB}. Using the fitted values of the distribution parameters of a data set related to a hospital pumps, \cite{Wang2011-mo_ASJBZB} have shown that when the failure and fault arriving characteristics of the items can be captured, it is recommended to use the DT model to forecast the spare part demand with a higher forecast accuracy. However, when such information is not available, then time series forecasting methods, such as those presented in \S\ref{sec:Parametric_methods_for_intermittent_demand_forecasting}, are recommended. The maintenance based forecasting is further discussed in \S\ref{sec:Predictive_maintenance}.

Given the life cycle of products, spare parts are associated with a risk of obsolescence. \cite{Molenaers2012-fz_ASJBZB} discussed a case study where 54\% of the parts stocked at a large petrochemical company had seen no demand for the last 5 years. \cite{Hinton1999-lm_ASJBZB} reported that the US Department of Defence was holding 60\% excess of spare parts, with 18\% of the parts (with a total value of \$1.5 billion) having no demand at all. To take into account the issue of obsolescence in spare parts demand forecasting, \cite{Teunter2011-dy_ASJBZB} have proposed the TSB method, which deals with linearly decreasing demand and sudden obsolescence cases. By means of an empirical investigation based on the individual demand histories of 8000 spare parts SKUs from the automotive industry and the Royal Air Force (RAF, UK), \cite{Babai2014-pg} have demonstrated the high forecast accuracy and inventory performance of the TSB method. Other variants of the Croston's method developed to deal with the risk of obsolescence in forecasting spare parts demand include the Hyperbolic-Exponential Smoothing method proposed by \cite{Prestwich2014-ed} and the modified Croston's method developed by \cite{Babai2019-np_ASJBZB}.
 
\subsubsection[Predictive maintenance (Juan Ramón Trapero Arenas)]{Predictive maintenance\protect\footnote{This subsection was written by Juan Ramón Trapero Arenas.}}
\label{sec:Predictive_maintenance}
A common classification of industrial maintenance includes three types of maintenance \citep{Montero_Jimenez2020-wf_JRTA}. Corrective maintenance refers to maintenance actions that occur after the failure of a component. Preventive maintenance consists of maintenance actions that are triggered after a scheduled number of units as cycles, kilometers, flights, etc. To schedule the fixed time between two preventive maintenance actions, the Weibull distribution is commonly used \citep{Baptista2018-lv_JRTA}. The drawbacks of preventive maintenance are related to the replacement of components that still have a remaining useful life; therefore, early interventions imply a waste of resources and too late actions could imply catastrophic failures. Additionally, the preventive intervention itself could be a source of failures too. Finally, predictive maintenance (PdM) complements the previous ones and, essentially, uses predictive tools to determine when actions are necessary \citep{Carvalho2019-nt_JRTA}. Within this predictive maintenance group, other terms are usually found in the literature as Condition-Based Maintenance and Prognostic and Health Management, \citep{Montero_Jimenez2020-wf_JRTA}.

The role of forecasting in industrial maintenance is of paramount importance. One application is to forecast spare parts (see \S\ref{sec:Spare_parts_forecasting}), whose demands are typically intermittent, usually required to carry out corrective and preventive maintenances \citep{Wang2011-mo_ASJBZB,Van_der_Auweraer2019-lq_JRTA}. On the other hand, it is crucial for PdM the forecast of the remaining useful time, which is the useful life left on an asset at a particular time of operation \citep{Si2011-ow_JRTA}. This work will be focused on the latter, which is usually found under the prognostic stage \citep{Jardine2006-qf_JRTA}.

The typology of forecasting techniques employed is very ample. \cite{Montero_Jimenez2020-wf_JRTA} classify them in three groups: physics-based models, knowledge-based models, and data-driven models. Physics-based models require high skills on the underlying physics of the application. Knowledge-based models are based on facts or cases collected over the years of operation and maintenance. Although, they are useful for diagnostics and provide explicative results, its performance on prognostics is more limited. In this sense, data-driven models are gaining popularity for the development of computational power, data acquisition, and big data platforms. In this case, data coming from vibration analysis, lubricant analysis, thermography, ultrasound, etc. are usually employed. Here, well-known forecasting models as VARIMAX/GARCH (see also \S\ref{sec:Statistical_and_econometric_models}) are successfully used \citep{Garcia2010-uz_JRTA,Cheng2012-ck_JRTA,Gomez_Munoz2014-ok_JRTA,Baptista2018-lv_JRTA}. State Space models based on the Kalman Filter are also employed \citep[][and \S\ref{sec:state_space_models}]{Pedregal2006-tj_JRTA,Pedregal2009-bx_JRTA}. 
Recently, given the irruption of the Industry 4.0, physical and digital systems are getting more integrated and Machine Learning/Artificial Intelligence are drawing the attention of practitioners and academics alike \citep{Carvalho2019-nt_JRTA}. In that same reference, it is found that the most frequently used Machine Learning methods in PdM applications were Random Forest, Artificial Neural Networks, Support Vector Machines and K-means.

\subsubsection[Reverse logistics (Aris A. Syntetos)]{Reverse logistics\protect\footnote{This subsection was written by Aris A. Syntetos.}}
\label{sec:Reverse_logistics}
As logistics and supply chain operations rely upon accurate demand forecasts (see also \S\ref{sec:Forecasting_in_the_supply_chain}), reverse logistics and closed loop supply chain operations rely upon accurate forecasts of returned items. Such items (usually referred as cores) can be anything from reusable shipping or product containers to used laptops, mobile phones or car engines. If some (re)manufacturing activity is involved in supply chains, it is both demand and returned items forecasts that are needed since it is net demand requirements (demand – returns) that drive remanufacturing operations. 

Forecasting methods that are known to work well when applied to demand forecasting, such as SES for example (see \S\ref{sec:exponential_smoothing_models}), do not perform well when applied to time-series of returns because they assume returns to be a process independent of sales. There are some cases when this independence might hold, such as when a recycler receives items sold by various companies and supply chains \citep{Goltsos2020-oj_ASJBZB}. In these cases, simple methods like SES applied on the time series of returns might prove sufficient. Typically though, returns are strongly correlated with past sales and the installed base (number of products with customers). After all, there cannot be a product return if a product has not first been sold. This lagged relationship between sales and returns is key to the effective characterisation of the returns process.

Despite the increasing importance of circular economy and research on closed loop supply chains, returns forecasting has not received sufficient attention in the academic literature \citep[notable contributions in this area include][]{Goh1986-qp_ASJBZB,Toktay2000-iy_ASJBZB,Toktay2003-fu_ASJBZB,De_Brito2009-ix_ASJBZB,Clottey2012-dh_ASJBZB}. The seminal work by \cite{Kelle1989-ey_ASJBZB} offers a useful framework to forecasting that is based on the degree of available information about the relationship between demand and returns. Product level (PL) information consists of the time series of sales and returns, alongside information on the time each product spends with a customer. The question then is how to derive this time to return distribution. This can be done through managerial experience, by investigating the correlation of the demand and the returns time series, or by serialising and tracking a subset (sample) of items. Past sales can then be used in conjunction with this distribution to create forecasts of returns. Serial number level (SL) information, is more detailed and consists of the time matching of an individual unit item's issues and returns and thus exactly the time each individual unit, on a serial number basis, spent with the customer. Serialisation allows for a complete characterisation of the time to return distribution. Very importantly, it also enables tracking exactly how many items previously sold remain with customers, providing time series of unreturned past sales. Unreturned past sales can then be extrapolated –- along with a time to return distribution -– to create forecasts of returns. 

\cite{Goltsos2019-gb_ASJBZB} offered empirical evidence in the area of returns forecasting by analysing a serialised data set from a remanufacturing company in North Wales. They found the Beta probability distribution to best fit times-to-return. Their research suggests that serialisation is something worthwhile pursuing for low volume products, especially if they are expensive. This makes a lot of sense from an investment perspective, since the relevant serial numbers are very few. However, they also provided evidence that such benefits expand in the case of high volume items. Importantly, the benefits of serialisation not only enable the implementation of the more complex SL method, but also the accurate characterisation of the returns process, thus also benefiting the PL method (which has been shown to be very robust).

\subsection{Economics and finance}
\label{sec:Economics_and_finance}

\subsubsection[Macroeconomic survey expectations (Michael P. Clements)]{Macroeconomic survey expectations\protect\footnote{This subsection was written by Michael P. Clements.}}
\label{sec:Macroeconomic_survey_expectations}
Macroeconomic survey expectations allow tests of theories of how agents form their expectations. Expectations play a central role in modern macroeconomic research \citep{MPC_Gali08}. Survey expectations have been used to test theories of expectations formation for the last 50 years. Initially the
Livingston survey data on inflationary expectations was used to test extrapolative or adaptive hypothesis, but the focus soon turned to testing whether expectations are formed rationally (see \citeauthor{MPC_Turn72}, \citeyear{MPC_Turn72}, for an early contribution). According to \cite[p.316]{MPC_Muth61}, rational expectations is the hypothesis that: `expectations, since they are informed predictions of
future events, are essentially the same as the predictions of the relevant economic theory.' This assumes all agents have access to all relevant information. Instead, one can test whether agents make efficient use of the information they possess. This is the notion of forecast efficiency \citep{MPC_Minc69b}, and can be tested by regressing the outturns on a constant and the forecasts of those outturns. Under forecast efficiency, the constant should be zero and the coefficient on the forecasts should be one. When the slope coefficient is not equal to one, the forecast errors will be systematically related to information available at the forecast origin, namely, the forecasts, and cannot be optimal. The exchange between \cite{MPC_Figl81,MPC_Figl83} and \cite{MPC_Kimb83} clarifies the role of partial information in testing forecast efficiency (that is, full information is not necessary), and shows that the use of the aggregate or consensus forecast in the individual realisation-forecast regression outlined above will give rise to a slope parameter less than one when forecasters are efficient but possess partial information. \cite{MPC_Zarn85}, \cite{MPC_Kean90} and \cite{MPC_Bonh01} consider pooling across individuals in the realisation-forecast regression, and the role of correlated shocks across individuals.

Recently, researchers considered why forecasters might not possess full-information, stressing informational rigidities: sticky information \citep[see, \textit{inter alia},][]{MPC_MankReis,MPC_MRW03}, and noisy information \citep[see, \textit{inter alia},][]{MPC_Wood01,MPC_Sims03}. \cite{MPC_Coib08,MPC_Coib15} test these models using aggregate quantities, such as mean errors and revisions.

Forecaster behaviour can be characterised by the response to new information (see also \S\ref{sec:judgmental_forecasting}). Over or under-reaction would constitute inefficiency. \cite{MPC_Broe18} and \cite{MPC_Bord18} find that agents over-react, generating a negative correlation between their forecast revision and error. The forecast is revised by more than is warranted by the new information (over-confidence regarding the value of the new information). \cite{MPC_Bord18} explain the over-reaction with a model of `diagnostic' expectations, whereas \cite{MPC_Fuhr18} finds `intrinsic inflation persistence': individuals under-react to new information, smoothing their responses to news.

The empirical evidence is often equivocal, and might reflect: the vintage of data assumed for the outturns; whether allowance is made for `instabilities' such as alternating over- and under-prediction \citep{MPC_Ross16} and the assumption of squared-error loss \citep[see, for example,][]{MPC_Patt_Timm06,MPC_ClemIAS}.

Research has also focused on the histogram forecasts produced by a number of macro-surveys. Density forecast evaluation techniques such as the probability integral transform\footnote{See, for example, \cite{MPC_Rose52}, \cite{MPC_Shep94}, \cite{MPC_KSC-98}, \cite{Diebold1998-ti_ET} and \cite{Berkowitz_2001_AY}.} 
have been applied to histogram forecasts, and survey histograms have been compared to benchmark forecasts \citep[see, for example,][]{MPC_Bao07,MPC_HallMitc,MPC_ClemUNINF}. Research has also considered uncertainty measures based on the histograms \cite{MPC_ClemEAEP}. \S\ref{sec:evaluating_probabilistic_forecasts} and \S\ref{sec:assessing_the_reliability_of_probabilistic forecats} also discuss the evaluation and reliability of probabilistic forecasts.  

\cite{MPC_Mansk06} and \cite{MPC_ClemCONSIST,MPC_ClemEXPL} considered the consistency between the point predictions and histogram forecasts. Reporting practices such as `rounding' have also been considered \citep{MPC_Bind17,MPC_ManskMolin,MPC_ClemRound}.

\cite{MPC_ClemPal19} reviews macroeconomic survey expectations.

\subsubsection[Forecasting GDP and inflation (Alessia Paccagnini)]{Forecasting GDP and inflation\protect\footnote{This subsection was written by Alessia Paccagnini.}}
\label{sec:Forecasting_GDP_and_Inflation}
As soon as Bayesian estimation of DSGEs became popular, these models have been employed in forecasting horseraces to predict the key macro variables, for example, Gross Domestic Product (GDP) and inflation, as discussed in \cite{MPC_Deln13}. The forecasting performance is evaluated using rolling or recursive (expanded) prediction windows \citep[for a discussion, see][]{CPV15_Alessia}. DSGEs are usually estimated using revised data, but several studies propose better results estimating the models using real-time data
\citep[see, for example,][]{MPC_Deln13,Wolters15_Alessia,KR15a_Alessia,CPV19_Alessia}.

The current DSGE model forecasting compares DSGE models to competitors (see \S\ref{sec:Forecasting_with_DSGE_Models} for an introduction to DSGE models). Among them, we can include the Random Walk (the naive model which assumes a stochastic trend), the Bayesian VAR models (Minnesota Prior à la \citeauthor{DLS84_Alessia}, \citeyear{DLS84_Alessia}; and Large Bayesian VAR à la \citeauthor{BGR10_Alessia}, \citeyear{BGR10_Alessia}), the Hybrid-Models (the DSGE-VAR à la \citeauthor{DNS04_Alessia}, \citeyear{DNS04_Alessia}; and the DSGE-Factor Augmented VAR à la \citeauthor{CFP09_Alessia}, \citeyear{CFP09_Alessia}), and the institutional forecasts (Greenbook, Survey Professional Forecasts, and the Blue Chip, as illustrated in \citeauthor{RG10_Alessia}, \citeyear{RG10_Alessia}).

Table \ref{tab:References_alessia} summarises the current DSGE forecasting literature mainly for the US and Euro Area and provided by estimating medium-scale models. As general findings, DSGEs can outperform other competitors, with the exception for the Hybrid-Models, in the medium and long-run to forecast GDP and inflation. In particular, \cite{SW07_Alessia} was the first empirical evidence of how DSGEs can be competitive with forecasts from Bayesian VARs, convincing researchers and policymakers in adopting DSGEs for prediction evaluations. As discussed in \cite{MPC_Deln13}, the accuracy of DSGE forecasts depends on how the model is able to capture low-frequency trends in the data. To explain the macro-finance linkages during the Great Recession, the Smets and Wouters model was also compared to other DSGE specifications including the financial sector. For example, \cite{MPC_Deln13}, \cite{KR15_Alessia}, \cite{GGKP16_Alessia}, and \cite{CPV19_Alessia} provide forecasting performance for DSGEs with financial frictions. This strand of the literature shows how this feature can improve the baseline Smets and Wouters predictions for the business cycle, in particular during the recent Great Recession. 

However, the Hybrid-Models always outperform the DSGEs thanks to the combination of the theory-based model (DSGE) and the statistical representation (VAR or Factor Augmented VAR), as illustrated by \cite{DNS04_Alessia} and \cite{CFP09_Alessia}.

Moreover, several studies discuss how prediction performance could depend on the parameters' estimation. \cite{KR15a_Alessia} suggest that updating DSGE model parameters only once a year is enough to have accurate and efficient predictions about the main macro variables.
 
\begin{table}[t]
\centering
\textbf{\caption{Alternative Competitors to DSGE Models}\label{tab:References_alessia}}
\medskip
\begin{tabular}{ p{4cm} p{10cm} }
 \hline 
 \textbf{Competitor} & \textbf{Reference}\\
 \hline
 Hybrid Models & 
US: \cite{DNS04_Alessia}, \cite{CFP09_Alessia}\\
Random Walk & US: \cite{GKR13_Alessia}, Euro Area: \cite{WCC10_Alessia}, \cite{SWW14_Alessia}\\
Bayesian VAR & US: \cite{SW07_Alessia}, \cite{GKR13_Alessia}, \cite{Wolters15_Alessia}, \cite{BP14_Alessia}, \cite{BP15a_Alessia}, \cite{BP15b_Alessia}, Euro Area: \cite{WCC10_Alessia} \\
Time-Varying VAR and Markov-Switching & US: \cite{BCPV16_Alessia}, Euro Area: \cite{BP16_Alessia}\\
Institutional Forecasts & US: \cite{RG10_Alessia}, \cite{KRS12_Alessia}, \cite{MPC_Deln13}, \cite{Wolters15_Alessia} \\
 \hline
\end{tabular}
\end{table}

\subsubsection[Forecasting unemployment (Jennifer L. Castle)]{Forecasting unemployment\protect\footnote{This subsection was written by Jennifer L. Castle.}}
\label{sec:Forecasting_UK_unemployment}
Unemployment has significant implications at both the micro and macro levels, influencing individual living standards, health and well-being, as well as imposing direct costs on the economy. Given its importance, policy-makers put unemployment at the heart of their economic plans, and as such require accurate forecasts to feed into economic policy decisions. Unemployment is described as a lagging indicator of the economy, with characteristics including business cycles and persistence. Despite this, forecasting the unemployment rate is difficult, because the data are highly non-stationary with abrupt distributional shifts, but persistence within regimes. In this section we focus on methods used to forecast the aggregate unemployment rate.

Unemployment is the outcome of supply and demand for labour, aggregated across all prospective workers, with labour demand derived from demand for goods and services. This implies a highly complex data generating process. Empirical forecasting models tend to simplify this relationship, with two approaches dominating the literature. The first is based on the \cite{JLC_Phillips58} curve capturing a non-linear relationship between nominal wage inflation and the unemployment rate, or the relation between unemployment and output described as Okun's \citeyear{JLC_Okun62} Law. The second uses the time-series properties of the data to produce statistical forecasts, such as univariate linear models (for example, ARIMA or unobserved component models; see \S\ref{sec:autoregressive_integrated_moving_average_models} and \S\ref{sec:state_space_models}), multivariate linear models (for example, VARMA or CVAR; see \S\ref{sec:Forecasting_with_many_variables}), various threshold autoregressive models (see \S\ref{sec:threshold_models}), Markov Switching models (see \S\ref{sec:Markov_switching_models}) and Artificial Neural Networks (see \S\ref{sec:neural_networks}).

The empirical literature is inconclusive as to the `best' forecasting models for unemployment, which varies by country, time period and forecast horizon. There is some evidence that non-linear statistical models tend to outperform within business cycle contractions or expansions, but perform worse across business cycles \citep[see, for example,][]{JLC_MontZarnTsayTiao98,JLC_Rothman98,JLC_KoopPotter99}, whereas \cite{JLC_Proietti03} finds that linear models characterised by higher persistence perform significantly better. Evidence of non-linearities is found by \cite{JLC_PeelSpeight00}, \cite{JLC_MilasRothman08} and \cite{JLC_Johnes99}, and \cite{JLC_GilAlana01} finds evidence of long-memory. \cite{JLC_BarnichonGarda16} applies a flow approach to unemployment forecasting and finds improvements, as does \cite{JLC_SmithJ11}. 

One approach that does yield accurate forecasts is to use a measure of profitability as the explanatory variable, assuming that unemployment will fall when hiring is profitable. \cite{JLC_HendUkinf01} proxies profitability ($\pi$) by the gap between the real interest rate (reflecting costs) and the real growth rate (reflecting the demand side), such that the unemployment rate rises when the real interest rate exceeds the real growth rate, and vice versa:
\begin{equation*}
\pi_{t} = \left(R_{L}-\Delta p -\Delta y\right)_{t}
\end{equation*}
where $R_{L}$ is the long-term interest rate, $\Delta p$ is a measure of inflation and $\Delta y$ is a measure of output growth. This is then embedded within a dynamic equilibrium correction model, using impulse indicator saturation \citep[IIS:][]{DFH_HendJoha07,johansen2009analysis_JJR} and step indicator saturation \citep[SIS:][]{castle2015detecting_JJR} to capture outliers, breaks and regime shifts, as well as allowing for any non-linearities using Taylor expansions for the regressors. The resulting forecasts perform well over the business cycle relative to alternative statistical models (also see \citealp{JLC_HendryIntroMacroMetrics15}, and \citealp*{JLC_CastleHendryMartinezVoxEU20}).

Forecasts from models of unemployment could be improved with either better economic theories of aggregate unemployment,\footnote{There are many relevant theories based on microfoundations, including search and matching, loss of skills, efficiency wages, and insider-outsider models, see \cite{JLC_LNJ91} for a summary.} or more general empirical models that tackle stochastic trends, breaks, dynamics, non-linearities and interdependence,\footnote{See \cite{JLC_HendryDoornikModelDiscovery} for an approach to jointly tackling all of these issues.} or better still, both. The COVID-19 pandemic and subsequent lockdown policies highlight just how important forecasts of unemployment are \citep{JLC_Castle2021}. 

\subsubsection[Forecasting productivity (Andrew B. Martinez)]{Forecasting productivity\protect\footnote{This subsection was written by Andrew B. Martinez.}}
\label{sec:Forecasting_productivity}
\begin{figure}[ht!]
\begin{center}
\includegraphics[width=16cm]{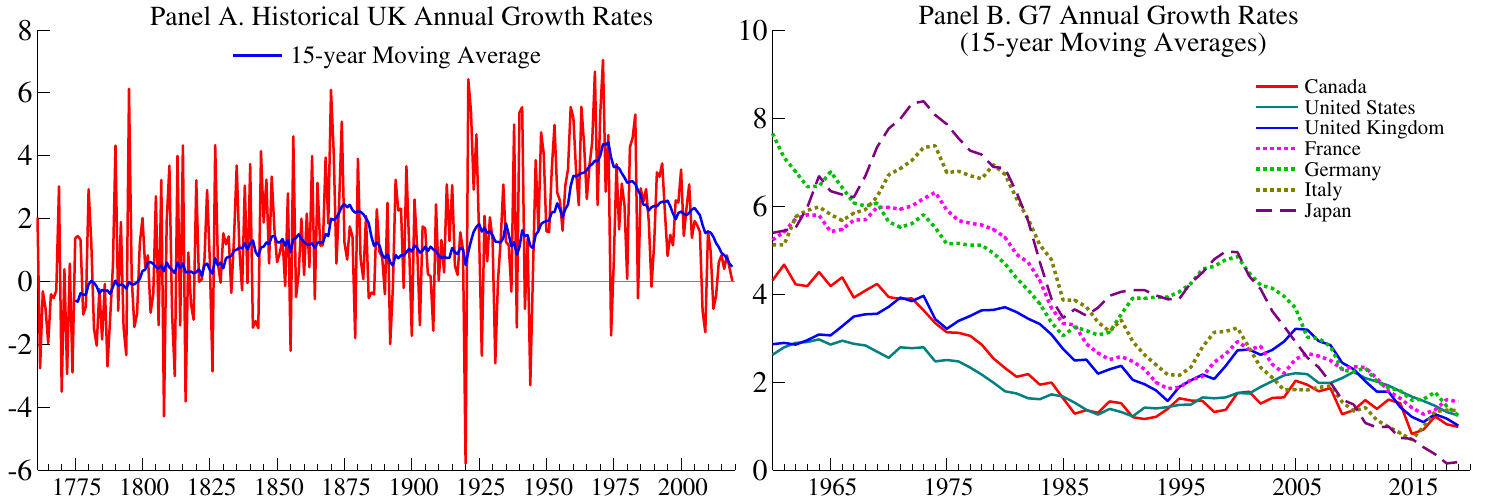} 
\caption{Productivity Growth (Output per total hours worked). Sources: Bank of England and Penn World Table Version 10.0.}
\label{fig:ProdFig1_ABM}
\end{center} 
\end{figure}

The growth of labour productivity, measured by the percent change in output per hours worked, has varied dramatically over the last 260 years. In the UK it ranged from -5.8\% at the onset of the 1920 Depression to just over 7\% in 1971; see panel A in figure \ref{fig:ProdFig1_ABM}. Productivity growth is very volatile and has undergone large historical shifts with productivity growth averaging around 1\% between 1800-1950 followed by an increase in the average annual growth to 3\% between 1950-1975. Since the mid-1970's productivity growth has gradually declined in many developed economies; see panel B of figure \ref{fig:ProdFig1_ABM}. In the decade since 2009, 2\% annual productivity growth was an upper bound for most G7 countries.

The most common approach for forecasting productivity is to estimate the trend growth in productivity using aggregate data. For example, \citet{gordon2003_ABM} considers three separate approaches for calculating trend labor productivity in the United States based on (\textit{i}) average historical growth rates outside of the business cycle, (\textit{ii}) filtering the data using the HP filter \citep{hodrick1997_ABM}, and (\textit{iii}) filtering the data using the Kalman filter \citep[see][]{Kalman1960_ABM}. The Office for Budget Responsibility (OBR) in the UK and the Congressional Budget Office (CBO) in the US follow similar approaches for generating its forecasts of productivity based on average historical growth rates as well as judgments about factors that may cause productivity to deviate from its historical trend in the short-term.\protect\footnote{See \protect\url{https://obr.uk/forecasts-in-depth/the-economy-forecast/potential-output-and-the-output-gap}. (Accessed: 2020-09-05)} Alternative approaches include forecasting aggregate productivity using disaggregated firm-level data \citep[see][and \S\ref{sec:Cross_sectional_hierarchical_forecasting}]{bartelsman2011_ABM,bartelsman2014_ABM} and using time-series models \citep[see][and \S\ref{sec:autoregressive_integrated_moving_average_models}]{vzmuk2018_ABM}.

In the last few decades there have been several attempts to test for time-varying trends in productivity and to allow for them. However, the focus of these approaches has been primarily on the United States \citep{hansen2001_ABM,roberts2001_ABM}, which saw a sharp rise in productivity growth in the 1990's that was not mirrored in other countries \citep{basu2003_ABM}. Test for shifts in productivity growth rates in other advanced economies did not find evidence of a changes in productivity growth until well after the financial crisis in 2007 \citep{benati2007_ABM,turner2011_ABM,glocker2018_ABM}. 

A more recent approach by \citet{Martinez2019_ABM} allows for a time-varying long-run trend in UK productivity. They show that are able to broadly replicate the OBR's forecasts using a quasi-transformed autoregressive model with one lag, a constant, and a trend. The estimated long-run trend is just over 2\% per year through 2007 Q4 which is consistent with the OBR's assumptions about the long-run growth rate of productivity \citep{obr2019_ABM}. However, it is possible to dramatically improve upon OBR's forecasts in real-time by allowing for the long-term trend forecast to adjust based on more recent historical patterns. By taking a local average of the last four years of growth rates, \citet{Martinez2019_ABM} generate productivity forecasts whose RMSE is on average more than 75\% smaller than OBR's forecasts extending five-years-ahead and is 84\% smaller at the longest forecast horizon.
 
\subsubsection[Fiscal forecasting for government budget surveillance (Diego J. Pedregal)]{Fiscal forecasting for government budget surveillance\protect\footnote{This subsection was written by Diego J. Pedregal.}}
\label{sec:Fiscal_forecasting_for_government_budget_surveillance}
Recent economic recessions have led to a renewed interest in fiscal forecasting, mainly for deficit and debt surveillance. This was certainly true in the case of the 2008 recession, and looks to become even more important in the current economic crisis brought on by the COVID-19 pandemic. This is particularly important in Europe, where countries are subject to strong fiscal monitoring mechanisms. Two main themes can be detected in the fiscal forecasting literature \citep{Leal_Diego}. First, investigate the properties of forecasts in terms of bias, efficiency and accuracy. Second, check the adequacy of forecasting procedures.

The first topic has its own interest for long, mainly restricted to international institutions \citep{Artis_Diego}. Part of the literature, however, argue that fiscal forecasts are politically biased, mainly because there is usually no clear distinction between political targets and rigorous forecasts \citep{Strauch_Diego,Frankel_Diego}. In this sense, the availability of forecasts from independent sources is of great value \citep{Jonung_Diego}. But it is not as easy as saying that independent forecasters would improve forecasts due to the absence of political bias, because forecasting accuracy is compromised by complexities of data, country-specific factors, outliers, changes in the definition of fiscal variables, etc. Very often some of these issues are known by the staff of organisations in charge of making the official statistics and forecasts long before the general public, and some information never leaves such institutions. So this insider information is actually a valuable asset to improve forecasting accuracy \citep{Leal_Diego}.

As for the second issue, namely the accuracy of forecasting methods, the literature can be divided into two parts, one based on macroeconomic models with specific fiscal modules that allows to analyse the effects of fiscal policy on macro variables and vice versa (see \cite{Favero_Diego} and references therein), and the other based on pure forecasting methods and comparisons among them. This last stream of research basically resembles closely what is seen in other forecasting areas: (\textit{i}) there is no single method outperforming the rest generally, (\textit{ii}) judgmental forecasting is especially important due to data problems (see \S\ref{sec:forecasting_with_judgment}), and (\textit{iii}) combination of methods tends to outperform individual ones, see \cite{Leal_Diego} and \S\ref{sec:combining_forecasts}.

Part of the recent literature focused on the generation of very short-term public finance monitoring systems using models that combine annual information with intra-annual fiscal data \citep{PEDREGAL2010_Diego} by time aggregation techniques (see \S\ref{sec:temporal_aggregation}), often set up in a SS framework (see \S\ref{sec:state_space_models}). The idea is to produce global annual end-of-year forecasts of budgetary variables based on the most frequently available fiscal indicators, so that changes throughout the year in the indicators can be used as early warnings to infer the changes in the annual forecasts and deviations from fiscal targets \citep{Antonio_Diego}.

The level of disaggregation of the indicator variables are established according to the information available and the particular objectives. The simplest options are the accrual National Accounts annual or quarterly fiscal balances running on their cash monthly counterparts. A somewhat more complex version is the previous one with all the variables broken down into revenues and expenditures. Other disaggregation schemes have been applied, namely by region, by administrative level (regional, municipal, social security, etc.), or by items within revenue and/or expenditure \citep[VAT, income taxes, etc.][]{PAREDES2014_Diego,paredes2020_Diego}.

Unfortunately, what is missing is a comprehensive and transparent forecasting system, independent of Member States, capable of producing consistent forecasts over time and across countries. This is certainly a challenge that no one has yet dared to take up. 

\subsubsection[Interest rate prediction (Massimo Guidolin \& Manuela Pedio)]{Interest rate prediction\protect\footnote{This subsection was written by Massimo Guidolin \& Manuela Pedio.}}
\label{sec:Interest_rate_prediction}
The (spot) rate on a (riskless) bond represents the ex-ante return (yield) to maturity which equates its market price to a theoretical valuation. Modelling and predicting default-free, short-term interest rates are crucial tasks in asset pricing and risk management. Indeed, the value of interest rate–sensitive securities depends on the value of the riskless rate. Besides, the short interest rate is a fundamental ingredient in the formulation and transmission of the monetary policy (see, for example, \S\ref{sec:Forecasting_with_DSGE_Models}). However, many popular models of the short rate (for instance, continuous time, diffusion models) fail to deliver accurate out-of-sample forecasts. Their poor predictive performance may depend on the fact that the stochastic behaviour of short interest rates may be time-varying (for instance, it may depend on the business cycle and on the stance of monetary policy). 

Notably, the presence of nonlinearities in the conditional mean and variance of the short-term yield influences the behaviour of the entire term structure of spot rates implicit in riskless bond prices. For instance, the level of the short-term rate directly affects the slope of the yield curve. More generally, nonlinear rate dynamics imply a nonlinear equilibrium relationship between short and long-term yields. Accordingly, recent research has reported that dynamic econometric models with regime shifts in parameters, such as Markov switching (MS; see \S\ref{sec:Markov_switching_models}) and threshold models (see \S\ref{sec:threshold_models}), are useful at forecasting rates.

The usefulness of MS VAR models with term structure data had been established since \cite{Hamilton1988-pl_MG} and \cite{Garcia1996-kc_MG}: single-state, VARMA models are overwhelmingly rejected in favour of multi-state models. Subsequently, a literature has emerged that has documented that MS models are required to successfully forecast the yield curve. \cite{Lanne2003-wq_MG} showed that a mixture of autoregressions with two regimes improves the predictions of US T-bill rates. \cite{Ang2002-px_MG} found support for MS dynamics in the short-term rates for the US, the UK, and Germany. \cite{Cai1994-nc_MG} developed a MS ARCH model to examine volatility persistence, reflecting a concern that it may be inflated by regimes. \cite{Gray1996-vg_MG} generalised this attempt to MS GARCH and reported improvements in pseudo out-of-sample predictions. Further advances in the methods and applications of MS GARCH are in \cite{Haas2004-qz_MG} and \cite{Smith2002-hi_MG}. A number of papers have also investigated the presence of regimes in the typical factors (level, slope, and convexity) that characterise the no-arbitrage dynamics of the term structure, showing the predictive benefits of incorporating MS \citep[see, for example,][]{Guidolin2019-jg_MG,Hevia2015-yf_MG}. 

Alternatively, a few studies have tried to capture the time-varying, nonlinear dynamics of interest rates using threshold models. As discussed by \cite{Pai1999-or_MP}, threshold models have an advantage compared to MS ones: the regimes are not determined by an unobserved latent variable, thus fostering interpretability. In most of the applications to interest rates, the regimes are determined by the lagged level of the short rate itself, in a self-exciting fashion. For instance, \cite{Pfann1996-xy_MP} explored nonlinear dynamics of the US short-term interest rate using a (self-exciting) threshold autoregressive model augmented by conditional heteroskedasticity (namely, a TAR-GARCH model) and found strong evidence of the presence of two regimes. More recently, also \cite{Gospodinov2005-jj_MP} used a TAR-GARCH to predict the short-term rate and showed that this model can capture some well-documented features of the data, such as high persistence and conditional heteroskedasticity.

Another advantage of nonlinear models is that they can reproduce the empirical puzzles that plague the expectations hypothesis of interest rates (EH), according to which it is a weighted average of short-term rates to drive longer-term rates \citep[see, for example,][]{Bansal2004-id_MG,Dai2007-ab_MG}. For instance, while \cite{Bekaert2001-zk_MG} show single-state VARs cannot generate distributions consistent with the EH, \cite{guidolin2009forecasts_DKB} find that the optimal combinations of lagged short and forward rates depend on regimes so that the EH holds only in some states.

As widely documented \citep[see, for instance,][]{Guidolin2018-od_MG}, the predictable component in mean rates is hardly significant. As a result, the random walk remains a hard benchmark to outperform as far as the prediction of the mean is concerned. However, density forecasts reflect all moments and the models that capture the dynamics of higher-order moments tend to perform best. MS models appear at the forefront of a class of non-linear models that produce accurate density predictions \citep[see, for example,][]{Hong2004-np_MG,Maheu2016-ta_MG}. Alternatively, \cite{Pfann1996-xy_MP} and more recently \cite{Dellaportas2007-rs_MP} estimated TAR models to also forecast conditional higher order moments and all report reasonable accuracy.

Finally, a literature has strived to fit rates not only under the physical measure, i.e., in time series, but to predict rates when MS enters the pricing kernel, the fundamental pricing operator. A few papers have assumed that regimes represent a new risk factor \citep[see, for instance,][]{Dai2003-mv_MG}. This literature reports that MS models lead to a range of shapes for nominal and real term structures \citep[see, for instance,][]{Veronesi1999-ws_MG}. Often the model specifications that are not rejected by formal tests include regimes \citep{Ang2008-lo_MG,Bansal2002-je_MG}. 

To conclude, it is worthwhile noting that, while threshold models are more interpretable, MS remain a more popular alternative for the prediction of interest rates. This is mainly due to the fact that statistical inference for threshold regime switching models poses some challenges, because the likelihood function is discontinuous with respect to the threshold parameters.
 
\subsubsection[House price forecasting (Alisa Yusupova)]{House price forecasting\protect\footnote{This subsection was written by Alisa Yusupova.}}
\label{sec:House_Price_Forecasting}
The boom and bust in housing markets in the early and mid 2000s and its decisive role in the Great Recession has generated a vast interest in the dynamics of house prices and emphasised the importance of accurately forecasting property price movements during turbulent times. International organisations, central banks and research institutes have become increasingly engaged in monitoring the property price developments across the world.\footnote{For instance, the International Monetary Fund recently established the Global Housing Watch, the Globalisation and Monetary Policy Institute of the Federal Reserve Bank of Dallas initiated a project on monitoring international property price dynamics, and the UK Housing Observatory initiated a similar project for the UK national and regional housing markets.} At the same time, a substantial empirical literature has
developed that deals with predicting future house price movements \citep[for a comprehensive survey see][]{GHYSELS20135_AY}. Although this literature concentrates almost entirely on the US~\citep[see, for example,][]{RapachStrauss_2009_AY,BorkMoller_2015_AY}, there are many other countries, such as the UK, where house price forecastability is of prime importance. Similarly to the US, in the UK, housing activities account for a large fraction of GDP and of households' expenditures; real estate property comprises a significant component of private wealth and mortgage debt constitutes a main liability of households \citep{ONS_2019_AY}. 

The appropriate forecasting model has to reflect the dynamics of the specific real estate market and take into account its particular characteristics. In the UK, for instance, there is a substantial empirical literature that documents the existence of strong spatial linkages between regional markets, whereby the house price shocks emanating from southern regions of the country, and in particular Greater London, have a tendency to spread out and affect neighbouring regions with a time lag \citep[see, for example,][\textit{inter alia}]{CookThomas_2003_AY, Holly_2010_AY,Antonakakis_CFG2018_AY}; see also \S\ref{sec:Functional_time_series_models} on forecasting functional data.

Recent evidence also suggests that the relationship between real estate valuations and conditioning macro and financial variables displayed a complex of time-varying patterns over the previous decades \citep{Aizenman_2013_AY}. Hence, predictive methods that do not allow for time-variation in both predictors and their marginal effects may not be able to capture the complex house price dynamics in the UK \citep[see][for a comparison of forecasting accuracy of a battery of static and dynamic econometric methods]{yusupova2019adaptive_AY}.

An important recent trend is to attempt to incorporate information from novel data sources (such as newspaper articles, social media, etc.) in forecasting models as a measure of expectations and perceptions of economic agents (see also \S\ref{sec:Text_based_forecasting}). It has been shown that changes in uncertainty about house prices impact on housing investment and real estate construction decisions~\citep{Cunningham2006_AY,BanksBOS2015_AY,OhY2019_AY}, and thus incorporating a measure of uncertainty in the forecasting model can improve the forecastability of real estate prices. For instance in the UK, the House Price Uncertainty (HPU) index \citep{HPU_AY}, constructed using the methodology outlined in \citet{BakerScott_2016_AY},\footnote{For a comparison of alternative text-based measures of economic uncertainty see \citet{Kapetanios2020_AY}.} was found to be important in predicting property price inflation ahead of the house price collapse of the third quarter of 2008 and during the bust phase \citep{yusupova2019adaptive_AY}. Along with capturing the two recent recessions (in the early 1990s and middle 2000s) this index also reflects the uncertainly related to the EU Referendum, Brexit negotiations and COVID-19 pandemic.
 
\subsubsection[Exchange rate forecasting (Michał Rubaszek)]{Exchange rate forecasting\protect\footnote{This subsection was written by Michał Rubaszek.}}
\label{sec:Exchange_rate_forecasting}
Exchange rates have long fascinated and puzzled researchers in international finance. The reason is that following the seminal paper of \cite{MeeseRogoff:1983JIE_MR}, the common wisdom is that macroeconomic models cannot outperform the random walk in exchange rate forecasting \citep[see][for a survey]{Rossi:2013JEL_MR}. This view is difficult to reconcile with the strong belief that exchange rates are driven by fundamentals, such as relative productivity, external imbalances, terms of trade, fiscal policy or interest rate disparity \citep{MacDonald:1998JIFMIM_MR, LeeEtAl:2013JMCB_MR, CouhardeEtAl:2017IE_MR}. These two contradicting assertions by the academic literature is referred to as ``exchange rate disconnect puzzle''.

The literature provides several explanations for this puzzle. First, it can be related to the forecast estimation error (see \S\ref{sec:Model_complexity}). The studies in which models are estimated with a large panels of data \citep{MarkSul:2001JIE_MR, EngelEtAl:2008NBER_MR, Ince:2014JIMF_MR}, long time series \citep{LothianTaylor:1996JPE_MR} or calibrated \citep{CaZorziRubaszek:2020JIMF_MR} deliver positive results on exchange rate forecastability. Second, there is ample evidence that the adjustment of exchange rates to equilibrium is non-linear \citep{TaylorPeel:2000JIMF_MR, CurranVelic:2019JIMF_MR}, which might diminish the out-of-sample performance of macroeconomic models \citep{KilianTaylor:2003JIE_MR, LopezSuarez:2011JIMF_MR}. Third, few economists argue that the role of macroeconomic fundamentals may be varying over time and this should be accounted for in a forecasting setting \citep{ByrneEtAl:2016JIMF_MR, BeckmanSchuster:2016JIMF_MR}. 

The dominant part of the exchange rate forecasting literature investigates which macroeconomic model performs best out-of-sample. The initial studies explored the role of monetary fundamentals to find that these models deliver inaccurate short-term and not so bad long-term predictions in comparison to the random walk \citep{MeeseRogoff:1983JIE_MR,Mark:1995AER_MR}. In a comprehensive study from mid-2000s, \citet{CheungEtAl:2005JIMF_MR} showed that neither monetary, uncovered interest parity (UIP) nor behavioural equilibrium exchange rate (BEER) model are able to outperform the no-change forecast. A step forward was made by \citet{MolodtsovaPapell:2009JIE_MR}, who proposed a model combining the UIP and Taylor rule equations and showed that it delivers competitive exchange rate forecasts. This result, however, has not been confirmed by more recent studies \citep{CheungEtAl:2019JIMF_MR, EngelEtAl:2019JIMF_MR}. In turn, \citet{CaZorziRubaszek:2020JIMF_MR} argue that a simple method assuming gradual adjustment of the exchange rate towards the level implied by the Purchasing Power Parity (PPP) performs well over shorter as well as longer horizon. This result is consistent with the results of \citet{CaZorziEtAl:2017JIE_MR} and \citet{EichenbaumEtAl:2017NBER_MR}, who showed that exchange rates are predictable within a general equilibrium DSGE framework (see \S\ref{sec:Forecasting_with_DSGE_Models}), which encompasses an adjustment of the exchange rate to a PPP equilibrium. Finally, \citet{CapEtAl:2019ECB_MR} discuss how extending the PPP framework for other fundamentals within the BEER framework is not helping in exchange rate forecasting. Overall, at the current juncture it might be claimed that ``exchange rate disconnect puzzle'' is still puzzling, with some evidence that methods based on PPP and controlling the estimation forecast error can deliver more accurate forecast than the random walk benchmark. A way forward to account for macroeconomic variables in exchange rate forecasting could be to use variable selection methods that allow to control for the estimation error (see \S\ref{sec:Variable_Selection}).

\subsubsection[Financial time series forecasting with range-based volatility models (Piotr Fiszeder)]{Financial time series forecasting with range-based volatility models\protect\footnote{This subsection was written by Piotr Fiszeder.}}
\label{sec:Financial_time_series_forecasting_with_range_based_volatility_models}
The range-based (RB) volatility models is a general term for the models constructed with high and low prices, and most often with their difference i.e., the price range. A short review and classification of such models is contained in \S\ref{sec:Low_and_high_prices_in_volatility_models}. From practical point of view, it is important that low and high prices are almost always available with daily closing prices for financial series. The price range (or its logarithm) is a significantly more efficient estimator of volatility than the estimator based on closing prices \citep{Alizadeh2002_PF}. Similarly the co-range (the covariance based on price ranges) is a significantly more efficient estimator of the covariance of returns than the estimator based on closing prices \citep{Brunetti2002_PF}. For these reasons models based on the price range and the co-range better describe variances and covariances of financial returns than the ones based on closing prices. 

Forecasts of volatility from simple models like moving average, EWMA, AR, ARMA based on the RB variance estimators are more accurate than the forecasts from the same models based on squared returns of closing prices \citep{Vipul2007_PF,Rajvanshi2015_PF}. Forecasts of volatility from the AR model based on the Parkinson estimator are more precise even than the forecasts from the standard GARCH models (see \S\ref{sec:arch_garch_models}) based on closing prices \citep{Li2011_PF}. 

In plenty of studies it was shown that forecasts of volatility of financial returns from the univariate RB models are more accurate than the forecasts from standard GARCH models based on closing prices (see, for example, \citeauthor{Mapa2003_PF}, \citeyear{Mapa2003_PF} for the GARCH-PARK-R model; \citeauthor{Chou2005_PF}, \citeyear{Chou2005_PF} for the CARR model; \citeauthor{Fiszeder2005_PF}, \citeyear{Fiszeder2005_PF} for the GARCH-TR model; \citeauthor{Brandt2006_PF}, \citeyear{Brandt2006_PF} for the REGARCH model; \citeauthor{Chen2008_PF}, \citeyear{Chen2008_PF} for the TARR model; \citeauthor{Lin2012_PF}, \citeyear{Lin2012_PF} for the STARR model; \citeauthor{Fiszeder2016_PF}, \citeyear{Fiszeder2016_PF} for the GARCH model estimated with low, high and closing prices during crisis periods; \citeauthor{Molnar2016_PF}, \citeyear{Molnar2016_PF} for the RGARCH model). 

The use of daily low and high prices in the multivariate volatility models leads to more accurate forecasts of covariance or covariance matrix of financial returns than the forecasts from the models based on closing prices (see, for example, \citeauthor{Chou2009a_PF}, \citeyear{Chou2009a_PF} for the RB DCC model; \citeauthor{Harris2010_PF}, \citeyear{Harris2010_PF} for the hybrid EWMA model; \citeauthor{Fiszeder2018_PF}, \citeyear{Fiszeder2018_PF} for the BEKK-HL model; \citeauthor{Fiszeder2019_PF}, \citeyear{Fiszeder2019_PF} for the co-range DCC model; \citeauthor{Fiszeder2019a_PF}, \citeyear{Fiszeder2019a_PF} for the DCC-RGARCH model).

The RB models were used in many financial applications. They lead for example to more precise forecasts of value-at-risk measures in comparison to the application of only closing prices (see, for example, \citeauthor{Chen2012_PF}, \citeyear{Chen2012_PF} for the threshold CAViaR model; \citeauthor{Asai2013a_PF}, \citeyear{Asai2013a_PF} for the HVAR model; \citeauthor{Fiszeder2019a_PF}, \citeyear{Fiszeder2019a_PF} for the DCC-RGARCH model; \citeauthor{Meng2020_PF}, \citeyear{Meng2020_PF} for scoring functions). The application of the multivariate RB models provides also the increase in the efficiency of hedging strategies (see, for example, \citeauthor{Chou2009a_PF}, \citeyear{Chou2009a_PF} for the RB DCC model; \citeauthor{Harris2010_PF}, \citeyear{Harris2010_PF} for the hybrid EWMA model; \citeauthor{Su2014_PF}, \citeyear{Su2014_PF} for the RB-MS-DCC model). Moreover, the RB volatility models have more significant economic value than the return-based ones in the portfolio construction (\citeauthor{Chou2010_PF}, \citeyear{Chou2010_PF} for the RB DCC model; \citeauthor{Wu2011_PF}, \citeyear{Wu2011_PF} for the RB-copula model). Some studies show that based on the forecasts from the volatility models with low and high prices it is possible to construct profitable investment strategies (\citeauthor{He2010_PF}, \citeyear{He2010_PF} for the VECM model; \citeauthor{Kumar2015_PF}, \citeyear{Kumar2015_PF} for the CARRS model).

\subsubsection[Copula forecasting with multivariate dependent financial times series (Feng Li)]{Copula forecasting with multivariate dependent financial times series\protect\footnote{This subsection was written by Feng Li.}}
\label{sec:Copula_forecasting_with_dependent_financial_times_series}
In this section, we focus on the practical advances on jointly forecasting multivariate financial time series with copulas. In the copula framework (see \S\ref{sec:bayesian_forecasting_with_copulas}), because marginal models and copula models are separable, point forecasts are straightforward with marginal models, but dependence information is ignored. A joint probabilistic forecast with copulas involves both estimations of the copula distribution and marginal models.

In financial time series, an emerging interest is to model and forecast the asymmetric dependence. A typical asymmetric dependence phenomenon is that two stock returns exhibit greater correlation during market downturns than market upturns. \citet{patton2006modelling_Feng} employs the asymmetric dependence between exchange rates with a time-varying copula construction with AR and GARCH margins. A similar study for measuring financial contagion with copulas allows the parameters of the copula to change with the states of the variance to identify shifts in the dependence structure in times of crisis \citep{rodriguez2007measuring_Feng}.

In stock forecasting, \citet{almeida2012efficient_Feng} employ a stochastic copula autoregressive model to model DJI and Nasdaq, and the dependence at the time is modelled by a real-valued latent variable, which corresponds to the Fisher transformation of Kendall's $\tau$. \citet{li2018improving_Feng} use a covariate-dependent copula framework to forecast the time varying dependence that improves both the probabilistic forecasting performance and the forecasting interpretability. Liquidity risk is another focus in finance. \citet{weiss2013forecasting_Feng} forecast three types of liquidity-adjusted intraday Value-at-Risk (L-IVaR) with a vine copula structure. The liquidity-adjusted intraday VaR is based on simulated portfolio values, and the results are compared with the realised portfolio profits and losses.

In macroeconomic forecasting, most existing reduced-form models for multivariate time series produce symmetric forecast densities. Gaussian copulas with skew Student's-\emph{t} margins depict asymmetries in the predictive distributions of GDP growth and inflation \citep{smith2016asymmetric_feng}. Real-time macroeconomic variables are forecasted with heteroscedastic inversion copulas \citep{smith2018inversion_Feng} that allow for asymmetry in the density forecasts, and both serial and cross-sectional dependence could be captured by the copula function \citep{loaiza2020real_Feng}.

Copulas are also widely used to detect and forecast default correlation, which is a random variable called \emph{time-until-default} to denote the survival time of each defaultable entity or financial instrument \citep{li2000default_Feng}. Then copulas are used in modelling the dependent defaults \citep{li2000default_Feng}, forecasting credit risk \citep{bielecki2013credit_Feng}, and credit derivatives market forecasting \citep{schonbucher2003credit_Feng}. A much large volume of literature is available for this specific area. See the aforementioned references therein. For particular applications in credit default swap (CDS) and default risk forecasting, see \cite{oh2018time_Feng} and \cite{li2019credit_Feng} respectively.

In energy economics, \citet{aloui2013time_Feng} employ the time-varying copula approach, where the marginal models are from ARMA($p$,$q$)–GARCH(1,1) to investigate the conditional dependence between the Brent crude oil price and stock markets in the Central and Eastern European transition economies. \citet{bessa2012time_Feng} propose a time-adaptive quantile-copula where the copula density is estimated with a kernel density forecast method. The method is applied to wind power probabilistic forecasting (see also \S\ref{sec:Wind_Power_orecasting}) and shows its advantages for both system operators and wind power producers. Vine copula models are also used to forecast wind power farms' uncertainty in power system operation scheduling. \citet{wang2017probabilistic_Feng} shows vine copulas have advantages of providing reliable and sharp forecast intervals, especially in the case with limited observations available.
 
\subsubsection[Financial forecasting with neural networks (Georgios Sermpinis)]{Financial forecasting with neural networks\protect\footnote{This subsection was written by Georgios Sermpinis.}}
\label{sec:Financial_forecasting_with_neural_networks}
Neural Networks (NNs; see \S\ref{sec:neural_networks}) are capable of successfully modelling non-stationary and non-linear series. This property has made them one of the most popular (if not the most) non-linear specification used by practitioners and academics in Finance. For example, 89\% of European banks use NNs to their operations \citep{European_Banking_Federation2019-sn} while 25.4\% of the NNs applications in total is in Finance \citep{Wong1995-zr_GS}. 

The first applications of NNs in Finance and currently the most widespread, is in financial trading. In the mid-80s when computational power became cheaper and more accessible, hedge fund managers started to experiment with NNs in trading. Their initial success led to even more practitioners to apply NNs and nowadays 67\% of hedge fund managers use NNs to generate trading ideas \citep{BarclayHedge2018-jx_GS}. A broad measure of the success of NNs in financial trading is provided by the Eurekahedge AI Hedge Fund Index\footnote{https://www.eurekahedge.com/Indices/IndexView/Eurekahedge/683/Eurekahedge-AI-Hedge-fund-Index (Accessed: 2020-09-01)} where it is noteworthy the 13.02\% annualised return of the selected AI hedge funds over the last 10 years. 

In academia, financial trading with NNs is the focus of numerous papers. Notable applications of NNs in trading financial series were provided by \cite{Kaastra1996-sm_GS}, \cite{Tenti1996-wc_GS}, \cite{Panda2007-ts_GS}, \cite{Zhang2008-kk_GS}, and \cite{Dunis2010-rx_GS}. The aim of these studies is to forecast the sign or the return of financial trading series and based on these forecasts to generate profitable trading strategies. These studies are closely related to the ones presented in \S\ref{sec:Forecasting_stock_returns} but the focus is now in profitability. 
The second major field of applications of NNs in Finance is in derivatives pricing and financial risk management. The growth of the financial industry and the provided financial services have made NNs and other machine learning algorithms a necessity for tasks such as fraud detection, information extraction and credit risk assessment \citep{Buchanan2019-ij_GS}. In derivatives pricing, NNs try to fill the limitations of the Black-Scholes model and are being used in options pricing and hedging. In academia notable applications of NNs in risk management are provided by \cite{Locarek-Junge1998-mz_GS} and \cite{Liu2005-lz_GS} and in derivatives by \cite{Bennell2004-vn_GS} and \cite{Psaradellis2016-ac_GS}. 

As discussed before, financial series due to their non-linear nature and their wide applications in practice seems the perfect forecasting data set for researchers that want to test their NN topologies. As a result, there are thousands of forecasting papers in the field of NNs in financial forecasting. However, caution is needed in interpretation of their results. NNs are sensitive to the choice of their hyperparameters. For a simple MLP, a practitioner needs to set (among others) the number and type of inputs, the number of hidden nodes, the momentum, the learning rate, the number of epochs and the batch size. This complexity in NN modelling leads inadvertently to the data snooping bias (see also \S\ref{sec:Statistical_tests_of_forecast_performance}). In other words, a researcher that experiments long enough with the parameters of a NN topology can have excellent in-sample and out-of-sample results for a series. However, this does not mean that the results of his NN can be generalised. This issue has led the related literature to be stained by studies cannot be extended in different samples. 

\subsubsection[Forecasting returns to investment style (Ross Hollyman)]{Forecasting returns to investment style\protect\footnote{This subsection was written by Ross Hollyman.}}
\label{sec:Forecasting_returns_to_investment_style}
Investment style or factor portfolios are constructed from constituent securities on the basis of a variety of a-priori observable characteristics, thought to affect future returns. For example a `Momentum' portfolio might be constructed with positive (`long') exposures to stocks with positive trailing 12-month returns, and negative (`short') exposure to stocks with negative trailing 12-month returns \citep[for full background and context, see, for example][]{Bernstein1995_RH,Haugen2010_RH}.\footnote{The website of Kenneth French is an excellent source of data on investment style factor data and research. http://mba.tuck.dartmouth.edu/pages/faculty/ken.french/data\_library.html} Explanations as to why such characteristics seem to predict returns fall in to two main camps: firstly that the returns represent a risk premium, earned by the investor in return for taking on some kind of (undiversifiable) risk, and secondly that such returns are the result of behavioural biases on the part of investors. In practice, both explanations are likely to drive style returns to a greater or lesser extent. Several such strategies have generated reasonably consistent positive risk-adjusted returns over many decades, but as with many financial return series, return volatility is large relative to the mean, and there can be periods of months or even years when returns deviate significantly from their long-run averages. The idea of timing exposure to styles is therefore at least superficially attractive, although the feasibility of doing so is a matter of some debate \citep{Arnott2016_RH,Asness2016_RH,Bender2018_RH}. Overconfidence in timing ability has a direct cost in terms of trading frictions and opportunity cost in terms of potential expected returns and diversification forgone.

A number of authors write on the general topic of style timing \citep[recent examples include][]{Hodges2017_RH,Dichtl2019_RH}, and several forecasting methodologies have been suggested, falling in to three main camps:
\begin{enumerate}[noitemsep]
 \item Serial Correlation: Perhaps the most promising approach is exploiting serial correlation in style returns. \cite{Tarun2019_RH} and \cite{Babu2020_RH} outline two such approaches and \cite{Ehsani2020} explore the relationship between momentum in factor portfolios and momentum in underlying stock returns. As with valuation spreads mentioned below, there is a risk that using momentum signals to time exposure to momentum factor portfolios risks unwittingly compounding exposure. A related strand of research relates (own) factor volatility to future returns, in particular for momentum factors \citep{Barroso2015_RH,Daniel2016_RH}.
 \item Valuation Spreads: Using value signals (aggregated from individual stock value exposures) to time exposure to various fundamental-based strategies is a popular and intuitively appealing approach \citep{Asness2016_RH}; however evidence of value added from doing so is mixed, and the technique seems to compound risk exposure to value factors.
 \item Economic \& Financial Conditions: \cite{Polk2020_RH} explore how economic and financial conditions affect style returns (an idea that dates back at least to \cite{Bernstein1995_RH} and references therein).
\end{enumerate}

Style returns exhibit distinctly non-normal distributions. On a univariate basis, most styles display returns which are highly negatively skewed and demonstrate significant kurtosis. The long-run low correlation between investment styles is often put forward as a benefit of style-based strategies, but more careful analysis reveals that non-normality extends to the co-movements of investment style returns; factors exhibit significant tail dependence. \cite{Christoffersen2013_RH} explores this issue, also giving details of the skew and kurtosis of weekly style returns. These features of the data mean that focusing solely on forecasting the mean may not be sufficient, and building distributional forecasts becomes important for proper risk management. \cite{Jondeau2007_RH} writes extensively on modelling non-gaussian distributions. 
 
\subsubsection[Forecasting stock returns (David E. Rapach)]{Forecasting stock returns\protect\footnote{This subsection was written by David E. Rapach.}}
\label{sec:Forecasting_stock_returns}
Theory and intuition suggest a plethora of potentially relevant predictors of stock returns. Financial statement data \citep[e.g.,][]{Chan2001_RAPACH,Yan2017_RAPACH} provide a wealth of information, and variables relating to liquidity, price trends, and sentiment, among numerous other concepts, have been used extensively by academics and practitioners alike to predict stock returns. The era of big data further increases the data available for forecasting returns. When forecasting with large numbers of predictors, conventional ordinary least squares (OLS) estimation is highly susceptible to overfitting, which is exacerbated by the substantial noise in stock return data (reflecting the intrinsically large unpredictable component in returns); see \S\ref{sec:machine_Learning_with_very_noisy_data}.

Over the last decade or so, researchers have explored methods for forecasting returns with large numbers of predictors. Principal component regression extracts the first few principal components (or factors) from the set of predictors; the factors then serve as predictors in a low-dimensional predictive regression, which is estimated via OLS (see \S\ref{sec:Forecasting_with_Big_Data}). Intuitively, the factors combine the information in the individual predictors to reduce the dimension of the regression, which helps to guard against overfitting. \cite{LudvigsonNg2007_DR} find that a few factors extracted from hundreds of macroeconomic and financial variables improve out-of-sample forecasts of the US market return. \cite{KellyPruitt2013_DR} and \cite{HuangJiangTuZhou2015_DR} use partial least squares \citep{Wold1966_DR} to construct target-relevant factors from a cross section of valuation ratios and a variety of sentiment measures, respectively, to improve market return forecasts.

Since \cite{BatesGranger1969_DR}, it has been known that combinations of individual forecasts often perform better than the individual forecasts themselves \citep[and \S\ref{sec:Forecast_combination_a_brief_review_of_statistical_approaches}]{Timmermann2006_DR}. \cite{RapachStraussZhou2010_DR} show that forecast combination can significantly improve out-of-sample market return forecasts. They first construct return forecasts via individual univariate predictive regressions based on numerous popular predictors from the literature \citep{GoyalWelch2008_DR}. They then compute a simple combination forecast by taking the average of the individual forecasts. \cite{RapachStraussZhou2010_DR} demonstrate that forecast combination exerts a strong shrinkage effect, thereby helping to guard against overfitting.

An emerging literature uses machine-learning techniques to construct forecasts of stock returns based on large sets of predictors. In an investigation of lead-lag relationships among developed equity markets, \cite{RapachStraussZhou2013_DR} appear to be the first to employ machine-learning tools to predict market returns. They use the elastic net \citep[ENet,][]{JLC_ZouHastie2005}, a generalisation of the popular least absolute shrinkage and selection operator \citep[LASSO,][]{JLC_Tibshir96}. The LASSO and ENet employ penalised regression to guard against overfitting in high-dimensional settings by shrinking the parameter estimates toward zero. \cite{ChincoClark-JosephYe2019_DR} use the LASSO to forecast high-frequency (one-minute-ahead) individual stock returns and report improvements in out-of-sample fit, while \cite{Rapach2018_HR} use the LASSO to improve monthly forecasts of industry returns.

Incorporating insights from \cite{DieboldShin2019_DR}, \cite{HanHeRapachZhou2020_DR} use the LASSO to form combination forecasts of cross-sectional stock returns based on a large number of firm characteristics from the cross-sectional literature \citep[e.g.,][]{HarveyLiuZhu2016_DR, McLeanPontiff2016_DR, HouXueZhang2020_DR}, extending the conventional OLS approach of \cite{HaugenBaker1996_DR}, \cite{Lewellen2015_DR}, and \cite{GreenHandZhang2017_DR}. \cite{RapachZhou2020_DR} and \cite{DongLiRapachZhou2020_DR} use the ENet to compute combination forecasts of the market return based on popular predictors from the time-series literature and numerous anomalies from the cross-sectional literature, respectively. Forecasting individual stock returns on the basis of firm characteristics in a panel framework, \cite{FreybergerNeuhierlWeber2020_DR} and \cite{GuKellyXiu2020_DR} employ machine-learning techniques -- such as the nonparametric additive LASSO \citep{HuangHorowitzWei2010_DR}, random forests \citep{Breiman2001_DR}, and artificial neural networks -- that allow for nonlinear predictive relationships.
 
\subsubsection[Forecasting crashes in stock markets (Philip Hans Franses)]{Forecasting crashes in stock markets\protect\footnote{This subsection was written by Philip Hans Franses.}}
\label{sec:Forecasting_crashes_in_stock_markets}
Time series data on financial asset returns have special features. Returns themselves are hard to forecast, while it seems that volatility of returns can be predicted. Empirical distributions of asset returns show occasional clusters of large positive and large negative returns. Large negative returns, that is, crashes seem to occur more frequently than large positive returns. Forecasting upcoming increases or decreases in volatility can be achieved by using variants of the Autoregressive Conditional Heteroskedasticity (ARCH) model \citep[][and \S\ref{sec:arch_garch_models}]{Engle1982_JJ,Bollerslev1986-zs_PHF} or realized volatility models \citep{Taylor1986-ir_PHF}. These models take (functions of) past volatility and past returns as volatility predictors, although also other explanatory variables can be incorporated in the regression.

An important challenge that remains is to predict crashes. \cite{Sornette2003-cf} summarises potential causes for crashes and these are computer trading, increased trading in derivatives, illiquidity, trade and budget deficits, and especially, herding behaviour of investors. Yet, forecasting the exact timing of crashes may seem impossible, but on the other hand, it may be possible to forecast the probability that a crash may occur within a foreseeable future. Given the herding behaviour, any model to use for prediction should include some self-exciting behaviour. For that purpose, \cite{Ait-Sahalia2015-nn_PHF} propose mutually exciting jump processes, where jumps can excite new jumps, also across assets or markets \citep[see also][]{Chavez-Demoulin2005-td_PHF}. Another successful approach is the Autoregressive Conditional Duration (ACD) model \citep{Engle1997-jd_PHF,Engle1998-js_PHF}, which refers to a time series model for durations between (negative) events.

An alternative view on returns' volatility and the potential occurrence of crashes draws upon the earthquake literature \citep{Ogata1978-fm_PHF,Ogata1988-lq_PHF}. The idea is that tensions in and across tectonic plates build up, until an eruption, and after that, tension starts to build up again until the next eruption. By modelling the tension-building-up process using so-called Hawkes processes \citep{Hawkes1971-wl_PHF,Hawkes2018-na_PHF,Hawkes1974-ri_PHF,Ozaki1979-qr_PHF}, one can exploit the similarities between earthquakes and financial crashes (see also \S\ref{sec:peak_over_the_threshold_POT}). \cite{Gresnigt2015-zt} take Hawkes processes to daily S\&P 500 data and show that it is possible to create reliable probability predictions of a crash occurrence within the next five days. \cite{Gresnigt2017-rx_PHF,Gresnigt2017-jq_PHF} further develop a specification strategy for any type of asset returns, and document that there are spillovers across assets and markets.

Given investor behaviour, past crashes can ignite future crashes. Hawkes processes are particularly useful to describe this feature and can usefully be implemented to predict the probability of nearby crashes. By the way, these processes can also be useful to predict social conflicts, as also there one may discern earthquake-like patterns. \cite{Van_den_Hengel2020-hs_PHF} document their forecasting power for social conflicts in Africa.

\subsection{Energy}
\label{sec:Energy}

\subsubsection[Building energy consumption forecasting and optimisation (Christoph Bergmeir \& Evangelos Spiliotis)]{Building energy consumption forecasting and optimisation\protect\footnote{This subsection was written by Christoph Bergmeir \& Evangelos Spiliotis.}}
\label{sec:Building_energy_consumption_forecasting}
In Europe, buildings account for 40\% of total energy consumed and 36\% of total $\text{CO}_2$ emissions  \citep{Patti2016}. Given that energy consumption of buildings is expected to increase in the coming years, forecasting electricity consumption becomes critical for improving energy management and planning by supporting a large variety of optimisation procedures.

The main challenge in electricity consumption forecasting is that building energy systems are complex in nature, with their behaviour depending on various factors related to the type (e.g., residential, office, entertainment, business, and industrial) and the end-uses (e.g., heating, cooling, hot water, and lighting) of the building, its construction, its occupancy, the occupants' behaviour and schedule, the efficiency of the installed equipment, and the weather conditions \citep{ZHAO20123586}. Special events, holidays, and calendar effects can also affect the behaviour of the systems and further complicate the consumption patterns, especially when forecasting at hourly or daily level (see \S\ref{sec:forecasting_for_multiple_seasonal_cycles}). As a result, producing accurate forecasts typically requires developing tailored, building-specific methods.

To deal with this task, the literature focuses on three main classes of forecasting methods, namely engineering, statistical, and ML \citep{MATDAUT20171108}. Engineering methods, typically utilised through software tools such as DOE-2, EnergyPlus, BLAST, and ESP-r, build on physical models that forecast consumption through detailed equations which account for the particularities of the building \citep{ALHOMOUD2001421, ZHAO20123586, FOUCQUIER2013272}. Statistical methods usually involve linear regression (see \S\ref{sec:time_series_regression_models}), ARIMA/ARIMAX (see \S\ref{sec:autoregressive_integrated_moving_average_models}), and exponential smoothing (see \S\ref{sec:exponential_smoothing_models}) models that forecast consumption using past consumption data or additional explanatory variables, such as weather or occupancy and calendar related information \citep{DEB2017902}. Finally, ML methods (see \S\ref{sec:machine_Learning}) typically involve neural networks (see \S\ref{sec:neural_networks}), support vector machines, and grey models that account for multiple non-linear dependencies between the electricity consumed and the factors influencing its value \citep{AHMAD2014102}. Till present, the literature has been inconclusive about which class of methods is the most appropriate, with the conclusions drawn being subject to the examined building type, data set used, forecasting horizon considered, and data frequency at which the forecasts are produced \citep{WEI2019106187}. To mitigate this problem, combinations of methods (see \S\ref{sec:combining_forecasts}) and hybrids (see \S\ref{sec:Hybrid_methods}) have been proposed, reporting encouraging results \citep{ZHAO20123586, MOHANDES201955}.

Other practical issues refer to data pre-processing. Electricity consumption data is typically collected at high frequencies through smart meters and therefore display noise and missing or extreme values due to monitoring issues (see \S\ref{sec:machine_Learning_with_very_noisy_data}). As a result, verifying the quality of the input data through diagnostics and data cleansing techniques (see \S\ref{sec:Anomaly_detection_and_time_series_forecasting} and \S\ref{sec:Robust_handling_of_outliers}), as well as optimising the selected time frames, are important for improving forecasting performance \citep{BOURDEAU2019101533}. Similarly, it is critical to engineer (see \S\ref{sec:Exogenous_variables_and_feature_engineering}) and select (see \S\ref{sec:Variable_Selection}) appropriate regressor variables which are of high quality and possible to accurately predict to assist electricity consumption forecasting. Finally, it must be carefully decided whether the bottom-up, the top-down or a combination method (see \S\ref{sec:Cross_sectional_hierarchical_forecasting}) will be used for producing reconciled forecasts at both building and end-use level \citep{KUSTER2017257}, being also possibly mixed with temporal aggregation approaches \citep[][but also \S\ref{sec:crosstemporal_hierarchies}]{Spiliotis2020-hj}.

Provided that accurate forecasts are available, effective energy optimisation can take place at a building level or across blocks of buildings (see \S\ref{sec:Collaborative_forecasting_in_the_energy_sector}) to reduce energy cost, improve network stability, and support efforts towards a carbon-free future, by exploiting smart grid, internet of things (IoT), and big data technologies along with recommendation systems \citep{MARINAKIS2020572}. 

An example for a typical application in this area is the optimisation of heating, ventilation, and air conditioning (HVAC) systems. The goal is to minimise the energy use of the HVAC system under the constraints of maintaining certain comfort levels in the building \citep{Marinakis2017}. Though this is predominantly an optimisation exercise, forecasting comes in at different points of the system as input into the optimisation, and many problems in this space involve forecasting as a sub-problem, including energy consumption forecasting, room occupancy forecasting, inside temperature forecasting, (hyper-local) forecasts of outside temperature, and air pressure forecasting for ventilation, among others. For instance, \cite{Krueger2004Predicting_CB} use a linear regression approach to predict inside temperatures in 3 houses in Brazil, and \cite{ruano2006prediction_CB} propose the use of a neural network to predict temperatures in a school building. \cite{madaus2020hyper_CB} predict hyper-local extreme heat events, combining global climate models and machine learning models. \cite{Jing2018air_CB} predict air pressure to tackle the air balancing problem in ventilation systems, using a support vector machine.

Predicting energy demand on a building/household level from smart meter data is an important research topic not only for energy savings. In the building space, \cite{Ahmad2017Trees_CB}, \cite{Touzani2018Gradient_CB}, and \cite{Wang2018Random_CB} predict building energy consumption of residential and commercial buildings using decision tree-based algorithms (random forests and gradient boosted trees) and neural networks to improve energy efficiency. 

A recent trend in forecasting are global forecasting models, built across sets of time series \citep{januschowski2020criteria_CB}. (Recurrent) neural networks \citep{bandara2020lstm_BRT, Hewamalage2019Recurrent_CB} are particularly suitable for this type of processing due to their capabilities to deal with external inputs and cold-start problems. Such capabilities are necessary if there are different regimes in the simulations under which to predict, an example of such a system for HVAC optimisation is presented by \cite{Godahewa2020Simulation_CB}.

More generally, many challenges in the space of building energy optimisation are classical examples of so-called ``predict then optimise'' problems \citep{demirovic2019predict_CB, elmachtoub2017smart_CB}. Here, different possible scenario predictions are obtained from different assumptions in the form of input parameters. These input parameters are then optimised to achieve a desired predicted outcome. As both prediction and optimisation are difficult problems, they are usually treated separately \citep{elmachtoub2017smart_CB}, though there are now recent works where they are considered together \citep{el2019generalization_CB, demirovic2019predict_CB}, and this will certainly be an interesting avenue for future research.

\subsubsection[Electricity price forecasting (Luigi Grossi \& Florian Ziel)]{Electricity price forecasting\protect\footnote{This subsection was written by Luigi Grossi \& Florian Ziel.}}
\label{sec:Electricity_price_forecasting}
\begin{figure}[ht!]
\resizebox{\textwidth}{!}{
\includegraphics{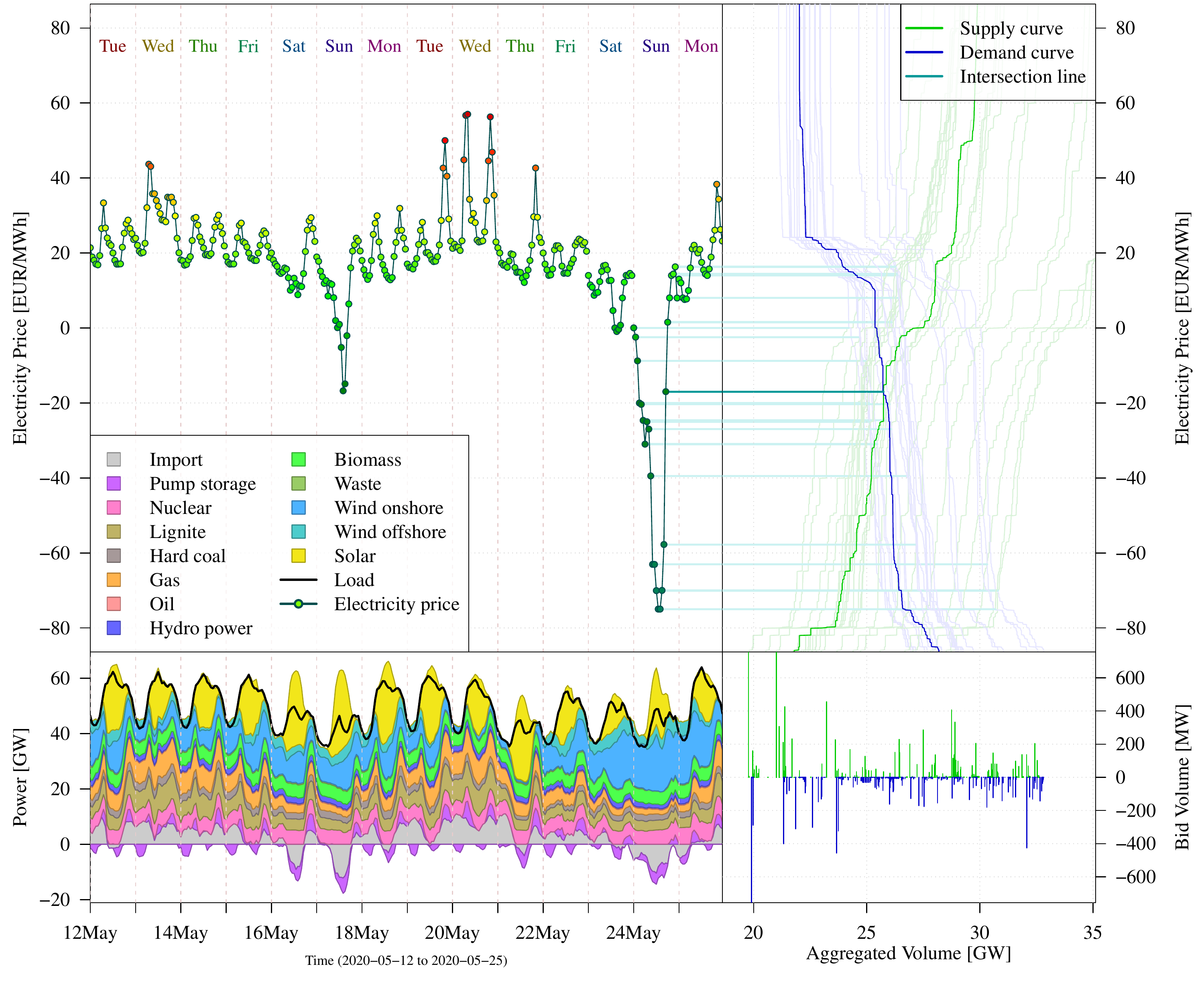}
}
\caption{Hourly German day-ahead electricity price data resulting from a two-sided auction (top left) with corresponding 24 sale/supply and purchase/demand curves for 24 May 2020 and highlighted curves for 17:00 (top right), power generation and consumption time series (bottom left), and bid structure of 24 May 2020 17:00 (bottom right).}
\label{fig:fig_epf_ziel}
\end{figure}

Forecasting electricity prices has various challenges that are highlighted in the detailed review papers by \cite{weron2014electricity_FZ}. Even though there are economically well motivated fundamental electricity price models, forecasting models based on evaluating historic price data are the dominating the academic literature. In recent years the focus on probabilistic forecasting grew rapidly, as they are highly relevant for many applications in energy trading and risk management, storage optimisation and predictive maintenance, \citep{ziel2018probabilistic_FZ, nowotarski2018recent_FZ}. Electricity price data is highly complex and is influenced by regulation. However, there is electricity trading based on auctions and on continuous trading. Many markets like the US and European markets organise day-ahead auctions for electricity prices, see figure \ref{fig:fig_epf_ziel}. Thus, we have to predict multivariate time series type data, \citep{ziel2018day_FZ}. In contrast, intraday markets usually apply continuous trading to manage short term variations due to changes in forecasts of renewable energy and demand, and outages \citep{kiesel2017econometric_FZ}.

The key challenge in electricity price forecasting is to address all potential characteristics of the considered market, most notably (some of them visible in figure \ref{fig:fig_epf_ziel}):
\begin{enumerate}[noitemsep]
 \item (time-varying) autoregressive effects and (in)stationarity,
 \item calendar effects (daily, weekly and annual seasonality, holiday effects, clock-change),
\item (time-varying) volatility and higher moment effects,
 \item price spikes (positive and negative), and
\item price clustering.
 \end{enumerate}

Some of those impacts can be explained by external inputs, that partially have to be predicted in advance:
\begin{enumerate}[noitemsep]
 \item load/demand/consumption (see \S\ref{sec:Load_forecasting}),
 \item power generation, especially from the renewable energy sources (RES) of wind and solar (see \S\ref{sec:Wind_Power_orecasting} and \S\ref{sec:Solar_power_forecasting}),
 \item relevant fuel prices (especially oil, coal, natural gas; see also  \S\ref{sec:Crude_oil_price_forecasting}),
 \item prices of emission allowances ($CO_2e$ costs),
 \item related power market prices (future, balancing and neighboring markets),
 \item availabilities of power plants and interconnectors,
 \item import/export flow related data, and
 \item weather effects (e.g. temperature due to cooling and heating and combined heat and power (CHP) effects; see also \S\ref{sec:Weather_Forecasting}).
\end{enumerate}
Note that other weather effects might be relevant as well, but should be covered from the fundamental point of view by the listed external inputs. Obvious examples are wind speed for the wind power prediction,  cloud cover for the solar power production and illumination effects in the electricity consumption.

\begin{figure}[ht!]
 \resizebox{\textwidth}{!}{
\includegraphics{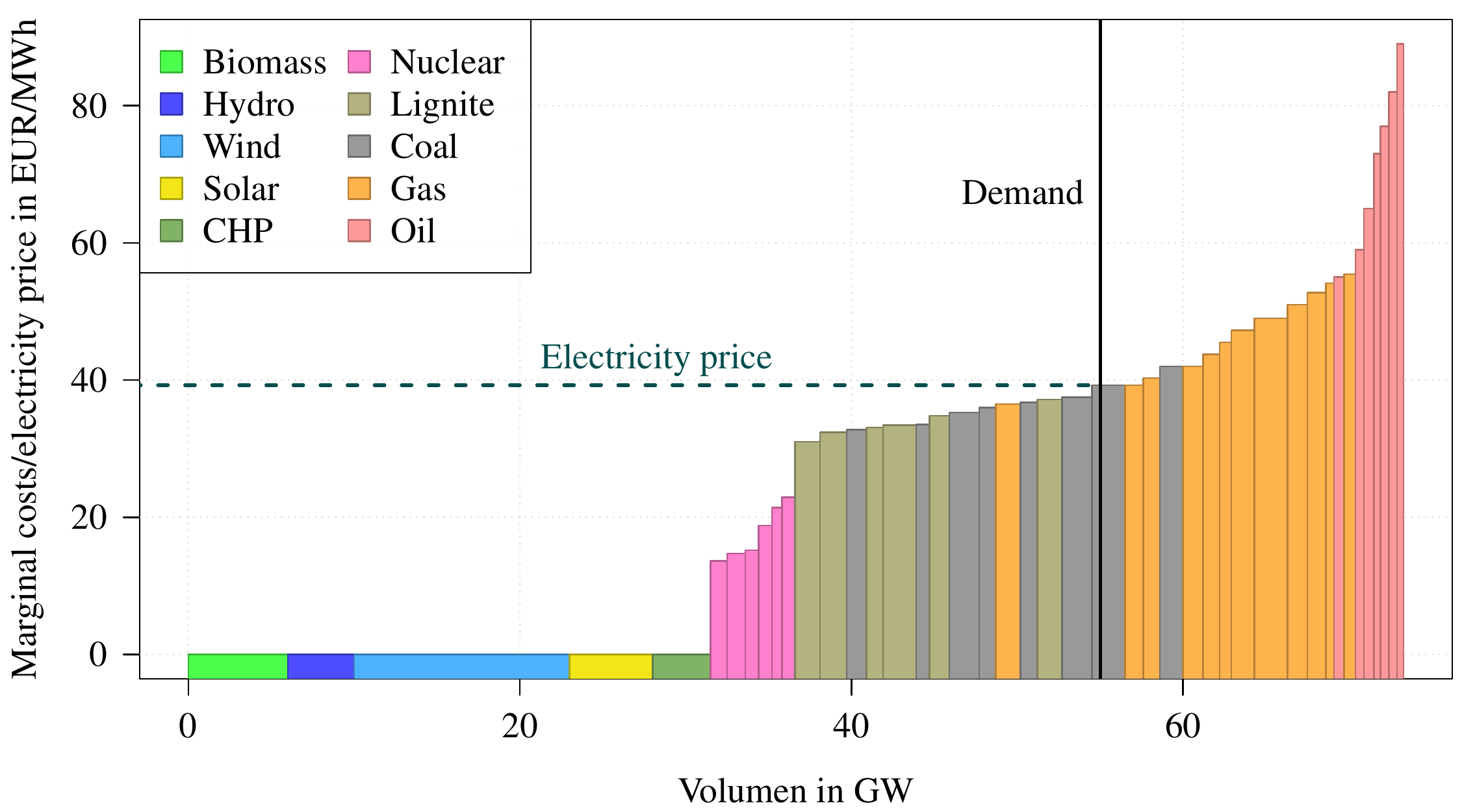}
}
\caption{Illustrative example of a supply stack model with inelastic demand for different power plant types, roughly covering the situation in Germany 2020.}
\label{fig:fig_supply_stack_model}
 \end{figure}

Many of those external effects may be explained by standard economic theory from fundamental electricity price models \citep{cludius2014merit, kulakov2021impact}. Even the simple supply stack model (merit order model), see figure \ref{fig:fig_supply_stack_model}, explains many features and should be kept in mind when designing an appropriate electricity price forecasting model.
 
In recent years, statistical and machine learning methods gained a lot of attraction in day-ahead electricity price forecasting. Even though the majority of effects is linear there are specific non-linear dependencies that can be explored by using non-linear models, especially neural networks \citep{dudek2016multilayer_FZ, lago2018forecasting_FZ, ugurlu2018electricity_FZ, marcjasz2019importance_FZ}. Of course this comes along with higher computational costs compared to linear models. \cite{fezzi2020size_FZ} illustrate that even simple linear models can give highly accurate forecasts, if correctly calibrated. However, there seems to be consensus that forecast
combination is appropriate, particularly for models that have different structures or different calibration window length \citep{gaillard2016additive_FZ, mirakyan2017composite_FZ, hubicka2018note_FZ}.

Another increasing stream of electricity price forecasting models do not focus on the electricity price itself, but the bid/sale/sell/supply and ask/sell/purchase/demand curves of the underlying auctions \citep[see figure \ref{fig:fig_epf_ziel}, but also][]{ziel2016electricity_FZ, kulakov2020x, shah2020forecasting_FZ, mestre2020forecasting_FZ}. This sophisticated forecasting problem allows more insights for trading applications and the capturing of price clusters.

In forecasting intraday markets the literature just started to grow quickly. As the aforementioned market characteristics get less distinct if information from day-ahead markets is taken into account appropriately. However, intraday prices are usually more volatile and exhibit more stronger price spikes. Thus, probabilistic forecasting is even more relevant \citep{janke2019forecasting_FZ,narajewski2020ensemble_FZ}. Recent studies showed that European markets are close to weak-form efficiency. Thus naive point forecasting benchmarks perform remarkably well \citep{oksuz2019neural_FZ,narajewski2020econometric_FZ, marcjasz2020beating_FZ}.

As pointed out above, predicting price spikes is particularly important in practice, due to the high impact in decision making problems which occur usually in extreme situations, see figure \ref{fig:fig_supply_stack_model}. Very high electricity prices are usually observed in connection to high demand and low renewable energy generation, sometimes together with sudden power plant failures. In contrast, negative price spikes occur in oversupply situation, when there is low demand but high penetration from wind and solar power. The presence of spikes is explored in two main streams in literature: spike forecasting and prediction of prices under normal regime through robust estimators.

Within the first set of papers, spikes are often modelled as one regime of non-linear models for time series. This approach is followed by \citet{Mount2006-fw_LG} focusing on regime-switching models with parameters driven by time-varying variables and by \citet{Becker2008-wu_LG} who adopt Markov switching models for spikes prediction. \cite{Christensen2009-rz_LG,Christensen2012-mv_LG} suggest treating and forecasting price spikes through Poisson autoregressive and discrete-time processes, respectively. \citet{Herrera2014-di_LG} use a Hawkes model combined with extreme events theory. Interregional links among different electricity markets are used by \citet{Clements2015-tt_LG} and \citet{Manner2016-yu_LG} to forecast electricity price spikes. A new procedure for the simulation of electricity spikes has been recently proposed by \citet{Muniain2020-dz} utilising bivariate jump components in a mean reverting jump diffusion model in the residuals.

The second stream of literature includes papers developing outlier detection methods or robust estimators to improve the forecasting performance of the models. \citet{MartinezAlvarez2011-ug_LG} tackle the issue of outlier detection and prediction defining ``motifs'', that is patches of units preceding observations marked as anomalous in a training set. \citet{Janczura2013-ny_LG} focus on the detection and treatment of outliers in electricity prices. A very similar approach, based on seasonal autoregressive models and outlier filtering, is followed by \citet{Afanasyev2019-cr_LG}. \citet{Grossi2019-pg_LG} introduced a procedure for the robust statistical prediction of electricity prices. The econometric framework is represented by the robust estimation of non-linear SETAR processes. A similar approach has been followed by \citet{Wang2020-mk_LG} using an outlier-robust machine learning algorithm.
 
\subsubsection[Load forecasting (Ioannis Panapakidis)]{Load forecasting\protect\footnote{This subsection was written by Ioannis Panapakidis.}}
\label{sec:Load_forecasting}
Load forecasting forms the basis where power system operation and planning builds upon. Based on the time horizon of the forecasts, load forecasting can be classified into very short-term (VSTLF), that refers to horizon from several minutes ahead up to 1 hour, short-term (STLF), that spans from 1 hour to 168 hours ahead, medium-term (MTLF), that spans from 168 hours to 1 year ahead and finally, and long-term (LTLF) that concerns predictions from 1 year to several years ahead. In VSTLF and STLF applications, the focus is on the sub-hourly or hourly load. In MTLF and LTLF, the variables of interest can be either monthly electricity peak load and total demand for energy. 

Inputs differ in the various horizons. In VSTLF and STLF, apart from meteorological data, day type identification codes are used. In LTLF, macroeconomic data are used since total demand of energy is influenced by the long-term modifications of the social and economic environments. Among the horizons, special attention is placed at STLF. This is reflected by the research momentum that have been placed in the load forecasting related literature by other researchers \citep{Hong2016-yf_IP}. Processes like unit commitment and optimal power flow rely on STLF \citep{Saksornchai2005-cz_IP,Bo2012-sw_IP}. Additionally, since competitive energy markets continually evolve, STLF becomes vital for new market entities such as retailers, aggregators, and prosumers for applications such as strategic bidding, portfolio optimisation, and tariff design \citep{Danti2017-zf_IP,Ahmad2019-mi_IP}. 

The models that can be found in the load forecasting related literature can in general categorised into three types: time-series, machine learning, and hybrid. Time-series models historically precede the others. Typical examples of this family are ARMA, ARIMA, and others (see also \S\ref{sec:autoregressive_integrated_moving_average_models}). In the machine learning models, the structure is usually determined via the training process. NNs are commonly used. Once a NN is sufficiently trained, it can provide forecasts for all types of forecasting horizons \citep{Hippert2001-lw}. The third category of models refers to the integration of two or more individual forecasting approaches (see also see \S\ref{sec:Hybrid_methods}). For instance, a NN can be combined with time series methods, with unsupervised machine learning algorithms, data transformation, and with meta-heuristics algorithms \citep{Bozkurt2017-mv_IP,Lopez2017-ni_IP,Lu2019-sm_IP,El-Hendawi2020-mf_IP}. 

Hybrid systems has been tested on validation data (through forecasting competitions), power system aggregated load, and application oriented tasks. \cite{Ma2021-la_IP} proposed an ensemble method based on a combination of various single forecasters on GEFCom2012 forecasting competition data that outperformed benchmark forecasters such as Theta method, NN, ARIMA, and others (see \S\ref{sec:Forecasting_competitions} for further discussions on forecasting competitions). For aggregated load cases, researchers focus on different countries and energy markets. \cite{Zhang2018-zd_IP} combined empirical mode decomposition (EMD), ARIMA, and wavelet neural networks (WNN) optimised by the fruit fly algorithm on Australian Market data and New York City data. Their approach was to separate the linear and nonlinear components from original electricity load; ARIMA is used for linear part while the WNN for the non-linear one.

\cite{Sideratos2020-wc_IP} proposed that a radial basis network that performs the initial forecasting could serve as input to a convolutional neural network that performs the final forecasting. The proposed model led to lower error compared to the persistence model, NN, and SVM. \cite{Semero2020-yj_IP} focused on the energy management of a microgrid located in China using EMD to decompose the load, adaptive neuro-fuzzy inference system (ANFIS) for forecasting and particle swarm intelligence (PSO) to optimize ANFIS parameters. The results show that the proposed approach yielded superior performance over four other methods. \cite{Faraji2020-sv_IP} proposed a hybrid system for the scheduling of a prosumer microgrid in Iran. Various machine learning algorithms provided load and weather forecasts. Through an optimisation routine, the best individual forecast is selected. The hybrid system displayed better accuracy from the sole application of the individual forecasters.

\subsubsection[Crude oil price forecasting (Xiaoqian Wang)]{Crude oil price forecasting\protect\footnote{This subsection was written by Xiaoqian Wang.}}
\label{sec:Crude_oil_price_forecasting}
Crude oil, one of the leading energy resources, has contributed to over one-third of the world's energy consumption \citep{alvarez2003symmetry_XW}. The fluctuations of the crude oil price have a significant impact on industries, governments as well as individuals, with substantial up-and-downs of the crude oil price bringing dramatic uncertainty for the economic and political development \citep{kaboudan2001compumetric_XW,cunado2005oil_XW}. Thus, it is critical to develop reliable methods to accurately forecast crude oil price movement, so as to guard against the crude oil market extreme risks and improve macroeconomic policy responses. However, the crude oil price movement suffers from complex features such as nonlinearity, irregularities, dynamics and high volatility \citep[][and also \S\ref{sec:arch_garch_models}]{kang2009forecasting_XW,alquist2013forecasting_XW,herrera2018forecasting_XW}, making the crude oil price forecasting still one of the most challenging forecasting problems.

Some prior studies have suggested that the crude oil price movement is inherently unpredictable, and it would be pointless and futile to attempt to forecast future prices, see \citet{miao2017influential_XW} for a detailed summary. These agnostics consider the naive no-change forecast as the best available forecast value of future prices. In recent years, however, numerous studies result in forecasts that are more accurate than naive no-change forecasts, making the forecasting activities of crude oil prices promising \citep{alquist2013forecasting_XW,baumeister2015high_XW}. 

Extensive research on crude oil price forecasting has focused predominantly on the econometric models, such as VAR, ARCH-type, ARIMA, and Markov models \citep[see, for example,][and \S\ref{sec:Statistical_and_econometric_models}]{mirmirani2004comparison_XW,agnolucci2009volatility_XW,mohammadi2010international_XW,e2010forecasting_XW}. 
In the forecasting literature, unit root tests (see \S\ref{sec:autoregressive_integrated_moving_average_models}) are commonly applied to examine the stationarity of crude oil prices prior to econometric modelling \citep{silvapulle1999relationship_XW,serletis2004testing_XW,rahman2012oil_XW}.
It is well-documented that crude oil prices are driven by a large set of external components, which are themselves hard to predict, including supply and demand forces, stock market activities, oil-related events (e.g., war, weather conditions), political factors, etc. In this context, researchers have frequently considered structural models (see \S\ref{sec:Forecasting_with_many_variables}), which relate the oil price movements to a set of economic factors. With so many econometric models, is there an optimal one? Recently, \citet{de2018forecasting_XW} proposed a SETAR model, allowing for predictive regimes changing after a detected threshold, and achieved performance improvements over six widely used econometric models. Despite their high computational efficiency, the econometric models are generally limited in the ability to nonlinear time series modelling.

On the other hand, artificial intelligence and machine learning techniques, such as belief networks, support vector machines (SVMs), recurrent neural networks (RNNs), and extreme gradient boosting (XGBoost), provided powerful solutions to recognise the nonlinear and irregular patterns of the crude oil price movement with high automation \citep[see, for example,][]{abramson1991using_XW,xie2006new_XW,mingming2012multiple_XW,gumus2017crude_XW}. However, challenges also exist in these techniques, such as computational cost and overfitting. In addition, a large number of studies have increasingly focused on the hybrid forecasting models (see also \S\ref{sec:Hybrid_methods}) based on econometrics models and machine learning techniques
\citep{jammazi2012crude_XW,he2012crude_XW,baumeister2015forecasting_XW,chiroma2015evolutionary_XW}, achieving improved performance. Notably, the vast majority of the literature has focused primarily on the deterministic prediction, with much less attention paid to the probabilistic prediction and uncertainty analysis. However, the high volatility of crude oil prices makes probabilistic prediction more crucial to reduce the risk in decision-making \citep{abramson1995probabilistic_XW,sun2018interval_XW}.
 
\subsubsection[Forecasting renewable energy technologies (Mariangela Guidolin)]{Forecasting renewable energy technologies\protect\footnote{This subsection was written by Mariangela Guidolin.}}
\label{sec:Forecasting_renewable_energy_technologies}

The widespread adoption of renewable energy technologies, RETs, plays a driving role in the transition to low-carbon energy systems, a key challenge to face climate change and energy security problems. Forecasting the diffusion of RETs is critical for planning a suitable energy agenda and setting achievable targets in terms of electricity generation, although the available time series are often very short and pose difficulties in modelling. According to \cite{rao:10_MG}, renewables' typical characteristics such as low load factor, need for energy storage, small size, high upfront costs create a competitive disadvantage, while \cite{Meade2015-lh-MG} suggested that renewable technologies are different from other industrial technological innovations because, in the absence of focused support, they are not convenient from a financial point of view. In this sense, policy measures and incentive mechanisms, such as feed-in tariffs, have been used to stimulate the market. As highlighted in \cite{Lee2017-py-MGa}, forecasting RETs requires to capture different socio-economic aspects, such as policy choices by governments, carbon emissions, macroeconomic factors, economic and financial development of a country, competitive strength of traditional energy technologies.

The complex and uncertain environment concerning RETs deployment has been faced in literature in several ways, in order to account for various determinants of the transition process. A first stream of research employed a bottom-up approach, where forecasts at a lower level are aggregated to higher levels within the forecasting hierarchy. For instance \cite{Park2016-ii-MG} realised a bottom-up analysis to study the optimum renewable energy portfolio, while \cite{Lee2017-am-MGb} performed a three-step forecasting analysis, to reflect the specificities of renewable sources, by using different forecasting methods for each of the sources considered. A similar bottom-up perspective was adopted in \cite{Zhang2020-ar-MG}, by conducting a multi-region study, to understand how multi-level learning may affect RETs dynamics, with the regionalised model of investment and technological development, a general equilibrium model linking a macro-economic growth with a bottom-up engineering-based energy system model.

The relative newness of RETs has posed the challenge of forecasting with a limited amount of data: in this perspective, several contributions applied the `Grey System' theory, a popular methodology for dealing with systems with partially unknown parameters \citep{Kayacan2010-tc-MG}. Grey prediction models for RETs forecasting were proposed in \cite{Tsai2017-jn-MG}, \cite{Lu2019-oz-MG}, \cite{Wu2019-kv-MG}, \cite{Moonchai2020-vt-MG}, and \cite{Liu2021-il-MG}.

Other studies developed forecasting procedures based on growth curves and innovation diffusion models (see \S\ref{sec:Innovation_diffusion_models}, \S\ref{sec:The_natural_law of growth_in_competition_Logistic_growth}, and \S\ref{sec:Synchronic_and_diachronic_competition}): from the seminal work by \cite{marchetti:79_MG}, contributions on the diffusion of RETs were proposed by \cite{guidolinmortarino:10_MG}, \cite{Dalla_Valle2011-fz-MG}, \cite{Meade2015-lh-MG}, \cite{Lee2017-py-MGa}, and \cite{bunea2020adoption_MG}. Forecasting the diffusion of renewable energy technologies was also considered within a competitive environment in \cite{Huh2014-ma-MG}, \cite{guidolin2016german_MG}, \cite{furlan2018forecasting_MG}, and \cite{guidolin2019transition_MG}. 

\subsubsection[Wind power forecasting (Jethro Browell)]{Wind power forecasting\protect\footnote{This subsection was written by Jethro Browell.}}
\label{sec:Wind_Power_orecasting}
Wind energy is a leading source of renewable energy, meeting 4.8\% of global electricity demand in 2018, more than twice that of solar energy \citep{IEA_windshare_2018_JB}. Kinetic energy in the wind is converted into electrical energy by wind turbines according to a characteristic `power curve'. Power production is proportion to the cube of the wind speed at low-to-moderate speeds, and above this is constant at the turbine's rated power. At very high or low wind speeds no power is generated. Furthermore, the power curve is influenced by additional factors including air density, icing, and degradation of the turbine's blades.

Forecasts of wind energy production are required from minutes to days-ahead to inform the operation of wind farms, participation in energy markets and power systems operations. However, the limited predictability of the weather (see also \S\ref{sec:Weather_Forecasting}) and the complexity of the power curve make this challenging. For this reason, probabilistic forecasts are increasing used in practice \citep{Bessa2017_JB}. Their value for energy trading is clear \citep{Pinson2007_JB}, but quantifying value for power system operation is extremely complex. Wind power forecasting may be considered a mature technology as many competing commercial offerings exist, but research and development efforts to produce novel and enhanced products is ongoing (see also \S\ref{sec:Forecasting_renewable_energy_technologies}).

Short-term forecasts (hours to days ahead) of wind power production are generally produced by combining numerical weather predictions (NWP) with a model of the wind turbine, farm or even regional power curve, depending on the objective. The power curve may be modelled using physical information, e.g. provided by the turbine manufacturer, in which case it is also necessary to post-process NWP wind speeds to match the same height-above-ground as the turbine's rotor. More accurate forecasts can be produced by learning the NWP-to-energy relationship from historic data when it is available. State-of-the-art methods for producing wind power forecasts leverage large quantities of NWP data to produce a single forecast \citep{Andrade2017_JB} and detailed information about the target wind farm \citep{Gilbert_2020_JB}. A number of practical aspects may also need to be considered by users, such as maintenance outages and requirements to reduce output for other reasons, such as noise control or electricity network issues.

Very short-term forecast (minutes to a few hours ahead) are also of value, and on these time scales recent observations are the most significant input to forecasting models and more relevant than NWP. Classical time series methods perform well (see \S\ref{sec:Statistical_and_econometric_models}), and those which are able to capture spatial dependency between multiple wind farms are state-of-the-art, notably vector autoregressive models and variants \citep{Cavalcante2016_JB,Messner2018_JB}. Care must be taken when implementing these models as wind power time series are bounded by zero and the wind farm's rated power meaning that errors may not be assumed to be normally distributed. The use of transformations is recommended (see also \S\ref{sec:BoxCox_Transformations}), though the choice of transformation depends on the nature of individual time series \citep{Pinson2012c_JB}.

Wind power forecasting is reviewed in detail in \cite{Zhang2014_JB}, \cite{Giebel2017_JB}, \cite{Hong2020_JB} and research is ongoing in a range of directions including:  improving accuracy and reducing uncertainty in short-term forecasting, extending forecast horizons to weeks and months ahead, and improving very short-term forecast with remote sensing and data sharing \citep[][and \S\ref{sec:Collaborative_forecasting_in_the_energy_sector}]{Sweeney2019_JB}.

\subsubsection[Wave forecasting (Jooyoung Jeon)]{Wave forecasting\protect\footnote{This subsection was written by Jooyoung Jeon.}}
\label{sec:Wave_forecasting}
Ocean waves are primarily generated by persistent winds in one direction. The energy thus propagated by the wind is referred to as wave energy flux and follows a linear function of wave height squared and wave period. Wave height is typically measured as significant wave height, the average height of the highest third of the waves. The mean wave period, typically measured in seconds, is the average time between the arrival of consecutive crests, whereas the peak wave period is the wave period at which the highest energy occurs at a specific point.

The benefit of wave energy is that it requires significantly less reserve compared to those from wind (see \S\ref{sec:Wind_Power_orecasting}) and solar (see \S\ref{sec:Solar_power_forecasting}) renewable energy sources \citep{hong2016probabilistic_FZ}. For example, the forecast error at one hour ahead for the simulated wave farms is typically in the range of 5–7\%, while the forecast errors for solar and wind are 17 and 22\% respectively \citep{Reikard2011_JJ}. Solar power is dominated by diurnal and annual cycles but also exhibits nonlinear variability due to factors such as cloud cover, temperature and precipitation. Wind power is dominated by large ramp events such as irregular transitions between states of high and low power. Wave energy exhibits annual cycles and is generally smoother although there are still some large transitions, particularly during the winter months. In the first few hours of forecasting wave energy, time series models are known to be more accurate than numerical wave prediction. Beyond these forecast horizons, numerical wave prediction models such as SWAN \citep[Simulating WAves Nearshore,][]{Booij1999} and WAVEWATCH III\textsuperscript{\textregistered} \citep{Tolman2008} are widely used.
As there is as yet no consensus on the most efficient model for harnessing wave energy, potential wave energy is primarily measured with energy flux, but the wave energy harnessed typically follows non-linear functions of wave height and wave period in the observations of the six different types of wave energy converters \citep{Reikard2015_JJ}.

To model the dependencies of wind speed, wave height, wave period and their lags, \cite{Reikard2011_JJ} uses linear regressions, which were then converted to forecasts of energy flux. \cite{Pinson2012_JJ} uses \citeauthor{Reikard2011_JJ}'s \citeyearpar{Reikard2011_JJ} regression model and log-normal distribution assumptions to produce probabilistic forecasts. \cite{Lopez-Ruiz2016_JJ} model the temporal dependencies of significant wave heights, peak wave periods and mean wave direction using a vector autoregressive model, and used them to produce medium to long term wave energy forecasts. \cite{Jeon2016_JJ} model the temporal dependencies of significant wave heights and peak wave periods using a bivariate VARMA-GARCH (see also \S\ref{sec:arch_garch_models}) to convert the two probabilistic forecasts into a probabilistic forecast of wave energy flux, finding this approach worked better than either univariate modelling of wave energy flux or bivariate modelling of wave energy flux and wind speed. \cite{Taylor2018_JJ} produce probabilistic forecasts for wave heights using a bivariate VARMA-GARCH model of wave heights and wind speeds, and using forecasts so as to optimise decision making for scheduling offshore wind farm maintenance vessels dispatched under stochastic uncertainty. On the same subject, \cite{Gilbert2020_JJ} use statistical post-processing of numerical wave predictions to produce probabilistic forecasts of wave heights, wave periods and wave direction and a logistic regression to determine the regime of the variables. They further applied the Gaussian copula to model temporal dependency but this did not improve their probabilistic forecasts of wave heights and periods.

\subsubsection[Solar power forecasting (Sonia Leva)]{Solar power forecasting\protect\footnote{This subsection was written by Sonia Leva.}}
\label{sec:Solar_power_forecasting}
Over the past few years, a number of forecasting techniques for photovoltaic (PV) power systems has been developed and presented in the literature. In general, the quantitative comparison among different forecast techniques is challenging, as the factors influencing the performance are numerous: the historical data, the weather forecast, the temporal horizon and resolution, and the installation conditions. A recent review by \cite{Sobri2018-qa_SL} presents a comparative analysis of previous works, also including statistical errors. However, since the conditions and metrics used in each work were different, the comparison is not very meaningful. \cite{Dolara2018-vh_SL} present relevant evaluation metrics for PV forecasting accuracy, while \cite{Leva2019-fs_SL} compare their effectiveness and immediate comprehension. In term of forecast horizon for PV power systems, intraday \citep{Nespoli2019-vg_SL} and the 24 hours of the next day \citep{Mellit2020-eb_SL} are considered the most important.

\begin{figure}[ht!]
\centering
\includegraphics[trim=120 110 120 110, clip, width=4.5in]{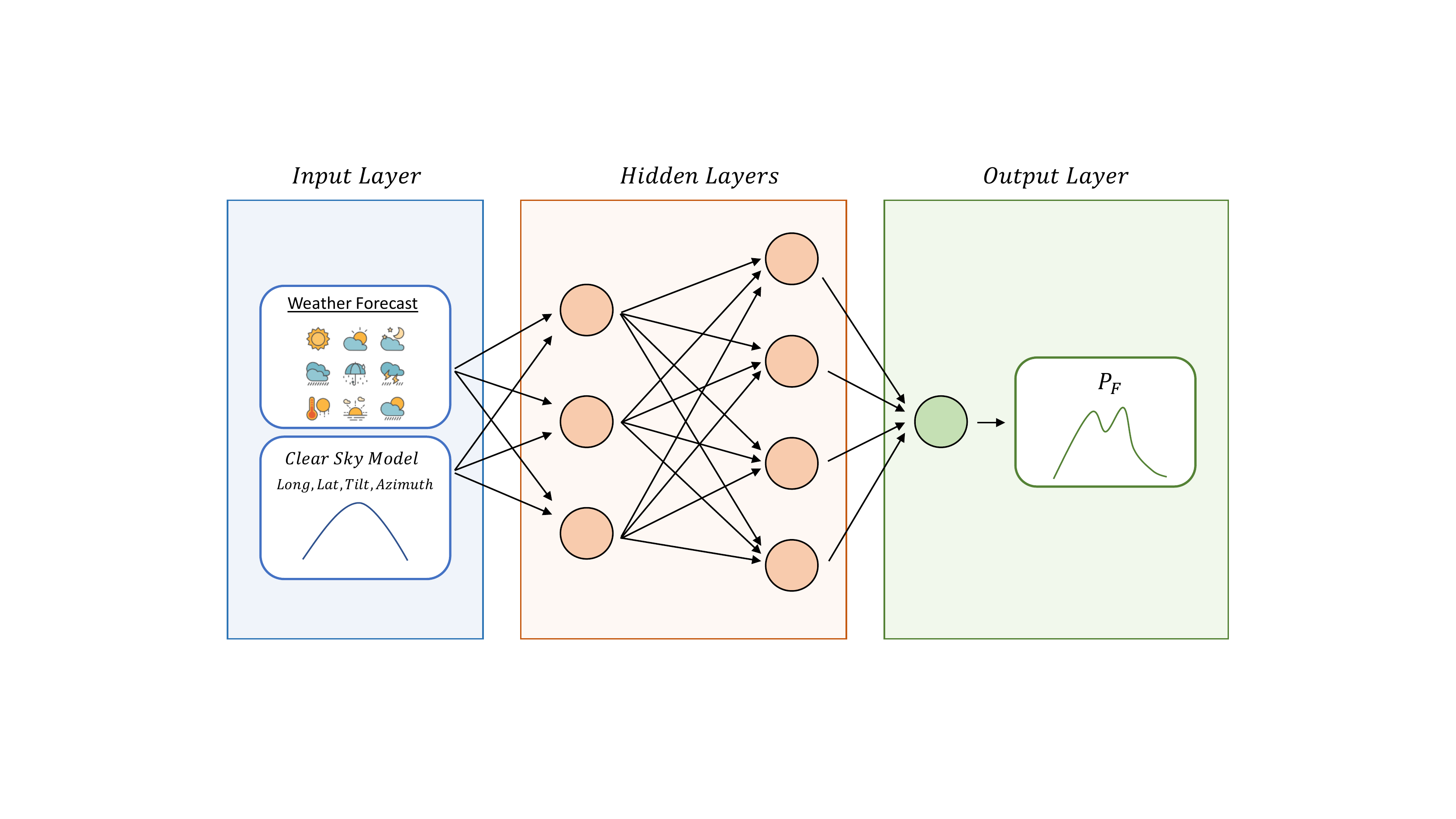}
\caption{Physical Hybrid Artificial Neural Network (PHANN) for PV power forecasting.}
\label{fig:fig_SL1}
\end{figure}

\begin{figure}[ht!]
\centering
\includegraphics[trim=250 75 250 165, clip, width=4 in]{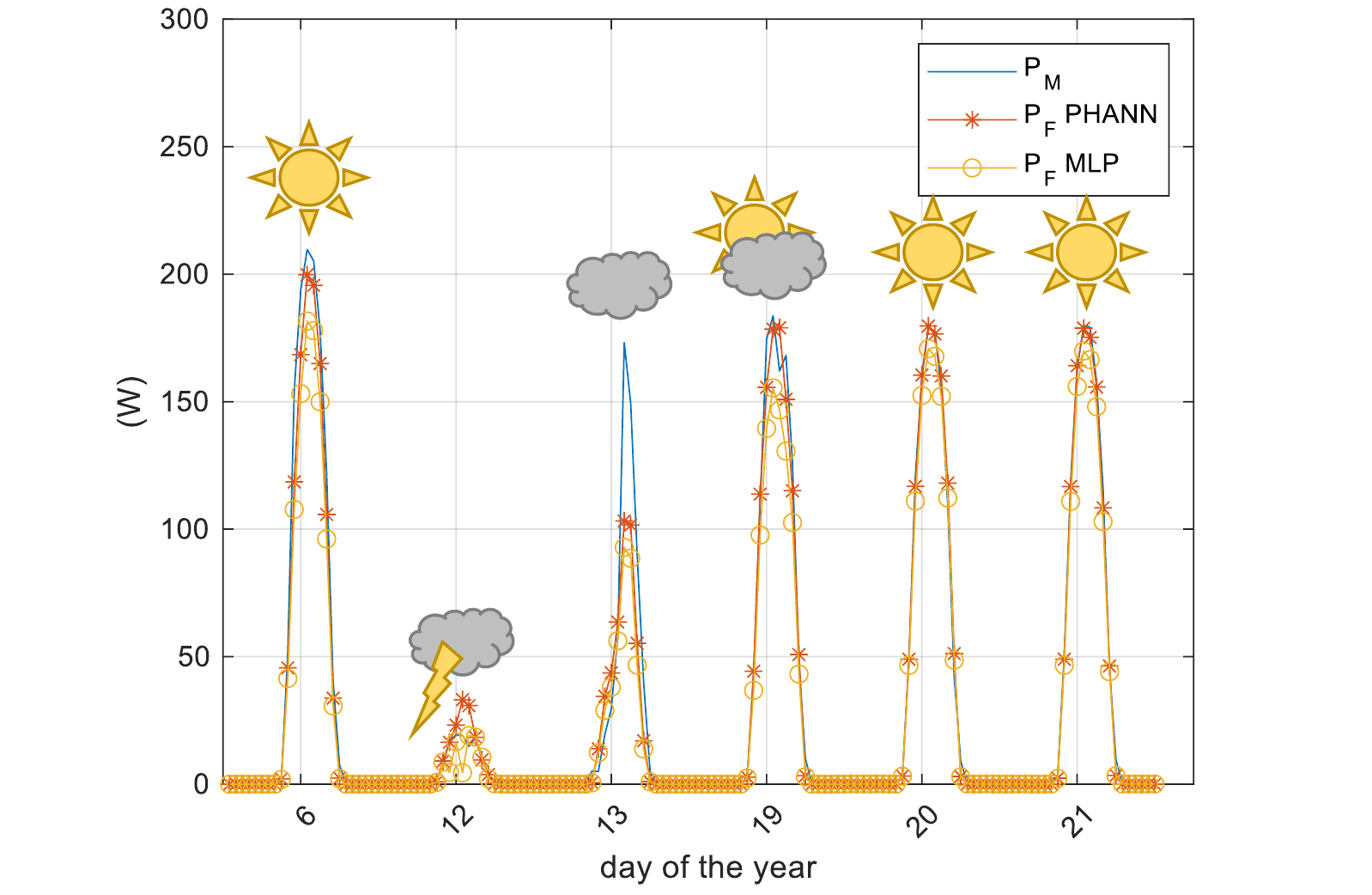}
\caption{Measured versus forecasted output power by MLP and PHANN methods.}
\label{fig:fig_SL2}
\end{figure}

\cite{Nespoli2019-vg_SL} compared two of the most widely used and effective methods for the forecasting of the PV production: a method based on Multi-Layer Perceptron (MLP) and a hybrid method using artificial neural network combined with clear sky solar radiation (see also \S\ref{sec:neural_networks} and \S\ref{sec:Hybrid_methods}). In the second case, the simulations are based on a feed-forward neural network (FFNN) but, among the inputs, the irradiation in clear sky conditions is provided. This method is called Physical Hybrid Artificial Neural Network (PHANN) and is graphically depicted in figure \ref{fig:fig_SL1} \citep{Dolara2015-vn_SL}. PHANN method demonstrates better performance than classical NN methods. Figure \ref{fig:fig_SL2} shows a comparison between the measured and forecasted hourly output power of the PV plant for both sunny and cloudy days. The PHANN method shows good forecasting performance, especially for sunny days.

\begin{figure}[ht!]
	\centering
	\includegraphics[trim=50 195 50 255, clip, width=0.47\textwidth]{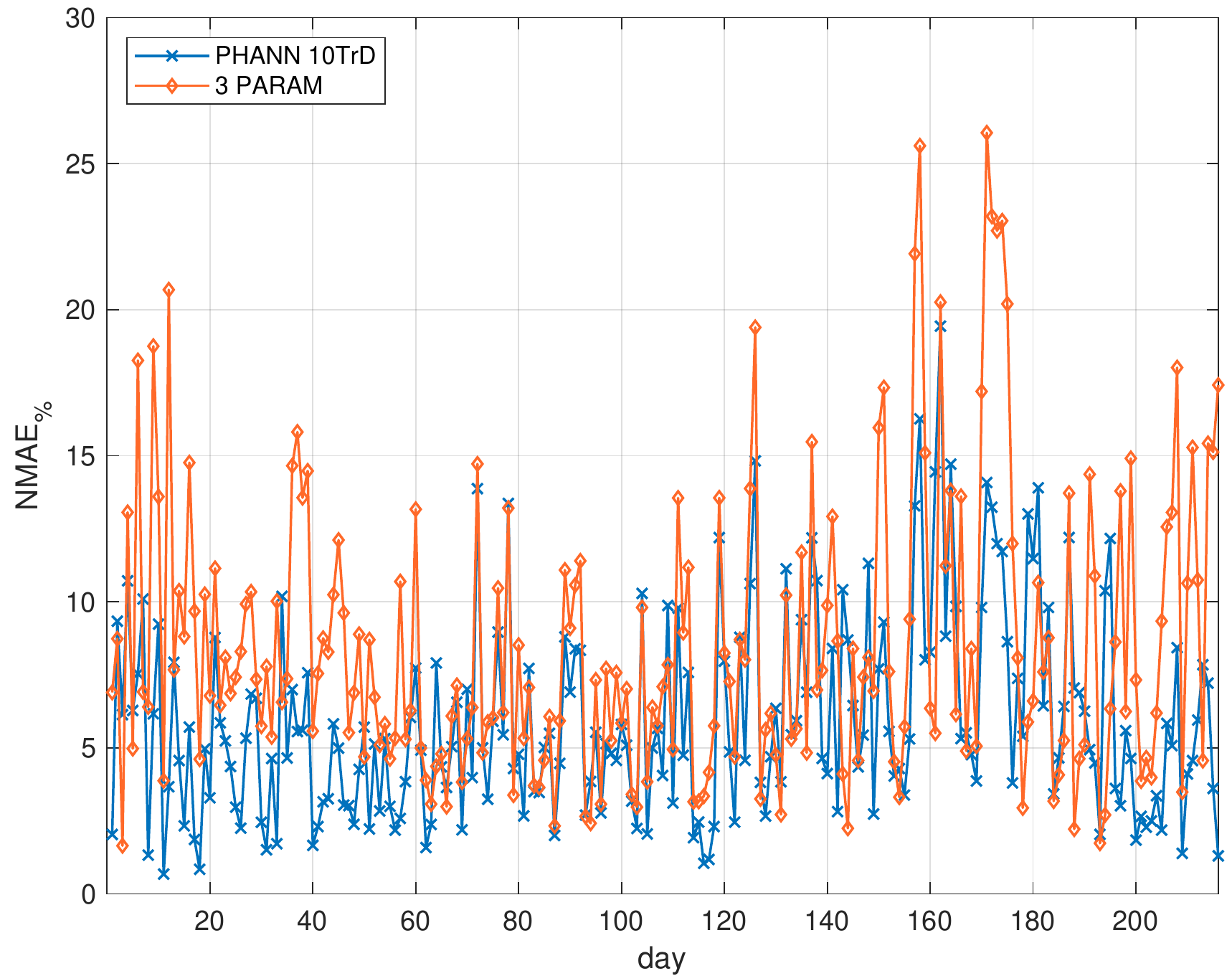}
	\hspace{5mm}
	\includegraphics[trim=50 195 50 255, clip, width=0.47\textwidth]{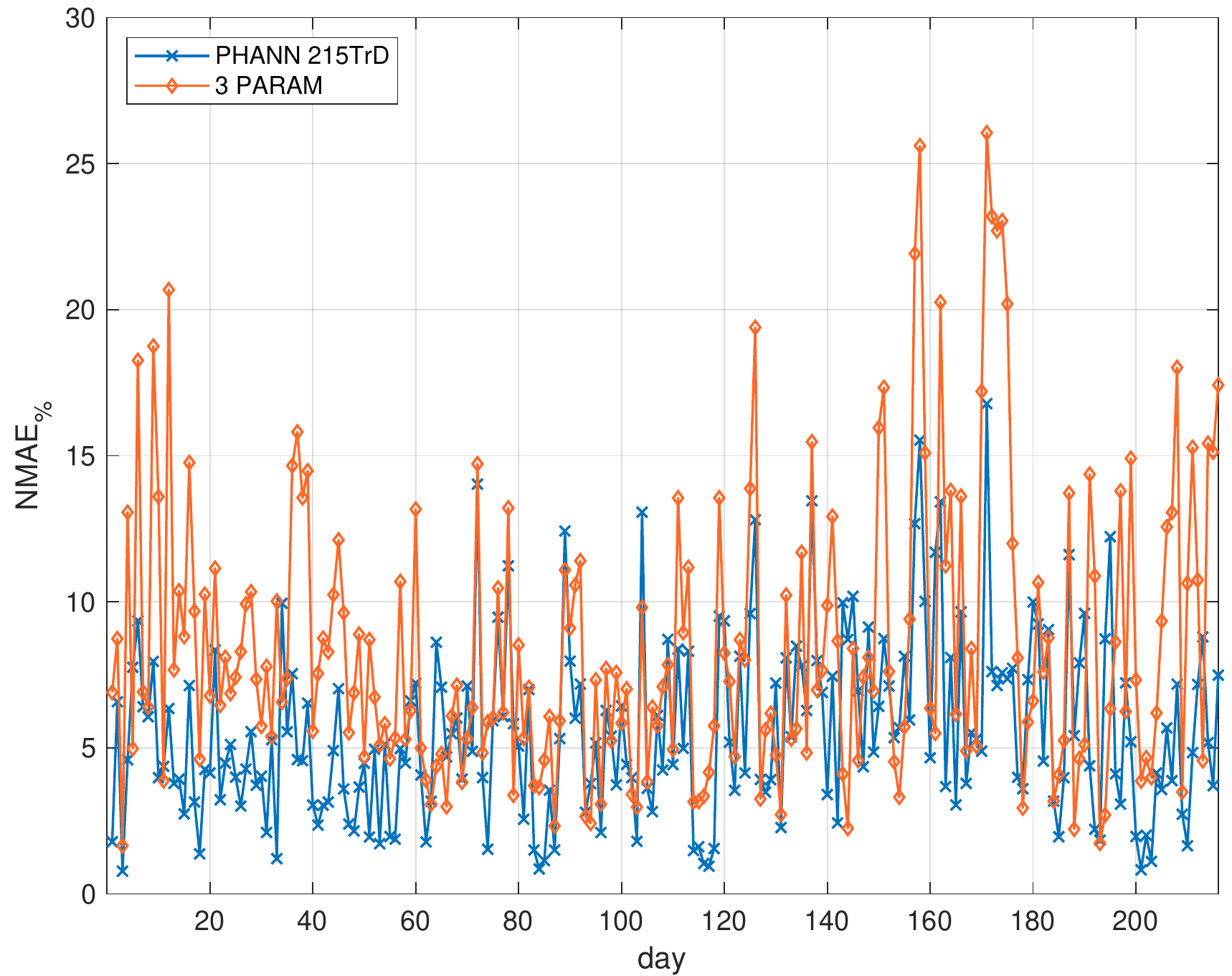}
	\caption{Daily NMAE\% of the PHANN method trained with 10 days (left) and with 215 days (right) compared with the three-parameters model.}
	\label{fig:fig_SL3}
\end{figure}

\cite{Ogliari2017-mm_SL} compared the PV output power day-ahead forecasts performed by deterministic (based on three and five parameters electric equivalent circuit) and stochastic hybrid (based on artificial neural network models) methods aiming to find the best performance conditions. In general, there is no significant difference between the two deterministic models, with the three-parameter approach being slightly more accurate. Figure \ref{fig:fig_SL3} shows the daily value of normalised mean absolute error (NMAE\%) for 216 days evaluated by using PHANN and three parameters electric circuit. The PHANN hybrid method achieves the best forecasting results, and only a few days of training can provide accurate forecasts.

\cite{Dolara2018-vh_SL} analysed the effect of different approaches in the composition of a training data-set for the day-ahead forecasting of PV power production based on NN. In particular, the influence of different data-set compositions on the forecast outcome has been investigated by increasing the size of the training set size and by varying the lengths of the training and validation sets, in order to assess the most effective training method of this machine learning approach. As a general comment on the reported results, it can be stated that a method that employs the same chronologically consecutive samples for training is best suited when the availability of historical data is limited (for example, in newly deployed PV plant), while training based on randomly mixed samples method, appears to be most effective in the case of a greater data availability. Generally speaking, ensembles composed of independent trials are most effective.

\subsubsection[Long-term simulation for large electrical power systems (Fernando Luiz Cyrino Oliveira)]{Long-term simulation for large electrical power systems\protect\footnote{This subsection was written by Fernando Luiz Cyrino Oliveira.}}
\label{sec:Long-term_simulation_for_large_electrical_power_systems}
In large electrical power systems with renewable energy dependence, the power generators need to be scheduled to supply the system demand \citep{De_Queiroz2016-pc}. In general, for modelling long-term renewables future behaviour, such as hydro, wind and solar photovoltaics (PV), stochastic scenarios should be included in the scheduling, usually in a dispatch optimisation problem under uncertainty – like described, for small systems, in \S\ref{sec:Building_energy_consumption_forecasting} and, for wave forecasting, in \S\ref{sec:Wave_forecasting}. Due to the complexity and uncertainly associated, this problem is, in general, modelled with time series scenarios and multi-stage stochastic approaches. \cite{De_Queiroz2016-pc} presented a review for hydrothermal systems, with a focus on the optimisation algorithms. \S\ref{sec:Wind_Power_orecasting} and \S\ref{sec:Solar_power_forecasting} explore the up-to-date methods for wind and PV solar power forecasting.

Here, we emphasise the importance of forecasting with simulation in the long-term renewable energy planning, especially in hydroelectric systems. In this context, due to the data spatial and temporal dependence structure, time series models are useful for future scenarios generation. Although the proposal could be forecasting for short-term planning and scheduling (as described in \S\ref{sec:Wind_Power_orecasting}, \ref{sec:Wave_forecasting}, and \S\ref{sec:Solar_power_forecasting}), simulation strategies are explored for considering and estimating uncertainty in medium and/or long-term horizons. 

According to \cite{Hipel1994-oz}, stochastic processes of natural phenomena, such as the renewables ones, are, in general, stationary. One of the main features of hydroelectric generation systems is the strong dependence on hydrological regimes. To deal with this task, the literature focuses on two main classes for forecasting/simulation streamflow data: physical and data-driven models \citep{Zhang2015-ql}. Water resources management for hydropower generation and energy planning is one of the main challenges for decision-makers. At large, the hydrological data are transformed into the so-called affluent natural energy, that is used for scenarios simulation and serve as input for the optimisation algorithms \citep{Oliveira2015-cb}. The current state-of-the-art models for this proposal are the periodic ones. \cite{Hipel1994-oz} presented a wide range of possibilities, but the univariate periodic autoregressive (PAR, a periodic extension version of the ones presented in \S\ref{sec:autoregressive_integrated_moving_average_models}) is still the benchmark, with several enhanced versions. The approach fits a model to each period of the historical data and the residuals are simulated to generate new future versions of the time series, considered stationary. Among many others, important variations and alternative proposals to PAR with bootstrap procedures (see bootstrap details in \S\ref{sec:Forecasting_with_bootstrap}), Bayesian dynamic linear models, spatial information and copulas versions (for copulas references, see \S\ref{sec:bayesian_forecasting_with_copulas}) are detailed in \cite{Souza2012-rq}, \cite{Lima2014-eo}, \cite{Lohmann2016-bm} and \cite{De_Almeida_Pereira2019-eh}, respectively. 

It is worth considering the need for renewables portfolio simulation. This led \cite{Pinheiro_Neto2017-ut} to propose a model to integrate hydro, wind and solar power scenarios for Brazilian data. For the Eastern United States, \cite{Shahriari2018-hl} add to the literature on the wind, solar and blended portfolios over several spatial and temporal scales. For China, \cite{Liu2020-lb} proposed a multi-variable model, with a unified framework, to simulate wind and PV scenarios to compensate hydropower generation. However, in light of the aforementioned, one of the key challenges and trends for renewable electrical power systems portfolio simulation are still related to the inclusion of exogenous variables, such as climate, meteorological, calendar and economic ones, as mentioned in \S\ref{sec:Electricity_price_forecasting}.

\subsubsection[Collaborative forecasting in the energy sector (Ricardo Bessa)]{Collaborative forecasting in the energy sector\protect\footnote{This subsection was written by Ricardo Bessa.}}
\label{sec:Collaborative_forecasting_in_the_energy_sector}
As mentioned in \S\ref{sec:Wind_Power_orecasting}, the combination of geographically distributed time series data, in a collaborative forecasting (or data sharing) framework, can deliver significant improvements in the forecasting accuracy of each individual renewable energy power plant. The same is valid for hierarchical load forecasting~\citep{Hong2019} and energy price forecasting (see \S\ref{sec:Electricity_price_forecasting}). A review of multivariate time series forecasting methods can be found in \S\ref{sec:Forecasting_with_many_variables} 2.3.11 and \S\ref{sec:bayesian_forecasting_with_copulas}. However, this data might have different owners, which are unwilling to share their data due to the following reasons: (\textit{i}) personal or business sensitive information, (\textit{ii}) lack of understanding about which data can and cannot be shared, and (\textit{iii}) lack of information about economic (and technical) benefits from data sharing.

In order to tackle these limitations, recent research in energy time series forecasting is exploring two alternative (and potentially complementary) pathways: (\textit{i}) privacy-preserving analytics, and (\textit{ii}) data markets.

The role of privacy-preserving techniques applied collaborative forecasting is to combine time series data from multiple data owners in order to improve forecasting accuracy and keep data private at the same time. For solar energy forecasting, \cite{Berdugo2011} described a method based on local and global analog-search that uses solar power time series from neighbouring sites, where only the timestamps and normalised weights (based on similarity) are exchanged and not the time series data.~\cite{Zhang2018} proposed, for wind energy forecasting with spatia-temporal data, a combination of ridge linear quantile regression and Alternating Direction Method of Multipliers (ADMM) that enables each data owner to autonomously solve its forecasting problem, while collaborating with the others to improve forecasting accuracy. However, as demonstrated by \cite{Goncalves2020}, the mathematical properties of these algorithms should be carefully analysed in order to avoid privacy breaches (i.e., when a third party recovers the original data without consent). 

An alternative approach is to design a market (or auction) mechanism for time series or forecasting data where the data owners are willing to sell their private (or confidential) data in exchange for an economic compensation~\citep{Agarwal2018}. The basic concept consists in pricing data as a function of privacy loss, but it can be also pricing data as a function of tangible benefits such as electricity market profit maximization.~\cite{Goncalves2020a} adapted for renewable energy forecasting the model described in~\cite{Agarwal2018}, by considering the temporal nature of the data and relating data price with the extra revenue obtained in the electricity market due to forecasting accuracy improvement. The results showed a benefit in terms of higher revenue resulting from the combination of electricity and data markets. With the advent of peer-to-peer energy markets at the domestic consumer level~\citep{Parag2016}, smart meter data exchange between peers is also expected to increase and enable collaborative forecasting schemes. For this scenario,~\cite{Yassine2015} proposed a game theory mechanism where a energy consumer maximizes its reward by sharing consumption data and a data aggregator can this data with a data analyst (which seeks data with the lowest possible price). 

Finally, promoting data sharing via privacy-preserving or data monetisation can also solve data scarcity problems in some use cases of the energy sector, such as forecasting the condition of electrical grid assets~\citep{Fan2020}. Moreover, combination of heterogeneous data sources (e.g., numerical, textual, categorical) is a challenging and promising avenue of future research in collaborative forecasting~\citep{Obst2019}.
 
\subsection{Environmental applications}
\label{sec:Environmental_applications}

\subsubsection[Forecasting two aspects of climate change (David F. Hendry)]{Forecasting two aspects of climate change\protect\footnote{This section was written by David F. Hendry.}}
\label{sec:Climate_Forecasting}
\begin{figure}[ht!]
\begin{center}
\includegraphics[width=14.5cm]{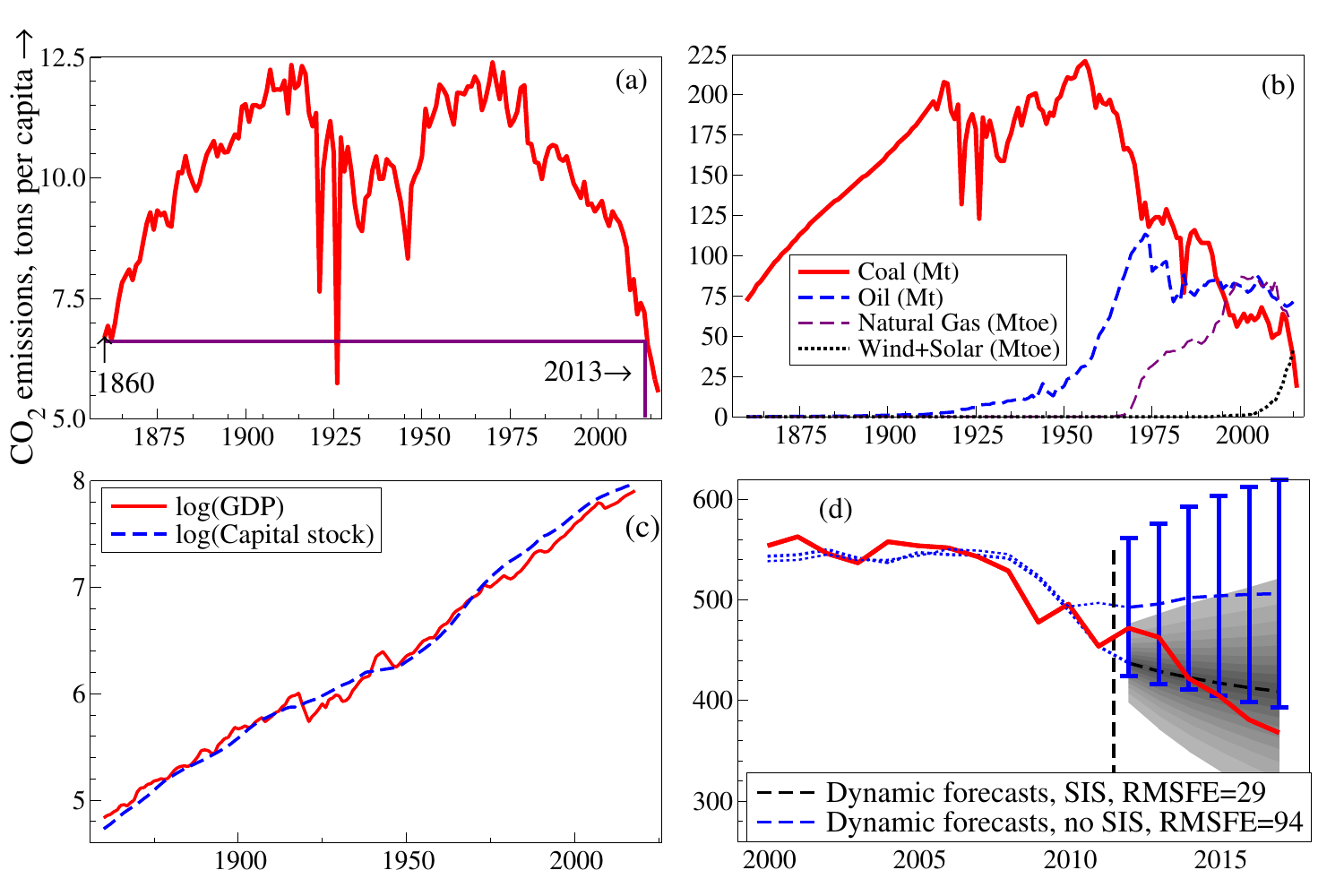} \vspace{-.2in}
\caption{(a) UK emissions, (b) energy sources in megatonnes (Mt) and megatonnes of oil equivalent (Mtoe), (c) economic variables, and (d) multi-step forecasts of CO2 emissions in Mt.}
\label{fig:hendry_UKCO2Figure1}
\end{center}
\end{figure}

First into the Industrial Revolution, the UK is one of the first out: in 2013
its per capita CO$_2$ emissions dropped below their 1860 level, despite per capita real incomes being around 7-fold higher \citep{DFH_HendUKCO2FIFO}. The model for forecasting UK CO$_2$ emissions was selected from annual data 1860-2011 on CO$_2$ emissions, coal and oil usage, capital and GDP, their lags and non-linearities (see \S\ref{sec:Weather_Forecasting} for higher frequency weather forecasts). Figures \ref{fig:hendry_UKCO2Figure1}(a) to \ref{fig:hendry_UKCO2Figure1}(c) show the non-stationary time series with strong upward then downward trends, punctuated by large outliers from world wars, miners strikes plus shifts from legislation and technological
change: \cite{DFH_CastleFTECE20}. Saturation estimation at 0.1\% using \textit{Autometrics} \citep{JLC_Door07Auto} retaining all other regressors, detected 4 step shifts coinciding with major policy interventions like the 2008 Climate Change Act, plus numerous outliers, revealing a cointegrated relation. The multi-step forecasts over 2012—2017 from a VAR in panel (d) of figure \ref{fig:hendry_UKCO2Figure1} show the advantage of using step-indicator saturation (SIS: \citealp{JLC_CDHPSIS15}).

\begin{figure}[ht!]
\begin{center}
\includegraphics[width=14.5cm]{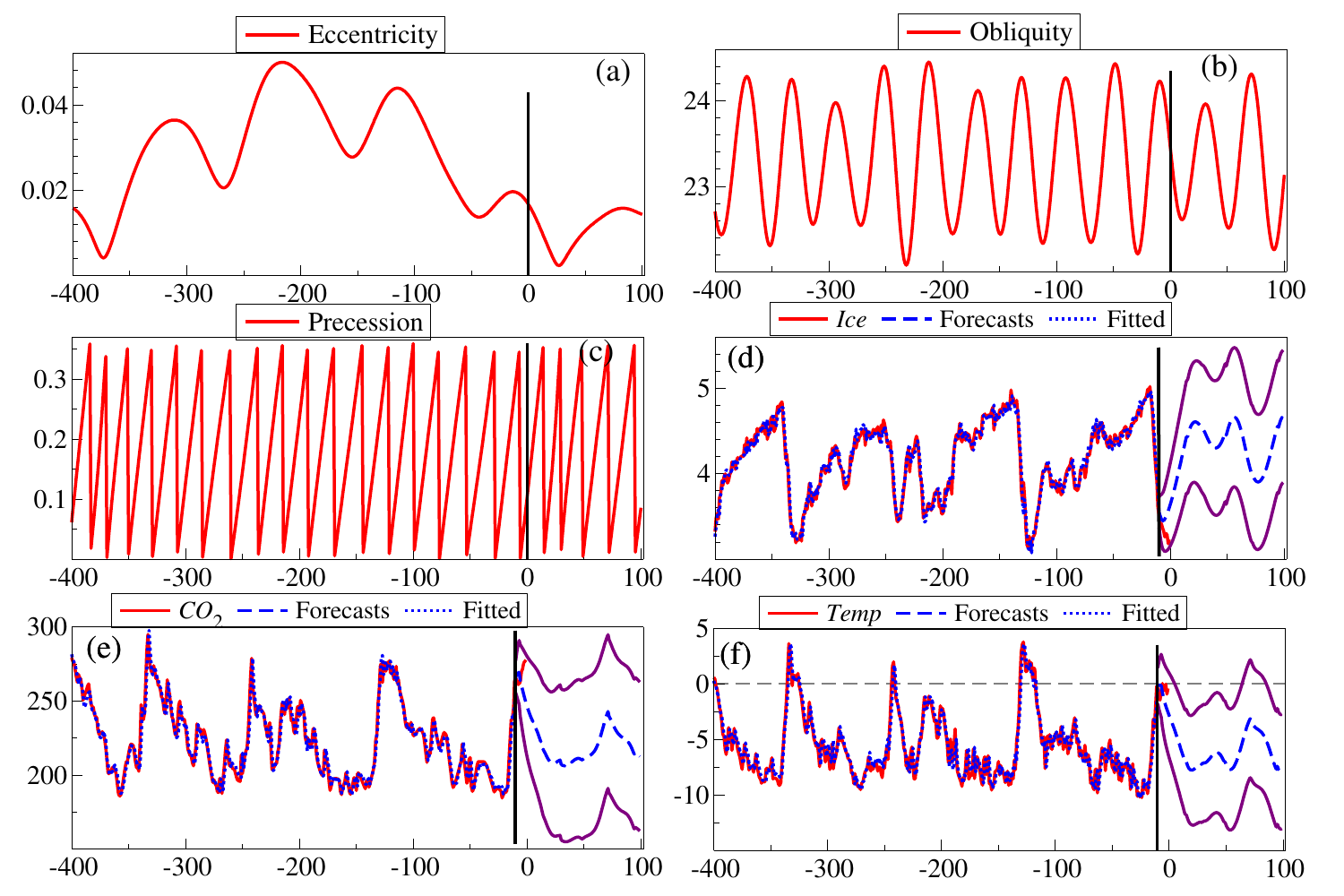} \vspace{-.2in}
\caption{Ice-Age data, model fits, and forecasts with endogenous CO\textsubscript{2}.}
\label{fig:hendry_IJFIceAge1}
\end{center}
\end{figure}

We formulated a 3-equation simultaneous model of atmospheric CO$_2$ and Antarctic Temperature and Ice volume over 800,000 years of Ice Ages in 1000-year frequency \citep{DFH_Paillard01,DFH_KaufJuse2013}. Driven by non-linear functions of eccentricity, obliquity, and precession (see panels (a), (b), and (c) of figure \ref{fig:hendry_IJFIceAge1} respectively), the model was selected with saturation estimation. 
Earth's orbital path is calculable into the future (\citealp{DFH_CrollJames} and
\citealp{DFH_Milankovitch41}), allowing 100,000 years of multi-step forecasts at endogenous emissions. Humanity has affected climate since 10 thousand years ago (kya: \citealp{DFH_Ruddiman05}), so we commence forecasts there. Forecasts over $-10$ to 100 with time series from 400kya in panels (d) to (f) of figure \ref{fig:hendry_IJFIceAge1} show paths within the ranges of past data $\pm$2.2SE \citep{DFH_PretisKauf18}.

\begin{figure}[ht!]
\begin{center}
\includegraphics[width=14.5cm]{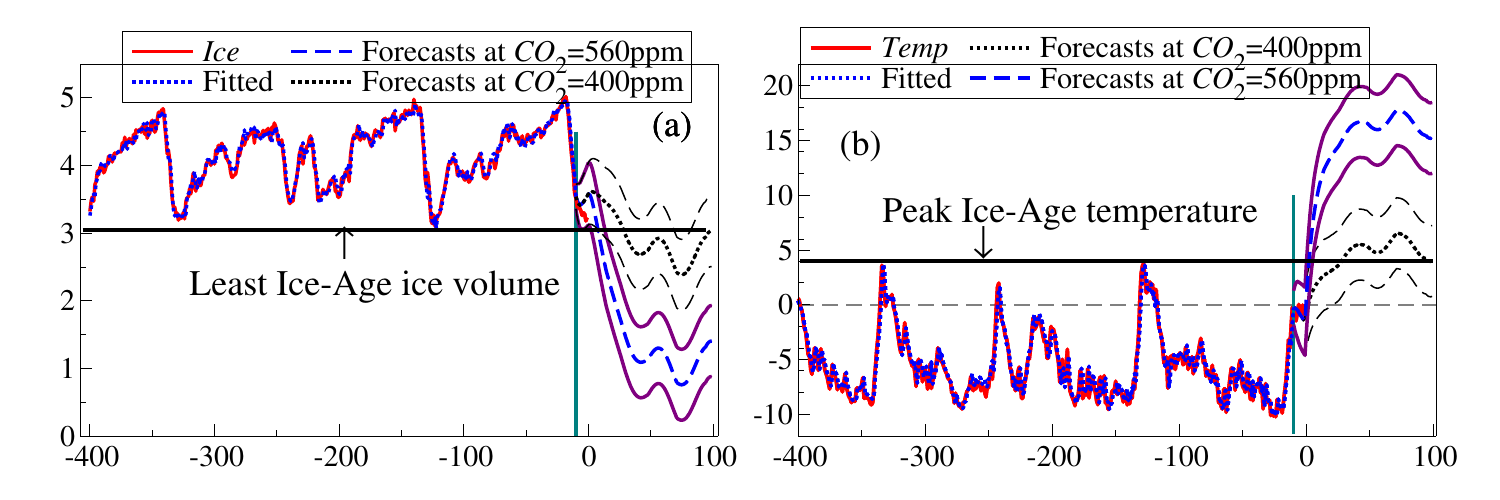} \vspace{-.2in}
\caption{Ice-Age simulations with exogenous CO\textsubscript{2}.}
\label{fig:hendry_IcAgeFigure2}
\end{center}
\end{figure}

Atmospheric CO$_2$ already exceeds 400ppm (parts per million), dramatically outside the
Ice-Age range \citep{DFH_SundKeel09}. Consequently, we conditionally forecast the next 100,000 years, simulating the potential climate for anthropogenic CO$_2$ \citep{DFH_CHIceAgeCO2} noting the `greenhouse' temperature is proportional to the logarithm of CO$_2$ (\citealp{DFH_Arrhen96}). The orbital drivers will continue to influence all three variables but that relation is switched off in the scenario for `exogenised' CO$_2$. The 110 dynamic forecasts conditional on 400ppm and 560ppm with $\pm$2SE bands are shown in figure \ref{fig:hendry_IcAgeFigure2}, panels (a) and (b) for Ice and Temperature respectively. The resulting global temperature rises inferred from these Antarctic temperatures would be dangerous, at more than 5$^{\circ}$C, with Antarctic temperatures positive for thousands of years \citep{DFH_VaksHender19,DFH_PretisKauf20}.

\subsubsection[Weather forecasting (Thordis Thorarinsdottir)]{Weather forecasting\protect\footnote{This subsection was written by Thordis Thorarinsdottir.}}
\label{sec:Weather_Forecasting}
The weather has a huge impact on our lives, affecting health, transport, agriculture (see also \S\ref{sec:The_Forecastability_of_agricultural_time_series}), energy use (see also \S\ref{sec:Energy}), and leisure. Since \citet{Bjerknes1904_TT} introduced hydrodynamics and thermodynamics into meteorology, weather prediction has been based on merging physical principles and observational information. Modern weather forecasting is based on numerical weather prediction (NWP) models that rely on accurate estimates of the current state of the climate system, including ocean, atmosphere and land surface. Uncertainty in these estimates is propagated through the NWP model by running the model for an ensemble of perturbed initial states, creating a weather forecast ensemble \citep{Buizza2018_TT, TothBuizza2019_TT}.

One principal concern in NWP modelling is that small-scale phenomena such as clouds and convective precipitation are on too small a scale to be represented directly in the models and must, instead, be represented by approximations known as parameterisations. Current NWP model development aims at improving both the grid resolution and the observational information that enters the models \citep{Bannister&2020_TT, Leuenberger&2020_TT}. However, for fixed computational resources, there is a trade-off between grid resolution and ensemble size, with a larger ensemble generally providing a better estimate of the prediction uncertainty. Recent advances furthermore include machine learning approaches (see \S\ref{sec:machine_Learning}) to directly model the small-scale processes, in particular cloud processes \citep[see, for example,][]{Gentine&2018_TT, Rasp&2018_TT}. 

Despite rapid progress in NWP modelling, the raw ensemble forecasts exhibit systematic errors in both magnitude and spread \citep{Buizza2018_TT}. Statistical post-processing is thus routinely used to correct systematic errors in calibration and accuracy before a weather forecast is issued; see \citet{Vannitsem&2018_TT} for a recent review but also \S\ref{sec:evaluating_probabilistic_forecasts} and \S\ref{sec:assessing_the_reliability_of_probabilistic forecats}. A fundamental challenge here is to preserve physical consistency across space, time and variables \citep[see, for example,][]{Moeller&2013_TT, Schefzik&2013_TT, Heinrich&2020_TT}. This is particularly important when the weather forecast is used as input for further prediction modelling, e.g., in hydrometeorology \citep{Hemri&2015_TT, Hemri2018_TT}. 

At time scales beyond two weeks, the weather noise that arises from the growth of the initial uncertainty, becomes large \citep{Royer1993_TT}. Sources of long-range predictability are usually associated with the existence of slowly evolving components of the earth system, including the El Ni\~no Southern Oscillation (ENSO), monsoon rains, the Madden Julian Oscillation (MJO), the Indian Ocean dipole, and the North Atlantic Oscillation (NAO), spanning a wide range of time scales from months to decades \citep{Vitart&2012_TT, Hoskins2013_TT}. It is expected that, if a forecasting system is capable of reproducing these slowly evolving components, they may also be able to forecast them \citep{vanSchaeybroeckVannitsem2018_TT}. The next step is then to find relationships between modes of low-frequency variability and the information needed by forecast users such as predictions of surface temperature and precipitation \citep{RoulinVannitsem2019_TT, Smith&2020_TT}. 

\subsubsection[Air quality forecasting (Claudio Carnevale)]{Air quality forecasting\protect\footnote{This subsection was written by Claudio Carnevale.}}
\label{sec:Air_quality_forecasting}
To preserve human health, European Commission stated in the Directive (2008/50/EC) that member states have to promptly inform the population when the particulate matter (PM) daily mean value exceeds (or is expected to exceed) the threshold of $50 \mu g/m^3$. Therefore, systems have been designed in order to produce forecasts for up to three days in advance using as input the measured value of concentration and meteorological conditions. These systems can be classified in (\textit{i}) data-driven models \citep[][and \S\ref{sec:data_driven_approaches}]{CARNEVALE2016_CC,STADLOBER2018_CC,CORANI2005_CC}, and (\textit{ii}) deterministic chemical and transport models \citep{HONORE2007_CC,MANDERS2009_CC}. In this section, a brief overview of the application of these systems to the high polluted area of Lombardy region, in Italy, will be presented.

\cite{CARNEVALE2018_CC} compared the results of three different forecasting systems based on neural networks, lazy learning models, and regression trees respectively. A single model has been identified for each monitoring station. In the initial configuration, only the last three PM measurements available were used to produce the forecast. In this configuration, the systems offered reasonable
performance, with correlation coefficients ranging from 0.6 (lazy learning method) to 0.75 (neural network). The work also demonstrated that the performance of the ensemble of the three systems was better than the best model for each monitoring station (see also \S\ref{sec:combining_forecasts} for further discussions on forecast combinations).

Starting from the results of this work, a second configuration was implemented, using as input also the wind speed measured in the meteorological monitoring station closest to the measurement point of PM. The researchers observed an improvement in all performance indices, with the median of the correlation for the best model (neural networks) increasing from 0.75 to 0.82 and the RMSE dropping from $15 \mu g/m^3$ to $7 \mu g/m^3$.

One of the main drawbacks of data-driven models for air quality is that they provide information only in the point where the measurements are available. To overcome this limitation, recent literature has presented mixed deterministic and data-driven approaches \cite[see, for example,][]{CARNEVALE2020_CC} which use the data assimilation procedure and offer promising forecasting performance. 

From a practical point of view, critical issues regarding forecasting air quality include:
\begin{itemize}[noitemsep]
  \item Information collection and data access: even if regional authorities have to publicly provide data and information related to air quality and meteorology, the measured data are not usually available in real-time and the interfaces are sometimes not automated;
  \item Data quantity: the amount of information required by air quality forecasting systems is usually large, in particular towards the definition of the training and validation sets;
  \item Non-linear relationships: the phenomenon of accumulation of pollutants in atmosphere is usually affected by strong nonlinearities, which significantly impact the selection of the models and their performance;
  \item Unknown factors: it is a matter of fact that the dynamic of pollutants in atmosphere is affected by a large number of non-measurable variables (such as meteorological variables or the interaction with other non-measurable pollutants), largely affecting the capability of the models to reproduce the state of the atmosphere.
\end{itemize}

\subsubsection[Forecasting and decision making for floods and water resources management (Ezio Todini)]{Forecasting and decision making for floods and water resources management\protect\footnote{This subsection was written by Ezio Todini.}}
\label{sec:Forecasting_and_decision_making_for_floods}
In Water Resources and Flood Risk Management, decision makers are frequently confronted with the need of taking the most appropriate decisions not knowing what will occur in the future. To support their decision-making under uncertainty, decision theory \citep{Berger1985-sm_ET,Bernardo1994-is_ET,DeGroot2004-xh_ET} invokes Bayesian informed decision approaches, which find the most appropriate decision by maximising (or minimising) the expected value of a ``utility function'', thus requiring its definition, together with the estimation of a ``predictive probability'' density \citep{Berger1985-sm_ET} due to the fact that utility functions are rarely linear or continuous. Consequently, their expected value does not coincide with the value assumed on the predicted ``deterministic'' expected value. Accordingly, overcoming the classical 18\textsuperscript{th} century ``mechanistic'' view by resorting into probabilistic forecasting approaches becomes essential (see also \S\ref{sec:Density_forecast_combinations}).

The failure of decision-making based on deterministic forecasts in the case of Flood Risk Management is easily shown through a simple example. At a river section, the future water level provided by a forecast is uncertain and can be described by a Normal distribution with mean 10 meters and standard deviation of 5 meters. Given a dike elevation of 10.5 meters, damages may be expected as zero if water level falls below the dike elevation and linearly growing when level exceeds it with a factor of $10^6$ dollars. If one assumes the expect value of forecast as the deterministic prediction to compute the damage the latter will result equal to zero, while if one correctly integrates the damage function times the predictive density the estimated expected damage will results into 6.59 millions of dollars and educated decisions on alerting or not the population or evacuating or not a flood-prone area can be appropriately taken (see also \S\ref{sec:Social_good}).

Water resources management, and in particular reservoirs management, aim at deriving appropriate operating rules via long term expected benefits maximisation. Nonetheless, during flood events decision makers must decide how much to preventively release from multi-purpose reservoirs in order to reduce dam failure and downstream flooding risks the optimal choice descending from trading-off between loosing future water resource vs the reduction of short term expected losses. 

This is obtained by setting up an objective function based on the linear combination of long and short term ``expected losses'', once again based on the available probabilistic forecast. 
This Bayesian adaptive reservoir management approach incorporating into the decision mechanism the forecasting information described by the short-term predictive probability density, was implemented on the lake Como since 1997 \citep{Todini1999-xc_ET,Todini2017-jm_ET} as an extension of an earlier original idea \citep{Todini1991-vu_ET}. This resulted into:
\begin{itemize}[noitemsep]
 \item a reduction of over 30\% of of the city of Como frequency;
 \item an average reduction of 12\% of the water deficit;
 \item an increase of 3\% in the electricity production.
\end{itemize}

Lake Como example clearly shows that instead of basing decisions on the deterministic prediction, the use of a Bayesian decision scheme, in which model forecasts describe the predictive probability density, increases the reliability of the management scheme by essentially reducing the probability of wrong decisions \citep{Todini2017-jm_ET,Todini2018-dj_ET}.

\subsection{Social good and demographic forecasting}
\label{sec:Social_good}

\subsubsection[Healthcare (Bahman Rostami-Tabar)]{Healthcare\protect\footnote{This section was written by Bahman Rostami-Tabar.}}
\label{sec:Heathcare}
There are many decisions that depend on the quality of forecasts in the health care system, from capacity planning to layout decisions to the daily schedules. In general, the role of forecasting in health care is to inform both clinical and non-clinical decisions. While the former concerns decisions related to patients and their treatments \citep{makridakis2019forecasting_BRT}, the latter involves policy/management, and supply chain decisions that support the delivery of high-quality care for patients. 

A number of studies refer to the use of forecasting methods to inform clinical decision making. These methods are used to screen high risk patients for preventative health care \citep{chen2015predicting_BRT,van2014predicting_BRT,santos2015new_BRT,uematsu2014development_BRT}, to predict mental health issues \citep{shen2017depression_BRT,tran2013integrated_BRT}, to assist diagnosis and disease progression \citep{ghassemi2015multivariate_BRT,ma2017dipole_BRT,pierce2010chest_BRT,qiao2019mnn_BRT}, to determine prognosis \citep{dietzel2010application_BRT,ng2007comparison_BRT}, and to recommend treatments for patients \citep{kedia2003predictors_BRT,scerri2006intermed_BRT,shang2019pre_BRT}. Common forecasting methods to inform clinical decisions include time series (see \S\ref{sec:exponential_smoothing_models}, \S\ref{sec:autoregressive_integrated_moving_average_models}, and \S\ref{sec:forecasting_for_multiple_seasonal_cycles}), regression (see \S\ref{sec:time_series_regression_models}), classification tree (see \S\ref{sec:clustering_based_forecasting}), neural networks (see \S\ref{sec:neural_networks}), Markov models (see \S\ref{sec:Markov_switching_models}) and Bayesian networks. These models utilise structured and unstructured data including clinician notes \citep{austin2016application_BRT,labarere2014derive_BRT} which makes the data pre-processing a crucial part of the forecasting process in clinical health care.

One of the aspects of the non-clinical forecasting that has received the most attention in both research and application is the policy and management. Demand forecasting is regularly used in Emergency Departments \citep{arora2020probabilistic_BRT,choudhury2020forecasting_BRT,khaldi2019forecasting_BRT,rostami2020anticipating_BRT}, ambulance services \citep{al2020empirical_BRT,setzler2009ems_BRT,Vile2011_BRT,zhou2016predicting_BRT} and hospitals with several different specialities \citep{mccoy2018assessment_BRT,ordu2019comprehensive_BRT,zhou2018time_BRT} to inform operational, tactical and strategic planning. The common methods used for this purpose include classical ARIMA and exponential smoothing methods, regression, singular spectrum analysis, Prophet, Double-Seasonal Holt-Winter, TBATS and Neural Networks. In public health, forecasting can guide policy and planning. Although it has a wider definition, the most attention is given to Epidemic forecasting (see also \S\ref{sec:Pandemics}). 

Forecasting is also used in both national and global health care supply chains, not only to ensure the availability of medical products for the population but also to avoid excessive inventory. Additionally, the lack of accurate demand forecast in a health supply chain may cost lives \citep{baicker2012saving_BRT} and has exacerbated risks for suppliers \citep{levine2008demand_BRT}. Classical exponential smoothing, ARIMA, regression and Neural Network models have been applied to estimate the drug utilisation and expenditures \citep{dolgin2010better_BRT,linner2020forecasting_BRT}, blood demand \citep{fortsch2016reducing_BRT}, hospital supplies \citep{gebicki2014evaluation_BRT,riahi2013three_BRT} and demand for global medical items \citep{amarasinghe2010forecasting_BRT,hecht2008demand_BRT,van2016demand_BRT}. It is important to note that, while the demand in a health care supply chain has often grouped and hierarchical structures \citep[][see also \S\ref{sec:Cross_sectional_hierarchical_forecasting}]{Mircetica2020_BRT}, this has not been well investigated and needs more attention.

\subsubsection[Epidemics and pandemics (Konstantinos Nikolopoulos \& Thiyanga S. Talagala)]{Epidemics and pandemics\protect\footnote{This subsection was written by Konstantinos Nikolopoulos \& Thiyanga S. Talagala.}}
\label{sec:Pandemics}
Pandemics and epidemics both refer to disease outbreaks. An epidemic is a disease outbreak that spreads across a particular region. A pandemic is defined as spread of a disease worldwide. Forecasting the evolution of a pandemic or an epidemic, the growth of cases and fatalities for various horizons and levels of granularity, is a complex task with raw and limited data -- as each disease outbreak type has unique features with several factors affecting the severity and the contagiousness. Be that as it may, forecasting becomes an paramount task for the countries to prepare and plan their response \citep{Nikolopoulos2020_YK}, both in healthcare and the supply chains \citep[][see also \S\ref{sec:Heathcare} and \S\ref{sec:Forecasting_in_the_supply_chain}]{Belien2012-sp_KN}.

Successful forecasting methods for the task include time-series methods (see \S\ref{sec:Statistical_and_econometric_models}), epidemiological and agent-based models (see \S\ref{sec:Agent_based_models}), metapopulation models, approaches in metrology \citep{Nsoesie2013-sj_KN}, machine and deep learning methods \citep{Yang2020-ll_KN}. \cite{Andersson2008-yc_KN} used regression models for the prediction of the peak time and volume of cases for a pandemic with evidence from seven outbreaks in Sweden. \cite{Yaffee2011-pi_KN} forecasted the evolution of the Hantavirus epidemic in USA and compared causal and machine-learning methods with time-series methods and found that univariate methods quite successful. \cite{Soebiyanto2010-rh_KN} used ARIMA models for successfully short-term forecasting of influenza weekly cases. \cite{shaman2012forecasting_BRT} used Kalman filter based SIR epidemiological models to forecast the peak time of influenza 6-7 weeks ahead.

For COVID-19, \cite{Petropoulos2020-qm_KN} applied a multiplicative exponential smoothing model (see also \S\ref{sec:exponential_smoothing_models}) for predicting global number of confirmed cases, with very successful results both for point forecasts and prediction intervals. This article got serious traction with 100,000 views and 300 citations in the first twelve months since its publication, thus evidencing the importance of such empirical investigations. There has been a series of studies focusing on predicting deaths in the USA and European countries for the first wave of the COVID-19 pandemic \citep{IHME_COVID-19-yz_KN,IHME_COVID-19-nx_KN}. Furthermore, \cite{Petropoulos2020covidIJF_KN} expanded their investigation to capture the continuation of both cases and deaths as well as their uncertainty, achieving high levels of forecasting accuracy for ten-days-ahead forecasts over a period of four months. Along the same lines, \cite{Doornik2020-es_KN} have been publishing real-time accurate forecasts of confirmed cases and deaths from mid-March 2020 onwards. Their approach is based on extraction of trends from the data using machine learning.

\cite{Pinson2020-va_KN} organised a debate between Taleb and Ioannidis on forecasting pandemics. \cite{Ioannidis2020-wk_KN} claim that forecasting for COVID-19 has by and large failed. However they give recommendations of how this can be averted. They suggest that the focus should be on predictive distributions and models should be continuously evaluated. Moreover, they emphasise the importance of multiple dimensions of the problem (and its impact). \cite{Taleb2020-ki_PC} discuss the dangers of using naive, empirical approaches for fat-tailed variables and tail risk management. They also reiterate the inefficiency of point forecasts for such phenomena.

Finally, \cite{Nikolopoulos2020-ua_KN} focused on forecast-driven planning, predicting the growth of COVID-19 cases and the respective disruptions across the supply chain at country level with data from the USA, India, UK, Germany, and Singapore. Their findings confirmed the excess demand for groceries and electronics, and reduced demand for automotive – but the model also proved that the earlier a lock-down is imposed, the higher the excess demand will be for groceries. Therefore, governments would need to secure high volumes of key products before imposing lock-downs; and, when this is not possible, seriously consider more radical interventions such as rationing.

Dengue is one of the most common epidemic diseases in tropical and sub-tropical regions of the world. Estimates of World Health Organisation reveals that about half of the world's population is now at risk for Dengue infection \citep{romero2019applying_Thiyanga}. \emph{Aedes aegypti}
and \emph{Aedes albopictus} are the principal vectors of dengue transmission and they are highly domesticated mosquitoes. Rainfall, temperature and relative humidity are thought of as important factors attributing towards the growth and dispersion of mosquito vectors and potential of dengue outbreaks \citep{banu2011dengue_Thiyanga}.

In reviewing the existing literature, two data types have been used to forecast dengue incidence: (\textit{i}) spatio-temporal data: incidence of laboratory-confirmed dengue cases among the clinically suspected patients \citep{naish2014climate_Thiyanga}, (\textit{ii}) web-based data: Google trends, tweets associated with Dengue cases \citep{de2017dengue_Thiyanga}.

SARIMA models (see also \S\ref{sec:autoregressive_integrated_moving_average_models}) have been quite popular in forecasting laboratory-confirmed dengue cases \citep{martinez2011predicting_Thiyanga,gharbi2011time_Thiyanga,promprou2006forecasting_Thiyanga}. \cite{chakraborty2019forecasting_Thiyanga} used a hybrid
model combining ARIMA and neural network autoregressive (NNAR) to forecast dengue cases. In light of biological relationships between climate and transmission of \emph{Aedes} mosquitoes, several studies have used additional covariates such as, rainfall, temperature, wind speed, and humidity to forecasts dengue incidence \citep{banu2011dengue_Thiyanga,naish2014climate_Thiyanga,talagala2015distributed_Thiyanga}. Poisson regression model has been widely used to forecast dengue incidence using climatic factors and lagged time between dengue incidence and weather variables \citep{hii2012forecast_Thiyanga,koh2018model_Thiyanga}. Several researchers looked at the use of Quasi‐Poisson and negative binomial regression models to accommodate over dispersion in the counts \citep{lowe2011spatio_Thiyanga,wang2014study_Thiyanga}. \cite{cazelles2005nonstationary_Thiyanga} used wavelet analysis to explore the dynamic of dengue incidence and wavelet coherence analyses was used to identify time and frequency specific association with climatic variables. \cite{de2017dengue_Thiyanga} took a different perspective and look at weekly tweets to forecast Dengue cases. \cite{rangarajan2019forecasting_Thiyanga} used Google trend data to forecast Dengue cases. Authors hypothesised that web query search related to dengue disease correlated with the current level of dengue cases and thus may be helpful in forecasting dengue cases.

A direction for future research in this field is to explore the use of spatio-temporal hierarchical forecasting (see \S\ref{sec:forecasting_by_aggregation}).
 
\subsubsection[Forecasting mortality (Clara Cordeiro \& Han Lin Shang)]{Forecasting mortality \protect\footnote{This subsection was written by Clara Cordeiro \& Han Lin Shang.}}
\label{sec:Forecasting_mortality_data}
Actuarial, Demographic, and Health studies are some examples where mortality data are commonly used. A valuable source of mortality information is the Human Mortality Database (HMD), a database that provides mortality and population data for 41 mainly developed countries. Additionally, at least five country-specific databases are devoted to subnational data series: Australian, Canadian, and French Human Mortality Databases, United States and Japan Mortality Databases. In some situations, the lack of reliable mortality data can be a problem, especially in developing countries, due to delays in registering or miscounting deaths \citep{checchi2005interpreting_CC}. Analysis of National Causes of Death for Action (ANACONDA) is a valuable tool that assesses the accuracy and completeness of data for mortality and cause of death by checking for potential errors and inconsistencies \citep{mikkelsen2020anaconda_CC}.

The analysis of mortality data is fundamental to public health authorities and policymakers to make decisions or evaluate the effectiveness of prevention and response strategies.
When facing a new pandemic, mortality surveillance is essential for monitoring the overall impact on public health in terms of disease severity and mortality \citep{setel2020mortality_CC,vestergaard2020excess_CC}. A useful metric is excess mortality and is the difference between the observed number of deaths and the expected number of deaths under ``normal'' conditions \citep{checchi2005interpreting_CC, aron2020measuring_CC}. Thus, it can only be estimated with accurate and high-quality data from previous years. Excess mortality has been used to measure the impact of heat events \citep{matte2016excess_CC,limaye2018climate_CC}, pandemic influenza \citep{nunes2011excess_CC,nielsen2013pooling_CC}, and nowadays COVID-19 \citep[][and \S\ref{sec:Pandemics}]{nogueira2020excess_CC,sinnathamby2020all_CC,owidcoronavirus_CC, SX21_HLS}, among others. Excess mortality data have been making available by the media publications \textit{The Economist}, \textit{The New York Times} and \textit{The Financial Times}. Moreover, a monitoring system of the weekly excess mortality in Europe has been performed by the EuroMOMO project \citep{vestergaard2020excess_CC}.

An essential use of mortality data for those individuals at age over 60 is in the pension and insurance industries, whose profitability and solvency crucially rely on accurate mortality forecasts to adequately hedge longevity risks \citep[see, e.g.,][]{SH20, SH20b}. Longevity risk is a potential systematic risk attached to the increasing life expectancy of annuitants, and it is an important factor to be considered when determining a sustainable government pension age \citep[see, e.g.,][for Australia]{HZS21_HLS}. The price of a fixed-term or lifelong annuity is a random variable, as it depends on the value of zero-coupon bond price and mortality forecasts. The zero-coupon bond price is a function of interest rate (see \S\ref{sec:Interest_rate_prediction}) and is comparably more stable than the retirees' mortality forecasts. 

Several methodologies were developed for mortality modelling and forecasting \citep{booth2008mortality_CC, janssen2018advances_CC}. These methods can be grouped into three categories: expectation, explanation, and extrapolation \citep{booth2008mortality_CC}. 

The expectation approach is based on the subjective opinion of experts (see also \S\ref{sec:panels_of_experts}), who set a long-run mortality target. Methods based on expectation make use of experts' opinions concerning future mortality or life expectancy with a specified path and speed of progression towards the assumed value \citep{CMI20}. The advantage of this approach is that demographic, epidemiological, medical, and other relevant information may be incorporated into the forecasts. The disadvantages are that such information is subjective and biased towards experts' opinions, and it only produces scenario-based (see \S\ref{sec:scenarios_forecasting}) deterministic forecasts \citep{AV90, WH04}.

The explanation approach captures the correlation between mortality and the underlying cause of death. Methods based on the explanation approach incorporate medical, social, environmental, and behavioural factors into mortality modelling. Example include smoking and disease-related mortality models. The benefit of this approach is that mortality change can be understood from changes in related explanatory variables; thus, it is attractive in terms of interpretability \citep{GV98}.

The extrapolative approach is considered more objective, easy to use and more likely to obtain better forecast accuracy than the other two approaches \citep{janssen2018advances_CC}. The extrapolation approach identifies age patterns and trends in time which can be then forecasted via univariate and multivariate time series models (see \S\ref{sec:Statistical_and_econometric_models}). In the extrapolation approach, many parametric and nonparametric methods have been proposed \citep[see, e.g.,][]{AS05, HU07, SBH11}. Among the parametric methods, the method of \cite{HP80} is well-known. Among the nonparametric methods, the Lee-Carter model \citep{lee1992modeling_CC}, Cairns-Blake-Dowd model \citep{CDB+09, DCB+10}, and functional data model \citep[][and \S\ref{sec:Functional_time_series_models}]{HU07}, as well as their extensions and generalisations are dominant. The time-series extrapolation approach has the advantage of obtaining a forecast probability distribution rather than a deterministic point forecast and, also, enable the determination of forecast intervals \citep{booth2008mortality_CC}.

\cite{janssen2018advances_CC} presents a review of the advances in mortality forecasting and possible future research challenges.
 
\subsubsection[Forecasting fertility (Joanne Ellison)]{Forecasting fertility\protect\footnote{This subsection was written by Joanne Ellison.}}
\label{sec:Forecasting_fertility}
Aside from being a driver of population forecasts (see \S\ref{sec:Models_for_population_processes}), fertility forecasts are vital for planning maternity services and anticipating demand for school places. The key challenge relates to the existence of, and interaction between, the quantum (how many?) and tempo (when?) components \citep{Booth2006-zp_JE}. This intrinsic dependence on human decisions means that childbearing behaviour is influenced by numerous factors acting at different levels, from individual characteristics to societal change \citep{Balbo2013-jh_JE}. An important methodological challenge for many low- and middle-income countries is fertility estimation, due to deficiencies in vital statistics caused by inadequate birth registration systems \citep{Moultrie2013-ce_JE,AbouZahr2015-az_JE,Phillips2018-eo_JE}. Such countries are often also in the process of transitioning from high to low fertility, which induces greater forecast uncertainty compared to low-fertility countries \citep{United_Nations_Development_Programme2019-kb_JE}.

A range of statistical models have been proposed to forecast fertility -- see \cite{Booth2006-zp_JE}, \cite{Bohk-Ewald2018-vb_JE}, and \cite{Shang2020-jr_JE} for reviews. The Lee-Carter model \citep{lee1992modeling_CC} originally developed for mortality forecasting (see \S\ref{sec:Forecasting_mortality_data}) has been applied to fertility \citep{Lee1993-pg_JE}, with extensions in functional data \citep{HU07} and Bayesian \citep{Wisniowski2015-bp_JE} contexts. Other notable extrapolative methods include the cohort-ARIMA model of \cite{De_Beer1985-ia_JE,De_Beer1990-rz_JE} -- see \S\ref{sec:autoregressive_integrated_moving_average_models} -- and the linear extrapolation method of \cite{Myrskyla2013-ds_JE}. Many parametric models have been specified to describe the shapes of fertility curves \citep{Brass1974-zf_JE,Hoem1981-fk_JE,Evans1986-hf_JE,Schmertmann2003-tl_JE}, with forecasts obtained through time series extrapolations of the parameters \citep{Congdon1990-fy_JE,Knudsen1993-eh_JE,De_Iaco2016-uw_JE}. Bayesian methods have been used to borrow strength across countries \citep[for example,][]{Alkema2011-zd_JE,Schmertmann2014-fk_JE}, with \cite{Ellison2020-hb_JE} developing a hierarchical model in the spirit of the latter. The top-down approach (see \S\ref{sec:Cross_sectional_hierarchical_forecasting}) of the former, which is used by the United Nations, projects the aggregate Total Fertility Rate (TFR) measure probabilistically \citep[also see][]{Tuljapurkar1999-ij_JE} before decomposing it by age. \cite{Hyppola1949-uy_JE} provide one of the earliest known examples of probabilistic fertility forecasting \citep{AS05}.

Little work has been done to compare forecast performance across this broad spectrum of approaches. The study of \cite{Bohk-Ewald2018-vb_JE} is the most comprehensive to date. Most striking is their finding that few methods can better the naive freezing of age-specific rates, and those that can differ greatly in method complexity (see also \S\ref{sec:Model_complexity}). A recent survey of fertility forecasting practice in European statistical offices \citep{Gleditsch2020-nf_JE} found that forecasts tend to be deterministic and make use of expert panels (see \S\ref{sec:panels_of_experts}). Expert elicitation techniques are gaining in sophistication, highlighted by the protocol of Statistics Canada \citep{Dion2020-kq_JE} which requests a full probability distribution of the TFR.

A promising avenue is the development of forecasting methods that incorporate birth order (parity) information, supported by evidence from individual-level analyses \citep[for example,][]{Fiori2014-ax_JE}. Another underexplored area is the integration of survey data into fertility forecasting models, which tend to use vital statistics alone when they are of sufficient quality \citep[see][for Bayesian fertility estimation with imperfect census data]{Rendall2009-ci_JE,Zhang2019-vn_JE}. Alternative data sources also have great potential. For example, \cite{Wilde2020-sd_JE} use Google data to predict the effect of COVID-19 on US fertility in the absence of vital statistics. Lastly, investigation of the possible long-term impacts of delayed motherhood in high-income countries, alongside developments in assisted reproduction technology such as egg freezing, is required \citep[see, for example,][]{Sobotka2018-ii_JE}.

\subsubsection[Forecasting migration (Jakub Bijak)]{Forecasting migration\protect\footnote{This subsection was written by Jakub Bijak.}}
\label{sec:Forecasting_migration}
Migration forecasts are needed both as crucial input into population projections (see \S\ref{sec:Models_for_population_processes}), as well as standalone predictions, made for a range of users, chiefly in the areas of policy and planning. At the same time, migration processes are highly uncertain and complex, with many underlying and interacting drivers, which evade precise conceptualisation, definitions, measurement, and theoretical description \citep{Bijak2020-tr_JB}. Given the high level of the predictive uncertainty, and the non-stationary character of many migration processes \citep{Bijak2010-cn_JB}, the current state of the art of forward-looking migration studies reflects therefore a shift from prediction to the use of forecasts as contingency planning tools (idem).

Reviews of migration forecasting methods are available in \cite{Bijak2010-dd_JB} and \cite{Sohst2020-le_JB}. The applications in official statistics, with a few exceptions, are typically based on various forms scenario-based forecasting with judgment (see \S\ref{sec:scenarios_forecasting}), based on pre-set assumptions \citep[for an example, see][]{Abel2018-fb_JB}. Such methods are particularly used for longer time horizons, of a decade or more, so typically in the context of applications in population projections, although even for such long time horizons calibrated probabilistic methods have been used as well \citep{Azose2016-kt_JB}. 

The mainstream developments in migration forecasting methodology, however, include statistical and econometric methods discussed in \S\ref{sec:Statistical_and_econometric_models}, such as time series models, both uni- and multivariate \citep[for example,][]{Gorbey1999-hy_JB,Bijak2010-dd_JB,Bijak2019-fm_JB}, econometric models \citep[for example,][]{Brucker2006-ex_JB,Cappelen2015-uc_JB}, Bayesian hierarchical models \citep{Azose2015-nn_JB}, and dedicated methods, for example for forecasting data structured by age \citep{Raymer2018-ca_JB}. In some cases, the methods additionally involve selection and combining forecasts through Bayesian model selection and averaging \citep[][see also \S\ref{sec:Variable_and_model_selection} and \S\ref{sec:combining_forecasts}]{Bijak2010-dd_JB}. Such models can be expected to produce reasonable forecasts (and errors) for up to a decade ahead \citep{Bijak2010-cn_JB}, although this depends on the migration flows being forecast, with some processes (e.g., family migration) more predictable than other (e.g., asylum). Another recognised problem with models using covariates is that those can be endogenous to migration (e.g., population) and also need predicting, which necessitates applying structured models to prevent uncertainty from exploding. 

The methodological gaps and current work in migration forecasting concentrate in a few key areas, notably including causal (mechanistic) forecasting based on the process of migrant decision making \citep{Willekens2018-mg_JB}; as well as early warnings and `nowcasting' of rapidly changing trends, for example in asylum migration \citep{Napierala2021-pq_JB}. In the context of early warnings, forays into data-driven methods for changepoint detection, possibly coupled with the digital trace and other high-frequency `Big data', bear particular promise. At the same time, coherent uncertainty description across a range of time horizons, especially in the long range \citep{Azose2015-nn_JB}, remains a challenge, which needs addressing for the sake of proper calibration of errors in the population forecasts, to which these migration components contribute. 

\subsubsection[Forecasting risk for violence and wars (Pasquale Cirillo)]{Forecasting risk for violence and wars\protect\footnote{This subsection was written by Pasquale Cirillo.}}
\label{sec:Forecasting_risk_for_violence_and_wars}
Can we predict the occurrence of WW3 in the next 20 years? Is there any trend in the severity of wars? 

The study of armed conflicts and atrocities, both in terms of frequency over time and the number of casualties, has received quite some attention in the scientific literature and the media \citep[e.g.,][]{Cederman2003-vl_PC,Friedman2015-sb_PC,Hayes2002-iv_PC,Norton-Taylor2015-aq_PC,Richardson1948-pq_PC,Richardson1960-hg_PC}, falling within the broader discussion about violence \citep{Berlinski2009-vd_PC,Goldstein2011-il_PC,Spagat2009-xk_PC}, with the final goal of understanding whether humanity is becoming less belligerent \citep{Pinker2011-eb_PC}, or not \citep{Braumoeller2019-ip_PC}.

Regarding wars and atrocities, the public debate has focused its attention on the so-called \textit{Long Peace Theory} \citep{Gaddis1989-pd_PC}, according to which, after WW2, humanity has experienced the most peaceful period in history, with a decline in the number and in the severity of bloody events. Scholars like \cite{Mueller2009-gf_PC,Mueller2009-gs_PC} and \cite{Pinker2011-eb_PC,Pinker2018-gg_PC} claim that sociological arguments and all statistics suggest we live in better times, while others like \cite{Gray2015-ed_PC,Gray2015-px_PC} and \cite{Mann2018-my_PC} maintain that those statistics are often partial and misused, the derived theories weak, and that war and violence are not declining but only being transformed. For Mann, the Long Peace proves to be ad-hoc, as it only deals with Western Europe and North America, neglecting the rest of the world, and the fact that countries like the US have been involved in many conflicts out of their territories after WW2.

Recent statistical analyses confirm Gray's and Mann's views: empirical data do not support the idea of a decline in human belligerence (no clear trend appears), and in its severity. Armed conflicts show long inter-arrival times, therefore a relative peace of a few decades means nothing statistically \citep{Cirillo2016-nd_PC}. Moreover, the distribution of war casualties is extremely fat-tailed \citep{Clauset2018-wd_PC,Clauset2018-xh_PC}, often with a tail exponent $\xi=\frac{1}
{\alpha}>1$ \citep{Cirillo2016-nd_PC}, indicating a possibly infinite mean, i.e., a tremendously erratic and unforeseeable phenomenon (see \S\ref{sec:Forecasting_under_fat_tails}). An only apparently infinite-mean phenomenon though \citep{Cirillo2019-tv_PC}, because no single war can kill more than the entire world population, therefore a finite upper bound exists, and all moments are necessarily finite, even if difficult to estimate. Extreme value theory \citep{Embrechts2013-ne_PC} can thus be used to correctly model tail risk and make prudential forecasts \citep[with many caveats like in][]{Scharpf2014-aj_PC}, while avoiding naive extrapolations \citep{Taleb2020-ki_PC}.

As history teaches \citep{Nye1990-oh_PC}, humanity has already experienced periods of relative regional peace, like the famous Paces Romana and Sinica. The present Pax Americana is not enough to claim that we are structurally living in a more peaceful era. The Long Peace risks to be another apophenia, another example of Texan sharpshooter fallacy \citep{Carroll2011-em_PC}.

Similar mistakes have been made in the past. \cite{buckle_2011_PC} wrote: ``that [war] is, in the progress of society, steadily declining, must be evident, even to the most hasty reader of European history. If we compare one country with another, we shall find that for a very long period wars have been becoming less frequent; and now so clearly is the movement marked, that, until the late commencement of hostilities, we had remained at peace for nearly forty years: a circumstance unparalleled [...] in the affairs of the world''. Sadly, Buckle was victim of the illusion coming from the Pax Britannica \citep{Johnston2008-qv_PC}: the century following his prose turned out to be the most murderous in human history.
 
\subsection{Systems and humans}
\label{sec:Systems_and_humans}

\subsubsection[Support systems (Vassilios Assimakopoulos)]{Support systems\protect\footnote{This subsection was written by Vassilios Assimakopoulos.}}
\label{sec:Suport_systems}
Forecasting in businesses is a complicated procedure, especially when predicting numerous, diverse series (see \S\ref{sec:Feature_based_time_series_forecasting}), dealing with unstructured data of multiple sources (see \S\ref{sec:Forecasting_with_Big_Data}), and incorporating human judgment \citep[][but also \S\ref{sec:forecasting_with_judgment}]{LIM1996339}. In this respect, since the early 80's, various Forecasting Support Systems (FSSs) have been developed to facilitate forecasting and support decision making \citep{Kusters2006599}. \cite{RYCROFT1993531} provides an early comparative review of such systems, while many studies strongly support their utilisation over other forecasting alternatives \citep{TASHMAN1991209, NadaSander2003}. 

In a typical use-case scenario, the FSSs will retrieve the data required for producing the forecasts, will provide some visualisations and summary statistics to the user, allow for data pre-processing, and then produce forecasts that may be adjusted according to the preferences of the user. However, according to \cite{FildesOrd2013}, effective FSS should be able to produce forecasts by combining relevant information, analytical models, judgment, visualisations, and feedback. To that end, FSSs must (\textit{i}) elaborate accurate, efficient, and automatic statistical forecasting methods, (\textit{ii}) enable users to effectively incorporate their judgment, (\textit{iii}) allow the users to track and interact with the whole forecasting procedure, and (\textit{iv}) be easily customised based on the context of the company.

Indeed, nowadays, most off-the-self solutions, such as SAP, SAS, JDEdwards, and ForecastPro, offer a variety of both standard and advanced statistical forecasting methods (see \S\ref{sec:Statistical_and_econometric_models}), as well as data pre-processing (see \S\ref{sec:Preprocessing_time_series_data}) and performance evaluation algorithms (see \S\ref{sec:evaluation_and_validation}). On the other hand, many of them still struggle to incorporate state-of-the-art methods that can further improve forecasting accuracy, such as automatic model selection algorithms and temporal aggregation (see also \S\ref{sec:temporal_aggregation}), thus limiting the options of the users \citep{PetropoulosFSS}. Similarly, although many FSSs support judgmental forecasts (see \S\ref{sec:judgmental_forecasting}) and judgmental adjustments of statistical forecasts (see \S\ref{sec:judgmental_adjustments_of_computer_based_forecasts}), this is not done as suggested by the literature, i.e., in a guided way under a well-organised framework. As a result, the capabilities of the users are restrained and methods that could be used to mitigate biases, overshooting, anchoring, and unreasonable or insignificant changes that do not rationalise the time wasted, are largely ignored \citep{Fildes2013290, Fildes2006-km}. 

Other practical issues of FSSs are related with their engine and interfaces which are typically designed so that they are generic and capable to serve different companies and organisations of diverse needs \citep{Kusters2006599}. From a developing and economic perspective, this is a reasonable choice. However, the lack of flexibility and customisability can lead to interfaces with needless options, models, tools, and features that may confuse inexperienced users and undermine their performance \citep{Fildes2006-km}. Thus, simple, yet exhaustive interfaces should be designed in the future to better serve the needs of each company and fit its particular requirements \citep{SpiliotisForward}. Ideally, the interfaces should be adapted to the strengths and weaknesses of the user, providing useful feedback when possible \citep{Goodwin2007391}. Finally, web-based FSSs could replace windows-based ones that are locally installed and therefore of limited accessibility, availability, and compatibility \citep{Asimakopoulos2013322}. Cloud computing and web-services could be exploited in that direction.
 
\subsubsection[Cloud resource capacity forecasting (Tim Januschowski)]{Cloud resource capacity forecasting\protect\footnote{This subsection was written by Tim Januschowski.}}
\label{sec:Cloud_resource_capacity_forecasting}
One of the central promises in cloud computing is that of elasticity. Customers of cloud computing services can add compute resources in real-time to meet and satisfy increasing demand and, when demand for a cloud-hosted application goes down, it is possible for cloud computing customers to down-scale. The benefit of the latter is particularly economically interesting during the current pandemic. Popular recent cloud computing offerings take this elasticity concept one step further. They abstract away the computational resources completely from developers, so that developers can build serverless applications. In order for this to work, the cloud provider handles the addition and removal of compute resources ``behind the scenes''. 

To keep the promise of elasticity, a cloud provider must address a number of forecasting problems at varying scales along the operational, tactical and strategic problem dimensions \citep{janusch18_TJ}. As an example for a strategic forecasting problems: where should data centres be placed? In what region of a country and in what geographic region? As an example for tactical forecasting problems, these must take into account energy prices (see \S\ref{sec:Electricity_price_forecasting}) and also, classic supply chain problems \citep{simchi2001_TJ}. After all, physical servers and data centres are what enables the cloud and these must be ordered and have a lead-time. The careful incorporation of life cycles of compute types is important (e.g., both the popularity of certain machine types and the duration of a hard disk). Analogous to the retail sector, cloud resource providers have tactical cold-start forecasting problems. For example, while GPU or TPU instances are still relatively recent but already well estabilished, the demand for quantum computing is still to be decided. In the class of operational forecasting problems, cloud provider can choose to address short-term resource forecasting problems for applications such as adding resources to applications predictively and make this available to customers \citep{Barr2018-lz_TJWEB}. The forecasting of the customer's spend for cloud computing is another example. For serverless infrastructure, a number of servers is often maintained in a ready state \citep{gias2020cocoa_TJ} and the forecasting of the size of this `warmpool' is another example. We note that cloud computing customers have forecasting problems that mirror the forecasting challenges of the cloud providers. Interestingly, forecasting itself has become a software service that cloud computing companies offer \citep{ Januschowski2018-fw_TJWEB,Poccia2019-ty_TJWEB,Liberty2020_TJ}

Many challenges in this application area are not unique to cloud computing. Cold start problems exist elsewhere for example. What potentially stands out in cloud computing forecasting problems may be the scale (e.g., there are a lot of physical servers available), the demands on the response time and granularity of a forecast and the degree of automation. Consider the operational forecasting problem of predictive scaling. Unlike in retail demand forecasting, no human operator will be able to control this and response times to forecasts are in seconds. It will be interesting to see whether approaches based on reinforcement learning \citep{Gamble2018-fd_TJWEB,Dempster2001} can partially replace the need to have forecasting models \citep{januschowski18_TJ}.

\subsubsection[Judgmental forecasting in practice (Shari De Baets, M.~Sinan~G\"on\"ul, \& Nigel Harvey)]{Judgmental forecasting in practice\protect\footnote{This subsection was written by Shari De Baets, M.~Sinan~G\"on\"ul, \& Nigel Harvey.}}
\label{sec:Judgmental_forecasting_in_practice}

Surveys of forecasting practice \citep{De_Baets2019-ir_NG} have shown that the use of pure judgmental forecasting by practitioners has become less common. About 40 years ago, \cite{Sparkes1984-ay_NG} found that company action was more likely to be influenced by judgmental forecasts than by any other type of forecast. In contrast, \cite{Fildes2015-ug_NG} found that only 15.6\% of forecasts in the surveyed companies were made by judgment alone. The majority of forecasts (55.6\%) were made using a combination of statistical and judgmental methods. In this section, we discuss forecasting using unaided judgment (pure judgmental forecasting; see also \S\ref{sec:judgmental_forecasting}), judgmental adjustments (judgment in combination with statistical models; see also \S\ref{sec:judgmental_adjustments_of_computer_based_forecasts}), and the role of judgment in forecasting support systems. 

On the first theme, the survey results discussed above beg the question of whether pure judgmental forecasting is still relevant and reliable. Answers here depend on the type of information on which the judgmental forecasts are based \citep[][see also \S\ref{sec:judgmental_forecasting}]{Harvey2007-xo_NG}. For instance, people have difficulty making cross-series forecasts, as they have difficulty learning the correlation between variables and using it to make their forecasts \citep{Harvey1994-gg_NG,Lim1996-ck_NG,Lim1996-yk_NG}. Additionally, they appear to take account of the noise as well as the pattern when learning the relation between variables; hence, when later using one of the variables to forecast the other, they add noise to their forecasts \citep{Gray1965-fv_NG}. Judgmental extrapolation from a single time series is subject to various effects. First, people are influenced by optimism. For example, they over-forecast time series labelled as `profits' but under-forecast the same series labelled as `losses' \citep{Harvey2013-sa_NG}. Second, they add noise to their forecasts so that a sequence of forecasts looks similar to (`represents') the data series \citep{Harvey1995-sx_NG}. Third, they damp trends in the data \citep{Eggleton1982-oi_NG,Harvey2013-sa_NG,Lawrence1989-tg_NG}. Fourth, forecasts from un-trended independent series do not lie on the series mean but between the last data point and the mean; this is what we would expect if people perceived a positive autocorrelation in the series \citep{Reimers2011-qn_NG}. These last two effects can be explained in terms of the under-adjustment that characterises use of the anchor-and-adjust heuristic: forecasters anchor on the last data point and adjust towards the trend line or mean – but do so insufficiently. However, in practice, this under-adjustment may be appropriate because real linear trends do become damped and real series are more likely to contain a modest autocorrelation than be independent \citep{Harvey2011-ku_NG}. We should therefore be reluctant to characterise these last two effects as biases. 

Given these inherent flaws in people's decision making, practitioners might be hesitant to base their predictions on judgment. However, the reality is that companies persist in incorporating judgment into their forecasting. Assertions that they are wrong to do so represent an over-simplified view of the reality in which businesses operate. Statistical models are generally not able to account for external events, events with low frequency, or a patchy and insufficient data history \citep{Armstrong1998-gh,Goodwin2002-eg,Hughes2001-my}. Hence, a balance may be found in the combination of statistical models and judgment (see \S\ref{sec:judgmental_adjustments_of_computer_based_forecasts}).

In this respect, judgmental adjustments to statistical model outputs are the most frequent form of judgmental forecasting in practice
\citep{Arvan2019-xy_SG,Eksoz2019-mh_PG,Lawrence2006-ey_SG,Petropoulos2016-uk_PG}. Judgmental adjustments give practitioners a quick and convenient way to incorporate their insights, their experience and the additional information that they possess into a set of statistical baseline forecasts. Interestingly, \cite{Fildes2009-tc} examined the judgmental adjustment applications in four large supply-chain companies and found evidence that the adjustments in a `negative' direction improved the accuracy more than the adjustments in a `positive' direction. This effect may be attributable to wishful thinking or optimism that may underlie positive adjustments. Adjustments that were `larger' in magnitude were also more beneficial in terms of the final forecast accuracy than `smaller' adjustments \citep{Fildes2009-tc}. This may simply be because smaller adjustments are merely a sign of tweaking the numbers, but large adjustments are carried out when there is a highly valid reason to make them. These findings have been confirmed in other studies \citep[see, for example,][]{Franses2009-ul_SG,Syntetos2009-dr_SG}.

What are the main reasons behind judgmental adjustments? \cite{Onkal2005-nr_PG} conducted a series of interviews and a survey on forecasting practitioners \citep{Gonul2009-pl_DILEK} to explore these. The main reasons given were (\textit{i}) to incorporate the practitioners’ intuition and experience about the predictions generated externally, (\textit{ii}) to accommodate sporadic events and exceptional occasions, (\textit{iii}) to integrate confidential/insider information that may have not been captured in the forecasts, (\textit{iv}) to hold responsibility and to gain control of the forecasting process, (\textit{v}) to incorporate the expectations and viewpoints of the practitioners, and (\textit{vi}) to compensate for various judgmental biases that are believed to exist in the predictions. These studies also revealed that forecasting practitioners are very fond of judgmental adjustments and perceive them as a prominent way of `completing' and `owning' the predictions that are generated by others. 

While the first three reasons represent the integration of an un-modelled component into the forecast, potentially improving accuracy, the other reasons tend to harm accuracy rather than improve it. In such cases, the forecast would be better off if left unadjusted. \cite{Onkal2005-nr_PG} and \cite{Gonul2009-pl_DILEK} report that the occasions when forecasters refrain from adjustments are (\textit{i}) when the practitioners are adequately informed and knowledgeable about the forecasting method(s) that are used to generate the baseline forecasts, (\textit{ii}) when there are accompanying explanations and convincing communications that provide the rationale behind forecast method selection, (\textit{iii}) when baseline predictions are supplemented by additional supportive materials such as scenarios and alternative forecasts, (\textit{iv}) when the forecasting source is believed to be trustworthy and reliable, and (\textit{v}) when organisational policy or culture prohibits judgmental adjustments. In these circumstances, the baseline forecasts are more easily accepted by practitioners and their adjustments tend to be less frequent. 

Ideally, a Forecast Support System (FSS; see \S\ref{sec:Suport_systems}) should be designed to ensure that it encourages adjustment or non-adjustment whichever is appropriate \citep{Fildes2006-km}. But how can this be achieved? The perceived quality and accessibility of a FSS can be influenced by its design. More on this can be found in the literature on the Technology Acceptance Model \citep{Davis1989-uo} and decision making 
\citep[for instance, by means of framing, visual presentation or nudging; e.g.,][]{Gigerenzer1996-jw,Kahneman1996-it,Payne1982-hq,Thaler2009-bb}. A number of studies have investigated the design aspects of FSS, with varying success. One of the more straightforward approaches is to change the look and feel of the FSS as well as its presentation style. \cite{Harvey1996-ll} found that trends were more easily discernible when the data was displayed in graphical rather than tabular format. Additionally, simple variations in presentation such as line graphs versus point graphs can alter accuracy \citep{Theocharis2018-xb}. The functionalities of the FSS can also be modified (see \S\ref{sec:judgmental_adjustments_of_computer_based_forecasts}). \cite{Goodwin2000-xp} investigated three ways of improving judgmental adjustment via changes in the FSS: a `no adjustment' default, requesting forecasters specify the size of an adjustment rather than give a revised forecast, and requiring a mandatory explanation for the adjustment. Only the default option and the explanation feature were successful in increasing the acceptance of the statistical forecast and so improving forecast accuracy.

\cite{Goodwin2011-mx} reported an experiment that investigated the effects of (\textit{i}) `guidance' in the form of providing information about when to make adjustments and (\textit{ii}) `restriction' of what the forecaster could do (e.g., prohibiting small adjustments). They found that neither restrictiveness nor guidance was successful in improving accuracy, and both were met with resistance by the forecasters. While these studies focused on voluntary integration, \cite{Goodwin2000-qo, Goodwin2002-eg} examined the effectiveness of various methods of mechanical integration and concluded that automatic correction for judgmental biases by the FSS was more effective than combining judgmental and statistical inputs automatically with equal or varying weights. Another approach to mechanical integration was investigated by \cite{Baecke2017-wj}. They compared ordinary judgmental adjustment with what they termed ``integrative judgment''. This takes the judgmental information into account as a predictive variable in the forecasting model and generates a new forecast. This approach improved accuracy. It also had the advantage that forecasters still had their input into the forecasting process and so the resistance found by \cite{Goodwin2011-mx} should not occur. Finally, it is worth emphasising that an effective FSS should not only improve forecast accuracy but should also be easy to use, understandable, and acceptable \citep[][see also \S\ref{sec:Trusting_model_and_expert_forecasts} and \S\ref{sec:Suport_systems}]{Fildes2006-km}. 

\subsubsection[Trust in forecasts (Dilek \"Onkal)]{Trust in forecasts\protect\footnote{This subsection was written by Dilek \"Onkal.}}
\label{sec:Trust_in_forecasts}
Regardless of how much effort is poured into training forecasters and developing elaborate forecast support systems, decision-makers will either modify or discard the predictions if they do not trust them (see also \S\ref{sec:judgmental_adjustments_of_computer_based_forecasts}, \S\ref{sec:Trusting_model_and_expert_forecasts}, \S\ref{sec:Suport_systems}, and \S\ref{sec:Judgmental_forecasting_in_practice}). Hence, trust is essential for forecasts to be actually used in making decisions \citep{Alvarado-Valencia2014-eb_DILEK,Onkal2019-uq_DILEK}.

Given that trust appears to be the most important attribute that promotes a forecast, what does it mean to practitioners? Past work suggests that trusting a forecast is often equated with trusting the forecaster, their expertise and skills so that predictions could be used without adjustment to make decisions \citep{Onkal2019-uq_DILEK}. It is argued that trust entails relying on credible forecasters that make the best use of available information while using correctly applied methods and realistic assumptions \citep{Gonul2009-pl_DILEK} with no hidden agendas \citep{Gonul2012-ku_DILEK}. Research suggests that trust is not only about trusting forecaster's competence; users also need to be convinced that no manipulations are made for personal gains and/or to mislead decisions \citep{Twyman_undated-zw_DILEK}.

Surveys with practitioners show that key determinants of trust revolve around (\textit{i}) forecast support features and tools (e.g., graphical illustrations, rationale for forecasts), (\textit{ii}) forecaster competence/credibility, (\textit{iii}) forecast combinations (from multiple forecasters/methods), and (\textit{iv}) forecast user’s knowledge of forecasting methods \citep{Onkal2019-uq_DILEK}.

What can be done to enhance trust? If trust translates into accepting guidance for the future while acknowledging and tolerating potential forecast errors, then both the providers and users of forecasts need to work as partners towards shared goals and expectations. Important pathways to accomplish this include (\textit{i}) honest communication of forecaster's track record and relevant accuracy targets \citep{Onkal2019-uq_DILEK}, (\textit{ii}) knowledge sharing \citep{Ozer2011-jx_DILEK,Renzl2008-lc_DILEK} and transparency of forecasting methods, assumptions and data \citep{Onkal2019-uq_DILEK}, (\textit{iii}) communicating forecasts in the correct tone and jargon-free language to appeal to the user audience \citep{Taylor1982-xz_DILEK}, (\textit{iv}) users to be supported with forecasting training \citep{Merrick2006-se_DILEK}, (\textit{v}) providing explanations/rationale behind forecasts \citep{Gonul2006-ia_DILEK,Onkal2008-ys_DILEK}, (\textit{vi}) presenting alternative forecasts under different scenarios (see \S\ref{sec:scenarios_forecasting}), and (\textit{vii}) giving combined forecasts as benchmarks \citep{Onkal2019-uq_DILEK}. 

Trust must be earned and deserved \citep{Maister2012-qb_DILEK} and is based on building a relationship that benefits both the providers and users of forecasts. Take-aways for those who make forecasts and those who use them converge around clarity of communication as well as perceptions of competence and integrity. Key challenges for forecasters are to successfully engage with users throughout the forecasting process (rather than relying on a forecast statement at the end) and to convince them of their objectivity and expertise. In parallel, forecast users face challenges in openly communicating their expectations from forecasts \citep{Gonul2009-pl_DILEK}, as well as their needs for explanations and other informational addendum to gauge the uncertainties surrounding the forecasts. Organisational challenges include investing in forecast management and designing resilient systems for collaborative forecasting.
 
\subsubsection[Communicating forecast uncertainty (Victor Richmond R. Jose)]{Communicating forecast uncertainty\protect\footnote{This subsection was written by Victor Richmond R. Jose.}}
\label{sec:Communicating_forecast_uncertainty}
Communicating forecast uncertainty is a critical issue in forecasting practice. Effective communication allows forecasters to influence end-users to respond appropriately to forecasted uncertainties. Some frameworks for effective communication have been proposed by decomposing the communication process into its elements: the communicator, object of uncertainty, expression format, audience, and its effect \citep{NRC2006_VRRJ,vanderbles2019_VRRJ}.

Forecasters have long studied part of this problem focusing mostly in the manner by which we express forecast uncertainties. \citet{gneiting2014_VRRJ} provides a review of recent probabilistic forecasting methods (see also \S\ref{sec:evaluating_probabilistic_forecasts} and \S\ref{sec:assessing_the_reliability_of_probabilistic forecats}). Forecasting practice however revealed that numeracy skills and cognitive load can often inhibit end-users from correctly interpreting these uncertainties \citep{joslyn2009b_VRRJ,raftery2016_VRRJ}. Attempts to improve understanding through the use of less technical vocabulary also creates new challenges. Research in psychology show that wording and verbal representation play important roles in disseminating uncertainty \citep{joslyn2009_VRRJ}. Generally forecasters are found to be consistent in their use of terminology, but forecast end-users often have inconsistent interpretation of these terms even those commonly used \citep{budescu1985_VRRJ,clark1990_VRRJ,ulkumen2016_VRRJ}. Pretesting verbal expressions and avoiding commonly misinterpreted terms are some easy ways to significantly reduce biases and improve comprehension.

Visualisations can also be powerful in communicating uncertainty. \citet{johnson1995_VRRJ} and \citet{spiegelhalter2011_VRRJ} propose several suggestions for effective communication (e.g., multiple-format use, avoiding framing bias, and acknowledging limitations), but also recognise the limited amount of existing empirical evidence. Some domain-specific studies do exist. For example, \cite{riveiro2014_VRRJ} showed uncertainty visualisation helped forecast comprehension in a homeland security context.

With respect to the forecaster and her audience, issues such as competence, trust, respect, and optimism have been recently examined as a means to improve uncertainty communication. \citet{fiske2014_VRRJ} discusses how forecast recipients often infer apparent intent and competence from the uncertainty provided and use these to judge trust and respect (see also \S\ref{sec:Trusting_model_and_expert_forecasts} and \S\ref{sec:Trust_in_forecasts} for discussion on trust and forecasting). This suggests that the amount of uncertainty information provided should be audience dependent \citep{politi2007_VRRJ,han2009_VRRJ}. \citet{raftery2016_VRRJ} acknowledges this by using strategies depending on the audience type (e.g., low-stakes user, risk avoider, etc.). \citet{fischhoff2014_VRRJ} suggests a similar approach by examining how people are likely to use the information (e.g., finding a signal, generating new options, etc.)

When dealing with the public, experts assert that communicating uncertainty helps users understand forecasts better and avoid a false sense of certainty \citep{morss2008_VRRJ}. Research however shows that hesitation to include forecast uncertainty exists among experts because it provides an opening for criticism and the possibility of misinterpration by the public \citep{fischhoff2012_VRRJ}. This is more challenging when the public has prior beliefs on a topic or trust has not been established. Uncertainty can be used by individuals to reinforce a motivated-reasoning bias that allows them to ``see what they want to see'' \citep{dieckmann2017_VRRJ}. Recent work however suggests that increasing transparency for uncertainty does not necessarily affect trust in some settings. \citet{vanderbles2020_VRRJ} recently showed in a series of experiments that people recognise greater uncertainty with more information but expressed only a small decrease in trust in the report and trustworthiness of the source.

\subsection{Other applications}
\label{sec:Other_applications}

\subsubsection[Tourism demand forecasting (Ulrich Gunter)]{Tourism demand forecasting\protect\footnote{This subsection was written by Ulrich Gunter.}}
\label{sec:Tourism_demand_forecasting}
As seen throughout 2020, (leisure) tourism demand is very sensitive to external shocks such as natural and human-made disasters, making tourism products and services extremely perishable \citep{Frechtling2001-ea_UG}. As the majority of business decisions in the tourism industry require reliable demand forecasts \citep{Song2008-lg_UG}, improving their accuracy has continuously been on the agenda of tourism researchers and practitioners alike. This continuous interest has resulted in two tourism demand forecasting competitions to date \citep{Athanasopoulos2011-lw,Song2021-ha_UG}, the current one with a particular focus on tourism demand forecasting during the COVID-19 pandemic (for forecasting competitions, see \S\ref{sec:Forecasting_competitions}). Depending on data availability, as well as on geographical aggregation level, tourism demand is typically measured in terms of arrivals, bed-nights, visitors, exports receipts, import expenditures, etc.

Since there are no specific tourism demand forecast models, standard univariate and multivariate statistical models, including common aggregation and combination techniques, etc., have been used in quantitative tourism demand forecasting \citep[see, for example,][for recent reviews]{Song2019-pc_UG,Jiao2019-hu_UG}. Machine learning and other artificial intelligence methods, as well as hybrids of statistical and machine learning models, have recently been employed more frequently.

Traditionally, typical micro-economic demand drivers (own price, competitors' prices, and income) and some more tourism-specific demand drivers (source-market population, marketing expenditures, consumer tastes, habit persistence, and dummy variables capturing one-off events or qualitative characteristics) have been employed as predictors in tourism demand forecasting \citep{Song2008-lg_UG}. One caveat of some of these economic demand drivers is their publication lag and their low frequency, for instance, when real GDP (per capita) is employed as a proxy for travellers' income.

The use of leading indicators, such as industrial production as a leading indicator for real GDP (see also \S\ref{sec:Forecasting_GDP_and_Inflation}), has been proposed for short-term tourism demand forecasting and nowcasting \citep{Chatziantoniou2016-ec_UG}. During the past couple of years, web-based leading indicators have also been employed in tourism demand forecasting and have, in general, shown improvement in terms of forecast accuracy. However, this has not happened in each and every case, thereby confirming the traded wisdom that there is no single best tourism demand forecasting approach \citep{Li2005-bp_UG}. Examples of those web-based leading indicators include Google Trends indices \citep{Bangwayo-Skeete2015-yk_UG}, Google Analytics indicators \citep{Gunter2016-wm_UG}, as well as Facebook `likes' \citep{Gunter2019-kp_UG}.

The reason why these expressions of interaction of users with the Internet have proven worthwhile as predictors in a large number of cases is that it is sensible to assume potential travellers gather information about their destination of interest prior to the actual trip, with the Internet being characterised by comparably low search costs, ergo allowing potential travellers to forage information \citep{Pirolli1999-sl_UG} with only little effort \citep{Zipf2016-wq_UG}. A forecaster should include this information in their own set of relevant information at the forecast origin \citep{Lutkepohl2005-xj_UG}, if taking it into account results in an improved forecast accuracy, with web-based leading indicators thus effectively Granger-causing \citep{Granger1969-uk_UG} actual tourism demand (see \S\ref{sec:Leading_indicators_and_Granger_causality}).

Naturally, tourism demand forecasting is closely related to aviation forecasting (see \S\ref{sec:Forecasting_for_airports}), as well as traffic flow forecasting (see \S\ref{sec:Traffic_flow_forecasting}). A sub-discipline of tourism demand forecasting can be found with hotel room demand forecasting. The aforementioned perishability of tourism products and services is particularly evident for hotels as a hotel room not sold is lost revenue that cannot be regenerated. Accurate hotel room demand forecasts are crucial for successful hotel revenue management \citep{Pereira2016-df_UG} and are relevant for planning purposes such as adequate staffing during MICE (i.e., Meetings, Incentives, Conventions, and Exhibitions/Events) times, scheduling of renovation periods during low seasons, or balancing out overbookings and ``no shows'' given constrained hotel room supply \citep{Ivanov2012_UG}.

Particularly since the onset of the COVID-19 pandemic in 2020, which has been characterised by global travel restrictions and tourism businesses being locked down to varying extents, scenario forecasting and other forms of hybrid and judgmental forecasting played an important role \citep[][see \S\ref{sec:scenarios_forecasting}]{Zhang2021-rl}, thereby highlighting an important limitation of quantitative tourism demand forecasting as currently practised. Based on the rapid development of information technology and artificial intelligence, \cite{Li2020-rj_UG}, however, envisage a ``super-smart tourism forecasting system'' \citep[][p. 264]{Li2020-rj_UG} for the upcoming 75 years of tourism demand forecasting. According to these authors, this system will be able to automatically produce forecasts at the micro level (i.e., for the individual traveller and tourism business) in real time while drawing on a multitude of data sources and integrating multiple (self-developing) forecast models.
 
\subsubsection[Forecasting for aviation (Xiaojia Guo)]{Forecasting for aviation\protect\footnote{This subsection was written by Xiaojia Guo.}}
\label{sec:Forecasting_for_airports}
Airports and airlines have long invested in forecasting arrivals and departures of aircrafts. These forecasts are important in measuring airspace and airport congestions, designing flight schedules, and planning for the assignment of stands and gates \citep{barnhart2004airline_XG}. Various techniques have been applied to forecast aircrafts' arrivals and departures. For instance, \cite{rebollo2014characterization_XG} apply random forests to predict air traffic delays of the National Airspace System using both temporal and network delay states as covariates. \cite{manna2017statistical_XG} develop a statistical model based on a gradient boosting decision tree to predict arrival and departure delays, using the data taken from the United States Department of Transportation \citep{ustransportation2020_XG}. \cite{rodriguez2019assessment_XG} develop a Bayesian Network model to predict flight arrivals and delays using the radar data, aircraft historical performance and local environmental data. There are also a few studies that have focused on generating probabilistic forecasts of arrivals and departures, moving beyond point estimates. For example, \cite{tu2008estimating_XG} develop a predictive system for estimating flight departure delay distributions using flight data from Denver International Airport. The system employs the smoothing spline method to model seasonal trends and daily propagation patterns. It also uses mixture distributions to estimate the residual errors for predicting the entire distribution. 

In the airline industry, accurate forecasts on demand and booking cancellations are crucial to revenue management, a concept that was mainly inspired by the airline and hotel industries \citep[][see also \S\ref{sec:Tourism_demand_forecasting} for a discussion on hotel occupancy forecasting]{lee1990airline_XG,mcgill1999revenue_XG}. The proposals of forecasting models for flight demand can be traced back to \cite{beckmann1958airline_XG}, where these authors demonstrate that Poisson and Gamma models can be applied to fit airline data. Then, the use of similar flights' short-term booking information in forecasting potential future bookings has been discussed by airline practitioners such as \cite{adams1987short_XG} at Quantas as well as \cite{smith1992yield_XG} at American Airlines. Regressions models (see \S\ref{sec:time_series_regression_models}) and time series models such as exponential smoothing (see \S\ref{sec:exponential_smoothing_models}) and ARIMA (see \S\ref{sec:autoregressive_integrated_moving_average_models}) have been discussed in \cite{sa1987reservations_XG}, \cite{wickham1995evaluation_XG}, and \cite{botimer1997select_XG}. There are also studies focusing on disaggregate airline demand forecasting. For example,  \cite{martinez1970automatic_XG} apply empirical probability distributions to predict bookings and cancellations of individual passengers travelling with Iberia Airlines. \cite{carson2011forecasting_XG} show that aggregating the forecasts of individual airports using airport-specific data could provide better forecasts at a national level. More recently, machine learning methods have also been introduced to generate forecasts for airlines. This can be seen in \cite{weatherford2003neural_XG} where they apply neural networks to forecast the time series of the number of reservations. Moreover, \cite{hopman2021machine_XG} show that an extreme gradient boosting model which forecasts itinerary-based bookings using ticket price, social media posts and airline reviews outperforms traditional time series forecasts. 

Forecasting passenger arrivals and delays in the airports have received also some attention in the literature, particularly in the past decade. \cite{wei2006aggregate_XG} build an aggregate demand model for air passenger traffic in a hub-and-spoke network. The model is a log-linear regression that uses airline service variables such as aircraft size and flight distance as predictors. \cite{barnhart2014modeling_XG} develop a multinomial logit regression model, designed to predict delays of US domestic passengers. Their study also uses data from the US Department of Transportation \citep{ustransportation2020_XG}. \cite{guo2020forecasting_XG} recently develop a predictive system that generates distributional forecasts of connection times for transfer passengers at an airport, as well as passenger flows at the immigration and security areas. Their approach is based on the application of regression trees combined with copula-based simulations. This predictive system has been implemented at Heathrow airport since 2017. 

With an increasing amount of available data that is associated with activities in the aviation industry, predictive analyses and forecasting methods face new challenges as well as opportunities, especially in regard to updating forecasts in real time. The predictive system developed by \cite{guo2020forecasting_XG} is able to generate accurate forecasts using real-time flight and passenger information on a rolling basis. The parameters of their model, however, do not update over time. Therefore, a key challenge in this area is for future studies to identify an efficient way to dynamically update model parameters in real time.
 
\subsubsection[Traffic flow forecasting (Alexander Dokumentov)]{Traffic flow forecasting\protect\footnote{This subsection was written by Alexander Dokumentov.}}
\label{sec:Traffic_flow_forecasting}
Traffic flow forecasting is an important task for traffic management bodies to reduce traffic congestion, perform planning and allocation tasks, as well as for travelling individuals to plan their trips. Traffic flow is complex spatial and time-series data exhibiting multiple seasonalities and affected by spatial exogenous influences such as social and economic activities and events, various government regulations, planned road works, weather, traffic accidents, etc. \citep{polson2017deep_ad}.

Methods to solve traffic flow forecasting problems vaguely fall into three categories.
The first uses parametric statistical methods such as ARIMA, seasonal ARIMA, space-time ARIMA, Kalman filters, etc. \citep[see, for example,][]{whittaker1997tracking_ad, vlahogianni2004short_ad, kamarianakis2005space_ad, vlahogianni2014short_ad}.
The second set of approaches uses purely of neural networks \citep{mena2020comprehensive_ad}.
The third group of methods uses various machine learning, statistical non-parametric techniques or mixture of them \citep[see, for example,][but also \S\ref{sec:neural_networks} and \S\ref{sec:machine_Learning} for an overview of NN and ML methods]{hong2011traffic_ad, zhang2016short_ad, zhang2017vehicle_ad}.

Although neural networks are probably the most promising technique for traffic flow forecasting \citep[see, for example,][]{polson2017deep_ad, do2019effective_ad}, statistical techniques, such as Seasonal-Trend decomposition based on Regression (STR, see \S\ref{sec:time_series_decomposition}), can outperform when little data is available or they can be used for imputation, de-noising, and other pre-processing before feeding data into neural networks which often become less powerful when working with missing or very noisy data.

\begin{figure}[ht!]
	\centering
	\includegraphics[width=0.72\textwidth]{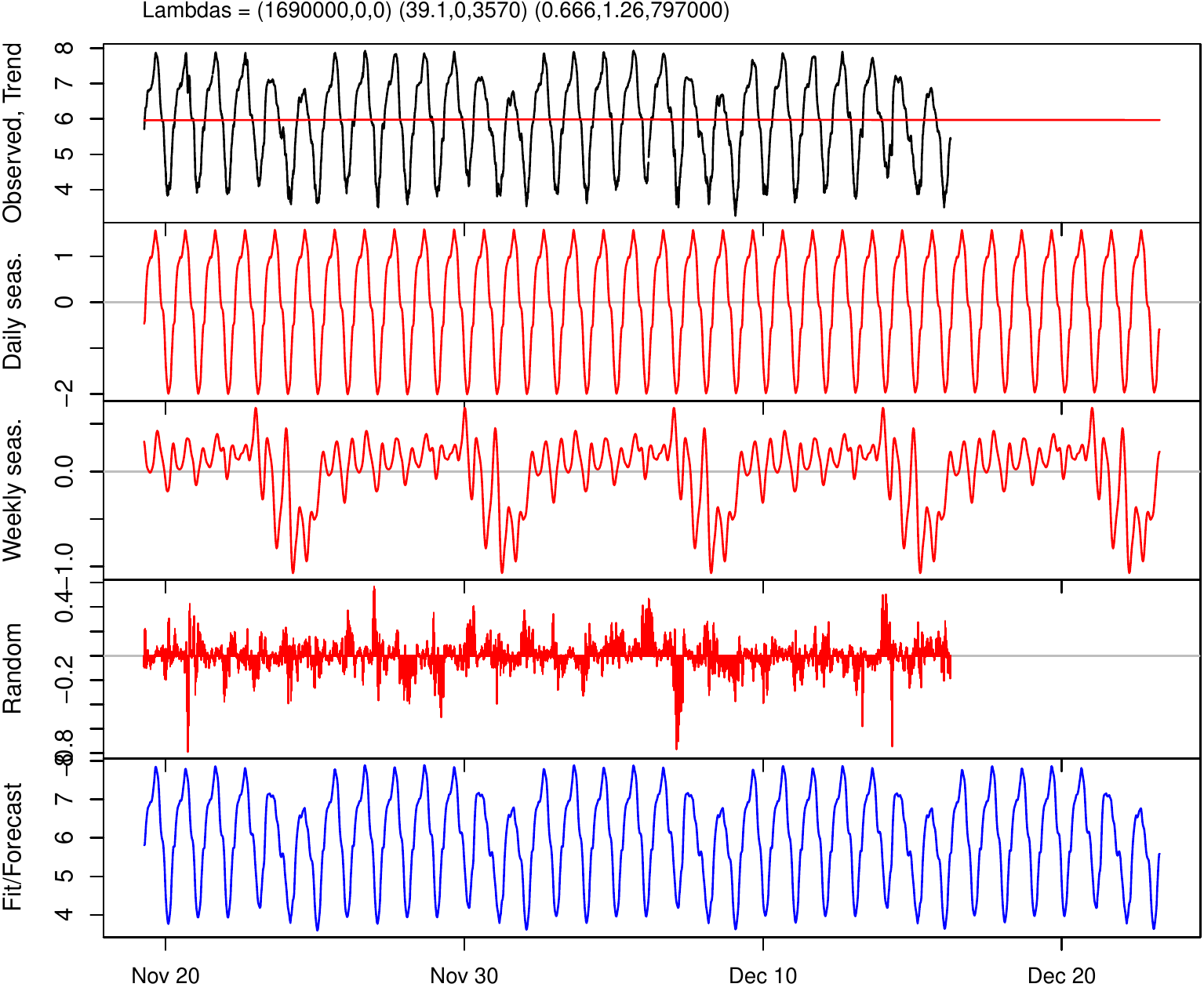}
	\caption{STR decomposition of the log transformed training data and the forecasts for the traffic flow data.}
	\label{fig:fig1_dokumentov}
\end{figure}

\begin{figure}[ht!]
	\centering
	\includegraphics[width=0.47\textwidth]{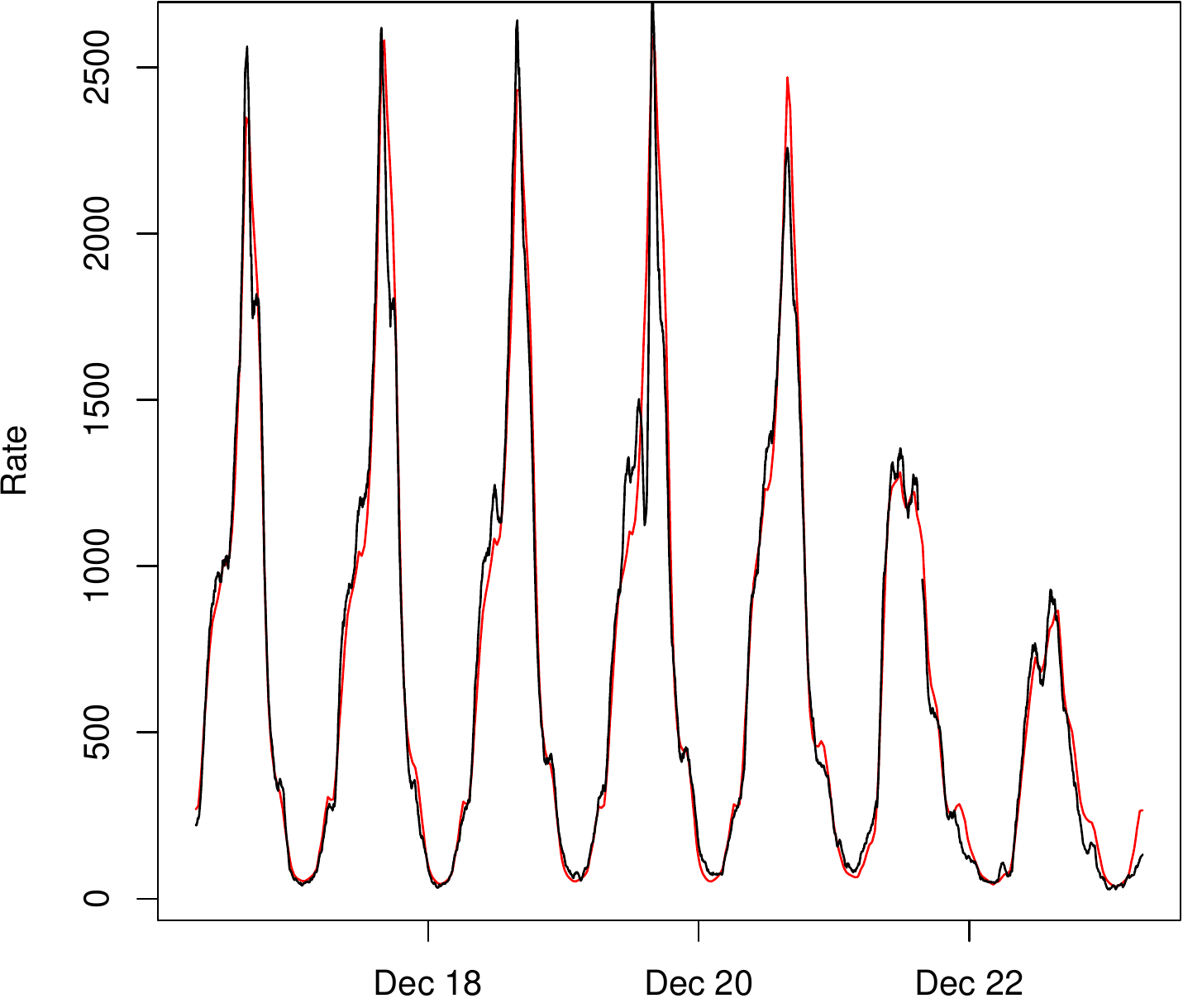}
	\hspace{5mm}
	\includegraphics[width=0.47\textwidth]{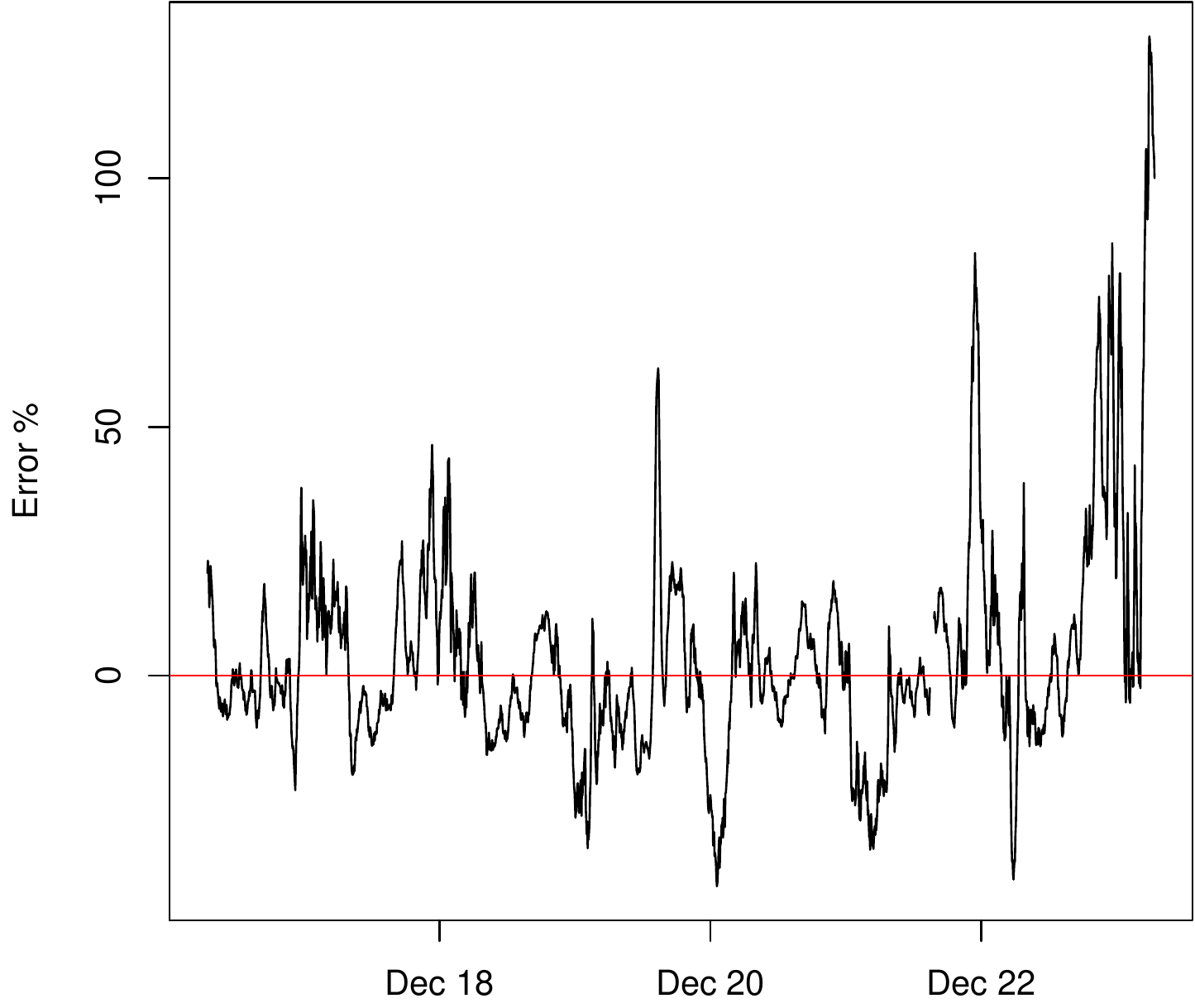}
	\caption{Left: forecast (red) and the test data (black); Right: the prediction error over time for the traffic flow data.}
	\label{fig:fig2_dokumentov}
\end{figure}

Traffic flow forecasting is illustrated below using vehicle flow rate data from road camera A1.GT.24538 on A1 highway in Luxembourg \citep{luxkaggledataset_ad} from 2019-11-19 06:44:00 UTC to 2019-12-23 06:44:00 UTC. Most of the data points are separated by 5 minutes intervals. Discarding points which do not follow this schedule leads to a data set where all data points are separated by 5 minutes intervals, although values at some points are missing. The data is split into training and test sets by setting aside last 7 days of data. As \cite{hou2014traffic_ad} and \cite{polson2017deep_ad} suggest, spatial factors are less important for long term traffic flow forecasting, and therefore they are not taken into account and only temporal data is used. Application of STR \citep{ad_dokumentov_2017} as a forecasting technique to the log transformed data leads to a forecast with Mean Squared Error 102.4, Mean Absolute Error 62.8, and Mean Absolute Percentage Error (MAPE) 14.3\% over the test set, outperforming Double-Seasonal Holt-Winters by 44\% in terms of MAPE.
The decomposition and the forecast obtained by STR are shown on figure \ref{fig:fig1_dokumentov} and the magnified forecast and the forecasting errors are on figure \ref{fig:fig2_dokumentov}.

\subsubsection[Call arrival forecasting (Devon K. Barrow)]{Call arrival forecasting\protect\footnote{This subsection was written by Devon K. Barrow.}}
\label{sec:Call_arrival_forecasting}
\begin{figure}[ht!]
	\centerline{\includegraphics[trim=70 720 70 70, clip,width=4.5in]{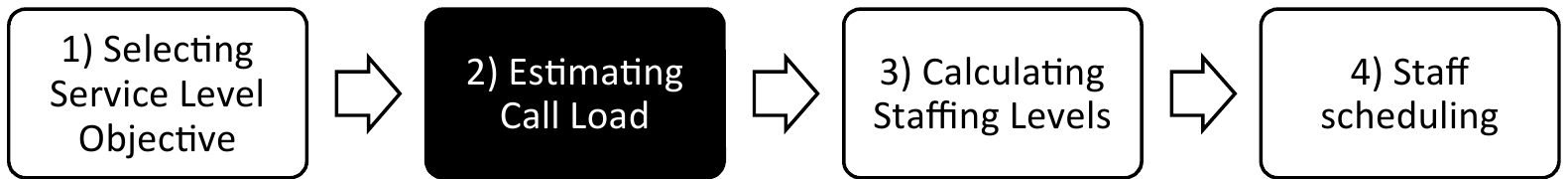}}
	\caption{The staffing decision process in call centres.}\label{fig:staff_devon}
\end{figure}

Forecasting of inbound call arrivals for call centres supports a number of key decisions primarily around staffing \citep{aksin2007modern_DKB}. This typically involves matching staffing level requirements to service demand as summarised in Figure \ref{fig:staff_devon}. To achieve service level objectives, an understanding of the call load is required in terms of the call arrivals \citep{gans2003telephone_DKB}. As such, forecasting of future call volume or call arrival rates is an important part of call centre management.

There are several properties to call arrival data. Depending on the level of aggregation and the frequency with which data is collected, e.g., hourly, call arrival data may exhibit intraday (within-day), intraweek, and intrayear multiple seasonal patterns \citep[and \S\ref{sec:forecasting_for_multiple_seasonal_cycles}]{avramidis2004modeling_DKB,brown2005statistical_DKB}. In addition, arrival data may also exhibit interday and intraday dependencies, with different time periods within the same day, or across days within the same week, showing strong levels of autocorrelation \citep{tanir1999call_DKB,brown2005statistical_DKB,shen2005analysis_DKB}. Call arrivals may also be heteroscedastic with variance at least proportional to arrival counts \citep{taylor2008comparison_DKB}, and overdispersed under a Poisson assumption having variance per time period typically much larger than its expected value \citep{jongbloed2001managing_DKB,avramidis2004modeling_DKB,steckley2005performance_DKB}. These properties have implications for various approaches to modelling and forecasting call arrivals. 

The first family of methods are time series methods requiring no distributional assumptions. Early studies employed auto regressive moving average (ARMA; see \S\ref{sec:autoregressive_integrated_moving_average_models}) models \citep{andrews1995ll_DKB,tandberg1995time_DKB,xu1999long_DKB,antipov2002forecasting_DKB}, exponential smoothing \citep[][see \S\ref{sec:exponential_smoothing_models}]{bianchi1993forecasting_DKB,bianchi1998improving_DKB}, fast Fourier transforms \citep{lewis2003application_DKB}, and regression \citep[][see \S\ref{sec:time_series_regression_models}]{tych2002unobserved_DKB}. The first methods capable of capturing multiple seasonality were evaluated by \cite{taylor2008comparison_DKB} and included double seasonal exponential smoothing \citep{taylor2003short_BRT} and multiplicative double seasonal ARMA (SARMA). Since then several advanced time series methods have been developed and evaluated \citep{taylor2010exponentially_DKB,de2011forecasting_DKB,taylor2012forecasting_BRT}, including artificial neural networks \citep{millan2013forecasting_DKB,pacheco2009neural_DKB,li2011research_DKB} and models for density forecasting \citep{taylor2012density_DKB}.

Another family of models relies on the assumption of a time-inhomogeneous Poisson process adopting fixed \citep{jongbloed2001managing_DKB,brown2005statistical_DKB,shen2008forecasting_DKB,taylor2012density_DKB} and mixed modelling \citep{avramidis2004modeling_DKB,aldor2009workload_DKB,ibrahim2013forecasting_DKB} approaches to account for the overdispersed nature of the data, and in some cases, interday and intraday dependence.

The works by \cite{weinberg2007bayesian_DKB} and \cite{soyer2008modeling_DKB} model call volumes from a Bayesian point of view. Other Bayesian inspired approaches have been adopted mainly for estimating various model parameters, but also allowing for intraday updates of forecasts \citep{landon2010modeling_DKB,aktekin2011call_DKB}.

A further class of approach addresses the dimensionality challenge related to high frequency call data using Singular Value Decomposition (SVD). \cite{shen2005analysis_DKB} and \cite{shen2008forecasting_DKB} use the same technique to achieve dimensionality reduction of arrival data, and to create a forecasting model that provides both interday forecasts of call volume, and an intraday updating mechanism. Several further studies have extended the basic SVD approach to realise further modelling innovations, for example, to forecast call arrival rate profiles and generate smooth arrival rate curves \citep{shen2007forecasting_DKB,shen2008interday_DKB,shen2009modeling_DKB}.
A more comprehensive coverage of different forecasting approaches for call arrival rate and volume can be found in a recent review paper by \cite{ibrahim2016modeling_DKB}.
 
\subsubsection[Elections forecasting (Jose M. Pav\'ia)]{Elections forecasting\protect\footnote{This subsection was written by Jose M. Pav\'ia.}}
\label{sec:Elections_forecasting}

With the exception of weather forecasts, there are few forecasts which have as much public exposure as election forecasts. They are frequently published by mass media, with their number and disclosure reaching a frenzy as the Election Day approaches. This explains the significant amount of methods, approaches and procedures proposed and the paramount role these forecasts play in shaping people's confidence in (soft/social) methods of forecasting.

The problem escalates because, regardless whether the goal of the election forecast is an attempt to ascertain the winner in two-choice elections (e.g., a referendum or a Presidential election) or to reach estimates within the margins of error in Parliamentary systems, the knowledge of the forecasts influences electors' choices \citep{Pavia2019-bx_JMP}. Election forecasts not only affect voters but also political parties, campaign organizations and (international) investors, who are also watchful of their evolution.

Scientific approaches to election forecasting include polls, information (stock) markets and statistical models. They can also be sorted by when they are performed; and new methods, such as social media surveillance (see also \S\ref{sec:Text_based_forecasting}), are also emerging \citep{Huberty2015-ug_JMP,Ceron2016-gq_JMP}. Probabilistic (representative) polls are the most commonly used instrument to gauge public opinions. The progressive higher impact of non-sampling errors \citep[coverage issues, non-response bias, measurement error:][]{Biemer2010-ks_JMP} is, however, severely testing this approach. Despite this, as \cite{Kennedy2017-la_JMP} show in a recent study covering 86 countries and more than 500 elections, polls are still powerful and robust predictors of election outcomes after adjustments \citep[see also][]{Jennings2020-yh_JMP}. The increasing need of post-sampling adjustments of probabilistic samples has led to a resurgence of interest in non-probabilistic polls \citep{Pavia2012-lt_JMP,Wang2015-xx_JMP,Elliott2017-ej_JMP}, abandoned in favour of probabilistic sampling in 1936, when Gallup forecasted Roosevelt's triumph over Landon using a small representative sample despite Literacy Digest failing to do so with a sample of near 2.5 million responses \citep{Squire1988-lw_JMP}.

A person knows far more than just her/his voting intention \citep{Rothschild2009-um_JMP} and when s/he makes a bet, the rationality of her/his prediction is reinforced because s/he wants to win. Expectation polls try to exloit the first issue \citep{Graefe2014-on_JMP}, while prediction markets, as efficient aggregators of information, exploit both these issues to yield election forecasts (see also \S\ref{sec:Wisdom_of_crowds} and \S\ref{sec:panels_of_experts}). Several studies have proven the performance of these approaches \citep{Wolfers2004-mq_JMP,Berg2008-um_JMP,Erikson2012-zb_JMP,Williams2016-xa_JMP}, even studying their links with opinion polls \citep{Brown2019-jf_JMP}. Practice has also developed econometric models \citep{Fair1978-bd_JMP} that exploit structural information available months before the election (e.g., the evolution of the economy or the incumbent popularity). Lewis-Beck has had great success in publishing dozens of papers using this approach \citep[see, e.g.,][]{Lewis-Beck2005-oa_JMP}.

Special mention also goes to Election-Day forecasting strategies, which have been systematically commissioned since the 1950s \citep{Mitofsky1991-np_JMP}. Exit (and entrance) polls \citep{Pavia2010-ae_JMP,Klofstad2012-uz_JMP}, quick-counts \citep{Pavia-Miralles2008-dt_JMP}, and statistical models \citep{Moshman1964-bb_JMP,Bernardo1984-zg_JMP,Pavia-Miralles2005-pj_JMP} have been used to anticipate outcomes on Election Day. Some of these strategies (mainly random quick-counts) can be also employed as auditing tools to disclose manipulation and fraud in weak democracies \citep{Scheuren2008-fw_JMP}.
 
\subsubsection[Sports forecasting (J. James Reade)]{Sports forecasting\protect\footnote{This subsection was written by J. James Reade.}}
\label{sec:Sports_forecasting}
Forecasting is inherent to sport. Strategies employed by participants in sporting contests rely on forecasts, and the decision by promoters to promote, and consumers to attend such events are conditioned on forecasts: predictions of how interesting the event will be. First in this section, we look at forecast competitions in sport, and following this we consider the role forecasts play in sporting outcomes.

Forecast competitions are common; see \S\ref{sec:Forecasting_competitions}. Sport provides a range of forecast competitions, perhaps most notably the competition between bookmakers and their customers -- betting. A bet is a contingent contract, a contract whose payout is conditional on specified future events occurring. Bets occur fundamentally because two agents disagree about the likelihood of that event occurring, and hence it is a forecast.

Bookmakers have been extensively analysed as forecasters; \cite{forrest2005oddssetters_JJR} evaluated biases in the forecasts implied by bookmaker odds over a period where the betting industry became more competitive, and found that relative to expert forecasts, bookmaker forecasts improved.

With the internet age, prediction markets have emerged, financial exchanges where willing participants can buy and sell contingent contracts. In theory, such decentralised market structures ought to provide the most efficient prices and hence efficient forecasts \citep{nordhaus1987forecasting_JJR}. A range of papers have tested this in the sporting context \citep{Gil2007-ao_JJR,Croxson2014-hh_JJR,Angelini2019-kg_JJR}, with conclusions tending towards a lack of efficiency.

judgmental forecasts by experts are commonplace too (see also \S\ref{sec:forecasting_with_judgment}); traditionally in newspapers, but more recently on television and online. \cite{reade2020strange_JJR} evaluate forecasts of scorelines from two such experts against bookmaker prices, a statistical model, and the forecasts from users of an online forecasting competition. \cite{singleton2019going_JJR} find that when forecasters in the same competition revise their forecasts, their forecast performance worsens. This forecasting competition is also analysed by \cite{butler2020expert_JJR} and \cite{reade2020strange_JJR}.

Sport is a spectacle, and its commercial success is conditioned on this fact. Hundreds of millions of people globally watch events like the Olympics and the FIFA World Cup -- but such interest is conditioned on anticipation, a forecast that something interesting will happen. A superstar is going to be performing, the match will be a close encounter, or it will matter a lot for a bigger outcome (the championship, say). These are the central tenets of sport economics back to \cite{neale1964peculiar_JJR} and \cite{rottenberg1956_JJR}, most fundamentally the `uncertainty of outcome hypothesis'.
A multitude of sport attendance prediction studies investigate this \citep[see, for example,][]{hart1975statistical_JJR,simmons2006new_JJR,sacheti2014uncertainty_JJR,coates2010week_JJR,vanours2021common_JJR}, and \cite{VANREETH2019810} considers this for forecasting TV audiences for the Tour de France.

Cities and countries bid to host large events like the World Cup based on forecasts regarding the impact of hosting such events. Forecasts that are often inflated for political reasons \citep{baade2016going_JJR}. Equally, franchise-based sports like many North American sports attract forecasts regarding the impact of a team locating in a city, usually resulting in public subsidies for the construction of venues for teams to play at \citep{coates1999growth_JJR}. Governments invest in sporting development, primarily to achieve better performances at global events, most notably the Olympics \citep{bernard2004wins_JJR}. 

Many sporting events themselves rely on forecasts to function; high jumpers predict what height they will be able to jump over, and free diving contestants must state the depth they will dive to. Less formally, teams will set themselves goals: to win matches, to win competitions, to avoid the `wooden spoon'. Here, forecast outcomes are influenced by the teams, and competitors, taking part in competitions and, as such, are perhaps less commonly thought of as genuine forecasts. Important works predicting outcomes range from \citet{dixon1997modelling_JRR} in soccer, to \citet{kovalchik2019calibration_JRR} for tennis, while the increasing abundance of data means that machine learning and deep learning methods are beginning to dominate the landscape. See, for example, \citet{maymin2019wage_JRR} and \citet{hubavcek2019exploiting_JRR} for basketball, and \citet{mulholland2019optimizing_JRR} for NFL.

\subsubsection[Forecasting for megaprojects (Konstantia Litsiou)]{Forecasting for megaprojects\protect\footnote{This subsection was written by Konstantia Litsiou.}}
\label{sec:Forecasting_for_megaprojects}
Megaprojects are significant activities characterised by a multi-organisation structure, which produces highly visible infrastructure or asset with very crucial social impacts \citep{Aaltonen2011-fj_Litsiou}. Megaprojects are complex, require huge capital investment, several stakeholders are identified and, usually a vast number of communities and the public are the receivers of the project's benefits. There is a need megaprojects especially those that deliver social and economic goods and create economic growth \citep{Flyvbjerg2003-wi_Litsiou}. Typical features of megaprojects include some or all the following: (\textit{i}) delivering a substantial piece of physical infrastructure with a life expectancy that spans across decades, (\textit{ii}) main contractor or group of contractors are privately owned and financed, and (\textit{iii}) the contractor could retain an ownership stake in the project and the client is often a government or public sector organisation \citep{Sanderson2012-mz_Litsiou}.

However, megaprojects are heavily laced with extreme human and technical complexities making their delivery and implementation difficult and often unsuccessful \citep{The_RFE_Working_Group_Report2015-ww_Litsiou,Merrow1988-ux_Litsiou}. This is largely due to the challenge of managing megaprojects including extreme complexity, increased risk, tight budget and deadlines, lofty ideals \citep{Fiori_Christine_undated-ft_Litsiou}. Due to the possibility and consequences of megaproject failure \citep{Misic2015-bk_Litsiou}, forecasting the outcomes of megaprojects is becoming of growing importance. In particular, it is crucial to identify and assess the risks and uncertainties as well as other factors that contribute to disappointing outcomes of megaprojects in order to mitigate them \citep{Flyvbjerg2003-wi_Litsiou,Miller2007-gz_Litsiou}.

Literature review in forecasting in megaprojects are scarce. However, there are a few themes that have emerged in the extant literature as characteristics of megaprojects that should be skilfully managed to provide a guideline for the successful planning and construction of megaprojects \citep{Fiori_Christine_undated-ft_Litsiou,Flyvbjerg2007-vy_Litsiou,Sanderson2012-mz_Litsiou}. \cite{Turner2012-iy_Litsiou} even claim that we cannot even properly define what success is. They argue that we need to reliable scales in order to predict multiple perspectives by multiple stakeholders over multiple time frames –- so definitely a very difficult long term problem. This could be done via a set of leading performance indicators that will enable managers of Megaprojects to forecast during project execution how various stakeholders will perceive success months or even years into the operation. At the very early stages of a project's lifecycle, a number of decisions must been taken and are of a great importance for the performance and successful deliverables/outcomes. \cite{Flyvbjerg2007-vy_Litsiou} stress the importance of the front-end considerations particularly for Megaprojects Failure to account for unforeseen events frequently lead to cost overruns. 

\cite{Litsiou2019-gy_Litsiou} suggest that forecasting the success of megaprojects is particularly a challenging and critical task due to the characteristics of such projects. Megaproject stakeholders typically implement impact assessments and/or cost benefit Analysis tools \citep{Litsiou2019-gy_Litsiou}. As \cite{Makridakis2010-pj_Litsiou} suggested, judgmental forecasting is suitable where quantitative data is limited, and the level of uncertainty is very high; elements that we find in megaprojects. By comparing the performance of three judgmental methods, unaided judgment, semi-structured analogies (sSA), and interaction groups (IG), used by a group of 69 semi-experts, \cite{Litsiou2019-gy_Litsiou} found that, the use of sSA outperforms unaided judgment in forecasting performance (see also \S\ref{sec:panels_of_experts}). The difference is amplified further when pooling of analogies through IG is introduced.

\subsubsection[Competing products (Renato Guseo)]{Competing products\protect\footnote{This subsection was written by Renato Guseo.}}
\label{sec:Competing_products}
Competition among products or technologies affects prediction due to local systematic deviations and saturating effects related to policies, and evolving interactions. The corresponding sales time series must be jointly modelled including the time varying reciprocal influence. Following the guidelines in subsection \S\ref{sec:Synchronic_and_diachronic_competition}, some examples are reported below.

Based on IMS-Health quarterly number of cimetidine and ranitidine packages sold in Italy, the CRCD model \citep{guseomortarino:12_RGuseo} was tested to evaluate a diachronic competition that produced substitution. Cimetidine is a histamine antagonist that inhibits the production of stomach acid and was introduced by Smith, Kline \& French in 1976. Ranitidine is an alternative active principle introduced by Glaxo in 1981 and was found to have far-improved tolerability and a longer-lasting action. The main effect in delayed competition is that the first compound spread fast but was suddenly outperformed by the new one principle that modified its stand-alone regime. \citet{guseomortarino:12_RGuseo} give some statistical and forecasting comparisons with the restricted Krishnan-Bass-Kummar Diachronic model (KBKD) by \citet{krishnanbasskumar:00_RGuseo}. Previous results were improved with the UCRCD model in \citet{guseomortarino:14_RGuseo} by considering a decomposition of word-of-mouth (WOM) effects in two parts: within-brand and cross-brand contributions. The new active compound exploited a large cross-brand WOM and a positive within-brand effect. After the start of competition, cimetidine experienced a negative WOM effect from its own adopters and benefited from the increase of the category's market potential driven by the antagonist. Forecasting is more realistic with the UCRCD approach and it avoids mistakes in long-term prediction.

Restricted and unrestricted UCRCD models were applied in Germany by \citet{guidolin2016german_MG} to the competition between nuclear power technologies and renewable energy technologies (wind and solar; see also \S\ref{sec:Forecasting_renewable_energy_technologies}, \S\ref{sec:Wind_Power_orecasting} and \S\ref{sec:Solar_power_forecasting}) in electricity production. Due to the `Energiewende' policy started around 2000, the substitution effect, induced by competition, is confirmed by the electricity production data provided by BP\footnote{https://www.bp.com/en/global/corporate/energy-economics/statistical-review-of-world-energy.html (Accessed: 2020-09-01)}. An advance is proposed in \citet{furlanmz:20_RGuseo} with three competitors (nuclear power, wind, and solar technologies) and exogenous control functions obtaining direct inferences that provide a deeper analysis and forecasting improvements in energy transition context.

Previous mentioned intersections between Lotka-Volterra approach and diffusion of innovations competition models suggested a more modulated access to the residual carrying capacity. The Lotka-Volterra with churn model (LVch) by \citet{guidolinguseo:15_RGuseo} represents `churn effects' preserving within and cross-brand effects in a synchronic context. 

An application of LVch model is discussed with reference to the competition/substitution between compact cassettes and compact discs for pre-recorded music in the US market. Obtained results of LVch outperform restricted and unrestricted UCRCD analyses. In this context the residual market is not perfectly accessible to both competitors and this fact, combined with WOM components, allows for better interpretation and forecasting especially in medium and long-term horizons.

A further application of the LVch model, Lotka-Volterra with asymmetric churn (LVac), is proposed in \citet{guidolinguseo:20_RGuseo}. It is based on a statistical reduction: The late entrant behaves as a standard \citet{bass:69_RGuseo} model that modifies the dynamics and the evolution of the first entrant in a partially overlapped market. The case study is offered by a special form of competition where the iPhone produced an inverse cannibalisation of the iPad. The former suffered a local negative interaction with some benefits: A long-lasting life cycle and a larger market size induced by the iPad.

A limitation in models for diachronic competition relates to high number of rivals, implying complex parametric representations with respect to the observed information. A second limitation, but also an opportunity, is the conditional nature of forecasting if the processes partially depend upon exogenous control functions (new policy regulations, new radical innovations, regular and promotional prices, etc.). These tools may be used to simulate the effect of strategic interventions, but a lack of knowledge of such future policies may affect prediction.
 
\subsubsection[Forecasting under data integrity attacks (Priyanga Dilini Talagala)]{Forecasting under data integrity attacks\protect\footnote{This subsection was written by Priyanga Dilini Talagala.}}
\label{sec:Safety_and_security}
Data integrity attacks, where unauthorized parties access protected or confidential data and inject false information using various attack templates such as ramping, scaling, random attacks, pulse and smooth-curve, has become a major concern in data integrity control in forecasting \citep{giani2013smart_PDT, sridhar2014model_PDT, yue2017integrated_PDT, singer2014cybersecurity_PDT}.

Several previous studies have given attention in anomaly detection pre-processing step in forecasting workflow with varying degree of emphasis. However, according to \cite{yue2017integrated_PDT}, the detection of  data integrity attacks is very challenging as such attacks are done by highly skilled adversaries in a coordinated manner without notable variations in the historical data patterns \citep{liang2019poisoning_PDT}. These attacks can cause  over-forecasts  that demand unnecessary expenses for the upgrade and maintenance, and can eventually lead to poor planning and business decisions \citep{luo2018benchmarking_PDT, luo2018robust_PDT, wu2020data_PDT}.  

Short-term load forecasting (see \S\ref{sec:Load_forecasting}) is one major field that are vulnerable to malicious data integrity attacks as many  power industry functions  such as  economic dispatch, unit commitment and automatic generation control  heavily depend on accurate load forecasts \citep{liang2019poisoning_PDT}. The cyberattack on U.S. power grid in 2018 is one such major incident related to the topic. According to the  study conducted by  \cite{luo2018benchmarking_PDT}, the widely used load forecasting models fail to produce reliable load forecast in the presence of such large malicious data integrity attacks. A submission to the Global Energy Forecasting Competition 2014 (GEFCom2014) incorporated an anomaly detection pre-processing step with a fixed anomalous threshold  to their load forecasting framework \citep{xie2016gefcom2014_PDT}. The method was later improved by \cite{luo2018real_PDT} by replacing the fixed threshold with a data driven anomalous threshold. \cite{sridhar2014model_PDT} also proposed a general framework to detect scaling and ramp attacks in  power systems. \cite{akouemo2016probabilistic_PDT} investigated the impact towards the gas load forecasting using hybrid approach based on Bayesian maximum likelihood classifier and a forecasting model. In contrast to the previous model based attempts,  \cite{yue2019descriptive_PDT} proposed a descriptive analytic-based approach to detect cyberattacks including long anomalous sub-sequences (see \S\ref{sec:Anomaly_detection_and_time_series_forecasting}), that are difficult to detect by the conventional anomaly detection methods. 

The problem of data integrity attacks is not limited to load forecasting. Forecasting fields such as elections forecasting (see \S\ref{sec:Elections_forecasting}), retail forecasting (see \S\ref{sec:Retail_sales_forecasting}), airline flight demand forecasting (see \S\ref{sec:Forecasting_for_airports}) and stock price forecasting \S\ref{sec:Forecasting_stock_returns}) are also vulnerable to data integrity attacks  \citep{SEAMAN2018822, luo2018benchmarking_PDT}. For instant, \cite{wu2020data_PDT}  explored the vulnerability of traffic modelling and forecasting in the presence of data integrity attacks with the aim of providing useful guidance for constrained network resource planning and scheduling.

However, despite of the increasing attention toward the topic, advancements in cyberattacks on critical infrastructure raise further data challenges. Fooling existing anomaly detection algorithms via novel cyberattack templates is one such major concern. In response to the above concern, \cite{liang2019poisoning_PDT} proposed a data poisoning algorithm that can fool existing load forecasting approaches with anomaly detection component while demanding further investigation into advanced anomaly detection methods. Further, adversaries can also manipulate other related input data without damaging the target data series. Therefore, further research similar to \citep{sobhani2020temperature_PDT} are required to handle such data challenges.
 
\subsubsection[The forecastability of agricultural time series (Dimitrios Thomakos)]{The forecastability of agricultural time series\protect\footnote{This subsection was written by Dimitrios Thomakos.}}
\label{sec:The_Forecastability_of_agricultural_time_series}
The forecasting of agricultural time series falls under the broader group of forecasting commodities, of which agricultural and related products are a critical subset. While there has been considerable work in the econometrics and forecasting literature on common factor models in general there is surprisingly little work so far on the application of such models for commodities and agricultural time series -- and this is so given that there is considerable literature in the linkage between energy and commodities, including agricultural products, their prices and futures prices, their returns and volatilities. Furthermore, a significant number of papers is fairly recent which indicates that there are many open avenues of future research on these topics, and in particular for applied forecasting. The literature on the latter connection can consider many different aspects in modelling as we illustrate below. We can identify two literature strands, a much larger one on the various connections of energy with commodities and the agricultural sector (and in this strand we include forecasting agricultural series) and a smaller one that explores the issue of common factors. 

An early reference of the impact of energy on the agricultural sector is \cite{Tewari1990-wz_DT} and then after a decade we find \cite{Gohin2010-wy_DT} on the long-run impact of energy prices on global agricultural markets. \cite{Byrne2013-me_DT} is an early reference for co-movement of commodity prices followed by \cite{Daskalaki2014-az_DT} on common factors of commodity future returns and then a very recent paper from \cite{Alquist2020-mb_DT} who link global economic activity with commodity price co-movement. The impact of energy shocks on US agricultural productivity was investigated by \cite{Wang2014-yw_DT} while \cite{Koirala2015-ke_DT} explore the non-linear correlations of energy and agricultural prices with \cite{Albulescu2020-zq_DT} exploring the latter issue further, the last two papers using copulas. \cite{Xiong2015-of_DT} is an early reference of forecasting agricultural commodity prices while \cite{Kyriazi2019-md_DT}, \cite{Wang2019-mm_DT}, and \cite{Li2020-el_DT} consider three novel and completely different approaches on forecasting agricultural prices and agricultural futures returns. \cite{Lopez_Cabrera2016-kj_DT} explore volatility linkages between energy and agricultural commodity prices and then \cite{Tian2017-ck_DT} start a mini-stream on volatility forecasting on agricultural series followed among others by the work of \cite{Luo2019-jh_DT} and of \cite{Degiannakis2020-ya_DT}. \cite{Nicola2016-tu_DT} examine the co-movement of energy and agricultural returns while \cite{Kagraoka2016-je_DT} and \cite{Lubbers2016-gx_DT} examine common factors in commodity prices. \cite{Wei_Su2019-gv_DT} and \cite{Pal2019-rt_DT} both investigate the linkages of crude oil and agricultural prices. Finally, \cite{Tiwari2020-ru_DT} examine the time-frequency causality between various commodities, including agricultural and metals. 

There is clearly room for a number of applications in the context of this recent research, such along the lines of further identifying and then using common factors in constructing forecasting models, exploring the impact of the COVID-19 crisis in agricultural production or that of climate changes on agricultural prices. 

\subsubsection[Forecasting in the food and beverage industry (Daniele Apiletti)]{Forecasting in the food and beverage industry\protect\footnote{This subsection was written by Daniele Apiletti.}}
\label{sec:Forecasting_food_beverage} 
Reducing the ecological impact and waste, and increasing the efficiency of the food and beverage industry are currently major worldwide issues. To this direction, efficient and sustainable management of perishable food and the control of the beverage quality is of paramount importance. A particular focus on this topic is placed on supply chain forecasting (see \S\ref{sec:Forecasting_in_the_supply_chain}), with advanced monitoring technologies able to track the events impacting and affecting the food and beverage processes \citep{elsevier2019reducing_DA}. Such technologies are typically deployed inside manufacturing plants, yielding to Industry 4.0 solutions \citep{ieee2018industry4_Da} that are enabled by state-of-the-art forecasting applications in smart factories. The transition from plain agriculture techniques to smart solutions for food processing is a trend that fosters emerging forecasting data-driven solutions in many parts of the world, with special attention to the sustainability aspects \citep{malaysia2012sustainable_DA}.

Various forecasting approaches have been successfully applied in the context of the food and beverage industry, from Monte Carlo simulations based on a shelf-life model \citep{elsevier2019reducing_DA}, to association rule mining (see \S\ref{sec:Association_rule_mining}) applied to sensor-based equipment monitoring measurements \citep{ap2020correlating_DA}, multi-objective mathematical models for perishable supply chain configurations, forecasting costs, delivery time, and emissions \citep{mdpi2021multi_DA}, and  intelligent agent technologies for network optimisation in the food and beverage logistics management \citep{elsevier2005foodBeverage_DA}.

We now focus on the case of forecasting the quality of beverages, and particularly coffee. Espresso coffee is among the most popular beverages, and its quality is one of the most discussed and investigated issues. Besides human-expert panels, electronic noses, and chemical techniques, forecasting the quality of espresso by means of data-driven approaches, such as association rule mining, is an emerging research topic \citep{ap2020correlating_DA, sustain2021coffee_DA, ap2020timeseries_DA}.

The forecasting model of the espresso quality is built from a real-world dataset of espresso brewing by professional coffee-making machines. Coffee ground size, coffee ground amount, and 
water pressure have been selected among the most influential external variables. The ground-truth quality evaluation has been performed for each shot of coffee based on three well-known quality variables selected by domain experts and measured by specific sensors: 
the extraction time, the average flow rate, and the espresso volume. An exhaustive set of more than a thousand coffees has been produced to train a model able to forecast the effect of non-optimal values on the espresso quality. 

For each variable considered, different categorical values are considered: ground size can be coarse, optimal, or fine; ground amount can be high, optimal, or low; brewing water pressure can be high, optimal, or low. The experimental setting of categorical variables enables the application of association rule mining (see \S\ref{sec:Association_rule_mining}), a powerful data-driven exhaustive and explainable approach \citep{han2011data_DA, KumarDMBook_DA}, successfully exploited in different application contexts \citep{acquaviva2015enhancing_DA, di2018metatech_DA}.

Several interesting findings emerged. If the water pressure is low, the amount of coffee ground is too high, and the grinding is fine, then we can forecast with confidence a low-quality coffee due to excessive percolation time. If the amount of coffee ground is low, the ground is coarse, and the pressure is high, then we can forecast a low-quality coffee due to excessive flow rate. Furthermore, the coarseness of coffee ground generates an excessive flow rate forecast, despite the optimal values of dosage and pressure, with very high confidence.

\subsubsection[Dealing with logistic forecasts in practice (Theodore Modis)]{Dealing with logistic forecasts in practice\protect\footnote{This subsection was written by Theodore Modis.}}
\label{sec:Dealing_with_logistic_forecasts_in_practice}

\begin{figure}[ht!]
\begin{center}
\includegraphics[trim=125 475 125 75, clip, width=4in ]{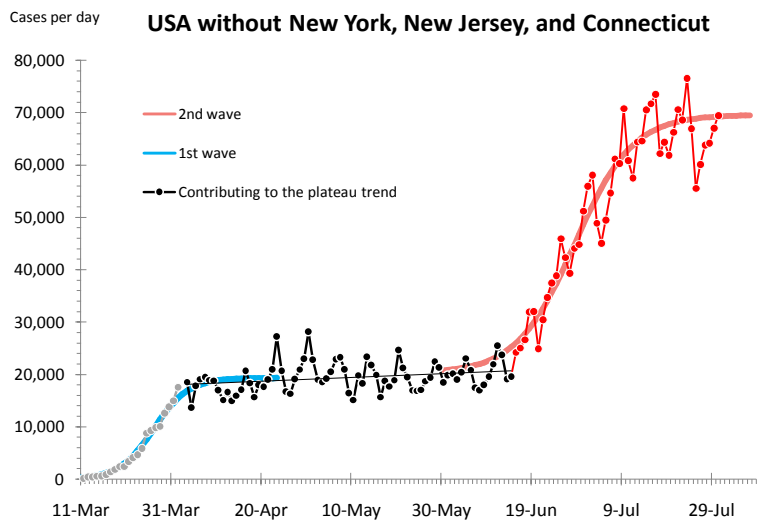}
\caption{Two logistic-growth steps during the early diffusion of COVID-19 in America (March to July, 2020).} 
\label{fig:modis_practice}
\end{center}
\end{figure}

The forecaster faces three major difficulties when using the logistic equation (S curve); see also \S\ref{sec:The_natural_law of growth_in_competition_Logistic_growth}. A first dilemma is whether he or she should fit an S curve to the cumulative number or to the number per unit of time. Here the forecaster must exercise wise judgment. What is the ``species'' and what is the niche that is being filled? To the frustration of business people there is no universal answer. When forecasting the sales of a new product it is often clear that one should fit the cumulative sales because the product's market niche is expected to eventually fill up. But if we are dealing with something that is going to stay with us for a long time (for example, the Internet or a smoking habit), then one should not fit cumulative numbers. At times this distinction may not be so obvious. For example, when COVID-19 first appeared many people (often amateurs) began fitting S curves to the cumulative number of infections (for other attempts on forecasting COVID-19, see \S\ref{sec:Pandemics}). Some of them were rewarded because indeed the diffusion of the virus in some countries behaved accordingly \citep{Debecker1994-zf}. But many were frustrated and tried to ``fix'' the logistic equation by introducing more parameters, or simply gave up on trying to use logistics with COVID 19. And yet, many cases (e.g., the US) can be illuminated by logistic fits but on the daily number of infections, not on the cumulative number. As of August 1, 2020, leaving out the three eastern states that had gotten things under control, the rest of the US displayed two classic S curve steps followed by plateaus (see figure \ref{fig:modis_practice}). The two plateaus reflect the number of infections that American society was willing to tolerate at the time, as the price to pay for not applying measures to restrict the virus diffusion.

The second difficulty in using the logistic equation has to do with its ability to predict from relatively early measurements the final ceiling. The crucial question is how early can the final ceiling be determined and with what accuracy. Some people claim that before the midpoint no determination of a final level is trustworthy \citep{Marinakis2021wp}. Forecasters usually abstain from assigning quantitative uncertainties on the parameters of their S curve forecasts mostly because there is no theory behind it. However, there is a unique study by \cite{Debecker2020wp} that quantifies the uncertainties on the parameters determined by logistic fits. The study was based on 35,000 S curve fits on simulated data, smeared by random noise and covering a variety of conditions. The fits were carried out via a $\chi^2$ minimisation technique. The study produced lookup tables and graphs for determining the uncertainties expected on the three parameters of the logistic equation as a function of the range of the S curve populated by data, the error per data point, and the confidence level required.

The third difficulty using the logistic equation comes from the fact that no matter what fitting program one uses, the fitted S curve will flatten toward a ceiling \textit{as early and as low} as it is allowed by the constraints of the procedure. As a consequence fitting programs may yield logistic fits that are often biased toward a low ceiling. Bigger errors on the data points accentuate this bias by permitting larger margins for the determination of the S curve parameters. To compensate for this bias the user must explore several fits with different weights on the data points during the calculation of the $\chi^2$. He or she should then favour the answer that gives the highest ceiling for the S curve (most often obtained by weighting more heavily the recent historical data points). Of course, this must be done with good justification; here again the forecaster must exercise wise judgment.

\subsection[The future of forecasting practice (Len Tashman)]{The future of forecasting practice\protect\footnote{This subsection was written by Len Tashman.}}
\label{sec:The_practice_of_forecasting_the_future}
\epigraph{Plus ça change, plus c’est la même chose.}{Jean-Baptiste Karr (1849)}

It would be a more straightforward task to make predictions about the future of forecasting practice if we had a better grasp of the present state of forecasting practice. For that matter, we lack even a common definition of forecasting practice. 
In a recent article, \cite{Makridakis2020foresight} lamented the failure of truly notable advances in forecasting methodologies, systems, and processes during the past decades to convince many businesses to adopt systematic forecasting procedures, leaving a wide swath of commerce under the guidance of ad hoc judgment and intuition. At the other extreme, we see companies with implementations that combine state-of-the-art methodology with sophisticated accommodations of computing time and costs, as well as consideration of the requirements and capabilities of a diverse group of stakeholders \citep{Yelland2019-mx-foresight}. So, it is not hyperbole to state that business forecasting practices are all over the place. What surely is hyperbole, however, are the ubiquitous claims of software providers about their products accurately forecasting sales, reducing costs, integrating functions, and elevating the bottom line \citep{Makridakis2020foresight,Sorensen2020-pq-foresight}.
For this section, we grilled a dozen practitioners and thought leaders (``the Group'') about developments playing out in the next decade of forecasting practice, and have categorised their responses:
\begin{itemize}[noitemsep]
 \item Nature of forecasting challenges;
 \item Changes in the forecasting toolbox;
 \item Evolution in forecasting processes such as integration of planning functions;
 \item Expectations of forecasters; and
 \item Scepticism about real change.
\end{itemize}

\noindent \textit{Forecasting Challenges}: Focusing on operations, the Group sees demand forecasting becoming ever more difficult due to product/channel proliferation, shorter lead times, shorter product histories, and spikes in major disruptions.
\begin{itemize}[noitemsep]
 \item Operational forecasts will have shorter forecast horizons to increase strategic agility required by business to compete, sustain, and survive. 
 \item New models will need to incorporate supply-chain disruption. Demand chains will need to be restarted, shortening historical data sets and making traditional models less viable due to limited history.
 \item Lead times will decrease as companies see the problems in having distant suppliers. Longer lead times make accurate forecasting more difficult.
\end{itemize}

\noindent \textit{Forecasting Tool Box}: Unsurprisingly, this category received most of the Group's attention. All predict greater reliance on AI/ML for automating supply-and-demand planning tasks and for reconciling discrepancies in hierarchical forecasting. Longer-horizon causal forecasting models will be facilitated by big data, social media, and algorithmic improvements by quantum computing. Post-COVID, we will see a greater focus on risk management/mitigation. The Cloud will end the era of desktop solutions.
\begin{itemize}[noitemsep]
 \item Quantum computers will improve algorithms used in areas like financial forecasting (e.g., Monte Carlo simulations), and will change our thinking about forecasting and uncertainty.
 \item Although social media is a tool for ``what's trending now'', new models will be developed to use social-media data to predict longer-term behaviour. Step aside Brown (exponential smoothing) and Bass (diffusion).
 \item Greater automation of routine tasks (data loading, scrubbing, forecast generation and tuning, etc.) through AI/ML-powered workflow, configurable limits, and active alerts. More black box under the hood, but more clarity on the dashboard.
 \item Greater focus on risk management/mitigation through what-if scenarios, simulations, and probabilistic forecasting.
\end{itemize}

\noindent \textit{Forecasting Processes and Functional Integration}: Systems will become more integrated, promoting greater collaboration across functional areas and coordination between forecast teams and those who rely upon them. Achieving supply-chain resilience will become as important as production efficiency, and new technology such as Alert and Root Cause Analysis systems will mitigate disruptions.
\begin{itemize}[noitemsep]
 \item S\&OP will expand from its home in operations to more fully integrate with other functions such as finance and performance management, especially in larger multinationals.
 \item The pandemic has forced firms to consider upping supply-chain resilience. Firms are building in capacity, inventory, redundancy into operations—somewhat antithetical to the efficiency plays that forecasting brings to the table.
 \item Forecasting will be more closely tied to Alert and Root Cause Analysis systems, which identify breakdowns in processes/systems contributing to adverse events, and prevent their recurrence.
\end{itemize}

\noindent \textit{Expectations of Forecasters}: Agreement was universal that the forecaster's job description will broaden and become more demanding, but that technology will allow some redirection of effort from producing forecasts to communicating forecasting insights.
\begin{itemize}[noitemsep]
 \item The interest around disease models increases our awareness of the strengths and weaknesses of mathematical models. Forecasters may need to become more measured in their claims, or do more to resist their models being exploited.
 \item We will see a transformation from demand planner to demand analyst, requiring additional skill sets including advanced decision making, data and risk analysis, communication, and negotiation.
 \item Professional forecasters will be rare except in companies where this expertise is valued. Fewer students are now educated or interested in statistical modelling, and time is not generally available for training.
 \item Forecasters will learn the same lesson as optimisation folks in the 1990s and 2000s: the importance of understanding the application area—community intelligence. 
\end{itemize}

\noindent \textit{Scepticism}: Many were sceptical about the current enthusiasm for AI/ML methods; disappointed about the slow adoption of promising new methods into software systems and, in turn, by companies that use these systems; and pessimistic about the respect given to and influence of forecasters in the company's decision making.
\begin{itemize}[noitemsep]
 \item While AI/ML are important additions to the forecaster's toolbox, they will not automatically solve forecasting issues. Problems include data hunger, capacity brittleness, dubious input data, fickle trust by users \citep{Kolassa2020-he-foresight}, and model bias.
 \item Practices in the next decade will look very similar to the present. Not that much has changed in the last decade, and academic developments are slow to be translated into practice.
 \item Politics, gaming, and the low priority given to forecasting are the prime drivers of practice, thus limiting interest in adopting new methodologies.
 \item None of the topical items (AI/ML, big data, demand sensing, new forecasting applications) will have much of an impact on forecasting practice. Forecasting departments hop from one trend to the other without making much progress towards better forecasting accuracy. 
 \item Software companies will struggle, despite good offerings. Most companies do not want to invest in excellent forecasting engines; whatever came with their ERP system is ``good enough''.
 \item Forecasting will continue to suffer from neglect by higher levels of management, particularly when forecasts are inconveniently contrary to the messages management hopes to convey.
\end{itemize}

Note finally that the COVID-19 pandemic has elevated practitioner concerns about disruptions to normal patterns as well as the fear of an increasingly volatile environment in which forecasts must be made. There are indications that companies will place more stress on judgmental scenarios, likely in conjunction with statistical/ML methods.

\clearpage

\section[Forecasting: benefits, practices, value, and limitations (Spyros Makridakis)]{Forecasting: benefits, practices, value, and limitations\protect\footnote{This subsection was written by Spyros Makridakis.}}
\label{sec:conclusions}
\epigraph{Mr. Buffett said his advice for the cash left to his wife was that 10 per cent should go to short-term government bonds and 90 per cent into a very low-cost S\&P 500 index fund.}

The purpose of this unique article is to provide an encyclopedic knowledge about the various aspects of forecasting. In this article, there are more than 140 sections and subsections, with more than 2,100 references, written by 80 of some of the best-known forecasting researchers and practitioners in the world, making it into a selective, encyclopedic piece covering, into a single source, a great deal of the available knowledge about the theory and practice of forecasting. We hope that this article will serve as an easy-to-use reference source. We aim to convert it into an online resource that will be regularly updated as new information becomes available.

But some people argue if there is any value in attempting to predict the future and if forecasting is any different than fortune telling, given the large numbers of mistaken forecasts made in the past, including our inability to accurately predict the progression of COVID-19 and its economic and human consequences? What is, therefore, the usefulness of a paper like the present one when crystal balling is not possible, and uncertainty reigns? It is the aim of this concluding article to set the record straight, explaining the benefits and practical value of forecasting while reporting its limitations too.

\textit{The Myriad of Forecasts}: All planning and the great majority of decisions we make require forecasting. Deciding what time to get up in the morning, not to be late for work implies a correct prediction of the commuting time to go to the office. Determining what to study is another decision requiring elaborate predictions about the demand for future jobs decades away. In the business world, firms must decide/forecast how many products to manufacture, the price they should be sold, how much money to spend on advertising and promotion, how much and in what type of new technologies to invests and a plethora of other future-oriented decisions requiring both predictions and assessing their inevitable uncertainty. Whether we like it or not, we have no choice but making these forecasts to benefit as much as possible from their value, knowing perfectly well that all predictions are uncertain while some may turn out to be wrong.

\textit{The Pervasiveness of Uncertainty}: Apart from some areas of hard sciences, all other forecasts are uncertain and must be accompanied with a measure of its magnitude, expressed as a prediction interval, or as a probability distribution around the most likely forecast. Although the value and usage of forecasts is clear, that of uncertainty is not. Worse, it becomes an unwelcome source of anxiety whose usefulness is misunderstood. Executives want to know the exact sales of their firm for next month to set up their production schedule. Instead, they are given prediction intervals (PIs) around such forecast and told that most of the time, sales will be within this interval, assuming the fluctuations follow some distributional assumptions. They argue that forecasting must decrease, not amplify, uncertainty and that the PIs are too wide and `uninformative' to be used for making practical business decisions. The trouble is that these PIs are based on past fluctuations and present the best estimation of future uncertainty, even if they seem too wide. Worse, empirical research has shown that they are too narrow, underestimating uncertainty often considerably.

\textit{Assessing Uncertainty and Dealing with its Implied Risks}: Uncertainty entails risks, requiring action to minimise their negative consequences. There are two kinds of uncertainty that can be illustrated by a commuting example. The first relates to fluctuations in the commuting time under normal driving conditions when there are no serious accidents, road works or major snowstorms. Such fluctuations are small and can be captured by a normal curve that allows to balance the risk of arriving earlier or later than the desired time. In the opposite case, uncertainty is fat-tailed and hard to estimate, as delays can be substantial depending upon the seriousness of the accident or that of the snowstorm while the risk of being early to work is smaller than being late. Moreover, such risk is substantially different when going to the airport to catch a flight, requiring starting much earlier than the average time it takes to go to the airport to minimise the risk of missing the flight.

\textit{More Accurate Ways of Forecasting and Assessing Uncertainty}: Extensive empirical research, including forecasting competitions, has shown that systematic approaches improve the accuracy of forecasting and the correct assessment of uncertainty resulting in substantial benefits when compared to ad-hoc judgmental alternatives \citep{Makridakis2020foresight}. The biggest advantage of such approaches is their ability to identify and estimate, in a mathematically optimal manner, past patterns and relationships that are subsequently extrapolated to predict their continuation, avoiding the over optimism and wishful thinking associated with judgmental approaches. At the same time, it must be clear that the accuracy of the forecasts and the correctness of uncertainty will depend on the established patterns/relationship not changing much during the forecasting period.

\textit{Using Benchmarks to Evaluate the Value of Forecasting}: The accuracy of the forecasts and the correct assessment of uncertainty must be judged not on their own but in comparison to some simple, readily available benchmarks. In stock market forecasts, for instance, the accuracy of predictions is compared to that of today’s price used as the forecast for future periods. Empirical comparisons have shown that such a benchmark beats the great majority of professional forecasters, hence Buffet’s advice in the epigram for his wife to invest in a low-cost index fund that selects stocks randomly. In weather forecasting, meteorologists are judged by the improvement of their forecasts over the naive prediction that tomorrow’s weather will be the same as today.

\textit{Concluding remark}: Accepting the advantages and limitations of systematic forecasting methods and most importantly avoiding any exaggerated expectations of what it can achieve is critical. Such methods do not possess any prophetic powers, they simply extrapolate established patterns and relationships to predict the future and assess its uncertainty. Their biggest advantage is their objectivity and ability for optimal extrapolation. Their biggest disadvantages are: (\textit{i}) the patterns and the relationships must remain fairly constant during the forecasting phase for the forecasts to be accurate, and (\textit{ii}) uncertainty must not be fat-tailed so it can be measured quantitatively. 

\clearpage

\section*{Disclaimer}
The views expressed in this paper are those of the authors and do not necessarily reflect the views of their affiliated institutions and organisations.

\section*{Acknowledgements}
Fotios Petropoulos would like to thank all the co-authors of this article for their very enthusiastic response and participation in this initiave. He would also like to thank Pierre Pinson for inviting this paper to be submitted to the \textit{International Journal of Forecasting}. The constructive comments and suggestions from this advisory board were vital in improving the paper. He also thanks Artur Tarassow for offering a list of Gretl's software functionalities.

Jakub Bijak's work received funding from the European Union's Horizon 2020 research and innovation programme, grant 870299 QuantMig: Quantifying Migration Scenarios for Better Policy.

Clara Cordeiro is partially financed by national funds through FCT – Fundação para a Ciência e a Tecnologia under the project UIDB/00006/2020.

Fernando Luiz Cyrino Oliveira acknowledges the support of the Coordination for the Improvement of Higher Level Personnel (CAPES) -- grant number 001, the Brazilian National Council for Scientific and Technological Development (CNPq) -- grant number 307403/2019-0, and the Carlos Chagas Filho Research Support Foundation of the State of Rio de Janeiro (FAPERJ) -- grant numbers 202.673/2018 and 211.086/2019.

Shari De Baets was funded by the FWO Research Foundation Flanders.

Joanne Ellison acknowledges the support of the ESRC FertilityTrends project (grant number ES/S009477/1) and the ESRC Centre for Population Change (grant number ES/R009139/1).

Piotr Fiszeder was supported by the National Science Centre project number 2016/21/B/HS4/00662 entitled ``Multivariate volatility models - the application of low and high prices''.

David T. Frazier has been supported by Australian Research Council (ARC) Discovery Grants DP170100729 and DP200101414, and ARC Early Career Researcher Award DE200101070.

Mariangela Guidolin acknowledges the support of the University of Padua, Italy, through the grant BIRD188753/18.

David F. Hendry gratefully acknowledges funding from the Robertson Foundation and Nuffield College.

Yanfei Kang acknowledges the support of the National Natural Science Foundation of China (number 11701022) and the National Key Research and Development Program (number 2019YFB1404600).

Stephan Kolassa would like to thank Tilmann Gneiting for some very helpful tips.

Gael M. Martin has been supported by Australian Research Council (ARC) Discovery Grants DP170100729 and DP200101414.

Alessia Paccagnini acknowledges the research support by COST Action ``Fintech and Artificial Intelligence in Finance - Towards a transparent financial industry'' (FinAI) CA19130.

Jose M. Pavía acknowledges the support of the Spanish Ministry of Science, Innovation and Universities and the Spanish Agency of Research, co-funded with FEDER funds, grant ECO2017-87245-R, and of Consellería d'Innovació, Universitats, Ciència i Societat Digital, Generalitat Valenciana -- grant number AICO/2019/053.

Diego J. Pedregal and Juan Ramon Trapero Arenas acknowledge the support of the European Regional Development Fund and Junta de Comunidades de Castilla-La Mancha (JCCM/FEDER, UE) under the project SBPLY/19/180501/000151 and by the Vicerrectorado de Investigación y Política Científica from UCLM through the research group fund program PREDILAB; DOCM 26/02/2020 [2020-GRIN-28770].

David E. Rapach thanks Ilias Filippou and Guofu Zhou for valuable comments.

J. James Reade and Han Lin Shang acknowledge Shixuan Wang for his constructive comments.

Michał Rubaszek is thankful for the financial support provided by the National Science Centre, grant No. 2019/33/B/HS4/01923 entitled ``Predictive content of equilibrium exchange rate models''.

\clearpage

\appendix

\section{List of acronyms}\label{sec:acronyms}

\noindent ABC: Approximate Bayesian Computation \\
ACC: Ant Colony Clustering \\
ACD: Autoregressive Conditional Duration \\
ADIDA: Aggregate–Disaggregate Intermittent Demand Approach \\
ADL: Autoregressive Distributed Lag \\
ADMM: Alternating Direction Method of Multipliers \\
AI: Artificial Intelligence \\
AIC: Akaike's Information Criterion \\
AICc: Akaike's Information Criterion corrected (for small sample sizes) \\
ANACONDA: Analysis of National Causes of Death for Action \\
ANFIS: Adaptive Neuro-Fuzzy Inference System \\
ANN: Artificial Neural Network \\
AO: Additive Outlier \\
AR: AutoRegressive (model) \\
ARX: AutoRegressive with eXogenous variables (model) \\ 
ARCH: AutoRegressive Conditional Heteroskedasticity \\
ARMA: AutoRegressive-Moving Average (model) \\
ARIMA: AutoRegressive Integrated Moving Average (model) \\
ARIMAX: AutoRegressive Integrated Moving Average with eXogenous variables (model) \\
B\&M: Brick and Mortar \\
BATS: Box-Cox transform, ARMA errors, Trend, and Seasonal components (model) \\
BEER: Behavioural Equilibrium Exchange Rate \\
BEKK: Baba-Engle-Kraft-Kroner GARCH \\
BEKK-HL: BEKK High Low \\
BIC: Bayesian Information Criterion \\
BLAST: Building Loads Analysis and System Thermodynamics \\
BM: Bass Model \\
BMC: Bootstrap Model Combination \\
BPNN: Back-Propagation Neural Network \\
CARGPR: Conditional AutoRegressive Geometric Process Range (model) \\
CARR: Conditional AutoRegressive Range (model) \\
CARRS: Conditional AutoRegressive Rogers and Satchell (model) \\
CBC: Choice Based Conjoint (analysis) \\
CBO: Congressional Budget Office \\
CDS: Credit Default Swap \\
CNN: Convolutional Neural Network \\
COVID-19: Coronavirus disease 2019 \\
CVAR: Cointegrated Vector AutoRegressive (model) \\
CAViaR: Conditional AutoRegressive Value At Risk \\
CL: Cross Learning \\
CPFR: Collaborative Planning, Forecasting, and Replenishment \\
CRCD: Competition and Regime Change Diachronic (model) \\
CRPS: Continuous Ranked Probability Score
CSR: Complete Subset Regression \\
CV: Cross-Validation \\
DA: Deterministic Annealing \\
DC: Distribution Centre \\
DCC: Dynamic Conditional Correlation \\
DCC-RGARCH: Range GARCH DCC \\
DFM: Dynamic Factor Model \\
DGP: Data Generating Process \\
DJI: Dow Jones Industrial
DM: Diebold-Mariano (test) \\
DSGE: Dynamic Stochastic General Equilibrium \\
DSHW: Double Seasonal Holt-Winters \\
DSTCC-CARR: Double Smooth Transition Conditional Correlation CARR \\
DT: Delay Time \\
EEG: ElectroEncephaloGram \\
EGARCH: Exponential GARCH \\
EH: Expectations Hypothesis \\
EMD: Empirical Mode Decomposition \\
ENet: Elastic Net \\
ENSO: El Ni\~no Southern Oscillation \\
ERCOT: Electric Reliability Council of Texas \\
ES: Expected Shortfall \\
ESP-r: Environmental Systems Performance -- research \\
ESTAR: Exponential STAR
ETS: ExponenTial Smoothing (or Error, Trend, Seasonality) \\
EVT: Extreme Value Theory \\
EWMA: Exponentially Weighted Moving Average \\
FAR: Functional AutoRegressive (model) \\
FASSTER: Forecasting with Additive Switching of Seasonality, Trend and Exogenous Regressors \\
FCM: Fuzzy C-Means \\
FIGARCH: Fractionally Integrated GARCH \\
FIS: Fuzzy Inference System \\
FFNN: Feed-Forward Neural Network \\
FFORMA: Feature-based FORecast Model Averaging \\
FMCG: Fast Moving Consumer Goods \\
FPCA: Functional Principal Component Analysis \\
FRB/EDO: Federal Reserve Board's Estimated, Dynamic, Optimisation-based (model) \\
FSS: Forecasting Support System \\
FVA: Forecast Value Added \\
GARCH: General AutoRegressive Conditional Heteroscedasticity \\
GARCH-PARK-R: GARCH PARKinson Range \\
GARCH-TR: GARCH True Range \\
GB: Givon-Bass (model) \\
GBM: Generalised Bass Model \\
GDP: Gross Domestic Product \\
GGM: Guseo-Guidolin Model (GGM) \\
GJR-GARCH: Glosten-Jagannathan-Runkle GARCH \\
GM: Generalised M-estimator \\
GMM: Generalised Methods of Moments \\
GPU: Graphics Processing Unit \\
GRNN: Generalised Regression Neural Network \\
HAR: Heterogeneous AutoRegressive (model) \\
HDFS: Hadoop Distributed File System \\
HMD: Human Mortality Database \\
HP: Hodrick-Prescott \\
HPU: House Price Uncertainty \\
HVAC: Heating, Ventilation, and Air Conditioning (system) \\
HAR: Heterogeneous AutoRegressive (model) \\
HQ: Hannan-Quinn \\
IEA: International Energy Agency \\
IG: Interaction Groups \\
iid: independent and identically distributed \\
IIS: Impulse Indicator Saturation \\
IO: Innovation Outlier \\
IT: Information Technology \\
KBKD: Krishnan-Bass-Kummar Diachronic (model) \\
KISS: Keep It Simple, Stupid (principle) \\
$k$NN: $k$ Nearest Neighbours \\
KPSS: Kwiatkowski–Phillips–Schmidt–Shin \\
L-IVaR: Liquidity-adjusted Intraday Value-at-Risk \\
LASSO: Least Absolute Shrinkage and Selection Operator \\
LH: Low and High \\
LLN: Law of Large Numbers \\
LN-CASS: Logit-Normal Continuous Analogue of the Spike-and-Slab \\
LS (or LPS): Logarithmic (Predictive) Score (log-score) \\
LSTAR: Logistic STAR \\
LSTM: Long Short-Term Memory Networks \\
LTLF: Long-Term Load Forecasting \\
LV: Lotka-Volterra (model) \\
LVac: Lotka-Volterra with asymmetric churn (model) \\
LVch: Lotka-Volterra with churn (model) \\
MAE: Mean Absolute Error \\
MAPE: Mean Absolute Percentage Error \\
MASE: Mean Absolute Scaled Error \\
MCMC: Markov Chain Monte Carlo \\
MFI: Marginal Forecast Interval \\
MICE: Meetings, Incentives, Conventions, and Exhibitions/Events
MIDAS: MIxed DAta Sampling \\
MJO: Madden Julian Oscillation \\
ML: Machine Learning \\
MLP: MultiLayer forward Perceptron \\
MLR: Multiple Linear Regression \\
MS: Markov Switching \\
MS VAR: Markov Switching VAR \\
MSARIMA: Multiple/Multiplicative Seasonal ARIMA \\
MSC: Multiple Seasonal Cycles \\
MSE: Mean Squared Error \\
MSRB: Markov-Switching Range-Based \\
MTA: Multiple Temporal Aggregation \\
MTLF: Medium-Term Load Forecasting \\
NAO: North Atlantic Oscillation \\
NLS: Nonlinear Least Squares \\
NLTK: Natural Language Toolkit \\
NMAE: Normalised Mean Absolute Error \\
NN: Neural Network \\
NNAR: Neural Network AutoRegressive \\
NOB: Non-Overlapping Blocks \\
NPF: New Product Forecasting \\
NWP: Numerical Weather Prediction \\
OB: Overlapping Blocks \\
OBR: Office for Budget Responsibility \\
ODE: Ordinary Differential Equations \\
OLS: Ordinary Least Squares \\
OWA: Overall Weighted Average \\
PAR: Periodic AutoRegressive (model) \\
PCA: Principal Components Analysis \\
pdf: probability density function \\
PdM: Predictive Maintenance \\
PFEM: Point Forecast Error Measure \\
PHANN: Physical Hybrid Artificial Neural Network \\
pHDR: predictive Highest Density Region \\
PI: Prediction Interval \\
PIT: Probability Integral Transform \\
PL: Product Level \\
PLS: Partial Least Squares \\
PM: Particulate Matter \\
POT: Peak Over Threshold \\
PPP: Purchasing Power Parity \\
PSO: Particle Swarm Intelligence \\
PV: PhotoVoltaic \\
$Q_\alpha$: quantile score or pinball loss for a level $\alpha \in (0, 1)$ \\
RAF: Royal Air Force (UK) \\
RB: Range-Based \\
RB-copula: Range-Based copula \\
RB-DCC: Range-Based DCC \\
RB-MS-DCC: Range-Based Markov-Switching DCC \\
RBF: Radial Basis Function \\
REGARCH: Range-Based Exponential GARCH \\
RET: Renewable Energy Technology \\
RGARCH: Range GARCH \\
RMSE: Root Mean Squared Error \\
RNN: Recurrent Neural Network \\
RR-HGADCC: Return and Range Heterogeneous General Asymmetric DCC \\
RTV: Real Time Vintage \\
SA: Structured Analogies \\
SARIMA: Seasonal AutoRegressive Integrated Moving Average (model) \\
SARIMAX: Seasonal AutoRegressive Integrated Moving Average with eXogenous variables \\
SARMA: Seasonal AutoRegressive Moving Average (model) \\
SARMAX: Seasonal AutoRegressive Moving Average with eXogenous variables \\
SBA: Syntetos-Boylan Approximation \\
SBC: Syntetos-Boylan-Croston (classification) \\
SEATS: Seasonal Extraction in ARIMA Time Series \\
SES: Simple (or Single) Exponential Smoothing \\
SETAR: Self-Exciting Threshold AutoRegressive (model) \\
SFI: Simultaneous Forecast Interval \\
SKU: Stock Keeping Unit \\
SGD: Stochastic Gradient Descent \\
SIS: Step Indicator Saturation \\
SL: Serial number Level \\
SMA: Simple Moving Average \\
sMAPE: symmetric Mean Absolute Percentage Error \\
SOM: Self-Organising Map \\
SS: State Space \\
sSA: semi-Structured Analogies \\
SSARIMA: Several Seasonalities (or State Space) ARIMA \\
STAR: Smooth Transition AutoRegressive (model) \\
STARR: Smooth Transition conditional AutoRegressive Range (model) \\
STL: Seasonal Trend decomposition using Loess \\
STLF: Short-Term Load Forecasting \\
STR: Seasonal-Trend decomposition based on Regression \\
SV: Stochastic Volatility \\
SVA: Stochastic Value Added \\
SVD: Singular Value Decomposition \\
SVM: Support Vector Machine \\
SWAN: Simulating WAves Nearshore \\
S\&OP: Sales and Operations Planning \\
S\&P: Standard \& Poor's \\
TAR: Threshold AutoRegressive (model) \\
TARMA: Threshold AutoRegressive Moving Average (model)
TARMASE: Threshold AutoRegressive Moving Average (model)\\
TARR: Range-Based Threshold conditional AutoRegressive (model) \\
TBATS: Exponential Smoothing state space model with Box-Cox transformation, ARMA errors, Trend and Seasonal components \\
TFR: Total Fertility Rate \\
TGARCH: Threshold GARCH \\
TMA: Threshold Moving Average (model) \\
TPU: Tensor Processing Unit \\
TRAMO: Time series Regression with ARIMA noise, Missing values and Outliers \\
TSB: Teunter-Syntetos-Babai (method) \\
UCRCD: Unbalanced Competition and Regime Change Diachronic (model) \\
UIP: Uncovered Interest Party \\
VaR: Value at Risk \\
VAR: Vector AutoRegressive (model) \\
VARX: VAR with eXogenous variables (model) \\
VARMA: Vector AutoRegressive Moving Average (model) \\
VAT: Value Added Tax \\
VARIMAX: Vector AutoRegressive Integrated Moving Average with eXogenous variables (model) \\
VECM (or VEC): Vector Error Correction Model \\
VEqCM: Vector Equilibrium-Correction Model \\
VSTLF: Very Short-Term Load Forecasting \\
WLS: Weighted Least Squares \\
WNN: Wavelet Neural Network \\
WOM: Word-Of-Mouth \\
WW2: World War 1 \\
WW2: World War 2 \\
WW3: World War 3 \\
XGBoost: eXtreme Gradient Boosting \\

\clearpage

\begin{landscape}
\section{Software}\label{sec:software}

\begin{center}
Table A.1: A list of indicative free or open-source packages, libraries, and toolboxes linking to the theory sections of this article. The authors assume no liability for the software listed below; interested users are strongly advised to read the respective documentations and licences terms.

\end{center}
\scriptsize
\begin{longtable}[l]{ p{.34\textwidth} p{.12\textwidth} p{.16\textwidth} p{.20\textwidth} p{.50\textwidth} }
\hline
\textbf{Related} \textbf{section} & \textbf{Software} & \textbf{Package/Library/Toolbox} & \textbf{Function(s)} & \textbf{Comments} \\
\hline
\S\ref{sec:BoxCox_Transformations}. Box-Cox transformations & R & \textit{forecast} & BoxCox; InvBoxCox; BoxCox.lambda; & Functions to transform the input variable using a Box-Cox transformation, reverse the transformation and find optimal parameters. \url{https://cran.r-project.org/package=forecast} \\
\S\ref{sec:time_series_decomposition}. Box-Cox transformations Time series decomposition	& R	& \textit{stats}	& decompose; stl	& Classical decomposition method (additive and multiplicative), and STL decomposition method. \url{https://stat.ethz.ch/R-manual/R-devel/library/stats/html/00Index.html} \\
	& R	& \textit{forecast}	& seasadj; seasonal; mstl; msts; tbats.components	& Tools for extracting components, and multiple seasonal decomposition methods. https://cran.r-project.org/package=forecast \\
	& R	& \textit{tsutils}	& decomp; seasplot	& Classical decomposition method, and functions for seasonal plots. \url{https://cran.r-project.org/package=tsutils}\\
	& R	& \textit{stR}	& AutoSTR; STR; heuristicSTR; plot.STR; seasadj.STR	& Seasonal-Trend decomposition based on Regression. \url{https://cran.r-project.org/package=stR} \\
	& R	& \textit{seasonal}	& seas	& Functions for X-11, SEATS, and X-13-ARIMA-SEATS decomposition methods. \url{https://cran.r-project.org/package=seasonal} \\
	& Gretl	& \textit{buys\_ballot}		& & Plot seasonal time series components. \url{http://gretl.sourceforge.net/} \\
& Gretl	& \textit{season\_plot}	& set\_season\_plot	& Plot seasonal time-series components. \url{http://gretl.sourceforge.net/}\\
& Gretl	& \textit{tsfcst} &	decompcl	& Classical time series decomposition. \url{http://gretl.sourceforge.net/}\\
& Gretl	& \textit{StrucTiSM}	& STSM\_components	& Decomposition using structural timeseries model. \url{http://gretl.sourceforge.net/} \\
\S\ref{sec:Anomaly_detection_and_time_series_forecasting}. Anomaly detection and time series forecasting	& R	& \textit{anomalize}	& time\_decompose; anomalize; time\_recompose	& A ``tidy'' workflow for detecting anomalies in data. \url{https://cran.r-project.org/package=anomalize}\\
	& R	& \textit{oddstream}	& find\_odd\_streams; extract\_tsfeatures; set\_outlier\_threshold	& Early detection of anomalous series within a large collection of streaming time series data. \url{https://cran.r-project.org/package=oddstream} \\
	& R	& \textit{tsoutliers}	& tso; locate.outliers.oloop; remove.outliers 	& Detection of outliers in time series such as Innovational outliers, additive outliers, level shifts, temporary changes and seasonal level shifts. \url{https://cran.r-project.org/package=tsoutliers}\\
	& R	& \textit{stray}	& find\_HDoutliers; find\_threshold; display\_HDoutliers	& Anomaly detection in high dimensional and temporal data. \url{https://cran.r-project.org/package=stray}\\
	& R	& \textit{forecast}	& tsoutliers; tsclean	& Provides some simple heuristic methods for identifying and correcting outliers. \url{https://cran.r-project.org/package=seasonal} \\
	& R	& \textit{OutliersO3}	& OutliersO3; O3plotM; O3plotT; O3prep	& Draws overview of outliers (O3) Plots. \url{https://cran.r-project.org/package=OutliersO3}\\
   & R	& CRAN Task View & Anomaly Detection with R		& Contains a list of R packages that can be used for anomaly detection. \url{https://github.com/pridiltal/ctv-AnomalyDetection}\\
	& Gretl	& \textit{tramolin}	& & Outlier detection/correction and missing data interpolation. \url{http://gretl.sourceforge.net/} \\
\hline
\textbf{Related} \textbf{section} & \textbf{Software} & \textbf{Package/Library/Toolbox} & \textbf{Function(s)} & \textbf{Comments} \\
\hline
\S\ref{sec:Robust_handling_of_outliers}. Robust handling of outliers in time series forecasting	& R	& \textit{gets}	& isat &	Function for running impulse and step indicator saturation. \url{https://cran.r-project.org/package=gets} \\
\S\ref{sec:exponential_smoothing_models}. Exponential smoothing models	& R	& \textit{forecast}	& ets; forecast.ets; ses; 	& Functions for simple exponential smoothing and automatic exponential smoothing modelling. \url{https://cran.r-project.org/package=forecast}\\
& R	& \textit{smooth}	& es	& Function for automatic exponential smoothing modelling. \url{https://cran.r-project.org/package=smooth}\\
& Gretl	& \textit{tsfcst}	& expsmpars	& Simple exponential smoothing minimising the sum of squared errors. \url{http://gretl.sourceforge.net/}\\
\S\ref{sec:time_series_regression_models}. Time-series regression models	& R	& \textit{stats}	& lm	& Fitting linear regression models. \url{https://stat.ethz.ch/R-manual/R-devel/library/stats/html/00Index.html} \\
& R	& \textit{leaps}	& regsubsets	& Functions for selecting linear regression models. \url{https://cran.r-project.org/package=leaps}\\
& R	& \textit{relaimpo}		& & Relative importance of regressors in linear models. \url{https://cran.r-project.org/package=relaimpo}\\
& R	& \textit{MASS}	& stepAIC 	& Choose a model by AIC in a stepwise algorithm. \url{https://cran.r-project.org/package=MASS}\\
& Gretl	& & ols, lad, midasreg	& Regression models with OLS, LAD, and MIDAS  with functionality for forecasting. \url{http://gretl.sourceforge.net/} \\
\S\ref{sec:Theta_method_and_models}. Theta method and models	& R	& \textit{forecast}	& thetaf	& Returns forecasts and prediction intervals for a theta method forecast.  \url{https://cran.r-project.org/package=forecast} \\
& R	& \textit{forecTheta}	& stheta; stm; otm; dstm; dotm	& Functions for forecasting univariate time series using Theta Models. \url{https://cran.r-project.org/package=forecTheta}\\
	& R	& \textit{tsutils}	& theta	& Estimate Theta method. \url{https://cran.r-project.org/package=tsutils}\\
& Gretl	& \textit{tsfcst}	& stheta	& Theta-method for univariate forecasting. \url{http://gretl.sourceforge.net/} \\
\S\ref{sec:autoregressive_integrated_moving_average_models}. Autoregressive integrated moving average (ARIMA) models	& R	& \textit{forecast}	& auto.arima; Arima; arfima; arima.errors; arimaorder	& Functions for fitting and forecasting with ARIMA models. \url{https://cran.r-project.org/package=forecast} \\
	& R	& \textit{smooth}	& auto.msarima; auto.ssarima; msarima; ssarima	& State-space and multiple seasonalities implementations of ARIMA models. \url{https://cran.r-project.org/package=smooth}\\
& Gretl		& & arima	& Functions for fitting and forecasting with SARIMAX models. \url{http://gretl.sourceforge.net/} \\
& Gretl	& \textit{auto\_arima}	& 	& Find best fitting SARIMAX model with functions for forecasting. \url{http://gretl.sourceforge.net/} \\
& Gretl	& \textit{armax}	&	& Automatically determine the best ARMAX model. \url{http://gretl.sourceforge.net/} \\
& R	& \textit{aTSA}	& adf.test	& Augmented Dickey-Fuller test. \url{https://cran.r-project.org/package=aTSA}\\
& R	& \textit{tseries}	& kpss.test	& Kwiatkowski-Phillips-Schmidt-Shin (KPSS) test. \url{https://cran.r-project.org/package=tseries}\\
& R	& \textit{forecast}	& ndiffs; nsdiffs	& Estimates the number of (seasonal) differences required to make a given time series stationary. \url{https://cran.r-project.org/package=forecast} \\
& Gretl	& & adf; adf-gls; kpss; levinlin &	Various unit-root tests for time-series and panel data. \url{http://gretl.sourceforge.net/} \\
\hline
\\ \\
\hline
\textbf{Related} \textbf{section} & \textbf{Software} & \textbf{Package/Library/Toolbox} & \textbf{Function(s)} & \textbf{Comments} \\
\hline
\S\ref{sec:autoregressive_integrated_moving_average_models}. Autoregressive integrated moving average (ARIMA) models (continued) & Gretl	& \textit{DHF\_test}	& & Package for Dickey-Hasza-Fuller seasonal Unit Root Test. \url{http://gretl.sourceforge.net/} \\
& Gretl	& \textit{DP}	& & Package for testing for a double unit root. \url{http://gretl.sourceforge.net/} \\
& Gretl	& \textit{GHegy}	& & Package Seasonal unit roots tests. \url{http://gretl.sourceforge.net/} \\
& Gretl	& \textit{Kapetanios}	& & Package for Kapetanios' unit root test with possible structural breaks. \url{http://gretl.sourceforge.net/} \\
& Gretl	& \textit{PPtest}	& & Package for running Phillips-Perron unit root test. \url{http://gretl.sourceforge.net/} \\
& Gretl	& \textit{VSG\_test}& & Package for test proposed by Ventosa-Santaulària and Gómez-Zaldívar. \url{http://gretl.sourceforge.net/} \\
\S\ref{sec:forecasting_for_multiple_seasonal_cycles}. Forecasting for multiple seasonal cycles	& R	& \textit{smooth}	& msarima; ssarima	& Functions for forecasting data with multiple seasonal cycles. \url{https://cran.r-project.org/package=smooth}\\
& R	& \textit{fable}	& model; forecast; fasster; ETS; ARIMA; TSLM	& Forecasting models for tidy time series. \url{https://cran.r-project.org/package=fable}\\
& R, Python	& \textit{prophet} & 	& Facebook's automatic forecasting procedure. \url{https://cran.r-project.org/package=prophet}\\
	& R	& \textit{tidymodels}		& & Collection of packages for modelling and machine learning using tidyverse principles. \url{https://cran.r-project.org/package=tidymodels}\\
	& R	& \textit{forecast}	& tbats; dshw	& Functions for forecasting data with multiple seasonal cycles. \url{https://cran.r-project.org/package=forecast} \\
	& R	& \textit{fable.prophet} 	& prophet; forecast	& A tidy R interface to the prophet forecasting procedure using fable. \url{https://github.com/mitchelloharawild/fable.prophet}\\
\S\ref{sec:state_space_models}. State-space models	& Matlab	& \textit{SSpace}		& & General modelling of linear, non-linear and non-Gaussian State Space systems. \url{https://github.com/djpedregal/SSpace} \\
& Matlab	& \textit{SSM}	& & 	General modelling of linear, non-linear and non-Gaussian State Space systems. \url{https://www.mathworks.com/help/econ/ssm-class.html}\\
	& Matlab	& \textit{SSMMATLAB}		& & A Set of MATLAB Programs for the Statistical Analysis of State Space Models. \url{https://github.com/vgomezenriquez/ssmmatlab}\\
	& Matlab	& \textit{E4}	& & 	A MATLAB toolbox for time series analysis in State Space form. \url{https://www.ucm.es/e-4} \\
	& R	& \textit{UComp}	& & 	Automatic identification of Unobserved Components models in State Space form. \url{https://cran.r-project.org/package=UComp}\\
  & R	& \textit{statespacer}	& & 	State Space modelling, mainly ARIMA and Basic Structural Models. \url{https://cran.r-project.org/package=statespacer}\\
  & R	& \textit{smooth}	& & 	Forecasting using single error State Space models. \url{https://cran.r-project.org/package=smooth}\\
	& R	& \textit{bssm}	& & 	Bayesian Inference of Non-Gaussian State Space Models. \url{https://cran.r-project.org/package=bssm}\\
	& R	& \textit{mssm}	& & 	Multivariate State Space models. \url{https://cran.r-project.org/package=mssm}\\
	& R	& \textit{KFAS}	& & 	Kalman Filter and Smoother for Exponential Family State Space Models. \url{https://cran.r-project.org/package=KFAS}\\
\hline
\\ \\
\hline
\textbf{Related} \textbf{section} & \textbf{Software} & \textbf{Package/Library/Toolbox} & \textbf{Function(s)} & \textbf{Comments} \\
\hline
\S\ref{sec:state_space_models}. State-space models (continued)	& R	& \textit{TSSS}	& & 	Time Series Analysis with State Space Model, based on the methods in Kitagawa (1993). \url{https://cran.r-project.org/package=TSSS}\\
	& R	& \textit{dlm}	& & 	Bayesian and Likelihood Analysis of Dynamic Linear Models (Gaussian State Space models). \url{https://cran.r-project.org/package=dlm}\\
	& Python	& \textit{statsmodels} & 	statespace	& Time Series Analysis by State Space Methods. \url{https://www.statsmodels.org/stable/index.html} \\
& Gretl	& & 	kfilter; ksmooth; kdsmooth; ksimul	& State Space Modeling functionality with function for forecasting. \url{http://gretl.sourceforge.net/}\\
& Gretl	& \textit{StrucTiSM}	& STSM\_fcast	& Harvey-style Structural Time Series Models with function for forecasting. \url{http://gretl.sourceforge.net/}\\
\S\ref{sec:Models_for_population_processes}. Models for population processes & R & dembase & & General-purpose tools for demographic analysis. \url{https://github.com/StatisticsNZ/dembase}\\
& R & demest & & Bayesian statistical methods for demography. \url{https://github.com/StatisticsNZ/demest}\\
& R & demlife & & Creating and working with life tables. \url{https://github.com/StatisticsNZ/demlife}\\
 & R & BayesPop & & Generating population projections for all countries of the world using several probabilistic components, such as total fertility rate and life expectancy. \url{https://cran.r-project.org/package=bayesPop}\\
& R & bayesTFR & & Making probabilistic projections of total fertility rate for all countries of the world, using a Bayesian hierarchical model. \url{https://cran.r-project.org/package=bayesTFR}\\
& R & bayesLife & & Making probabilistic projections of life expectancy for all countries of the world, using a Bayesian hierarchical model. \url{https://cran.r-project.org/package=bayesLife}\\
& Spreadsheet & DAPPS & & Demographic Analysis and Population Projection System: Standalone spreadsheet-based software for demographic estimation and projections, prepared by the US Census Bureau. \url{https://www.census.gov/data/software/dapps.Overview.html}\\
\S\ref{sec:Forecasting_with_many_variables}. Forecasting with many variables	& R	& \textit{gets}	& getsm, getsv, isat, isatvar	& Package that implements general to specific model selection, indicator saturation, with functionality for forecasting. \url{https://cran.r-project.org/package=gets}\\
& R	& \textit{vars}	& & 	Functions and routines for VAR Modelling. \url{https://cran.r-project.org/package=vars}\\
& Gretl		& & var; system	& Fitting system-models with functionality for forecasting. \url{http://gretl.sourceforge.net/}\\
\S\ref{sec:Functional_time_series_models}. Functional time series models	& R	& \textit{ftsa}	& ftsm; farforecast; T\_stationarity	& Functional time series analysis. \url{https://cran.r-project.org/package=ftsa}\\
\S\ref{sec:arch_garch_models}. ARCH/GARCH models	& R	& \textit{tseries} & garch		& Fit GARCH models to time series. \url{https://cran.r-project.org/package=tseries}\\
& Python	& \textit{PyFlux}	& & 	Time series analysis and prediction tools that focus on autoregressive methods (ARIMA, ARCH, GARCH, etc). \url{https://pyflux.readthedocs.io/en/latest/index.html}\\
& Gretl		& & arch, garch	& Fit (G)ARCH models to time series. \url{http://gretl.sourceforge.net/}\\
\hline
\textbf{Related} \textbf{section} & \textbf{Software} & \textbf{Package/Library/Toolbox} & \textbf{Function(s)} & \textbf{Comments} \\
\hline
\S\ref{sec:arch_garch_models}. ARCH/GARCH models (continued) & Gretl	& \textit{gig}	& gig\_estimate; gig\_var\_fcast	& Estimate various types of GARCH models. \url{http://gretl.sourceforge.net/} \\
\S\ref{sec:Markov_switching_models}. Markov switching models	& R	& \textit{MSwM}	& 	& Fitting Markov switching models. \url{https://cran.r-project.org/package=MSwM}\\
	& R	& \textit{NHMSAR}	& & 	Non-homogeneous Markov switching autoregressive models. \url{https://cran.r-project.org/package=NHMSAR}\\
\S\ref{sec:threshold_models}. Threshold models	& R	& \textit{TAR}	& & 	Bayesian modelling of autoregressive threshold time series models. \url{https://cran.r-project.org/package=TAR}\\
	& R	& \textit{TSA} & 	tar; star & 	Functions for threshold models (and general time series analysis). \url{https://cran.r-project.org/package=TSA}\\
& Gretl	& \textit{Threshold\_Panel}	& THRESH\_SETUP	& Hansen's panel threshold model. \url{http://gretl.sourceforge.net/}\\
& Gretl	& SETAR		& & Estimation of a SETAR model. \url{http://gretl.sourceforge.net/}\\
\S\ref{sec:Forecasting_with_DSGE_Models}. Forecasting with DSGE models	& R	& \textit{BMR}	& forecast	& Bayesian Macroeconometrics in R (BMR) is a package for estimating and forecasting Bayesian VAR and DSGE. \url{https://github.com/kthohr/BMR} \\
	& Matlab/GNU Octave	& \textit{Dynare} 	& & 	Software platform for solving, estimating, and making forecasts with DSGE. \url{https://www.dynare.org/}\\
\S\ref{sec:Innovation_diffusion_models}. Innovation diffusion models	& R	& \textit{DIMORA}	& & 	Estimation of Bass Model, Generalised Bass Model, GGM, UCRCD. \url{https://cran.r-project.org/package=DIMORA} \\
	& R	& \textit{diffusion}	& diffusion	& Various diffusion models to forecast new product growth. Currently the package contains Bass, Gompertz and Gamma/Shifted Gompertz curves. \url{https://cran.r-project.org/package=diffusion} \\
\S\ref{sec:The_natural_law of growth_in_competition_Logistic_growth}. The natural law of growth in competition	& R	& \textit{LS2Wstat}	& scurve	& An S curve function between two constant values. \url{https://cran.r-project.org/package=LS2Wstat} \\
\S\ref{sec:Estimation_and_representation_of_uncertainty}. Estimation and representation of uncertainty	& R	& \textit{hdrcde}	& cde	& Conditional kernel density estimation to produce marginal distributions (uncertainty forecasts). \url{https://cran.r-project.org/package=hdrcde}\\	
& R	& \textit{gamlss}	& gamlss & 	Semi-parametric models for uncertainty forecasting. \url{https://cran.r-project.org/package=gamlss}\\
	& R	& \textit{gamboostLSS}	& mboostLSS & 	Semi-parametric component-wise gradient boosting models for uncertainty forecasting. \url{https://cran.r-project.org/package=gamboostLSS}\\
	& Python & 	\textit{scikit-learn} & 	GradientBoostingRegressor; RandomForestQuantileRegressor	& Machine learning models (gradient boosting trees and random forests) for quantile forecasting. \url{https://scikit-learn.org/stable/}\\
	& R	& \textit{quantreg}	& rq; lprq; nlqr	& Estimation and inference methods for models of conditional quantiles. \url{https://cran.r-project.org/package=quantreg} \\
	& R	& \textit{EnvStats}	& FcnsByCatPredInts; pointwise 	& Functions for computing prediction intervals and simultaneous prediction intervals. \url{https://cran.r-project.org/package=EnvStats}\\
	& R	& \textit{rmgarch}	& dccfit-methods; dccforecast-methods	& Multivariate GARCH Models (e.g., forecasting covariance matrix). \url{https://cran.r-project.org/package=rmgarch}\\
\S\ref{sec:Forecasting_under_fat_tails}. Forecasting under fat tails	& R	& \textit{FatTailsR}	& & 	Functions for Kiener distributions and fat tails. \url{https://cran.r-project.org/package=FatTailsR} \\
\S\ref{sec:bayesian_forecasting_with_copulas}. Bayesian forecasting with copulas &	R	& \textit{VineCopula}		& & Statistical analysis of vine copula models. \url{https://cran.r-project.org/package=VineCopula}\\
&	R	& \textit{cdcopula} & & Covariate-dependent copula models. \url{https://github.com/feng-li/cdcopula} \\
\hline
\textbf{Related} \textbf{section} & \textbf{Software} & \textbf{Package/Library/Toolbox} & \textbf{Function(s)} & \textbf{Comments} \\
\hline
\S\ref{sec:bayesian_forecasting_with_copulas}. Bayesian forecasting with copulas (continued)  &	R	& \textit{FactorCopula}		& & Factor Copula Models for Mixed Continuous and Discrete Data. \url{https://cran.r-project.org/package=FactorCopula} \\
\S\ref{sec:Leading_indicators_and_Granger_causality}. Leading indicators and Granger causality		& R		& \textit{lmtest}		& grangertest 		& Test for Granger causality. \url{https://cran.r-project.org/package=lmtest} \\
& Gretl		& & var; omit	& Standard Granger-causality test. \url{http://gretl.sourceforge.net/}\\
& Gretl	& \textit{BreitungCandelonTest}		& & Breitung-Candelon test of frequency-wise Granger (non-)causality. \url{http://gretl.sourceforge.net/}\\
\S\ref{sec:Variable_Selection}. Variable Selection		& R		& \textit{glmnet}		& 		& Generalised linear models with Lasso or elastic net regularisation. \url{https://cran.r-project.org/package=glmnet} \\
& Gretl		& & omit & 	Sequential removing of variables to a model. \url{http://gretl.sourceforge.net/}\\
& Gretl	& \textit{addlist}		& & Sequential addition of variables to a model. \url{http://gretl.sourceforge.net/}\\
& Gretl	& \textit{regls}		& & Add-on for regularised least squares such as Ridge, Lasso and Elastic-Net. \url{http://gretl.sourceforge.net/}\\
& Gretl	& \textit{fsboost}	& fsreg	& Forward-stagewise boosted regression estimates with functionality for forecasting. \url{http://gretl.sourceforge.net/}\\
\S\ref{sec:Model_Selection}. Model Selection	& R	& \textit{gets}		& & Functions for automatic general to specific model selection. \url{https://cran.r-project.org/package=gets} \\
\S\ref{sec:Cross_validation_for_time_series_data}. Cross-validation for time-series data		& R		& \textit{forecast}		& CVar		& k-fold Cross-Validation applied to an autoregressive model. \url{https://cran.r-project.org/package=forecast} \\
& Gretl		& & fcast	& Forecasting command with functionality for recursive-window forecasts. \url{http://gretl.sourceforge.net/}\\
\S\ref{sec:Forecast_combination_a_brief_review_of_statistical_approaches}. Forecast combination: a brief review of statistical approaches		& R		& \textit{forecastHybrid}			& 	& Functions for ensemble time series forecasts. \url{https://cran.r-project.org/package=forecastHybrid} \\
\S\ref{sec:Forecasting_on_distributed_systems}. Forecasting on distributed systems		& Database		& \textit{InfluxDB} 			& 	& Scalable datastore for metrics, events, and real-time analytics. \url{https://www.influxdata.com/time-series-database/}\\
		& Database		& \textit{OpenTSDB} 		& 	& 	A scalable, distributed Time Series Database. \url{http://opentsdb.net/}\\
		& Database		& \textit{RRDtool} 		& 	& 	A program for easily maintaining a database of time-series data. \url{https://oss.oetiker.ch/rrdtool/}\\
		& Database		& \textit{Timely} 			& 	& A time series database application that provides secure access to time series data. \url{https://code.nsa.gov/timely/}\\
		& Python, R \& Spark		& \textit{darima} 		& 	& Implementations of distributed ARIMA models on Spark platform. \url{https://github.com/xqnwang/darima}\\
\S\ref{sec:Agent_based_models}. Agent-based models		& R	& 	\textit{SpaDES}			& 	& Spatially explicit discrete event simulation models. \url{https://cran.r-project.org/package=SpaDES}\\
\S\ref{sec:Feature_based_time_series_forecasting}. Feature-based time series forecasting 	& 	R		& \textit{tsfeatures}		& tsfeatures		& Methods for extracting various features from time series data. \url{https://cran.r-project.org/package=tsfeatures}\\
		& Python		& \textit{tsfresh}		& 	& 	Calculates a large number of time series characteristics, the so called features. Further the package contains methods to evaluate the explaining power and importance of such characteristics for regression or classification tasks. \url{https://tsfresh.readthedocs.io/en/latest/} \\
\hline
\\ \\ \\
\hline
\textbf{Related} \textbf{section} & \textbf{Software} & \textbf{Package/Library/Toolbox} & \textbf{Function(s)} & \textbf{Comments} \\
\hline
\S\ref{sec:Feature_based_time_series_forecasting}. Feature-based time series forecasting (continued) & Matlab		& \textit{hctsa} 		& 	& 	Code framework that enables the extraction of thousands of time-series features from a time series (or a time-series dataset). It also provides a range of tools for visualising and analysing the resulting time-series feature matrix. \url{https://github.com/benfulcher/hctsa}\\
	& 	Python		& \textit{pyopy}			& 	& Python binding for hctsa. \url{https://github.com/strawlab/pyopy} \\
		& R		& \textit{fforma}		& 	& 	Tools for forecasting using a model combination approach. It can be used for model averaging or model selection. It works by training a `classifier' that learns to select/combine different forecast models. \url{https://github.com/pmontman/fforma} \\
& R		& \textit{gratis}			& 	& Efficient algorithms for generating time series with diverse and controllable characteristics, which can be used as the training data in feature-based time series forecasting. \url{https://cran.r-project.org/package=gratis} \\
		& R		& \textit{seer}				& & Implementations of a novel framework for forecast model selection using time series features; FFORMS (Feature-based FORecast Model Selection). \url{https://cran.r-project.org/package=seer} \\
\S\ref{sec:Forecasting_with_bootstrap}. Forecasting with bootstrap		& R		& \textit{bootstrap}			& 	& Various bootstrapping functions. \url{https://cran.r-project.org/package=bootstrap} \\
& Gretl	& \textit{uniFCextensions}	& uniFCboot	& Estimate an interval forecast without assuming Gaussian innovations. \url{http://gretl.sourceforge.net/}\\
\S\ref{sec:Bagging_for_Time_Series}. Bagging for time series forecasting		& R		& \textit{forecast}		& baggedETS		& Returns forecasts and other information for bagged ETS models. \url{https://cran.r-project.org/package=forecast} \\
		& R		& \textit{tshacks} 		& baggedClusterETS; treatedETS		& Returns forecasts for bagged Cluster and Treated ETS models. \url{https://github.com/tiagomendesdantas/tshacks} \\
\S\ref{sec:neural_networks}. Neural Networks		& Python, MXNet \& PyTorch		& \textit{GluonTS} 		& 	& 	Framework for building deep learning based models including a number of pre-built models such as feed-forward neural networks. \url{https://github.com/awslabs/gluon-ts} \\
	& R		& \textit{nnfor}		& mlp; elm		& Time Series Forecasting with Neural Networks. \url{https://cran.r-project.org/package=nnfor} \\
	& Python		& \textit{neural prophet }			& 	& Reimplementation of prophet in PyTorch, and extensions to it. \url{https://github.com/ourownstory/neural\_prophet} \\
	& Python		& 			& 	& RNNs for forecasting in Tensorflow. \url{https://github.com/HansikaPH/time-series-forecasting} \\
	& R	& ANN2	& neuralnetwork	& Artificial Neural Networks. \url{https://cran.r-project.org/package=ANN2} \\
	& R	& nnet	& nnet	& Feed-Forward Neural Networks and Multinomial Log-Linear Models. \url{https://cran.r-project.org/package=nnet} \\
	& R	& forecast	& nnetar	& Auto-regressive Neural Network for time series forecasting. \url{https://cran.r-project.org/package=forecast} \\
\S\ref{sec:Deep_Probabilistic_Forecasting_models}. Deep Probabilistic models		& Python, MXNet \& PyTorch		& \textit{GluonTS} 			& 	& Framework for building deep learning based models including a number of pre-built models such as DeepAR, DeepState models and NBEATS. \url{https://github.com/awslabs/gluon-ts} \\
		& Python, PyTorch		& \textit{PyTorchTS} 		&	& Clone of GluonTS in PyTorch. \url{https://github.com/zalandoresearch/pytorch-ts} \\
\S\ref{sec:machine_Learning}. Machine learning	& R	& \textit{RSNNS}	& mlp; rbf; dlvq; elman; jordan; som	& Neural Networks using the Stuttgart Neural Network Simulator (SNNS). \url{https://cran.r-project.org/package=RSNNS} \\
\hline
\textbf{Related} \textbf{section} & \textbf{Software} & \textbf{Package/Library/Toolbox} & \textbf{Function(s)} & \textbf{Comments} \\
\hline
\S\ref{sec:machine_Learning}. Machine learning (continued)	& R	& \textit{rpart}	& rpart; prune	& Recursive partitioning and regression trees. \url{https://cran.r-project.org/package=rpart} \\
	& R	& \textit{caret}		& & Classification and regression training. \url{https://cran.r-project.org/package=caret} \\
	& R	& \textit{e1071}	& svm 	& Misc ML functions of the Department of Statistics, Probability Theory Group.  \url{https://cran.r-project.org/package=e1071} \\
	& R	& \textit{kernlab}	& gausspr 	& Gaussian processes for regression and classification. \url{https://cran.r-project.org/package=kernlab} \\
	& R	& \textit{brnn}	& brnn	& Bayesian Regularisation for Feed-Forward Neural Networks. \url{https://cran.r-project.org/package=brnn} \\
	& R	& \textit{grnn}	& grnn	& General regression neural network. \url{https://cran.r-project.org/package=grnn} \\
	& R	& \textit{randomForest}	& randomForest	& Breiman and Cutler's Random Forests for Classification and Regression. \url{https://cran.r-project.org/package=randomForest} \\
    & R	& \textit{gbm}	& gbm	& Generalised Boosted regression models. \url{https://cran.r-project.org/package=gbm} \\
	& R	& \textit{neuralnet}	& neuralnet	& Training of simple Neural Networks. \url{https://cran.r-project.org/package=neuralnet} \\
	& Python	& \textit{Tensorflow}	& & A framework, developed by Google, offering tools for designing, building, and deploying ML models. \url{https://tensorflow.org/api\_docs/python/tf} \\
	& Python	& \textit{Keras API}	& & 	A deep learning API built on top of Tensorflow. It provides high level blocks for building and training NN models. \url{https://keras.io/} \\
	& R	& \textit{Tensorflow}		& & R Interface to Tensorflow (https://tensorflow.rstudio.com/). \url{https://tensorflow.rstudio.com/} \\
	& R	& \textit{deepnet}		& & Deep learning toolkit. \url{https://cran.r-project.org/package=deepnet} \\
	& R	& \textit{h2o}		& & R Interface for the 'H2O' Scalable Machine Learning Platform. \url{https://cran.r-project.org/package=h2o} \\
	& R	& \textit{Apache MXNet}		& & A flexible library for deep learning. \url{https://mxnet.apache.org/} \\
	& Python	& \textit{scikit-learn}	& & Ordinary Least Squares, Ridge regression, Lasso, Bayesian Regression, Generalized Linear Regression, Stochastic Gradient Descent and Polynomial regression, Support Vector Machines, Nearest Neighbors, Gaussian Processes, Decision Trees, Ensemble methods (Forests of randomised trees, AdaBoost and Gradient Tree Boosting), Multi-layer Perceptrons. \url{https://scikit-learn.org/stable/} \\
	& Python	& \textit{CNTK}		& & A framework, developed by Microsoft, that provides tools for building ML and DL models. \url{https://docs.microsoft.com/en-us/cognitive-toolkit/} \\
	& Python	& \textit{PyTorch}		& & A framework, developed by Facebook, for building ML and DL models. \url{https://pytorch.org/} \\
\S\ref{sec:clustering_based_forecasting}. Clustering-based forecasting	& R	& \textit{tsfknn}	& tsfknn	& Time Series Forecasting Using Nearest Neighbours. \url{https://cran.r-project.org/package=tsfknn} \\
\S\ref{sec:Hybrid_methods}. Hybrid methods	& Python	& \textit{ESRNN-GPU}	& 	& A GPU-enabled version of the hybrid model used by the winner of M4 competition. \url{https://github.com/damitkwr/ESRNN-GPU} \\
\hline
\textbf{Related} \textbf{section} & \textbf{Software} & \textbf{Package/Library/Toolbox} & \textbf{Function(s)} & \textbf{Comments} \\
\hline
\S\ref{sec:Parametric_methods_for_intermittent_demand_forecasting}. Parametric methods for intermittent demand forecasting	& R	& \textit{tsintermittent}	& crost; tsb	& Parametric forecasting methods for intermitternt demand. \url{https://cran.r-project.org/package=tsintermittent} \\
	& R	& \textit{forecast}	& croston	& Forecasts for intermittent demand using Croston's method. \url{https://cran.r-project.org/package=forecast} \\
\S\ref{sec:non_parametric_intermittent_demand_methods}. Non-parametric intermittent demand methods	& R	& \textit{tsintermittent}	& imapa	& MAPA for intermittent demand data. \url{https://cran.r-project.org/package=tsintermittent} \\
\S\ref{sec:classification_methods}. Classification methods	& R	& \textit{tsintermittent}	& idclass	& Time series categorisation for intermittent demand. \url{https://cran.r-project.org/package=tsintermittent} \\
	& R	& \textit{tsutils}	& abc; xyz; abcxyz	& Classification functions and routines. \url{https://cran.r-project.org/package=tsutils} \\
\S\ref{sec:Text_based_forecasting}. Forecasting with text information	& Python	& \textit{NLTK}		& & The Natural Language Toolkit in Python. \url{https://www.nltk.org/} \\
	& Python	& \textit{SpaCy}		& & An open source library for advanced Natural Language Processing in Python. \url{https://spacy.io/} \\
\S\ref{sec:Cross_sectional_hierarchical_forecasting}. Cross-sectional hierarchical forecasting	& R	& \textit{hts}		& & Functions and routines for hierarchical and grouped time series forecasting. \url{https://cran.r-project.org/package=hts} \\
\S\ref{sec:temporal_aggregation}. Temporal aggregation	& R	& \textit{MAPA}	& mapa; mapasimple	& Functions and wrappers for using the Multiple Aggregation Prediction Algorithm (MAPA) for time series forecasting. \url{https://cran.r-project.org/package=MAPA} \\
& R	& \textit{thief}	& thief	& Temporal Hierarchical Forecasting. \url{https://cran.r-project.org/package=thief} \\
	& R	& \textit{tsintermittent}	& imapa	& MAPA for intermittent demand data with automatic model selection based on the PK classification. \url{https://cran.r-project.org/package=tsintermittent} \\
\S\ref{sec:ecological_inference_forecasting}. Ecological inference forecasting	& R	& \textit{ei}	& ei	& Returns local and global forecasts of inner cells in $2 \times 2$ tables. \url{https://cran.r-project.org/package=ei} \\
& R	& \textit{eiPack}	& ei.MD.bayes; ei.reg; ei.reg.bayes	& Returns local and global forecasts of inner cells in R$\times$C tables under a Multinomial Dirichlet model or using ecological regression. \url{https://cran.r-project.org/package=eiPack} \\
	& R	& \textit{lphom} 	& lphom; tslphom; nslphom	& Returns forecasts of inner cells of a R$\times$C table using linear programming optimisation. \url{https://CRAN.R-project.org/package=lphom} \\
    & R	& \textit{eiCompare} 	& ei\_est\_gen; ei\_good; ei\_rxc	& Returns forecasts of inner cells of a  $R \times C$  tables using iterative versions of $2 \times 2$ methods and the Multinomial Dirichlet model. \url{https://cran.r-project.org/package=eiCompare} \\
\S\ref{sec:point_forecast_accuracy_measures}. Point, interval, and pHDR forecast error measures	& R	& forecast	& accuracy	& Accuracy measures for a forecast model. \url{https://cran.r-project.org/package=forecast} \\
\S\ref{sec:evaluating_probabilistic_forecasts}. Evaluating probabilistic forecasts	& R	& \textit{scoringRules}		& & Scoring rules for parametric and simulated distribution forecasts. \url{https://cran.r-project.org/package=scoringRules} \\
	& R	& \textit{verification}	& crps	& Continuous ranked probability score. \url{https://cran.r-project.org/package=verification} \\
\S\ref{sec:Statistical_tests_of_forecast_performance}. Statistical tests of forecast performance	& R	& \textit{forecast}	& dm.test	& Diebold-Mariano test for predictive accuracy. \url{https://cran.r-project.org/package=forecast} \\
	& R	& \textit{tsutils}	& nemenyi	& Nonparametric multiple comparisons (Nemenyi test). \url{https://cran.r-project.org/package=tsutils} \\
\hline
\\ \\
\hline
\textbf{Related} \textbf{section} & \textbf{Software} & \textbf{Package/Library/Toolbox} & \textbf{Function(s)} & \textbf{Comments} \\
\hline
\S\ref{sec:Statistical_tests_of_forecast_performance}. Statistical tests of forecast performance (continued) & Gretl	& \textit{FEP}	& doMZtest; doHPtest; doEKTtest; doPTtest; doDLtest; doDMtest; doGWtest; doCWtest	& Various statistical tests on forecast unbiasedness, efficiency, asymmetric loss and directional changes. \url{http://gretl.sourceforge.net/}\\
& Gretl	& \textit{DiebMar}	& & 	Diebold-Mariano test. \url{http://gretl.sourceforge.net/}\\

\hline
\label{tab:software}
\end{longtable}
\normalsize

\clearpage

\section{Data sets}\label{sec:datasets}

\begin{center}
Table A.2: A list of indicative publicly available data sets.
\end{center}
\scriptsize
\begin{longtable}{ p{.36\textwidth} p{.36\textwidth} p{.60\textwidth} }
\hline
\textbf{Related section} & \textbf{Description} & \textbf{Link} \\
\hline
\S\ref{sec:Deep_Probabilistic_Forecasting_models}. Deep Probabilistic Models & Data for wikipedia page views, Dominicks retail, electricity consumption, traffic lane occupation. & 	\url{https://gluon-ts.mxnet.io/api/gluonts/gluonts.dataset.repository.datasets.html} \\
\S\ref{sec:Text_based_forecasting}. Forecasting with text Information & Movie reviews data provided by the Stanford NLP group. & 	\url{https://nlp.stanford.edu/sentiment/code.html} \\
\S\ref{sec:ecological_inference_forecasting}. Ecological inference forecasting	& Party registration in South-East North Carolina (eiPack R package).	 & \url{https://www2.ncleg.net/RnR/Redistricting/BaseData2001} \\
& ei.Datasets: Real Datasets for Assessing Ecological Inference Algorithms.	 & \url{https://cran.csiro.au/web/packages/ei.Datasets/index.html} \\
\S\ref{sec:Forecasting_competitions}. Forecasting competitions & 	Data for the M, M2, M3 and M4 forecasting competitions.	 & \url{https://forecasters.org/resources/time-series-data/} \\
& 	Time Series Competition Data (R package) & 	\url{https://github.com/robjhyndman/tscompdata} \\
& 	Mcomp: Data for the M and M3 forecasting competitions (R package). & 	\url{https://cran.r-project.org/package=Mcomp} \\
 & 	M4comp2018: Data for the M4 forecasting competition (R package). & 	\url{https://github.com/carlanetto/M4comp2018} \\
& 	Data for the M4 forecasting competition (csv files). & 	\url{https://github.com/Mcompetitions/M4-methods/tree/master/Dataset} \\
 & 	Tcomp: Data from the 2010 Tourism forecasting competition (R package) & 	\url{https://cran.r-project.org/package=Tcomp} \\
& Data for the M5 forecasting competition (csv files). & 	\url{https://github.com/Mcompetitions/M5-methods/tree/master/Dataset} \\
\S\ref{sec:Forecasting_for_inventories}. Forecasting for inventories	& Grupo Bimbo Inventory Demand.	& \url{https://www.kaggle.com/c/grupo-bimbo-inventory-demand} \\
\S\ref{sec:Retail_sales_forecasting}. Forecasting in retail	& Rossmann Store Sales.	& \url{https://www.kaggle.com/c/rossmann-store-sales} \\
& Corporación Favorita Grocery Sales Forecasting.	& \url{https://www.kaggle.com/c/favorita-grocery-sales-forecasting} \\
& Walmart Recruiting -- Store Sales Forecasting.	& \url{https://www.kaggle.com/c/walmart-recruiting-store-sales-forecasting} \\
& Walmart Recruiting II: Sales in Stormy Weather.	& \url{https://www.kaggle.com/c/walmart-recruiting-sales-in-stormy-weather} \\
& Store Item Demand Forecasting Challenge.	& \url{https://www.kaggle.com/c/demand-forecasting-kernels-only} \\
& Online Product Sales.	& \url{https://www.kaggle.com/c/online-sales} \\
\S\ref{sec:Predictive_maintenance}. Predictive maintenance & Robot Execution Failures. & 	\url{https://archive.ics.uci.edu/ml/datasets/Robot+Execution+Failures} \\
& Gearbox Fault Detection.	& \url{https://c3.nasa.gov/dashlink/resources/997/} \\
& Air Pressure System Failure at Scania Trucks. & 	\url{https://archive.ics.uci.edu/ml/datasets/IDA2016Challenge} \\
 & Generic, Scalable and Decentralised Fault Detection for Robot Swarms & 	\url{https://zenodo.org/record/831471\#\#.WwQIPUgvxPY} \\
& Wind turbine data (e.g., failures). & 	\url{https://opendata.edp.com/pages/homepage/} \\
\S\ref{sec:Economics_and_finance}. Economics and finance	& Two Sigma Financial Modelling Challenge.	& \url{https://www.kaggle.com/c/two-sigma-financial-modeling/overview} \\
& Financial, economic, and alternative data sets, serving investment professionals.	& \url{https://www.quandl.com/} \\
\S\ref{sec:Forecasting_GDP_and_Inflation}. Forecasting GDP and Inflation & Repository website with Dynare codes and data sets to estimate different DSGE models and use them for forecasting. & 	\url{https://github.com/johannespfeifer/dsge\_mod} \\
& Data set for Macroeconomic variables for US economy. & 	\url{https://fred.stlouisfed.org/} \\
& Data set for Macroeconomic variables for OECD economy. & 	\url{https://data.oecd.org/} \\
\hline
\textbf{Related section} & \textbf{Description} & \textbf{Link} \\
\hline
\S\ref{sec:Forecasting_GDP_and_Inflation}. Forecasting GDP and Inflation (continued) & Data set for Macroeconomic variables for EU economy. & 	\url{https://ec.europa.eu/eurostat/data/database} \\
\S\ref{sec:House_Price_Forecasting}. House price forecasting	& Zillow Prize: Zillow’s Home Value Prediction (Zestimate).	& \url{https://www.kaggle.com/c/zillow-prize-1} \\
& Sberbank Russian Housing Market.	& \url{https://www.kaggle.com/c/sberbank-russian-housing-market} \\
& Western Australia Rental Prices.	& \url{https://www.kaggle.com/c/deloitte-western-australia-rental-prices} \\
\S\ref{sec:Forecasting_returns_to_investment_style}. Forecasting returns to investment style	& Algorithmic Trading Challenge.	& \url{https://www.kaggle.com/c/AlgorithmicTradingChallenge} \\
\S\ref{sec:Forecasting_stock_returns}. Forecasting stock returns	& The Winton Stock Market Challenge.	& \url{https://www.kaggle.com/c/the-winton-stock-market-challenge} \\
& The Big Data Combine Engineered by BattleFin.	& \url{https://www.kaggle.com/c/battlefin-s-big-data-combine-forecasting-challenge/data} \\
\S\ref{sec:Energy}. Energy	& VSB Power Line Fault Detection.	& \url{https://www.kaggle.com/c/vsb-power-line-fault-detection} \\
& ASHRAE -- Great Energy Predictor III	& \url{https://www.kaggle.com/c/ashrae-energy-prediction} \\
\S\ref{sec:Load_forecasting}. Hybrid machine learning system for short-term load forecasting	& Global Energy Forecasting Competition 2012 -- Load Forecasting.	& \url{https://www.kaggle.com/c/global-energy-forecasting-competition-2012-load-forecasting} \\
\S\ref{sec:Wind_Power_orecasting}. Wind power forecasting	& Global Energy Forecasting Competition 2012 -- Wind Forecasting.	& \url{https://www.kaggle.com/c/GEF2012-wind-forecasting} \\
\S\ref{sec:Solar_power_forecasting}. Solar power forecasting & 	Power measurements from a PV power plant and grid of numerical weather predictions. & 	\url{https://doi.org/10.25747/edf8-m258} \\
& AMS 2013-2014 Solar Energy Prediction Contest.	& \url{https://www.kaggle.com/c/ams-2014-solar-energy-prediction-contest} \\
&	SolarTechLab data set. &	\url{https://ieee-dataport.org/open-access/photovoltaic-power-and-weather-parameters} \\
\S\ref{sec:Long-term_simulation_for_large_electrical_power_systems}. Long-term simulation for large electrical power systems & 	Brazilian National Electric Systems Operator (hydro, solar, wind, nuclear and thermal generation data). & 	\url{http://www.ons.org.br/Paginas/resultados-da-operacao/historico-da-operacao/geracao\_energia.aspx} \\
\S\ref{sec:Collaborative_forecasting_in_the_energy_sector}. Collaborative forecasting in the energy sector	& Solar power time series from 44 small-scale PV in Évora, Portugal.	& \url{https://doi.org/10.25747/gywm-9457} \\
	& Australian Electricity Market Operator (AEMO) 5 Minute Wind Power Data.	& \url{https://doi.org/10.15129/9e1d9b96-baa7-4f05-93bd-99c5ae50b141}\\
	& Electrical energy consumption data from domestic consumers.	& \url{data.london.gov.uk/dataset/smartmeter-energy-use-data-in-london-households} \\
	& Electric vehicles charging data (arrivals, departures, current, voltage, etc.).	& \url{https://eatechnology.com/consultancy-insights/my-electric-avenue/} \\
	& Wind power plant data and numerical weather predictions from CNR (France).	& \url{https://challengedata.ens.fr/challenges/34} \\
\S\ref{sec:Weather_Forecasting}. Weather forecasting	& How Much Did It Rain?	& \url{https://www.kaggle.com/c/how-much-did-it-rain-ii} \\
\S\ref{sec:Air_quality_forecasting}. Air quality forecasting	& EMC Data Science Global Hackathon (Air Quality Prediction).	& \url{https://www.kaggle.com/c/dsg-hackathon/overview} \\
\S\ref{sec:Social_good}. Social good and demographic forecasting	& LANL Earthquake Prediction.	& \url{https://www.kaggle.com/c/LANL-Earthquake-Prediction} \\
\S\ref{sec:Heathcare}. Healthcare	& Flu Forecasting.	& \url{https://www.kaggle.com/c/genentech-flu-forecasting} \\
& West Nile Virus Prediction.	& \url{https://www.kaggle.com/c/predict-west-nile-virus} \\
\S\ref{sec:Pandemics}. Epidemics and pandemics	& COVID-19 Data Repository by the Center for Systems Science and Engineering (CSSE) at Johns Hopkins University.	& \url{https://github.com/CSSEGISandData/COVID-19} \\
\S\ref{sec:Forecasting_mortality_data}. Forecasting mortality & 	Human Mortality Database. & 	\url{https://www.mortality.org} \\
& 	EuroMOMO. & 	\url{https://www.euromomo.eu/} \\
& 	The Economist. & 	\url{https://github.com/TheEconomist/covid-19-excess-deaths-tracker} \\
& 	The New York Times. & 	\url{https://github.com/Financial-Times/coronavirus-excess-mortality-data} \\
\hline
\textbf{Related section} & \textbf{Description} & \textbf{Link} \\
\hline
\S\ref{sec:Forecasting_mortality_data}. Forecasting mortality data (continued) & 	The Financial Times. & 	\url{https://github.com/nytimes/covid-19-data/tree/master/excess-deaths} \\
& 	ANACONDA- Quality assessment of mortality data. & 	\url{https://crvsgateway.info/anaconda} \\
& 	Australian Human Mortality Database. & 	\url{https://demography.cass.anu.edu.au/research/australian-human-mortality-database} \\
& 	Canadian Human Mortality Database. & 	\url{http://www.bdlc.umontreal.ca/CHMD/} \\
& 	French Human Mortality Database. & 	\url{https://frdata.org/fr/french-human-mortality-database} \\
\S\ref{sec:Forecasting_fertility}. Forecasting fertility & Human Fertility Database: fertility data for developed countries with complete birth registration based on official vital statistics. & \url{https://www.humanfertility.org/} \\
& World Fertility Data: UN's collection of fertility data based on additional data sources such as surveys. & \url{https://www.un.org/development/desa/pd/data/world-fertility-data} \\
\S\ref{sec:Forecasting_migration}. Forecasting migration & Integrated Modelling of European Migration (IMEM) Database, with estimates of migration flows between 31 European countries by origin, destination, age and sex, for 2002–2008. & \url{https://www.imem.cpc.ac.uk/} \\
& QuantMig data inventory: meta-inventory on different sources of data on migration and its drivers, with European focus. & \url{https://quantmig.eu/data\_inventory/} \\
& Bilateral international migration flow estimates for 200 countries \cite{Abel2019-rm}. & \url{https://doi.org/10.1038/s41597-019-0089-3} \\ 
& UN World Population Prospects: UN global population estimates and projections, including probabilistic & \url{https://population.un.org/wpp/} \\
\S\ref{sec:Other_applications}. Other applications	& Forecast Eurovision Voting.	& \url{https://www.kaggle.com/c/Eurovision2010} \\
& Reducing Commercial Aviation Fatalities.	& \url{https://www.kaggle.com/c/reducing-commercial-aviation-fatalities} \\
& Porto Seguro's Safe Driver Prediction.	& \url{https://www.kaggle.com/c/porto-seguro-safe-driver-prediction} \\
& Recruit Restaurant Visitor Forecasting.	& \url{https://www.kaggle.com/c/recruit-restaurant-visitor-forecasting} \\
& Restaurant Revenue Prediction.	& \url{https://www.kaggle.com/c/restaurant-revenue-prediction} \\
& Coupon Purchase Prediction.	& \url{https://www.kaggle.com/c/coupon-purchase-prediction} \\
& Bike Sharing Demand	& \url{https://www.kaggle.com/c/bike-sharing-demand} \\
& Google Analytics Customer Revenue Prediction.	& \url{https://www.kaggle.com/c/ga-customer-revenue-prediction} \\
 & Santander Value Prediction Challenge.	& \url{https://www.kaggle.com/c/santander-value-prediction-challenge} \\
& Santander Customer Transaction Predictio.n	& \url{https://www.kaggle.com/c/santander-customer-transaction-prediction} \\
& Acquire Valued Shoppers Challenge	& \url{https://www.kaggle.com/c/acquire-valued-shoppers-challenge} \\
 & Risky Business	& \url{https://www.kaggle.com/c/risky-business} \\
& Web Traffic Time Series Forecasting.	& \url{https://www.kaggle.com/c/web-traffic-time-series-forecasting} \\
& A repository of data sets, including time series ones, that can be used for benchmarking forecasting methods in various applications of interest. &	\url{https://github.com/awesomedata/awesome-public-dataset}s \\
& WSDM -- KKBox's Churn Prediction Challenge.	& \url{https://www.kaggle.com/c/kkbox-churn-prediction-challenge/overview} \\
& Homesite Quote Conversion	& \url{https://www.kaggle.com/c/homesite-quote-conversion} \\
& Liberty Mutual Group: Property Inspection Prediction.	& \url{https://www.kaggle.com/c/liberty-mutual-group-property-inspection-prediction} \\
& Liberty Mutual Group -- Fire Peril Loss Cost.	& \url{https://www.kaggle.com/c/liberty-mutual-fire-peril} \\
& A set of more than 490,000 time series (micro, macro, demographic, finance, other) to download.	& \url{http://fsudataset.com/} \\
\hline
\textbf{Related section} & \textbf{Description} & \textbf{Link} \\
\hline
\S\ref{sec:Other_applications}. Other applications (continued) & A repository of data sets, including time series ones, that can be used for benchmarking forecasting methods in various applications of interest.	& \url{https://github.com/awesomedata/awesome-public-datasets} \\
\S\ref{sec:Tourism_demand_forecasting}. Tourism demand forecasting & 	TourMIS: Database with annual and monthly tourism time series (e.g., arrivals and bednights) covering European countries, regions, and cities (free registration required). & 	\url{https://www.tourmis.info/index\_e.html} \\
& Tourism Forecasting. 	& \url{https://www.kaggle.com/c/tourism1; https://www.kaggle.com/c/tourism2} \\
\S\ref{sec:Forecasting_for_airports}. Forecasting for aviation & 	Airline and Airport performance data provided by the U.S. Department of Transportation & \url{https://www.transtats.bts.gov/} \\
\S\ref{sec:Traffic_flow_forecasting}. Traffic flow forecasting	& New York City Taxi Fare Prediction.	& \url{https://www.kaggle.com/c/demand-forecasting-kernels-only} \\
& RTA Freeway Travel Time Prediction	& \url{https://www.kaggle.com/c/RTA} \\
& ECML/PKDD 15: Taxi Trip Time Prediction.	& \url{https://www.kaggle.com/c/pkdd-15-taxi-trip-time-prediction-ii} \\
& BigQuery-Geotab Intersection Congestion.	& \url{https://www.kaggle.com/c/bigquery-geotab-intersection-congestion/overview} \\
& Traffic volume counts collected by DOT for New York Metropolitan Transportation Council.	& \url{https://data.cityofnewyork.us/Transportation/Traffic-Volume-Counts-2014-2019-/ertz-hr4r} \\
\S\ref{sec:Elections_forecasting}. Elections forecasting & 	New Zealand General Elections -- Official results and statistics.	 & \url{https://www.electionresults.govt.nz} \\
& 	Spanish Elections -- Official results and statistics.	 & \url{https://dataverse.harvard.edu/dataverse/SEA} \\
\S\ref{sec:Sports_forecasting}. Sports forecasting	& 	NFL Big Data Bowl.	& 	\url{https://www.kaggle.com/c/nfl-big-data-bowl-2020} \\
\S\ref{sec:Safety_and_security}. Forecasting under data integrity attacks	& 	Microsoft Malware Prediction.	& 	\url{https://www.kaggle.com/c/microsoft-malware-prediction} \\
\hline
\label{tab:datasets}
\end{longtable}
\end{landscape}
\normalsize

\clearpage

\bibliographystyle{elsarticle-harv}
\bibliography{refs}

\end{document}